\documentclass[12pt, a4paper, twoside]{book}
\usepackage{a4wide}

\usepackage[Sonny]{fncychap}
\usepackage{fancyhdr}
\usepackage[latin1]{inputenc}
\usepackage[dutch,english]{babel}
\usepackage[small,bf,flushleft]{caption}
\usepackage{amsmath}
\usepackage{amsfonts}
\usepackage{amssymb}
\usepackage{graphicx}
\usepackage[nottoc]{tocbibind}
\usepackage{emptypage}
\usepackage{braket}
\usepackage{caption}
\usepackage{subcaption}
\usepackage{multirow}
\usepackage{verbatim}
\usepackage{rotating}
\usepackage{bbm}
\usepackage[usenames,dvipsnames]{color}
\usepackage[colorlinks=true, urlcolor=blue, citecolor=OliveGreen]{hyperref}
\usepackage[numbers, sort&compress]{natbib}
\usepackage{algorithm}
\usepackage[noend]{algpseudocode}
\usepackage{appendix}
\usepackage{tabularx}

\usepackage{tikz}
\usetikzlibrary{shapes.callouts}
\tikzset{
  level/.style   = { thick, black },
  axes/.style = { thick, black }
}

\hypersetup{
    pdfauthor={Pieter W. Claeys},
    pdftitle={Richardson-Gaudin models and broken integrability},
    pdfkeywords={Integrability, Bethe ansatz, Richardson-Gaudin models},
    pdfsubject={},
    bookmarksopen=true,
    bookmarksnumbered=true,
    unicode=true,
}		

\usepackage{pdfpages}
\usepackage{multirow}

\author{Pieter W. Claeys}
\title{Richardson-Gaudin models and broken integrability}

\usepackage{makeidx}
\makeindex

\usepackage{color}



\usepackage{epigraph}

\newcommand*{\tbd}[1]{{\color{red}[TBD: #1]}}
\newcommand*{\optional}[1]{{\color{blue} #1 }}

\DeclareMathOperator{\per}{per}
\DeclareMathOperator{\sgn}{sgn}

\newcommand\undermat[2]{
  \makebox[0pt][l]{$\smash{\underbrace{\phantom{%
    \begin{matrix}#2\end{matrix}}}_{\text{$#1$}}}$}#2}

\begin{document}


\includepdf[pages={1}]{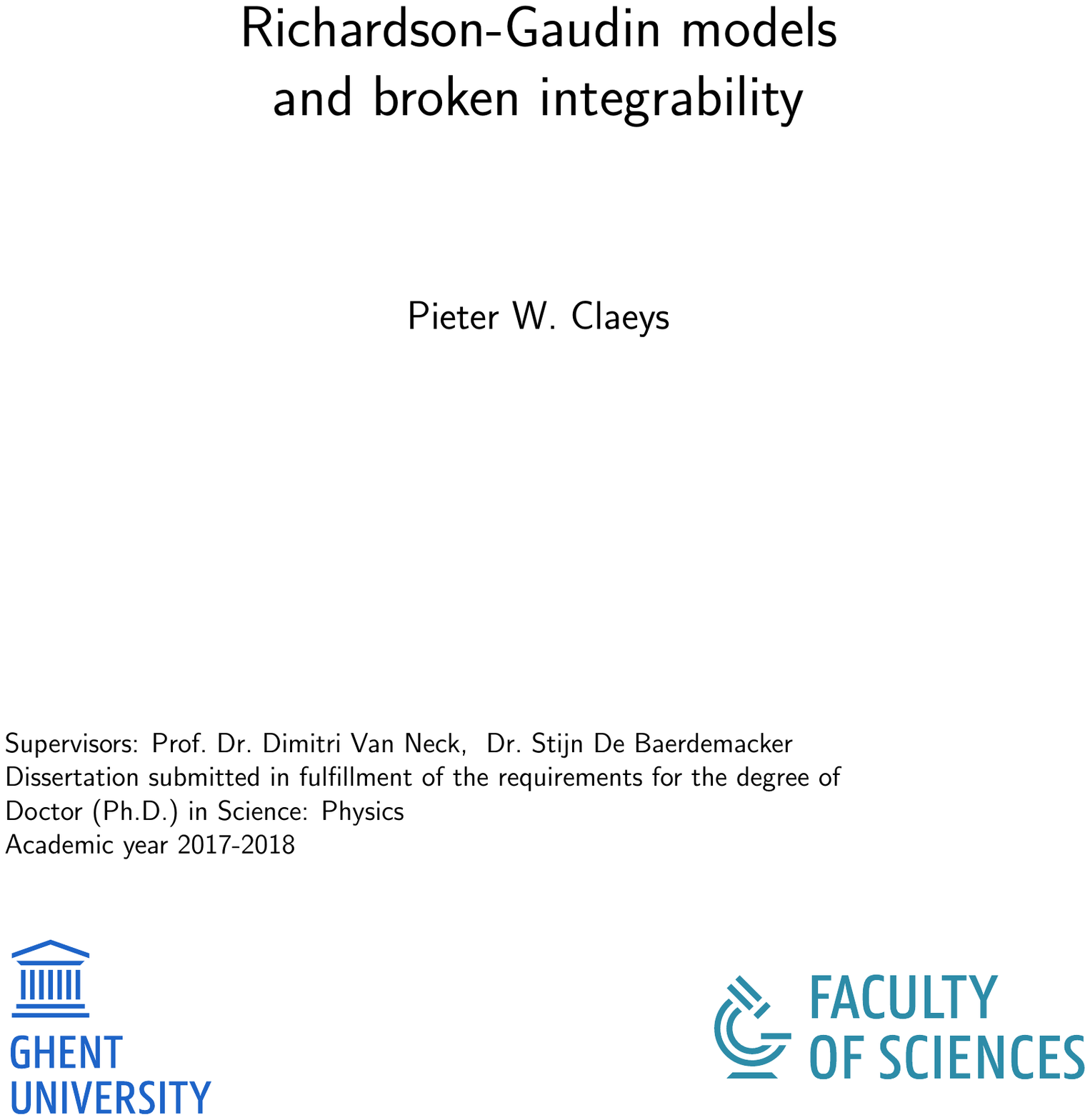}

\frontmatter


\thispagestyle{empty}
\setlength{\tabcolsep}{25pt} 

\vspace*{\fill}

\textbf{Supervisors}

\vspace{0.5cm}

\setlength{\tabcolsep}{25pt} 
\begin{tabular}{p{7.5cm} p{5cm}}
Prof. Dr. Dimitri Van Neck & Ghent University \\
Dr. Stijn De Baerdemacker & Ghent University \\
\end{tabular}

\vspace{0.5cm}

\textbf{Members of the examination committee}

\vspace{0.5cm}

\setlength{\tabcolsep}{25pt} 
\begin{tabular}{p{7.5cm} p{5cm}}
Prof. Dr. Jean-S\'ebastien Caux & University of Amsterdam \\
Prof. Dr. Alexandre Faribault & University of Lorraine \\
Prof. Dr. Michiel Wouters & University of Antwerp \\
Prof. Dr. Hendrik De Bie & Ghent University \\
Prof. Dr. Patrick Bultinck & Ghent University \\
Prof. Dr. Philippe Smet (Chair) & Ghent University \\
Dr. Laurens Vanderstraeten & Ghent University \\

\end{tabular}
\vspace*{\fill}

\vspace{0.5cm}

\hspace{-\parindent}The research described in this thesis was carried out at the Center for Molecular Modeling, Ghent University, and at the Institute of Physics, University of Amsterdam. This research was funded by a Ph.D. fellowship and a travel grant for a long stay abroad from the Research Foundation Flanders (FWO Vlaanderen).

\hspace{-\parindent}
\begin{flushleft}
\includegraphics[width=0.25\textwidth]{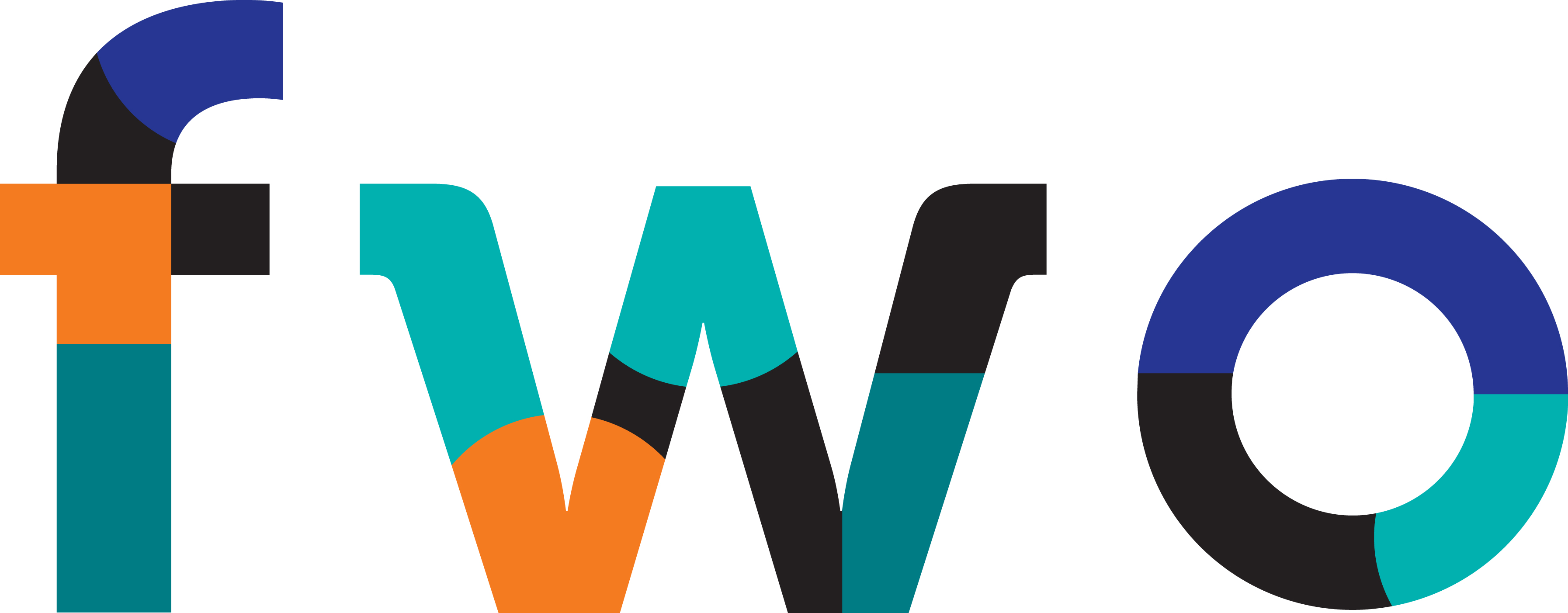}
\end{flushleft}



\chapter*{Acknowledgements}
\addcontentsline{toc}{chapter}{Acknowledgements}

While the four years of a Ph.D. might seem like a long time, there are some people that made these years fly by. Each of them contributed to this thesis in ways big or small, in many cases without even realizing. These brief acknowledgements are too short to properly thank any of them, but none of them should go unmentioned.

First of all, my thanks go to my supervisor Dimitri Van Neck for his support, for providing me with the opportunity to pursue a Ph.D., and for giving me the freedom to follow my own interests. Without him, this thesis would not have been possible.

The one person that simply cannot be overlooked in these acknowledgements is my second supervisor, Stijn De Baerdemacker. Thanks for the guidance, the infinite enthusiasm, and the many ideas and discussions. At each point in these years, your viewpoint proved to be invaluable, and I hope to have absorbed some of your ideas on research and physics throughout these years. Words cannot describe how grateful and proud I am to have you as a supervisor and a friend.

A good supervisor is invaluable, so I consider myself extremely lucky to have had not one, not two, but three of them throughout these years -- Jean-S\'ebastien Caux, thank you for having me in Amsterdam. The year I spent at Science Park was one of the most fun, motivating and interesting experiences of this Ph.D., in no small part because of you. Your passion for all things scientific continues to inspire me.

None of this research occurred in isolation, and I am grateful to the many colleagues at the CMM who made it a pleasure to come to `work' each day. Brecht, Sebastian, Ward, Klaas and Caitlin, it was a pleasure sharing an office and/or daily coffee breaks with each and every one of you. Further down the corridor, a special thanks to Sven and Ruben, for having been there from the start. Remaining in Zwijnaarde, Simon, it's always a pleasure talking physics with you. Wim, thanks for keeping everything running.

Again, it was a pleasure to have not one, but two groups where I felt at home. Eoin, Enej, Sangwoo, Shan, and Moos, thanks for being great office-mates and making the departure from Amsterdam so difficult. Vladimir, it was a pleasure discussing and collaborating with you. Your scientific knowledge never ceases to impress me. 

Speaking of Amsterdam, two people deserve a special mention in these acknowledgements. Sergio, you immediately made me feel welcome in Amsterdam, and I am grateful to have had you by my side as a friend ever since. Ana, thanks for the many extensive conversations and for giving me a daily break to look forward to.

Of course, there is more to life than physics (if at times only barely). Thanks to Sam, Melissa, and Kevin for the many entertaining game and movie nights, and thanks to Arne and Amber for making Ghent so much more enjoyable. Thanks to Lise for the friendship and support these past few years. Thanks to Rick, Thomas, and Pim for the extraordinary pasta-evenings, both in Amsterdam and Ghent, and to the entire Das Mag-group for being the most inspiring group of people I've had the pleasure of being part of. Bert, Michiel, Tine, Laura, Lisa, Ine, Sara, and Jolene, thanks for being part of my life from even before I became aware of quantum physics.

Briefly looking forward instead of backward, it's my pleasure to acknowledge Anatoli Polkovnikov. Collaborating at Boston University is one of the things I most look forward to the coming year.

It almost goes without saying that this thesis would not have been possible without the continuous support of my parents and my brother, Robbert. Any acknowledgement is necessarily an understatement, but I would not be who I am today without them. Even if it became increasingly more unintelligible what I was doing, their support remained unquestionable and I could not have wished for better.

\begin{flushright}
Pieter,

Ghent 2018
\end{flushright}


\chapter*{List of papers}
\addcontentsline{toc}{chapter}{List of papers}

\begin{enumerate}

\item Pieter W. Claeys, Stijn De Baerdemacker, Mario Van Raemdonck, and Dimitri Van Neck, \textit{The Dicke model as the contraction limit of a pseudo-deformed Richardson-Gaudin model}, Journal of Physics:
Conference Series \textbf{597}, 012025 (2015), \newline
\href{http://dx.doi.org/10.1088/1742-6596/597/1/012025}{\nolinkurl{doi:10.1088/1742-6596/597/1/012025}}

\item Pieter W. Claeys, Stijn De Baerdemacker, Mario Van Raemdonck, and Dimitri Van Neck, \textit{Eigenvalue-based method and form-factor determinant representations for integrable XXZ Richardson-Gaudin models}, Physical Review B \textbf{91}, 155102 (2015), \newline
\href{http://dx.doi.org/10.1103/PhysRevB.91.155102}{\nolinkurl{doi:10.1103/PhysRevB.91.155102}}

\item Pieter W. Claeys, Stijn De Baerdemacker, Mario Van Raemdonck, and Dimitri Van Neck, \textit{Eigenvalue-based determinants for scalar products and form factors in Richardson-Gaudin integrable models coupled to a bosonic mode}, Journal of Physics A: Mathematical and Theoretical \textbf{48}, 425201 (2015), \newline
\href{http://dx.doi.org/10.1088/1751-8113/48/42/425201}{\nolinkurl{doi:10.1088/1751-8113/48/42/425201}}

Selected as Publisher's pick.

\item Guillaume Acke, Stijn De Baerdemacker, Pieter W. Claeys, Mario Raemdonck, Ward Poelmans, Dimitri Van Neck, and Patrick Bultinck, \textit{Maximum probability domains for Hubbard models}, Molecular
Physics \textbf{114}, 1392-1405 (2016), \newline
\href{http://dx.doi.org/10.1080/00268976.2016.1153742}{\nolinkurl{doi:10.1080/00268976.2016.1153742}}

\item Pieter W. Claeys, Stijn De Baerdemacker, and Dimitri Van Neck, \textit{Read-Green resonances in a topological superconductor coupled to a bath}, Physical Review B \textbf{93}, 220503(R) (2016), \newline
\href{http://dx.doi.org/10.1103/PhysRevB.93.220503}{\nolinkurl{doi:10.1103/PhysRevB.93.220503}}

\item Pieter W. Claeys, Dimitri Van Neck, and Stijn De Baerdemacker, \textit{Inner products in integrable Richardson-Gaudin models}, SciPost Physics \textbf{3}, 28 (2017), \newline
\href{http://dx.doi.org/10.21468/SciPostPhys.3.4.028}{\nolinkurl{doi:10.21468/SciPostPhys.3.4.028}}

\item Pieter W. Claeys, Jean-S\'ebastien Caux, Dimitri Van Neck, and Stijn De Baerdemacker, \textit{Variational
method for integrability-breaking Richardson-Gaudin models}, Physical Review B \textbf{96}, 155149 (2017), \newline
\href{http://dx.doi.org/10.1103/PhysRevB.96.155149}{\nolinkurl{doi:10.1103/PhysRevB.96.155149}}

\item Pieter W. Claeys and Jean-S\'ebastien Caux, \emph{Breaking the integrability of the Heisenberg model through periodic driving}, ArXiv e-prints (2017), \newline
\href{http://arxiv.org/abs/1708.07324}{\nolinkurl{arXiv:1708.07324}}

\item Stijn De Baerdemacker, Pieter W. Claeys, Jean-S\'ebastien Caux,  Dimitri Van Neck, and Paul W. Ayers, \emph{Richardson-Gaudin Configuration-Interaction for nuclear pairing correlations}, ArXiv e-prints (2017), \newline
\href{http://arxiv.org/abs/1712.01673}{\nolinkurl{arXiv:1712.01673}}

\item Pieter W. Claeys, Stijn De Baerdemacker, Omar El Araby, and Jean-S\'ebastien Caux, \emph{Spin polarization through Floquet resonances in a driven central spin model}, ArXiv e-prints (2017), \newline
\href{http://arxiv.org/abs/1712.03117}{\nolinkurl{arXiv:1712.03117}}

\item Eyzo Stouten, Pieter W. Claeys, Jean-S\'ebastien Caux, and Vladimir Gritsev, \emph{Integrability and duality in spin chains}, ArXiv e-prints (2017), \newline
\href{http://arxiv.org/abs/1712.09375}{\nolinkurl{arXiv:1712.09375}}

\item Sergio Enrique Tapias Arze, Pieter W. Claeys, Isaac P\'erez Castillo, and Jean-S\'ebastien Caux, \emph{Out-of-equilibrium phase transitions induced by Floquet resonances in a periodically quench-driven XY spin chain}, ArXiv e-prints (2018), \newline
\href{http://arxiv.org/abs/1804.10226}{\nolinkurl{arXiv:1804.10226}}

\end{enumerate}


\pdfbookmark{\contentsname}{Contents}
\tableofcontents

\mainmatter


\chapter{Introduction} 
\label{Chap_Intro}

\setlength\epigraphwidth{.9\textwidth}
\epigraph{\emph{Now we know how the electrons and light behave. But what can I call it? If I say they behave like particles I give the wrong impression; also if I say they behave like waves. They behave in their own inimitable way, which technically could be called a quantum mechanical way. They behave in a way that is like nothing that you have seen before. Your experience with things that you have seen before is incomplete. The behavior of things on a very tiny scale is simply different. An atom does not behave like a weight hanging on a spring and oscillating. Nor does it behave like a miniature representation of the solar system with little planets going around in orbits. Nor does it appear to be somewhat like a cloud or fog of some sort surrounding the nucleus. It behaves like nothing you have seen before.}}{{Richard P. Feynman}}

\setlength\epigraphwidth{.48\textwidth}
\epigraph{\emph{The troubles came, I saved what I could save \\
A thread of light, a particle, a wave}}{{Leonard Cohen}}

Despite being one of the great scientific achievements of the last century, quantum physics still has the reputation of being notoriously difficult and counter-intuitive. In a way, this shouldn't be surprising -- whereas human intuition is gained through day-to-day experience, quantum mechanics involves precisely those phenomena occurring on scales different from those encountered in our daily lives. Examples include the behaviour of electrons in atoms and molecules, where distances become smaller than the nanoscale, and the sudden appearance of superconductivity if the temperature of some materials are brought close enough to absolute zero. This might make it seem as if any research in this area is merely of theoretical interest, from the extremely small to the extremely cold. Yet the influence of quantum mechanics on our current lives cannot be overstated, lying at the heart of electronics in our computers and a large part of modern technology.

The full theory of quantum physics was developed during the early 20th century, with pioneering roles being played by physicists such as Schr\"odinger, Heisenberg, Pauli and Dirac (among many others). Ever since, this theory has withstood the test of experiment time after time and has now reached a respectable level of maturity. On the most basic level, quantum physics can be considered mostly `complete'. The backbone of science is the assumption that the workings of nature, of our bodies,... follow a same set of scientific laws. Quantum mechanics provides us with exactly this, presenting a way of capturing the relevant laws of nature in a strict mathematical framework. Apart from some pathological situations, we feel that the physical laws are known, counter-intuitive as they may seem.

However, the story doesn't end here. While it is possible to write down the underlying laws as mathematical equations, this does not guarantee that we are able to solve them or extract physical information. In fact, this is rarely the case. The complexity of all involved equations grows quickly with system size, prohibiting solutions except for the smallest of systems. E.g. in quantum chemistry, it is only possible to solve the hydrogen atom, containing a single electron and proton. Systems containing more particles can no longer be solved exactly, and we often have to resort to approximate methods. This is what is known as the \emph{quantum many-body problem}. As Paul Dirac noted in 1929 \cite{dirac_quantum_1929}:

\begin{quotation}
``The underlying physical laws necessary for the mathematical theory of a large part of physics and the whole of chemistry are thus completely known, and
the difficulty is only that the exact application of these laws leads to equations much too complicated to be soluble. It therefore becomes desirable that approximate practical methods of applying quantum mechanics should be developed, which can lead to an explanation of the main features of complex atomic systems without too much computation.''
\end{quotation}

This would be no problem if it was simply possible to predict the behaviour of large systems by generalizing the behaviour of small systems. Instead of obtaining exact quantitative predictions, it would be possible to settle for qualitative ones by simply extrapolating the known behaviour of small systems. Unfortunately (or fortunately), this is not the case. When it comes to large systems, we are completely lost. In much the same way that it is impossible to predict the existence of waves and the sea from the study of a single water molecule, the behaviour of systems containing many interacting particles is infinitely more rich and interesting than the behaviour of systems containing few particles. Even if it is exactly known in what way particles interact, their macroscopic properties are rarely a direct reflection of their microscopic properties. The whole system is greater than the sum of its parts or, in Phil Anderson's words, `\emph{More is different}' \cite{anderson_more_1972}.

\section{The quantum many-body problem}
 
Postponing all technicalities to later chapters, it is already possible to gain some intuition for the peculiarities of quantum mechanics and the difficulties of the quantum many-body problem from a simple example. 

Suppose we have a system which acts as the quantum mechanical equivalent of a simple coin. Performing a measurement on this system can be seen as flipping this coin, and will return either heads (here denoted $\ket{\uparrow}$) or tails ($\ket{\downarrow}$). Quantum mechanically, the properties of such a system are encoded in the \emph{wave function} $\ket{\Psi}$.  Before flipping a coin, it is impossible to predict what the result will be. The only thing we can say is how \emph{probable} each outcome will be. The wave function then provides a convenient way of writing down all possible outcomes of a measurement, combined with the probability of each outcome, as
\begin{equation}\label{intro:qmb:psi}
\ket{\Psi} = C_{\uparrow}\ket{\uparrow}+C_{\downarrow}\ket{\downarrow}.
\end{equation}
Here $C_{\uparrow}$ and $C_{\downarrow}$ are two (complex) numbers, with the wave function telling us that the probability of the coin flip returning heads is $|C_{\uparrow}|^2$, and the probability of the coin flip returning tails is $|C_{\downarrow}|^2$. This already contains a crucial aspect of quantum theory -- it is a fundamentally \emph{probabilistic} theory. It is impossible to tell for certain what the outcome of any measurement will be, it is only possible to say how probable any possible outcome is. This is not in any way a shortcoming of the theory, but rather a fundamental property of nature.

Remarkably, such a toy system already plays an important role in the description of interacting electrons. From classical mechanics, it is well understood that electrons carry properties such as mass, electric charge, (angular) momentum,... In order to explain the energy spectra of atoms containing multiple electrons, Pauli proposed that electrons also possess a further, intrinsic, two-valued property \cite{pauli_uber_1925}. This was later termed electron spin by Kronig, Uhlenbeck and Goudsmit \cite{uhlenbeck_ersetzung_1925}, where it is now possible to distinguish between spin-up electrons ($\ket{\uparrow}$) and spin-down electrons ($\ket{\downarrow}$). 

The wave function (\ref{intro:qmb:psi}) can thus be seen as the wave function for a single electron, making abstraction of all properties except electron spin. By simply writing down the wave function for multiple particles, the origin of the quantum many-body problem as stated by Dirac can be made clear. For two electrons, each electron can again be in two different states, resulting in four possible outcomes and a wave function which can be written as
\begin{equation}
\ket{\Psi} = C_{\uparrow\uparrow}\ket{\uparrow\uparrow}+C_{\uparrow\downarrow}\ket{\uparrow\downarrow}+C_{\downarrow\uparrow}\ket{\downarrow\uparrow}+C_{\downarrow\downarrow}\ket{\downarrow\downarrow}.
\end{equation}
Adding a single particle to the system effectively doubles the amount of terms in the wave function. Adding more and more particles, a system containing $L$ particles would then lead to a wave function containing $2^L$ terms. In other words, the number of terms grows exponentially with system size. In order to fully appreciate exponential growth, it should be realized that 64 particles would lead to $2^{64}=18.446.744.073.709.551.616$ possible outcomes\footnote{The number 64 is not chosen accidentally. A famous legend about the invention of chess goes that its inventor, as a reward, asked his ruler for a total amount of wheat corresponding to that placed on a chessboard when a single grain of wheat is placed on the first square, two on the second, four on the third,... until the 64th square. The ruler then laughs it off as a meager prize, before eventually realizing that such a reward far exceeds his country's resources.}. We cannot possibly hope to obtain all these coefficients, and the wave function becomes untractable for large system sizes, leading to what is known as the `exponential wall'.

\section{Symmetry and the Schr\"odinger equation}

If the wave function is the fundamental object in the description of any system, a natural question to ask is how it can be obtained. The answer to this question is provided by the Schr\"odinger equation \cite{schrodinger_quantisierung_1926}, discovered by Erwin Schr\"odinger in 1925, which can be written down in a surprisingly concise manner as
\begin{equation}
\hat{H} \ket{\Psi} = E \ket{\Psi}.
\end{equation}
In this equation $\hat{H}$ is the Hamiltonian, a mathematical operator containing information about the interactions between all involved particles (e.g. electromagnetic interactions among electrons). For any physical system, it is known how to construct the Hamiltonian by following a strict set of rules, and the main goal in quantum many-body physics then consists of solving this equation for both the wave function $\ket{\Psi}$ and the energy $E$.

Circumventing the exponential wall, there are two main ways of tackling this problem. First, and as predicted by Dirac, there currently exists a wealth of approximate methods for solving the Schr\"odinger equation \cite{chaikin_principles_2000,sakurai_modern_2010,dickhoff_many-body_2005,helgaker_molecular_2014,orus_practical_2014}. In practice, this often corresponds to restricting the wave function in some way, imposing a specific structure on the coefficients in Eq. (\ref{intro:qmb:psi}). The success of any approach is then judged by how well the proposed structure of the wave function matches that of the exact solution. In this way Hartree-Fock (HF) \cite{hartree_wave_1928,hartree_wave_1928-1,fock_naherungsmethode_1930} and Bardeen-Cooper-Schrieffer (BCS) \cite{cooper_bound_1956,bardeen_microscopic_1957,bardeen_theory_1957} mean-field theory have proven to be remarkably successful in the description of atoms and molecules and superconductivity respectively. Both theories are based on the assumption that the specific details of the interparticle interactions can be captured in a mean-field or collective interaction, giving rise to remarkably simple and tractable wave functions.

However, it is not always necessary to resort to approximate methods. A second approach exploits the presence of \emph{symmetry} in physical systems. While the ancient Greeks already recognized the importance of symmetry, this concept gained in importance in the study of physics simultaneous with the advent of quantum mechanics. Usually, symmetry is thought of as being a property of objects -- a circle is more symmetric than a square. However, in the same way that objects can exhibit symmetries, physical theories can exhibit symmetry, and some theories are more symmetrical than others. 

The basic idea can be easily formulated --  `\emph{a thing is symmetrical if there is something you can do to it so that after you have finished doing it, it looks the same as before}' \cite{weyl_symmetry_2015}. It is impossible to tell if a circle has been rotated, and it is similarly impossible to tell if e.g. two electrons in an atom have been exchanged. Both these features are marks of specific symmetries. Remarkably, symmetry allows us to make \emph{exact} statements about physical systems, independent of system size. From a purely practical point of view, for a symmetric system the coefficients in the wave function will be subject to symmetry constraints and the number of relevant parameters in Eq. (\ref{intro:qmb:psi}) can be massively reduced \cite{mcweeny_symmetry_2002}. As an example, consider a two-electron system in which the two electrons cannot be distinguished. It can then be expected that the probability of finding the first electron in a specific state ($\ket{\uparrow}$ or $\ket{\downarrow}$) equals the probability of finding the second electron in the same state. The four simplest wave functions $\ket{\Psi}$ with this property are given by
\begin{equation}
\ket{\uparrow \uparrow}, \qquad \ket{\downarrow \downarrow}, \qquad \frac{1}{\sqrt{2}} \ket{\uparrow \downarrow}+\frac{1}{\sqrt{2}} \ket{\downarrow \uparrow}, \qquad \frac{1}{\sqrt{2}} \ket{\uparrow \downarrow}-\frac{1}{\sqrt{2}} \ket{\downarrow \uparrow},
\end{equation}
which can be considered the building blocks for any two-electron system purely from symmetry considerations. This can be quantified by defining an \emph{exchange operator} $\hat{P}$, exchanging the role of the two electrons. If the electrons in these wave functions are exchanged, this results in
\begin{align}
&\hat{P}  \ket{\uparrow \uparrow} = \ket{\uparrow \uparrow}, \qquad \hat{P}  \ket{\downarrow \downarrow} = \ket{\downarrow \downarrow},  \qquad \hat{P} \left( \frac{1}{\sqrt{2}} \ket{\uparrow \downarrow}+\frac{1}{\sqrt{2}} \ket{\downarrow \uparrow} \right) =  \frac{1}{\sqrt{2}} \ket{\uparrow \downarrow}+\frac{1}{\sqrt{2}} \ket{\downarrow \uparrow}, \nonumber \\
&\hat{P}\left(  \frac{1}{\sqrt{2}} \ket{\uparrow \downarrow}-\frac{1}{\sqrt{2}} \ket{\downarrow \uparrow} \right)  =  -\left(\frac{1}{\sqrt{2}} \ket{\uparrow \downarrow}-\frac{1}{\sqrt{2}} \ket{\downarrow \uparrow}\right).
\end{align}
The first three wave functions remain invariant under particle exchange, while the fourth gains a minus sign. A value $+1$ can now be associated with the first three wave functions, where $\hat{P}\ket{\Psi} = +\ket{\Psi}$, and a value $-1$ can be associated with the fourth wave function, where $\hat{P}\ket{\Psi} = - \ket{\Psi}$. Since these particles are indistinguishable, this value cannot be changed by the Hamiltonian, and this value is what is known as a \emph{conserved quantity}. This is a universal property -- symmetries always give rise to conservation laws \cite{noether_invariante_1918}, which are in turn reflected in the wave function. 

It should be noted that the relation between the two approaches of approximate methods and symmetry is a rather intricate one, since symmetry principles often guide the construction of approximate wave functions. The Slater determinant underlying HF theory \cite{slater_simplification_1951} can be thought of as the simplest wave function taking into account Pauli's exclusion principle, which is a direct consequence of the indistinguishability of electrons. Conversely, the BCS wave function explicitly had to break particle-number symmetry in order to explain superconductivity. Historically, there has always been a strong interplay between symmetry and the development of better approximate methods.

\section{Integrability}

Closely related to the concept of symmetry is that of \emph{integrability}. Where the symmetries mentioned so far seem quite intuitive, the exact statements we can make are also quite restricted, and are often not enough to completely characterize the physical system. Compared to these, integrable models are characterized by a far more mathematical symmetry which allows the Schr\"odinger equation to be solved \emph{exactly}, even for larger system sizes. These models all have in common that their wave function can be written as a \emph{Bethe ansatz} wave function, providing a highly efficient way of writing down the exact many-body wave function. The underlying symmetries are then reflected in the existence of an extensive set of conservation laws for integrable systems.

The field originated in 1931 with Hans Bethe's solution of the Heisenberg chain \cite{bethe_zur_1931}, describing a linear chain of two-level atoms interacting with their nearest neighbours \cite{heisenberg_zur_1928}. The structure of the wave function introduced in this seminal work is known as the \emph{coordinate Bethe ansatz}. Bethe's solution was later extended by Yang and Yang \cite{yang_one-dimensional_1966,yang_one-dimensional_1966-1,yang_one-dimensional_1966-2}, and it was realized that this solution could be connected to Baxter's results for the six-vertex model, arising in classical two-dimensional statistical mechanics \cite{baxter_exactly_2007}. The fundamental relation underlying these results is now known as the \emph{Yang-Baxter equation}. Within condensed matter physics, such exactly solvable models are generally obtained in the context of one-dimensional (1D) systems. The framework since developed in order to deal with these systems has been termed the \emph{Algebraic Bethe Ansatz} (ABA) \cite{korepin_quantum_1993}.

In parallel with these developments, Richardson obtained an exact solution for the so-called reduced BCS model in the field of nuclear physics \cite{richardson_restricted_1963,richardson_exact_1964,richardson_numerical_1966,richardson_pairing_1977}. Even though this work was published in the 1960s, it was largely overlooked by the community until resurfacing in the study of ultrasmall metallic grains \cite{sierra_exact_2000,von_delft_spectroscopy_2001}. It then became clear that there is an intimate connection between Richardson's solution and a class of integrable models known as Gaudin magnets, obtained by Michel Gaudin \cite{gaudin_diagonalisation_1976,gaudin_bethe_2014}. The precise nature of this connection was clarified by Cambiaggio \emph{et al.} \cite{cambiaggio_integrability_1997}, effectively establishing the integrability of the Richardson model and strengthening the connection between exact solvability and the existence of conservation laws. Following crucial works by Amico \emph{et al.} \cite{amico_integrable_2001} and Dukelsky \emph{et al.} \cite{dukelsky_class_2001}, these combined results have since led to a class of systems known as \emph{Richardson-Gaudin integrable models}, the main objects of study in this thesis. These were later also incorporated within the ABA and their connection to the classical Yang-Baxter equation was established  \cite{amico_bcs_2001,zhou_superconducting_2002,von_delft_algebraic_2002,links_algebraic_2003}. As such, Richardson-Gaudin models provide a specific class of integrable models which can be solved exactly using Bethe ansatz techniques. 

At all points, it should be kept in mind that integrability places strong constraints on the model under study. Similar to symmetric models, models need quite a bit of fine-tuning in order to be integrable. This can be seen as a shift in mentality -- whereas usually the model is considered to be exact and the solution approximate, now the models are approximate and their solutions are exact. However, in recent years there has been a move towards using techniques and concepts from integrability in the study of non-integrable models. This can be done either by adding more realistic terms to the Hamiltonian and treating these in an approximate (perturbative) manner \cite{delfino_non-integrableqft_1996, controzzi_mass_2005,delfino_non-integrable_1998,pozsgay_characterization_2006,delfino_decay_2006,groha_spinon_2017}, or by approximating wave functions of non-integrable models using Bethe ansatz techniques \cite{krauth_bethe_1991,kiwata_hirohito_bethe-ansatz_1994,okunishi_magnetization_1999,vanderstraeten_scattering_2015}. The majority of these results essentially build on the same idea  -- while integrable models are interesting in their own right, the rich framework of integrability also provides a convenient toolbox for the treatment of more involved systems. This will be one of the key ideas throughout this work. 


\section{Outlook and outline of the thesis}
 
The first goal of this thesis is then to provide some insight in the overarching structure of Richardson-Gaudin models and their integrability, with special focus on the structure of the Bethe ansatz wave function. The obtained framework is subsequently applied to integrable models in various physical contexts. 

The second goal is to investigate how methods from integrability can be extended towards non-integrable models. Here, the focus lies on models which are in some sense `close to integrability', where it is shown how the Bethe ansatz is still able to accurately model wave functions of non-integrable models in two different settings.

This thesis is structured as follows. The first half focuses on the theoretical aspects of (Richardson-Gaudin) integrability. Chapter \ref{Chap_RGmodels} provides an introduction to classical and quantum integrability, with special attention paid to Richardson-Gaudin models and the Bethe ansatz. In Chapter \ref{chap:EVB}, a framework is presented for the numerical solution of the resulting Bethe equations starting from the conserved charges of these models. This approach hints at an underlying duality in Richardson-Gaudin models, which is then extended to the calculation of inner products and correlation functions in Chapter \ref{chap:innerproducts}. While these first chapters focus on models based on the $su(2)$ algebra, it is shown in Chapter \ref{chap:contraction} how the results from these chapters can be extended to models containing a bosonic degree of freedom through a deformation of the underlying algebra, concluding the first half of this thesis. The second half of this thesis then presents applications of the outlined framework in different physical settings. A first application is presented in Chapter \ref{chap:readgreen}, where an integrable model describing the interaction of a topological superconductor with a bath is investigated and the exact Bethe ansatz solution is presented and compared with mean-field theory. Chapters \ref{chap:VarRG} and \ref{chap:floquet} investigate the use of the Bethe ansatz wave function in the context of integrability-breaking. In Chapter \ref{chap:VarRG}, the use of the Bethe ansatz as a variational ansatz is investigated and applied to models where the integrability of a given Hamiltonian is explicitly broken, and a method is presented for obtaining the low-lying spectrum of such models. Chapter \ref{chap:floquet} then studies the effects of periodic driving in integrable Richardson-Gaudin systems, implicitly breaking integrability, and it is shown how the properties of integrability and the Bethe ansatz can be used to describe adiabatic transitions in driven systems. It should be emphasized that, while the methods outlined in these chapters are applied to specific physical systems, the underlying principles can be applied to all Richardson-Gaudin systems. Chapter \ref{chap:concl} is reserved for conclusions.


\part{Richardson-Gaudin models}

\chapter{Richardson-Gaudin integrability} 
\label{Chap_RGmodels}


\setlength\epigraphwidth{.3\textwidth}
\epigraph{\emph{Reality favors symmetry.}}{Jorge Luis Borges}


In this chapter, a self-contained introduction to Richardson-Gaudin (RG) integrable models and their solution by Bethe ansatz is presented, setting the stage for the later chapters of this thesis. This is done through the framework of the Generalized Gaudin Algebra (GGA) \cite{ortiz_exactly-solvable_2005}, highlighting the algebraic properties underlying integrability. After a brief discussion of the concept of integrability in classical and quantum mechanics, it is shown how a GGA can be used to systematically construct integrable models and obtain their exact eigenstates using Bethe ansatz techniques. Non-interacting or free models can arguably be considered to be the simplest integrable models, arising as a particular limit of Richardson-Gaudin models. Since these already exhibit most of the crucial features of integrability, the connection with such non-interacting models is made throughout.

The introduction is initially kept purely algebraic, and it is then shown how different realizations of the GGA can be used to construct three commonly-encountered integrable models: the central spin model, the reduced BCS Hamiltonian, and the $p_x+ip_y$-wave pairing Hamiltonian. These correspond to different realizations of the GGA and the underlying $su(2)$-algebras, with a somewhat increasing level of complexity. Some physical context is given for these models, after which their integrability and Bethe ansatz eigenstates are presented.

\section{Classical and quantum integrability}
\subsection{Classical integrability}

Within classical mechanics, the notion of integrability is a well-defined one directly connected to the dynamics of a given system. Classically, a physical system can be described in terms of canonical variables $(\vec{q},\vec{p})=(q_1, \dots q_L,p_1, \dots p_L)$, for which the equations of motion follow from a classical Hamiltonian $\mathcal{H}(\vec{q},\vec{p})$ as
\begin{equation}
 \frac{d q_i}{dt}   = \frac{\partial }{\partial p_i}  \mathcal{H}(\vec{q},\vec{p}), \qquad  \frac{d p_i}{dt} = -\frac{\partial }{\partial q_i}\mathcal{H}(\vec{q},\vec{p}).
\end{equation}
A system with $L$ degrees of freedom (and hence a $2L$-dimensional phase space) is said to be \emph{Liouville-integrable} if the system possesses $L$ independent integrals of motion in involution \cite{laurent-gengoux_liouville_2013}. What does this tell us? The symplectic structure allows for the definition of Poisson brackets $\{\cdot,\cdot\}$, and the equations of motion for any physical observable $\mathcal{O}(\vec{q},\vec{p},t)$ can be recast as
\begin{equation}
\frac{d}{dt}{\mathcal{O}} = \{\mathcal{O},\mathcal{H}\}+\frac{\partial \mathcal{O}}{\partial t}, \qquad \textrm{with}\qquad  \{\mathcal{O},\mathcal{H}\} = \sum_{i=1}^L \frac{\partial \mathcal{O} }{\partial q_i} \frac{\partial \mathcal{H}}{\partial p_i} - \frac{\partial \mathcal{H} }{\partial q_i} \frac{\partial \mathcal{O}}{\partial p_i},
\end{equation}
where the dependence of both $\mathcal{O}$ and $\mathcal{H}$ on $(\vec{q},\vec{p})$ has been made implicit. These fully determine the dynamics, and  for an observable $\mathcal{O}(\vec{q},\vec{p})$ (with no explicit time-dependence) to be in involution with the Hamiltonian means that it Poisson-commutes with the Hamiltonian as $\{\mathcal{O},\mathcal{H}\}=0$, leading to $\dot{\mathcal{O}}=0$. Such an operator is then also known as a \emph{constant of motion} or a \emph{conserved charge}, since its numerical value remains constant during all dynamics. Integrable systems are now characterized by a maximal set of $L$ such conserved charges $\mathcal{Q}_1, \dots, \mathcal{Q}_L$, which are similarly in involution as $\{\mathcal{Q}_i, \mathcal{Q}_j\}=0$. 

Since the Hamiltonian itself is a trivial conserved quantity, the demand that this set is maximal implies that the Hamiltonian itself cannot be independent from these conserved charges, and can always be written as a function of them as $\mathcal{H}(\mathcal{Q}_1, \dots, \mathcal{Q}_L)$. Taking these variables as new canonical variables, the problem of time evolution reduces to finding the canonical conjugate variables $\mathcal{P}_1, \dots \mathcal{P}_L$. Following the Liouville-Arnol'd theorem \cite{arnold_mathematical_1989}, this can be done in a purely algebraic manner. Once these have been obtained, Hamilton's equations of motion immediately follow as
\begin{equation}
\frac{d}{dt}{\mathcal{Q}}_i = \frac{\partial \mathcal{H}}{\partial \mathcal{P}_i} = 0, \qquad \frac{d}{dt}{\mathcal{P}}_i = -\frac{\partial \mathcal{H}}{\partial \mathcal{Q}_i} = \textrm{Cst.}
\end{equation}
This allows the differential equations of motion to be explicitly integrated, resulting in the denomination of \emph{integrable} models. 

A simple example consists of a Hamiltonian with quadratic interaction terms
\begin{equation}
\mathcal{H} = \sum_{i=1}^L \frac{p_i^2}{2 m_i} + \frac{1}{2}\sum_{i,j=1}^L q_i V_{ij} q_j,
\end{equation}
where the integrability can be made explicit by diagonalizing the mass-weighted interaction matrix $\tilde{V}_{ij}=V_{ij}/\sqrt{m_i m_j}$, leading to eigenvalues $\omega_1, \dots , \omega_L$. A canonical transformation to the \emph{normal modes}
\begin{align}
\tilde{q}_i = \sum_{j=1}^L U_{ij}  \sqrt{ \omega_i m_j }  q_j, \qquad \tilde{p}_i = \sum_{j=1}^L U_{ji} \frac{p_j}{\sqrt{\omega_i m_j}} , \qquad \text{with} \qquad   U \tilde{V} U^T= \omega^2,
\end{align}
leads to a Hamiltonian expressed in the new canonical coordinates as
\begin{equation}
\mathcal{H}= \sum_{i=1}^L \frac{\omega_i}{2}\left(\tilde{p}_i^2+\tilde{q}_i^2\right)  = \sum_{i=1}^L \frac{\omega_i}{2}\mathcal{Q}_i .
\end{equation}
This reduces the Hamiltonian to a non-interacting or free one, which can be considered the simplest example of an integrable system. The conserved charges $\mathcal{Q}_i = \tilde{p}_i^2+\tilde{q}_i^2$ can be interpreted as single-particle Hamiltonians for the normal modes, and all dynamics have been decoupled. Solving the equations of motions has effectively been reduced to the algebraic problem of diagonalizing the mass-weighted interaction matrix.

\subsection{Quantum integrability}

Somewhat surprisingly, there exists no straightforward extension of this concept to quantum integrability \cite{weigert_problem_1992,links_algebraic_2003,clemente-gallardo_towards_2009,caux_remarks_2011}. One of the key problems is the definition of `integrals of motion' in the quantum case. Quantizing a Hamiltonian problem in the Heisenberg picture corresponds to replacing observables by operators and Poisson brackets by commutators, leading to equations of motion
\begin{equation}
\frac{d}{dt} \hat{O} = \frac{1}{i\hbar} [ \hat{O},\hat{H}] + \frac{\partial}{\partial t} \hat{O}.
\end{equation}
A natural suggestion would then be to define quantum integrability as the existence of a set of conserved charges satisfying
\begin{equation}
[\hat{Q}_i,\hat{H}] = [\hat{Q}_i, \hat{Q}_j]=0, \qquad \forall i,j =1 \dots L.
\end{equation}
Whereas this demand was extremely strict in classical mechanics, the opposite is now the case -- this definition would lead to all Hamiltonians being quantum integrable. According to the spectral theorem any Hermitian operator can be expanded as
\begin{equation}
\hat{H} = \sum_{i} \lambda_i \hat{P}_i,
\end{equation}
with $\lambda_i$ the eigenvalues and $\hat{P}_i$ the projector on the corresponding eigenspace. Trivially, we then have $[\hat{H},\hat{P}_i]=[\hat{P}_j, \hat{P}_i]=0$, and any Hamiltonian possesses a set of conserved charges.

This definition is then only sound if some structure can be imposed on these conserved charges. This can be done through the notion of \emph{ergodicity} \cite{caux_remarks_2011} -- roughly speaking a system is ergodic if, starting from an arbitrary state, time evolution `samples' the entire phase space. The time average of any physical observable can then be reduced to its ergodic average over the full phase space. In classical mechanics the existence of conserved charges from integrability strongly restricts time evolution, leading to \emph{non-ergodic} time evolution. Quantum mechanically, ergodicity can also be defined, but only in the limit of infinitely large systems. Nevertheless, demanding that the conserved charges lead to loss of ergodicity in this limit excludes the projectors as conserved charges, but includes the classes of both Yang-Baxter integrable \cite{korepin_quantum_1993} and many-body localized systems \cite{abanin_recent_2017} in lattice models\footnote{In these models, the loss of ergodicity is guaranteed by the (quasi-)local nature of the conserved charges.}. Another, closely related, class is that of Richardson-Gaudin integrable models, the main focus of this thesis. In all classes the notion of a maximal set of conserved charges is typically replaced by the demand that the number of conserved charges scales extensively with system size, leading to loss of ergodicity in infinitely large systems. 

Quantum integrability now has some striking consequences. Compared to the exactly-solvable dynamics in classical integrability, the dynamics of a quantum system are encoded in the eigenstates and eigenvalues of the Hamiltonian. Integrability now generally allows all these eigenstates and eigenvalues to be exactly determined using Bethe ansatz techniques. This can also be connected to the notion of thermalization -- whereas statistical physics uses ergodicity to predict the long-time (equilibrium) behaviour following classical dynamics, quantum systems necessitate the Eigenstate Thermalization Hypothesis (ETH), imposing some structure on the eigenstates \cite{dalessio_quantum_2016}.


It is also worth remarking that an often used criterion to numerically distinguish integrable and non-integrable systems is an investigation of the statistical properties of the energy spectrum. For non-integrable models the energy spectrum is expected to follow the Wigner-Dyson statistics associated with Gaussian Orthogonal Ensembles from Random Matrix Theory \cite{dalessio_quantum_2016}, whereas Berry and Tabor conjectured that the spectra of integrable models behave Poissonian \cite{berry_level_1977}. This could then also be connected to the existence of conserved charges, since it was argued that a number of conserved charges scaling at least logarithmically with system  size leads to Poissonian statistics \cite{scaramazza_integrable_2016}. While this can also be directly connected to the concept of ergodicity, exceptions to this rule have been found (see e.g. \cite{relano_stringent_2004}).

\section{Richardson-Gaudin models}

Non-interacting, or free, theories can arguably be considered to be the simplest case of integrable models. Not only do they satisfy the requirement of non-ergodicity, they are also known to be integrable in the classical limit. As such, these models form an ideal starting point for the study of RG integrability. Note that ``non-interacting'' does not imply that these are trivial, e.g. the 1D transverse field Ising model exhibits a phase transition and can be mapped to a non-interacting model after a Fourier transform and a Jordan-Wigner transformation \cite{pfeuty_one-dimensional_1970}. 

As mentioned before, the Poisson structure is replaced by a commutator structure, commonly encoded in the definition of a Lie algebra (see also Appendix \ref{RG:App:su2}). As an example, for a system containing $L$ spins labeled $i=1 \dots L$ (and as such assumed distinguishable), each separate spin corresponds to a realization of the $su(2)$ algebra generated by $\{S^+, S^-, S^z\}$, leading to a set of generators satisfying\footnote{From now on, we will drop the hat notation for operators.}
\begin{equation}
[S^z_i,S^{+}_j]=\delta_{ij}S^{+}_i, \qquad [S^z_i,S^-_j]=-\delta_{ij}S^-_i, \qquad [S^{+}_i,S^{-}_j]=2\delta_{ij}S^z_i.
\end{equation}
The simplest non-interacting Hamiltonian that can be written in this way is
\begin{equation}\label{RG:RG:nonint}
{H} = \sum_{i=1}^L H_i = \sum_{i=1}^L \frac{\omega_i}{2} S_i^z,
\end{equation}
with $\omega_i$ free variables that can be interpreted as magnetic fields applied on spin $i$ along the $z$-axis. The conserved charges here are simply given as ${Q}_i = {S}_i^z$, proportional to the single-particle Hamiltonians, and the total Hamiltonian is a simple function of the conserved charges as ${H} = \sum_{i=1}^L \omega_i {Q}_i/2$. Suppose we wish to introduce interactions in the model while still keeping the commutative structure. This can be done by proposing a set of interacting conserved charges, for which the non-interacting models are returned as a zeroth-order expansion in the interaction strength $g$ \cite{dukelsky_class_2001,amico_integrable_2001,sdb_saclay_2016},
\begin{equation}\label{RG:RG:com}
{Q}_i = {S}_i^z + g \sum_{j \neq i}^L \left[ X_{ij} \left(S_i^{+}S_j^-+ S_i^- S_j^{+}\right)+Z_{ij} S_i^z S_j^z \right].
\end{equation}
These reduce to the non-interacting conserved charges in the limit $g \to 0$, and the interactions are parametrized by a set of $X$- and $Z$-variables. In the limit $g \to \infty$, these return the class of \emph{XXZ Gaudin magnets} \cite{gaudin_bethe_2014}. The commutativity condition $[Q_i,Q_j]=0$, a necessary requirement for these operators to act as conserved charges, will be fulfilled if the following conditions hold
\begin{align}\label{RG:RG:GaudinXXZ}
&X_{ij}+X_{ji} = 0, \qquad Z_{ij}+Z_{ji} = 0, \qquad \qquad \forall i \neq j,\\
&X_{ij}X_{jk}-X_{ik}(Z_{ij}+Z_{jk}) = 0, \qquad \qquad \forall i \neq j \neq k.
\end{align}
These are also known as the \emph{Gaudin equations}. Obtaining a class of conserved charges corresponds to solving these equations. Gaudin already mentioned three classes of solutions, where each class considers $X_{ij}$ and $Z_{ij}$ as odd functions of some arbitrary parameters $\epsilon_i-\epsilon_j$, reducing the Gaudin equations to functional relations.
\begin{enumerate}
\item \textbf{The rational model}
\begin{equation}\label{RG:RG:rat}
X_{ij} = \frac{1}{\epsilon_i-\epsilon_j}, \qquad Z_{ij} = \frac{1}{\epsilon_i-\epsilon_j}
\end{equation}
\item \textbf{The trigonometric model}
\begin{equation}\label{RG:RG:trig}
X_{ij} = \frac{1}{\sin(\epsilon_i-\epsilon_j)}, \qquad Z_{ij}=\cot(\epsilon_i-\epsilon_j)
\end{equation}
\item \textbf{The hyperbolic model}
\begin{equation}\label{RG:RG:hyp}
X_{ij} = \frac{1}{\sinh(\epsilon_i-\epsilon_j)}, \qquad Z_{ij}=\coth(\epsilon_i-\epsilon_j)
\end{equation}
\end{enumerate}

Any linear combination of these constants of motion then gives rise to a Richardson-Gaudin integrable Hamiltonian ${H} = \sum_{i=1}^L\omega_i Q_i/2$ , where the conserved charges are explicitly known. More involved solutions for XYZ-models in terms of elliptic functions are also known \cite{sklyanin_algebraic_1996,gould_elliptic_2002,esebbag_elliptic_2015}, although these will not be studied in this thesis.

\section{Generalized Gaudin algebra}

The previous section can be formalized by introducing a \emph{Generalized Gaudin Algebra} (GGA) \cite{ortiz_exactly-solvable_2005}. Symmetry properties of a Hamiltonian are often captured by constructing eigenstates as specific irreducible representations  (irreps) of corresponding Lie algebras, and the GGA builds on this idea in order to provide an algebraic framework for Richardson-Gaudin integrable models. In many ways, it can also be seen as a rewriting of the Algebraic Bethe Ansatz (ABA) \cite{faddeev_how_1996}/Quantum Inverse Scattering Method (QISM) \cite{korepin_quantum_1993} that is better suited to the models under study \cite{amico_integrable_2001,amico_applications_2001,von_delft_algebraic_2002,di_lorenzo_quasi-classical_2002,links_algebraic_2003,dunning_exact_2010,cirilo_antonio_algebraic_2014,cirilo_antonio_algebraic_2015}.

Due to the direct connection with the non-interacting models in the $g \to 0$ limit, where the Lie algebra and its representations can be used to construct conserved charges and exact eigenstates, many properties of the GGA strongly resemble those of the usual $su(2)$ algebra (see Appendix \ref{RG:App:su2}). A GGA is defined by operators $S^x(u), S^y(u), S^z(u)$ satisfying commutation relations
\begin{align}
[S^x(u),S^y(v)] &= i \left( Y(u,v) S^z(u)-X(u,v) S^z(v) \right), \label{RG:GGA:GGA_1}\\
[S^y(u),S^z(v)] &= i \left( Z(u,v) S^x(u)-Y(u,v) S^x(v) \right), \label{RG:GGA:GGA_2}\\
[S^z(u),S^x(v)] &= i \left( X(u,v) S^y(u)-Z(u,v) S^y(v) \right), \label{RG:GGA:GGA_3}\\
[S^{\kappa}(u),S^{\kappa}(v)] &=0, \qquad \kappa =x,y,z,\label{RG:GGA:GGA_4}
\end{align}
with $u,v \in \mathbb{C}$. This is an infinite-dimensional Lie algebra, highly reminiscent of the $su(2)$ algebra, characterized by three functions $X(u,v)$, $Y(u,v)$ and $Z(u,v)$. These can be demanded to be antisymmetric under exchange of the variables $u$ and $v$. Given such an algebra, a continuous family of mutually commuting operators can be defined as 
\begin{align}\label{RG:GGA:Cas}
\mathbb{S}^2(u) &= S^x(u)^2+S^y(u)^2+S^z(u)^2,
\end{align}
where $u$ is also termed the spectral parameter. Integrability now comes into play by noting that it follows from the commutation properties of the GGA that
\begin{equation}
[\mathbb{S}^2(u),\mathbb{S}^2(v)]=0,\  \ \forall u,v \in \mathbb{C}.
\end{equation}
These operators generate a continuous set of commuting operators, leading to a continuous set of conserved charges. Note that although these resemble the Casimir operator of $su(2)$, they do not act as Casimir operators for the GGA since they do not commute with its generators.

However, not any set of antisymmetric functions can be used to generate a GGA. In order to guarantee consistency in the definition of any Lie algebra, the generators have to satisfy a set of Jacobi identities such as
\begin{equation}
[S^x(u),[S^x(v),S^y(w)]]+[S^y(w),[S^x(u),S^x(v)]]+[S^x(v),[S^y(w),S^x(u)]]=0.
\end{equation}
These can all be combined in a set of consistency equations for the $X$-, $Y$- and $Z$-functions as
\begin{equation}\label{RG:GGA:GaudinXYZ}
X(u,v) Y(v,w) + Y(w,u) Z(u,v) + Z(v,w)X(w,u) = 0.
\end{equation}
These provide the continuous equivalent of the Gaudin equations, with any solution to these equations again defining a GGA and a resulting class of Richardson-Gaudin models. Compared to the general theory of integrability, these play the role of the usual Yang-Baxter equations \cite{yang_exact_1967,baxter_exactly_2007}.

\section{XXZ models}
The similarity of both sets of Gaudin equations (\ref{RG:RG:GaudinXXZ}) and (\ref{RG:GGA:GaudinXYZ}) can be made explicit by setting $X_{ij} = Y_{ij}= X(\epsilon_i,\epsilon_j)$ and $Z_{ij} = Z(\epsilon_i,\epsilon_j)$, where the variables $\epsilon_i$ are also sometimes referred to as inhomogeneities. The advantage of this parametrization (leading to the class of XXZ models) is that the structure of the Bethe ansatz wave functions is explicitly known, as will be shown in the following section. From the Gaudin equations, it can be shown that the $X$- and $Z$-functions also satisfy
\begin{equation}\label{GGA:XXZ:Gamma}
X(u,v)^2-Z(u,v)^2 = \Gamma, \qquad \forall u,v \in \mathbb{C},
\end{equation}
with $\Gamma$ a constant. This can be used to write down Gaudin equations purely in terms of $Z$-functions as \cite{ortiz_exactly-solvable_2005}
\begin{equation}\label{GGA:XXZ:GaudinZ}
Z(u,v)Z(v,w)+Z(w,u)Z(u,v) + Z(v,w)Z(w,u) = \Gamma.
\end{equation}
The XXZ-parametrizations also allows for a clearer connection with the $su(2)$ algebra. While it is possible for general XYZ models to define raising and lowering operators as
\begin{equation}
S^+(u) = S^x(u)+ i S^y(u), \qquad S^-(u) = S^x(u) - i S^y(u),
\end{equation}
these generally do not commute among themselves, resulting in major complications. For XXZ models the commutation relations simplify to
\begin{align}
&\left[S^{z}(u),S^{\pm}(v)\right] = \pm \left(X(u,v) S^{\pm}(u) - Z(u,v) S^{\pm}(v)\right), \\
&\left[S^{-}(u),S^{+}(v)\right] = - 2 X(u,v) \left(S^z(u)-S^z(v)\right), \\
&\left[S^{\pm}(u),S^{\pm}(v)\right] = 0.
\end{align}

\section{Bethe ansatz}
From the existence of raising and lowering operators with known commutator structure, it is possible to write down a \emph{Bethe ansatz} wave function as
\begin{equation}
\ket{v_1 \dots v_N} = S^+(v_1) S^+(v_2) \cdots S^+(v_N) \ket{0} =  \prod_{a=1}^N S^+(v_a) \ket{0},
\end{equation}
defined as a product of raising operators depending on (possibly complex) parameters $\{v_1 \dots v_N\}$, also known as \emph{rapidities} or \emph{Bethe roots}, acting on a vacuum state $\ket{0}$. This vacuum is assumed to satisfy the usual properties of a lowest-weight representation
\begin{equation}
\mathbb{S}^2(u)\ket{0} = F_2(u)\ket{0}, \qquad S^z(u)\ket{0} = F_z(u)\ket{0}, \qquad S^-(u) \ket{0}=0.
\end{equation}
This structure is a common property of Bethe ansatz-solvable models, where the two crucial elements are (i) the existence of generalized raising/lowering operators with known commutator structure and (ii) the existence of a vacuum or reference state annihilated by the generalized lowering or raising operators. This first property is generally guaranteed by construction, either through the GGA or the ABA, whereas the existence of a vacuum state will be model-dependent. 

Due to the commutativity of $\mathbb{S}^2(u)$ at different values of the spectral parameter $u$, the conserved charges have a common set of eigenstates. Simultaneous diagonalization then corresponds to obtaining the eigenstates of a single $\mathbb{S}^2(u)$. Because of the clear product structure of the Bethe ansatz, the action of $\mathbb{S}^2(u)$ on such a Bethe state can be calculated as
\begin{align}\label{RG:BA:S2u_comm}
&\mathbb{S}^2(u) \left(\prod_{a=1}^N S^+(v_a) \right) \ket{0} = \sum_{a=1}^N \sum_{b=a+1}^N \left(\prod_{c \neq a,b}^N S^+(v_c) \right) \left[\left[\mathbb{S}^2(u),S^+(v_a)\right],S^+(v_b)\right]\ket{0} \nonumber \\
& \qquad \qquad  + \sum_{a=1}^N  \left(\prod_{b \neq a}^N S^+(v_b) \right) \left[\mathbb{S}^2(u),S^+(v_a)\right] \ket{0} +  \left(\prod_{a=1}^N S^+(v_a) \right) \mathbb{S}^2(u) \ket{0}.
\end{align}
It is easily shown that $[[[\mathbb{S}^2(u),S^+(v_a)],S^+(v_b)],S^+(v_c)]=0$, and as such no higher-order commutators will arise. The necessary single and double commutators can be found from the GGA, and incorporating these in the total expression for the action of $\mathbb{S}^2(u)$ on a Bethe state results in
\begin{align}
&\left[\mathbb{S}^2(u) -F_2(u) \right]\ket{v_1 \dots v_N} = -\sum_{a=1}^N \left[\Gamma+Z(u,v_a)\left(2F_z(u)-\sum_{b \neq a}^N Z(u,v_b)\right)\right]\ket{v_1 \dots v_N} \nonumber \\
&\qquad \qquad \qquad +2\sum_{a=1}^N X(u,v_a) \left[F_z(v_a)+\sum_{b\neq a}^N Z(v_b,v_a)\right] \ket{v_1 \dots v_a \rightarrow u \dots v_N}.
\end{align}
Two contributions can be clearly distinguished -- a diagonal part proportional to the original state $\ket{v_1 \dots v_N}$, and a set of non-diagonal states $\ket{v_1 \dots v_a \rightarrow u \dots v_N}$, where a single rapidity $v_a$ has been replaced by the spectral parameter $u$. If these additional terms would vanish, this would reduce to an exact eigenvalue equation. Luckily, some freedom has been left in the definition of the Bethe state. So far, the variables $\{v_1 \dots v_N\}$ have been chosen arbitrarily, but if they are chosen to satisfy the so-called \emph{Bethe} or \emph{Richardson-Gaudin equations}
\begin{equation}
F_z(v_a)+\sum_{b\neq a}^N Z(v_b,v_a) = 0, \qquad \forall a=1 \dots N,
\end{equation}
the unwanted off-diagonal contributions vanish and the resulting Bethe state is an exact eigenstate. Note how, as could be expected, these equations are independent of the spectral parameter $u$. Due to the commutativity condition $[\mathbb{S}^2(u),\mathbb{S}^2(v)]=0$ the Bethe states are eigenstates at each values of the spectral parameter, and the Bethe equations should hence be independent of $u$. In analogy with the non-interacting models, the Bethe equations can also be seen a set of self-consistency equations determining the normal modes.

In this way, the Bethe ansatz presents an alternative to direct diagonalization of the Hamiltonian matrix with an exceptional advantage. Instead of scaling exponentially with system size $L$, this solution method scales \emph{linearly} with the number of excitations $N$. This linear scaling allows calculation of the eigenvalues and eigenstates for systems where the classical approach of diagonalizing the Hamiltonian quickly proves to be impossible.

\section{Spin models from a GGA}
So far, the construction remained quite general and no explicit mention of spin models has been made. Richardson-Gaudin models can then be obtained by constructing a specific realization of the generators of a GGA in terms of interacting spins. A spin-$s$ particle can be realized through the irrep spanned by $\ket{s, m_s}$, with $m_s = -s, -s+1 , \dots, s$. Since only commutator properties have been used so far, the integrability is independent of the chosen spin representation.
In order to make the connection with the previously-presented conserved charges for interacting spin models (\ref{RG:RG:com}), a specific representation in terms of these $\oplus_{i=1}^L su(2)$ generators can be introduced as
\begin{equation}
S^{\pm}(u) = \sum_{i=1}^L X(u,\epsilon_i) S_i^{\pm}, \qquad S^z(u) = -\frac{1}{g}-\sum_{i=1}^L Z(u, \epsilon_i) S_i^z.
\end{equation}
Here $g$ and $\epsilon_1, \dots, \epsilon_L$ are arbitrary real parameters, and these operators satisfy the commutation relations of the GGA (\ref{RG:GGA:GGA_1}-\ref{RG:GGA:GGA_4}) by construction. Explicitly writing out 
\begin{equation}
\mathbb{S}^2(u) = \frac{1}{2}\left(S^+(u)S^-(u)+S^-(u)S^+(u)\right)+S^z(u)^2
\end{equation}
then returns the conserved operators previously introduced as 
\begin{equation}
\mathbb{S}^2(u) = \frac{2}{g}\sum_{i=1}^L Z(u,\epsilon_i)Q_i -\Gamma \left(\sum_{i=1}^LS_i^z\right)^2 + \textrm{Cst.}
\end{equation}
The chosen parametrization for $X$ and $Z$ now determines the type of interactions present in the conserved charges
\begin{equation}\label{RG:spin:Qi}
Q_i = S_i^z+g \sum_{j \neq i}^L \left[ \frac{1}{2}X(\epsilon_i,\epsilon_j)(S_i^+S_j^-+S_i^-S_j^+)+Z(\epsilon_i,\epsilon_j) S_i^z S_j^z \right].
\end{equation}
For the interacting spin RG models, the necessary vacuum state follows from the choice of irreducible representation of $su(2)$, and it can easily be checked that the lowest weight state $\ket{0}=\otimes_{i=1}^L \ket{s_i,-s_i}$ satisfies the properties of a vacuum state. A solution to the Gaudin equations (\ref{RG:GGA:GaudinXYZ}) then allows for the definition of Bethe ansatz eigenstates 
\begin{equation}\label{RG:spin:BA}
\ket{v_1 \dots v_N} = \prod_{a=1}^N \left(\sum_{j=1}^L X(\epsilon_j,v_a) S_j^+ \right)\ket{0}
\end{equation}
leading to the eigenvalue equation (see Appendix \ref{RG:App:S2u})
\begin{equation}\label{RG:spin:eigen}
Q_i \ket{v_1 \dots v_N} = -s_i \left[1+g \sum_{a=1}^N Z(\epsilon_i,v_a)-g \sum_{j \neq i}^L Z(\epsilon_i,\epsilon_j)s_j \right]\ket{v_1 \dots v_N},
\end{equation}
provided the rapidities satisfy the Bethe equations
\begin{equation}\label{RG:spin:BAE}
\frac{1}{g}+ \sum_{i=1}^L Z(\epsilon_i, v_a) s_i-\sum_{b \neq a}^N Z(v_b,v_a) = 0, \qquad a=1 \dots N.
\end{equation}

\section{Physical realizations}

The main challenge is now to obtain an integrable Hamiltonian with a clear physical interpretation. This is often done on an ad hoc basis, where the freedom in the GGA generally results in models with quite a lot of freedom (for integrable models). In this section, three exemplary Richardson-Gaudin models will be discussed: the central spin model (Section \ref{RG:subsec:cs}), the reduced BCS Hamiltonian (Section \ref{RG:subsec:redBCS}), and the $p_x+ip_y$-wave pairing Hamiltonian (Section \ref{RG:subsec:pip}). These all correspond to different realizations of both the GGA and the underlying $su(2)$ algebra, and can be considered as models with an increasing level of complexity.

\subsection{The central spin model}
\label{RG:subsec:cs}
\begin{figure}
\begin{center}
\includegraphics[width=0.8\textwidth]{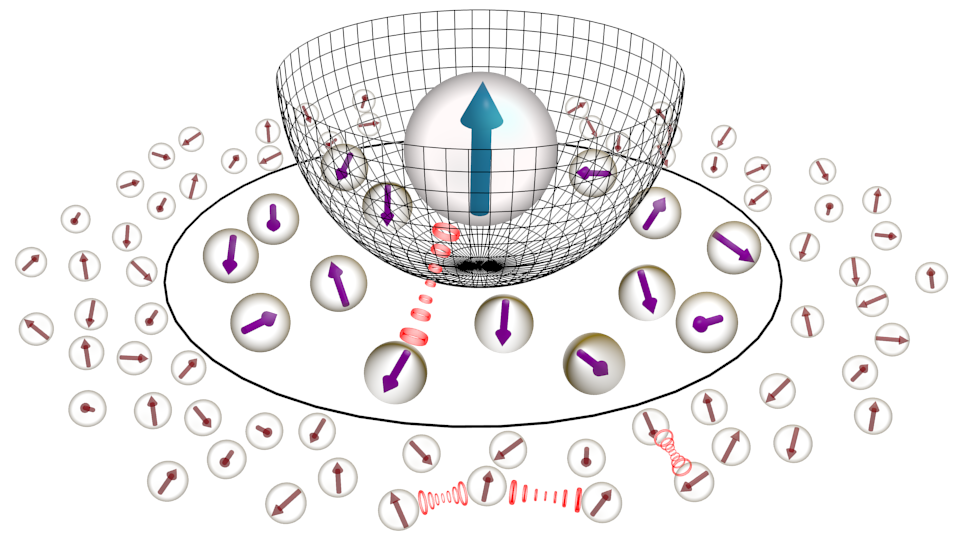}
\caption{Graphical representation of the central spin interacting with a bath of surrounding spins, reproduced from Ref. \cite{van_den_berg_competing_2014}. \label{fig:RG:centralspin}}
\end{center}
\end{figure}

The central spin model is the original realization of the Gaudin magnet \cite{gaudin_diagonalisation_1976,gaudin_bethe_2014}, describing the interaction of a single spin on which a magnetic field $B_z$ is applied along the $z$-axis with a bath of surrounding spins. The single central spin then experiences both the external magnetic field and the collective field created by the bath spins, also known as the Overhauser field. This can be modelled by a central spin Hamiltonian
\begin{equation}
H_{cs} = B_z S_{c}^z + \sum_{j \neq c}^L A_j \vec{S}_{c} \cdot \vec{S}_j,
\end{equation}
with the interaction strengths $A_j$ dependent on the specific model and the central spin denoted as $\vec{S}_c$. These are commonly taken to be $A_j = \exp\left[-(j-1)/L\right]$, corresponding to a quantum dot in a 2D Gaussian envelope \cite{coish_hyperfine_2004}. Such a model is important in the study of quantum dots \cite{coish_hyperfine_2004,van_den_berg_competing_2014}, solid-state nuclear magnetic resonance (NMR) \cite{bendall_broadband_1995,hediger_adiabatic_1995,hwang_broadband_1997}, and models the nitrogen-vacancy defect in diamond, which has been proposed as a promising qubit system due to its long decoherence time \cite{balasubramanian_ultralong_2009}. In most realizations, the additional bath spins represent the hyperfine interaction of the spin with its environment \cite{schliemann_electron_2003}, as illustrated in Figure \ref{fig:RG:centralspin}. Successful application of this model as a qubit then requires a thorough understanding of the dynamics and decoherence properties, leading to a wide range of studies \cite{yuzbashyan_solution_2005,bortz_exact_2007,bortz_spin_2007,bortz_dynamics_2010,erbe_different_2010,schliemann_spins_2010,barnes_nonperturbative_2012,stanek_dynamics_2013,faribault_integrability-based_2013}. 

The underlying GGA is the rational or XXX Gaudin algebra, where the solutions to the Gaudin equations are explicitly given by rational functions (\ref{RG:RG:rat}) as
\begin{equation}
X(u,v) = \frac{1}{u-v}, \qquad Z(u,v) = \frac{1}{u-v},
\end{equation}
where $X(u,v) = Z(u,v)$, which leads to conserved charges as\footnote{Note that the interaction term is closely related to the exchange operator $P$ for spin-$1/2$ models, where $\vec{S}_i \cdot \vec{S}_j = P/2-1/4$, tying back to the introductory chapter.}
\begin{equation}
Q_i = S_i^z + g \sum_{j \neq i}^L \frac{1}{\epsilon_i-\epsilon_j}\left[\frac{1}{2}\left(S_i^+S_j^-+S_i^-S_j^+\right)+S_i^z S_j^z\right] \equiv S_i^z + g \sum_{j \neq i}^L \frac{\vec{S}_i \cdot \vec{S}_j}{\epsilon_i-\epsilon_j}.
\end{equation}
The most straightforward way of obtaining an integrable Hamiltonian from these conserved charges is by simply selecting a single one and promoting it to be a Hamiltonian. The central spin Hamiltonian is then proportional to a single conserved charge $Q_c$ with $g=B_z^{-1}$, $\epsilon_c=0$ and  $A_j = -\epsilon_j^{-1}$. The freedom of choice in the variables $\epsilon_j$ then corresponds to the freedom to choose the inhomogeneous interaction strengths $A_j$. 

The eigenstates are given by
\begin{equation}
\ket{v_1 \dots v_N} =\prod_{a=1}^N \left(\sum_{j=1}^L \frac{S_j^+}{\epsilon_j-v_a}\right)\ket{0} \propto \prod_{a=1}^N  \left(S_c^++\sum_{j \neq c}^L \frac{A_j v_a}{1+A_j v_a} S_j^+\right)\ket{0},
\end{equation}
where the vacuum state is given by the lowest-weight state $\ket{0} = \otimes_{i=1}^L \ket{s_i,-s_i}$. This is an eigenstate provided the rapidities satisfy the Bethe equations
\begin{equation}\label{RG:cs:BAE}
B_z^{-1}+\sum_{j=1}^L \frac{s_j}{\epsilon_j-v_a}-\sum_{b \neq a}^N \frac{1}{v_b-v_a}=0, \qquad a=1 \dots N,
\end{equation}
with an eigenvalue of 
\begin{equation}
Q_i \ket{v_1 \dots v_N} = -s_i\left[1 + g \sum_{a=1}^N \frac{1}{\epsilon_i-v_a} - g \sum_{j \neq i}^L \frac{s_j}{\epsilon_i-\epsilon_j} \right] \ket{v_1 \dots v_N}.
\end{equation}

\subsection{The reduced BCS model}
\label{RG:subsec:redBCS}
\begin{figure}
\begin{center}
\begin{tikzpicture}[scale=0.85]
  \draw[level] (0,0) -- (1.5,0);
  \draw[level] (3,0) -- (4.5,0);
  \draw[level] (6,0) -- (7.5,0);
  \draw[level] (9,0) -- (10.5,0);
  \draw[level] (12,0) -- (13.5,0);
  \draw[level] (1.5,1) -- (3,1);
  \draw[level] (4.5,1) -- (6,1);
  \draw[level] (7.5,1) -- (9,1);
  \draw[level] (10.5,1) -- (12,1);
  \draw[level] (3,2) -- (4.5,2);
  \draw[level] (6,2) -- (7.5,2);
  \draw[level] (9,2) -- (10.5,2);
  \draw[level] (4.5,3) -- (6,3);
  \draw[level] (7.5,3) -- (9,3);
  \draw[level] (6,4) -- (7.5,4);
  \draw[axes] (-0.5,0) --  (-0.5,4);
  \node[rotate=90] at (-1.5,2) {Seniority $\nu$};
  \draw[axes] (14.0,0) --  (14.0,4);
  \node[rotate=90] at (15.5,2) {Quasi-spin $d$};
  \foreach \x in {0,...,4}
     		\draw[axes] (-0.5,\x) -- (-0.4,\x)
			node[left] {\x};
   \draw[axes] (13.9,0) -- (14.0,0) node[right] {2};
   \draw[axes] (13.9,1) -- (14.0,1) node[right] {3/2};
   \draw[axes] (13.9,2) -- (14.0,2) node[right] {1};
   \draw[axes] (13.9,3) -- (14.0,3) node[right] {1/2};
   \draw[axes] (13.9,4) -- (14.0,4) node[right] {0};
   
   \node at (0.2,-0.75){$n=0$};
   \node at (0.,-1.25){$\mu=-2$};   
   \node at (2.25,-0.75){$1$};
   \node at (2.25,-1.25){$-3/2$};   
   \node at (3.75,-0.75){$2$};
   \node at (3.75,-1.25){$-1$};     
   \node at (5.25,-0.75){$3$};
   \node at (5.25,-1.25){$-1/2$};   
   \node at (6.75,-0.75){$4$};
   \node at (6.75,-1.25){$0$};   
   \node at (8.25,-0.75){$5$};
   \node at (8.25,-1.25){$1/2$};
   \node at (9.75,-0.75){$6$};
   \node at (9.75,-1.25){$1$};
   \node at (11.25,-0.75){$7$};
   \node at (11.25,-1.25){$3/2$};   
   \node at (12.75,-0.75){$8$};
   \node at (12.75,-1.25){$2$};
\end{tikzpicture}
\caption{Schematic overview of all states in the seniority-coupling scheme for $j=7/2$ and $\Omega_j=2j+1=8$. The connection between the seniority $\nu$, the quasi-spin $d$, the total number of fermions $n$ and the irrep label $\mu$ is shown. Decreasing the seniority results in a larger possible occupation of fermion pairs and a larger quasi-spin. Based on a similar scheme in Ref. \cite{rowe_fundamentals_2010}.\label{fig:RG:seniority}}
\end{center}
\end{figure}
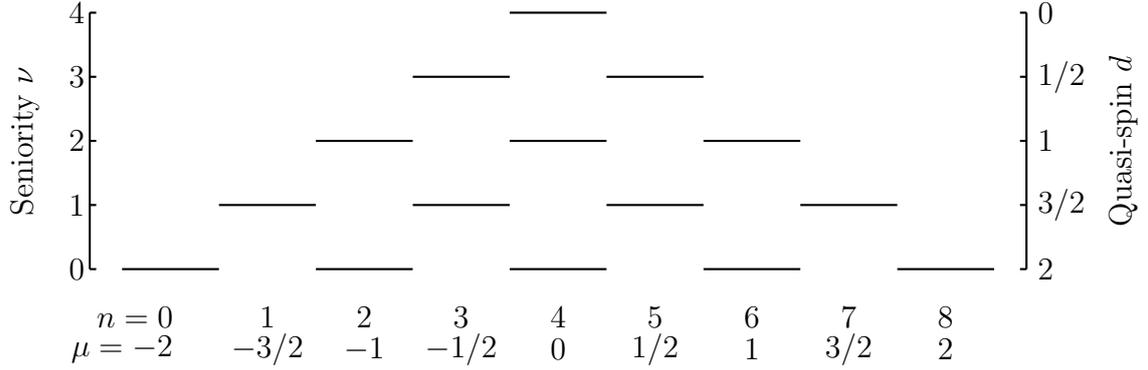

Richardson's exact solution was obtained in the context of nuclear superfluidity, and one of the major applications of RG integrability remains the treatment of strong pairing correlations (for an excellent review, see \cite{dukelsky_colloquium_2004}). As discovered by Bardeen, Cooper and Schrieffer, the mechanism underlying superfluidity/superconductivity is the formation of electron (Cooper) pairs \cite{bardeen_theory_1957}. The reduced BCS Hamiltonian describing fermion pair scattering can be written down as
\begin{equation}
{H}_{BCS} = \sum_{j m} \epsilon_j a^{\dagger}_{jm}a_{jm} + \frac{g}{4} \sum_{j j' m m'} (-1)^{j+j'+m+m'} a^{\dagger}_{jm}a^{\dagger}_{j -m} a_{j' m'} a_{j' -m'},
\end{equation}
in which $g$ is the level-independent pairing interaction. Suppose we have $n$ fermions moving in a set of $L$ single-particle states $i$ with angular momentum $j_i$ and $\epsilon_i$ the $\Omega_i=2j_i+1$ fold degenerate single-particle energies. The connection between spin models and fermion pairing models can be realized through the \emph{quasi-spin} realization of the $su(2)$ algebra \cite{heyde_nuclear_1994}. Pairing happens at the level of the total angular momentum, which leads to the definition of pair creation and annihilation operators
\begin{equation}
S_i^+=\sum_{m_i>0} a^\dag_{m_i}a^\dag_{\bar{m}_i},\quad {S}^-_i=\sum_{m_i>0} {a}_{\bar{m}_i}{a}_{m_i},
\end{equation}
where we have only indicated the index over which the summation runs.  The bar notation $\bar{m}_i$ denotes the time-reversed partner of $m_i$, with a phase correction $a^\dag_{j_i m_i}=(-)^{j_i-m_i}a^\dag_{j_i -m_i}$ in order to respect good angular momentum tensorial properties.  With this notation, the particle-number operators can be written as
\begin{equation}
n_i=\sum_{m_i>0}(a_{m_i}^\dag a_{m_i}+{a}_{\bar{m}_i}^\dag {a}_{\bar{m}_i}) = 2 \left(S_i^z+\frac{\Omega_i}{4}\right),
\end{equation}
again only summing over the relevant index. These fermion operators $\{S_i^+,S_i^-, S_i^z\}$ again constitute an $su(2)$ algebra, known as the \emph{quasi-spin} algebra \cite{heyde_nuclear_1994}. It is convenient to introduce the seniority quantum number $\nu_i$, which counts the number of unpaired fermions, and the related quasi-spin  pairing quantum number $d_i=\frac{1}{4}\Omega_i-\frac{1}{2}\nu_i$, which denotes (half of) the maximum allowed number of pairs in a level \cite{talmi_simple_1993}. The connection between these different quantum numbers is illustrated in Figure \ref{fig:RG:seniority}. When discussing spin models, we will use the notation $\ket{s,m_s}$ to denote the representations, whereas we will use the notation $\ket{d_i, \mu_i}$ in the context of fermion pairing models.

The reduced BCS Hamiltonian can then be rewritten as
\begin{equation}
H_{BCS} = \sum_{i=1}^L 2\epsilon_i \left(S_i^z+\frac{\Omega_i}{4}\right) + g \sum_{i,j=1}^L S_i^+ S_j^-.
\end{equation}
As discovered by Cambiaggio \emph{et al.}\ \cite{cambiaggio_integrability_1997}, this Hamiltonian supports a complete set of conserved charges
\begin{equation}
Q_i = S_i^z + g \sum_{j \neq i}^L \frac{1}{\epsilon_i-\epsilon_j}\left[\frac{1}{2} \left(S_i^+ S_j^-+S_i^- S_j^+\right)+ S_i^z S_j^z \right],
\end{equation}
which are again the conserved charges of the rational model. In fact this can be understood by noting that the reduced BCS Hamiltonian can be rewritten as\footnote{The constant contains Casimir operators and $\sum_{i=1}^L S_i^z$, both of which are symmetries of the system and can be replaced by their expectation values.}
\begin{equation}
H_{BCS} = \sum_{i=1}^L 2\epsilon_i Q_i + \textrm{Cst.}
\end{equation}
This then allows the eigenstates of this model to be obtained as Bethe ansatz states
\begin{equation}\label{RG:phys:BA_BCS}
\ket{v_1 \dots v_N} = \prod_{a=1}^N\left(\sum_{i=1}^L\frac{S_i^+}{\epsilon_i-v_a}\right)\ket{0} =  \prod_{a=1}^N\left(\sum_{\substack{i=1 \\ m_i>0}}^L\frac{a^\dag_{m_i}a^\dag_{\bar{m}_i}}{\epsilon_i-v_a}\right)\ket{0}.
\end{equation} 
The state $\ket{0}$ is now the pair vacuum state $\otimes_{i=1}^L\ket{d_i,-d_i}$, meaning that it contains no paired particles\footnote{The `blocking effect' allows for the decoupling of the unpaired particles from the paired fermions.}. The rapidities again have to form a solution to the set of Bethe equations
\begin{equation}\label{RG:phys:BAE_BCS}
\frac{1}{g}+\sum_{j=1}^L \frac{d_j}{\epsilon_j-v_a}-\sum_{b \neq a}^N \frac{1}{v_b-v_a}=0, \qquad a=1 \dots N.
\end{equation}
As soon as the RG equations have been solved, the energy of the associated eigenstate is readily given by
\begin{equation}
E=2 \sum_{a=1}^N v_a + \sum_{i=1}^L\epsilon_i \nu_i,
\end{equation}
giving an interpretation of half the pair energy to a rapidity $v_a$. It was for this model that Richardson obtained the exact eigenstates, by generalizing the wave function structure for a single pair to a product-like wave function and using this as a variational ansatz \cite{richardson_restricted_1963,richardson_exact_1964}. The availability of both an exact solution and the usual BCS mean-field (approximate) wave function then presents two ways of investigating superconductivity. This phenomenon was originally understood through the BCS mean-field approach, explicitly breaking particle-number symmetry. The latter could be restored by projecting the mean-field wave function on a sector with definite number of particles, giving rise to the projected BCS wave function. The analogy with Richardson's solution can then be made by noting that both approaches share a same product structure \cite{dussel_cooper_2007,sandulescu_accuracy_2008}. Some more intuition in the solutions to the Bethe equations and the connection to the BCS wave function can be gained by mapping these to an electrostatic problem (the so-called electrostatic analogy \cite{dukelsky_electrostatic_2002,roman_large-n_2002}). In this way, it was also shown how the Bethe equations can be rephrased as the BCS mean-field equations in the thermodynamic limit. Because of the success of the BCS mean-field approach, Richardson's solution was largely overlooked until reappearing in the study of ultrasmall superconducting grains \cite{braun_fixed-n_1998}. In these grains, the number fluctuations inherent in the BCS mean-field wave function can no longer be overlooked, and it proved to be necessary to obtain a more accurate method. Richardson's solution then succeeded in describing the crossover from the superconducting regime to the pairing fluctuation regime \cite{sierra_exact_2000,von_delft_spectroscopy_2001,crouzeix_energy_2011,combescot_bcs_2013,combescot_granules_2016}.


\subsection{The $p_x+ip_y$-wave pairing Hamiltonian} 
\label{RG:subsec:pip}
Richardson's original solution has since been generalized to more general pairing interactions in a variety of ways \cite{ibanez_exactly_2009,skrypnyk_spin_2009,skrypnyk_non-skew-symmetric_2009,dunning_exact_2010,rombouts_quantum_2010,van_raemdonck_exact_2014,lukyanenko_boundaries_2014}. Whereas the interaction in the reduced BCS model was assumed to be isotropic (leading to it being termed an $s$-wave pairing Hamiltonian), more involved interactions are possible ($p$-wave, $d$-wave,...). The physical importance of this is that, by allowing more general pairing interactions, it is possible to obtain phases with a non-trivial topology. From the study of such pairing interactions, it was e.g. shown how topological superconductivity arose from a chiral $p_x+ip_y$-wave interaction by breaking time-reversal symmetry \cite{read_paired_2000,ryu_topological_2010}. Such pairing interactions are believed to occur naturally \cite{mackenzie_superconductivity_2003,xia_high_2006,maeno_evaluation_2011,kallin_chiral_2016} and have also been argued to be technologically achievable \cite{zhang_px+ipy_2008}. A major interest in these systems originates from their topological properties and the subsequent potential for quantum computation \cite{tewari_quantum_2007,sau_generic_2010,sarma_majorana_2015}. 

\begin{figure}
\begin{center}
\includegraphics{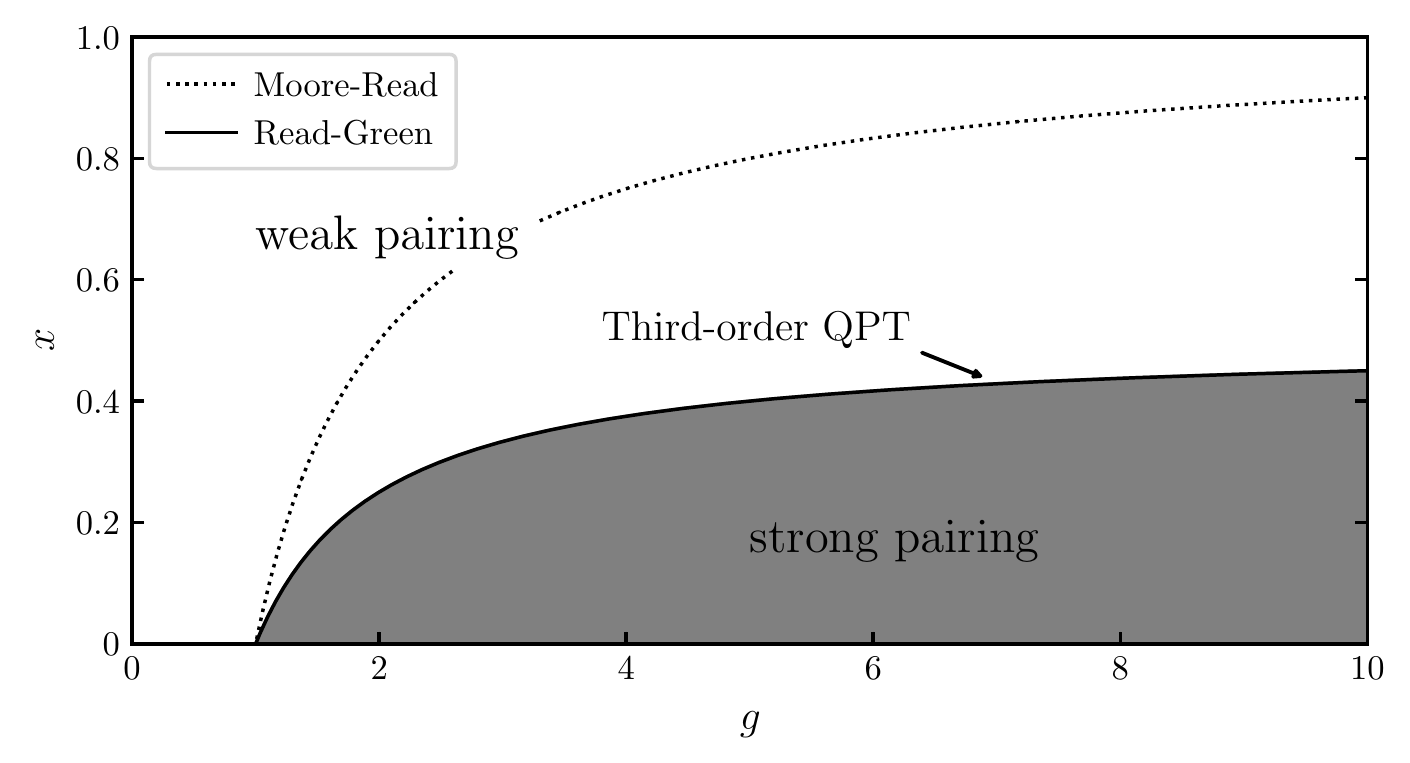}
\caption{Phase diagram of the $p_x+ip_y$-wave pairing Hamiltonian in terms of the fermion density $x=N/L$ and the interaction strength $g=GL$ in the thermodynamic limit. A quantum phase transition (QPT) occurs at the Read-Green line between a topologically non-trivial (weak-pairing) and a topologically trivial (strong-pairing) phase. At the Moore-Read line, all rapidities condense to 0 and the total energy vanishes. Based on a similar figure in Ref. \cite{rombouts_quantum_2010}. \label{fig:RG:phasepxipy}}
\end{center}
\end{figure}

Staying in the context of fermion pair interactions, the $p_x + i p_y$-wave pairing Hamiltonian\footnote{Also called chiral $p$-wave pairing Hamiltonian or $p+ip$-wave pairing Hamiltonian.} describes a chiral interaction between two-dimensional fermions with momentum $\vec{k} = (k_x, k_y)$. The main point of interest in this model is its topological phase transition between two superconducting states with different topologies (either trivial or non-trivial), as illustrated in Figure \ref{fig:RG:phasepxipy}. From the exact solution, it was shown how the topological phase transition is reflected in the Read-Green points for finite systems \cite{rombouts_quantum_2010}. At these points, it is possible to reach excited states through a fixed number of zero-energy pair excitations. When a single zero-energy excitation is allowed, this corresponds to a vanishing chemical potential and the topological phase transition is recovered in the thermodynamic limit of the Richardson-Gaudin solution \cite{ibanez_exactly_2009,rombouts_quantum_2010}. This exact solution has also led to a criterion for the characterization of topological superconductivity in finite systems \cite{ortiz_many-body_2014,ortiz_what_2016}.

Taking $a_{\mathbf{k}}$ and $a_{\mathbf{k}}^{\dagger}$ to denote annihilation/creation operators for two-dimensional spinless fermions of mass $m$ and momentum $\mathbf{k}=(k_x,k_y)$, the Hamiltonian reads
\begin{align}\label{RG:pip:ham}
H_{p+ip} = \sum_{\mathbf{k}}\frac{|\mathbf{k}|^2}{2m}a^{\dagger}_{\mathbf{k}}a_{\mathbf{k}}+\frac{G}{4m}\sum_{\mathbf{k},\mathbf{k}'}(k_x+ik_y)(k_x'-ik_y')a^{\dagger}_{\mathbf{k}}a^{\dagger}_{-\mathbf{k}}a_{-\mathbf{k}'}a_{\mathbf{k}'},
\end{align}
in which $G$ is a dimensionless interaction constant. The fermion pair interactions can again be captured in a quasi-spin algebra, where the coupling now occurs between fermions with opposite momenta $\pm \mathbf{k}$. Introducing a phase $\exp(i \phi_{\mathbf{k}}) = (k_x+ik_y)/|\mathbf{k}|$, an $su(2)$ algebra is generated by
\begin{align}
S^{+}_{\mathbf{k}}=e^{i \phi_{\mathbf{k}}}a^{\dagger}_{\mathbf{k}}a^{\dagger}_{-\mathbf{k}}, \qquad S^{-}_{\mathbf{k}}=e^{-i \phi_{\mathbf{k}}}a_{-\mathbf{k}}a_{\mathbf{k}}, \qquad S_{\mathbf{k}}^z = \frac{1}{2}(a^{\dagger}_{\mathbf{k}}a_{\mathbf{k}} + a^{\dagger}_{-\mathbf{k}}a_{-\mathbf{k}}-1),
\end{align} 
restricted to spin-$1/2$ representations. Taking $\epsilon_{\mathbf{k}}^2 = |\mathbf{k}|^2/m$ and labeling the allowed momenta $\pm \mathbf{k}$ with integers $i=1 \dots L$, the $p_x+ip_y$-wave pairing Hamiltonian can be recast as
\begin{align}
H=\sum_{i=1}^L \epsilon_{i} \left(S_i^z+\frac{1}{1}\right)+G \sum_{i, j =1}^L \sqrt{\epsilon_{i} \epsilon_j} S^{+}_{i}S_j^-.
\end{align}
As shown by Iba\~nez \emph{et al.} \cite{ibanez_exactly_2009} and Rombouts \emph{et al.} \cite{rombouts_quantum_2010}, the conserved charges of this model are given by
\begin{align}
Q_i = S_i^z + g \sum_{j \neq i}^L\left[\frac{\sqrt{\epsilon_i \epsilon_j}}{\epsilon_i-\epsilon_j}\left(S_i^+ S_j^-+S_i^- S_j^+\right)+ \frac{\epsilon_i + \epsilon_j}{\epsilon_i-\epsilon_j}S_i^z S_j^z \right],
\end{align}
corresponding to an XXZ Gaudin algebra generated by
\begin{equation}
X(u,v) = 2\frac{\sqrt{uv}}{u-v}, \qquad Z(u,v) = \frac{u+v}{u-v}.
\end{equation}
Constructing the Hamiltonian from these conserved charges is slightly more involved, since the interaction constants $g$ and $G$ differ. Taking the usual linear combination of charges results in
\begin{align}
\sum_{i=1}^L \epsilon_i Q_i = \sum_{i=1}^L \epsilon_i S_i^z \left(1-g+g\sum_{j=1}^L S_j^z\right) + g \sum_{i,j=1}^L \sqrt{\epsilon_i \epsilon_j} S_i^+ S_j^-  + \textrm{Cst.}
\end{align}
Since $\sum_{j=1}^L Q_j = \sum_{j=1}^L S_j^z = S^z$ is a symmetry of the system, which can be interpreted as the total number of (spin or fermion pair) excitations, any calculation can be restricted to a specific symmetry sector in which $S^z$ takes a definite value. This has already been incorporated in the Bethe ansatz wave function, which contains a fixed number of excitations $S^z \ket{v_1 \dots v_N} = \left[N-\sum_{j=1}^L s_j \right]\ket{v_1 \dots v_N}$. Without loss of generality this operator can be replaced by its expectation value, leading to
\begin{equation}
H_{p+ip} = \left[1-g+g(N-\sum_{j=1}^L s_j) \right]^{-1} \left(\sum_{i=1}^L \epsilon_i Q_i\right) + \textrm{Cst.}
\end{equation}
The relation between both coupling constants is then given by 
\begin{equation}
G^{-1} = g^{-1} - 1 +N - \sum_{j=1}^L s_j,
\end{equation}
where the implicit dependence of the interaction strength on the particle number will turn out to play an important role, as discussed in later chapters. The eigenstates are then given by
\begin{equation}\label{RG:pip:BA}
\ket{v_1 \dots v_N} = \prod_{a=1}^N\left(\sum_{i=1}^L 2 \frac{\sqrt{\epsilon_i v_a}}{\epsilon_i-v_a}S_i^+\right)\ket{0} \propto \prod_{a=1}^N\left(\sum_{\mathbf{k}} \frac{k_x+ik_y}{|\mathbf{k}|^2/m-v_a}a^{\dagger}_{\mathbf{k}}a^{\dagger}_{-\mathbf{k}}\right)\ket{0},
\end{equation} 
where the vacuum is again the pair vacuum $\ket{0}=\otimes_{i=1}^L \ket{d_i,-d_i}$. The Bethe equations can then either be written as
\begin{equation}
\frac{1}{g}+ \sum_{i=1}^L d_i \frac{\epsilon_i+v_a}{\epsilon_i-v_a} -  \sum_{b \neq a}^N \frac{v_b+v_a}{v_b-v_a} = 0, \qquad a =1 \dots N,
\end{equation}
or in terms of the coupling constant $G$ as 
\begin{equation}
\frac{1+G^{-1}}{2 v_a}+ \sum_{i=1}^L \frac{d_i }{\epsilon_i-v_a} -  \sum_{b \neq a}^N \frac{1}{v_b-v_a} = 0, \qquad a =1 \dots N.
\end{equation}
This model will be discussed in more detail in later chapters, with special attention paid to its various symmetries and their influence on the phase diagram.

\begin{subappendices}
\makeatletter
\@openrightfalse
\makeatother
\chapter*{Appendices}
\section{The $su(2)$ algebra - Lie algebraic structure}
\label{RG:App:su2}
An $su(2)$ algebra is generated by operators $\{S^x,S^y,S^z\}$ with $[S^a,S^b]= i\epsilon_{abc}S^c$, which is commonly recast by introducing the operators
\begin{equation}
S^+ = S^x+i S^y, \qquad S^- = S^x - i S^y,
\end{equation}
with $\{S^+,S^-,S^z\}$ now satisfying
\begin{equation}
[S^z,S^{+}]=S^{+}, \qquad [S^z,S^-]=-S^-, \qquad [S^{+},S^{-}]=2 S^z.
\end{equation}
The quadratic Casimir operator commuting with the generators is given by
\begin{equation}
\mathcal{C}[su(2)] = \frac{1}{2}(S^+S^-+S^-S^+)+S^z S^z.
\end{equation}
The operator $S^z$ acts as a Cartan operator, and representations of the algebra can be constructed using common eigenstates of $\mathcal{C}[su(2)]$ and $S^z$, with the eigenvalues of $S^z$ also referred to as \emph{weights}. A finite-dimensional irreducible representation of dimension $(2s+1)$ is spanned by the states $\ket{s,m_s}$, with $m_s = -s, -s+1, \dots,s$ and $s$ taking (half-)integer values. These representations are characterized by a highest-weight state $\ket{s,s}$ and a lowest-weight state $\ket{s,-s}$. The action of the $su(2)$ generators on these states is given by
\begin{align}
\mathcal{C}[su(2)] \ket{s, m_s} &= s(s+1) \ket{s, m_s}, \qquad S^z \ket{s, m_s} = m_s \ket{s, m_s}, \nonumber \\
S^+ \ket{s, m_s} &= \sqrt{(s-m_s)(s+m_s+1)}\ket{s, m_s+1}, \nonumber \\
S^- \ket{s, m_s} &= \sqrt{(s+m_s)(s-m_s+1)}\ket{s, m_s-1}.
\end{align}

\section{Obtaining the conserved charges from the GGA}
\label{RG:App:S2u}
In this Appendix, the action of $\mathbb{S}^2(u)$ on a Bethe state will be discussed in the case of interacting spin models. Performing the full expansion of $\mathbb{S}^2(u)$ results in
\begin{equation}\label{RG:AppGGA:Expansion}
\mathbb{S}^2(u) = \frac{2}{g}\sum_{i=1}^L Z(u,\epsilon_i)Q_i -\Gamma \left(\sum_{i=1}^LS_i^z\right)^2 + \sum_{i=1}^L X(u,\epsilon_i)^2 \mathcal{C} \left[su(2)_i\right] +\frac{1}{g^2}.
\end{equation}
Starting from its action on a Bethe state
\begin{align}
&\left[\mathbb{S}^2(u) -F_2(u) \right]\ket{v_1 \dots v_N} = -\sum_{a=1}^N \left[\Gamma+Z(u,v_a)\left(2F_z(u)-\sum_{b \neq a}^N Z(u,v_b)\right)\right]\ket{v_1 \dots v_N} \nonumber \\
&\qquad \qquad \qquad +2\sum_{a=1}^N X(u,v_a) \left[F_z(v_a)+\sum_{b\neq a}^N Z(v_b,v_a)\right] \ket{v_1 \dots v_a \rightarrow u \dots v_N},
\end{align}
and introducing the spin-$s_i$ representations, this can be rewritten as
\begin{align}
\mathbb{S}^2(u)& \ket{v_1 \dots v_N} =-\frac{2}{g}\sum_{i=1}^L Z(u,\epsilon_i)s_i\left[1+g \sum_{a=1}^N Z(\epsilon_i,v_a) -g \sum_{j \neq i}^L Z(\epsilon_i,\epsilon_j)s_j\right]\ket{v_1 \dots v_N} \nonumber \\
&+\left[-\Gamma \left(N-\sum_{i=1}^L s_i\right)^2 + \sum_{i=1}^L X(u,\epsilon_i)^2s_i(s_i+1)+\frac{1}{g^2}\right]\ket{v_1 \dots v_N} \nonumber \\
&+2 \sum_{a=1}^N Z(u,v_a) \left[\frac{1}{g}+\sum_{i=1}^L Z(\epsilon_i,v_a)s_i-\sum_{b \neq a}^N Z(v_b,v_a)\right]\ket{v_1 \dots v_N} \nonumber \\
&-2\sum_{a=1}^N X(u,v_a) \left[\frac{1}{g}+\sum_{i=1}^L Z(\epsilon_i,v_a)s_i-\sum_{b\neq a}^N Z(v_b,v_a)\right] \ket{v_1 \dots v_a \rightarrow u \dots v_N}.
\end{align}
The first three terms contribute to the eigenvalue of $\mathbb{S}^2(u)$, where the first term returns the eigenvalue of $Q_i$, the second term is the evaluation of the constant obtained in the expansion in Eq. (\ref{RG:AppGGA:Expansion}), and the third term will cancel if the Bethe equations are satisfied. Note that in this way, the Bethe equations are equivalent to the demand that the poles in the eigenvalue, or the terms proportional to $Z(u,v_a)$, vanish if $u \to v_a$. This is a common property of integrable models, and can also be used to derive the Bethe equations without explicit knowledge of the Bethe state.

\end{subappendices}
\makeatletter
\@openrighttrue
\makeatother



\chapter{An eigenvalue-based framework}
\label{chap:EVB}

\setlength\epigraphwidth{.8\textwidth}
\epigraph{\emph{Without any underlying symmetry properties, the job of proving interesting results becomes extremely unpleasant. The enjoyment of one's tools is an essential ingredient of successful work. }}{{Donald Knuth}}

In the previous chapter it was shown how integrability can be used to obtain exact Bethe ansatz states, circumventing the exponential scaling of the Hilbert space. In order to fully exploit the Bethe ansatz, efficient methods now have to be devised for a numerical solution of the Bethe equations. Unfortunately, these equations are highly non-linear and often give rise to singularities, making a straightforward numerical solution challenging. This was already noted by Richardson in an exploratory numerical study \cite{richardson_numerical_1966}, and a variety of methods have since been introduced as a way of resolving this difficulty \cite{rombouts_solving_2004,dominguez_solving_2006,de_baerdemacker_richardson-gaudin_2012,van_raemdonck_exact_2014,pan_extended_2011,guan_heine-stieltjes_2012,marquette_generalized_2012,pan_heine-stieltjes_2013,lerma_h._lipkinmeshkovglick_2013,guan_numerical_2014,qi_exact_2015,pan_exact_2017,pogosov_probabilistic_2012}.



In this chapter, we show how it is possible to efficiently solve the Bethe equations for Richardson-Gaudin models. Each eigenstate is defined in terms of a set of rapidities, but can also be characterized by its eigenvalues for the full set of conserved charges. Instead of obtaining these eigenvalues after solving the Bethe equations, we show how it is possible obtain the eigenvalues directly, and only afterwards extract the rapidities. This avoids the singularities plagueing the original Bethe equations, and can be connected to the structure of the conserved charges. The method has the further advantage that overlaps and correlation coefficients can be immediately expressed purely in terms of these eigenvalues, as will be discussed in great detail in the next chapter. The present chapter is largely based on Ref. \cite{claeys_eigenvalue-based_2015}.

\section{Singular points}
\begin{figure}
\begin{center}
\includegraphics{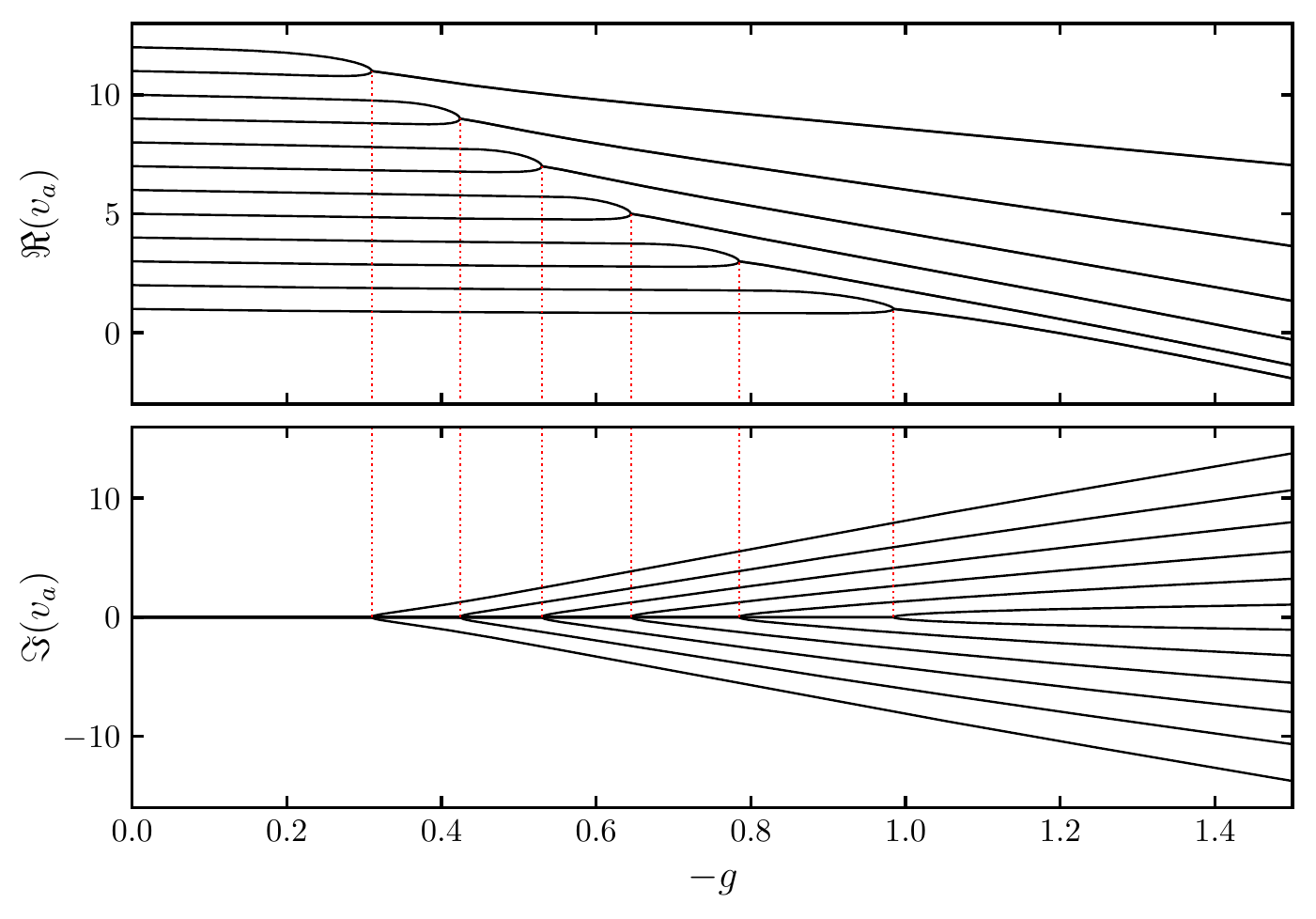}
\caption{Real and imaginary part of the rapidities $\{v_a\}=\{v_1 \dots v_N\}$ for the ground state of a picket-fence reduced BCS model ($\epsilon_i=i,i=1 \dots L$) at half-filling with $L=24$ and $N=12$. The red dotted lines mark the singular points where two real rapidities combine to form a complex conjugate pair.\label{fig:EVB:gsBCS}}
\end{center}
\end{figure}

The difficulties posed by a straightforward numerical solution of the Bethe equations can be easily illustrated. In Figure \ref{fig:EVB:gsBCS}, the rapidities are given for the ground state of the reduced BCS model, satisfying the Bethe equations (\ref{RG:phys:BAE_BCS})
\begin{equation}\label{EVB:sp:RGeq}
\frac{1}{g}+\frac{1}{2}\sum_{i=1}^L \frac{1}{\epsilon_i-v_a} - \sum_{b \neq a}^N \frac{1}{v_b-v_a}=0, \qquad \forall a=1 \dots N.
\end{equation}
The model is parametrized as a picket-fence model with equal level spacing, where $\epsilon_i = i$ \cite{hirsch_fully_2002}. The rapidities are presented for varying interaction strength $g$ starting from the non-interacting limit $g=0$. 
Several limiting behaviours can be observed. For $g \to 0$, the rapidities are purely real and collapse to the single-particle energies $\epsilon_i$. For $|g| \to \infty$, all rapidities are part of a complex conjugate pair, and diverge with increasing $|g|$. In between, an intermediate regime can be observed where real rapidities combine to form complex conjugate pairs. This occurs at the so-called \emph{singular points} \cite{richardson_numerical_1966}. At these points, the Bethe equations (\ref{EVB:sp:RGeq}) exhibit singular behaviour since $v_b-v_a \to 0$. Satisfying the Bethe equations then also requires $\epsilon_i-v_a \to 0$ in order to cancel this divergence, as can also be observed in Figure \ref{fig:EVB:gsBCS}. Despite these singularities, all physical observables vary smoothly along the singular points, so these cannot be interpreted as an indicator for e.g. a phase transition. Such singularities naturally hamper a straightforward solution of the Bethe equations. Solving these equations conventionally involves the targeting of specific eigenstates (e.g. the ground state), where the non-interacting limit $g \to 0$ provides a convenient way of labeling states since the solutions are explicitly known as $v_a = \epsilon_{i(a)}+\mathcal{O}(g)$. Adiabatically increasing $g$ up to the desired value while iteratively updating the solution would result in the targeted eigenstate, but this typically involves transitions across several singular points where iterative solution schemes are known to fail. 


\section{An eigenvalue-based numerical method}
Despite the seemingly singular behaviour of the Bethe wave function, all physical observables vary smoothly along these singular points. The constants of motion provide a specific set of physical observables for which the dependence on the rapidities is explicitly known. For a system containing only doubly-degenerate levels (spin-$1/2$ and all $\epsilon_i$ distinct), these can be written down as
\begin{equation}
Q_i = \left(S_i^z+\frac{1}{2}\right) + g \sum_{j \neq i}^L \left[\frac{1}{2}X(\epsilon_i,\epsilon_j)\left(S_i^+ S_j^-+S_i^- S_j^+\right)+ Z(\epsilon_i,\epsilon_j) \left(S_i^z S_j^z - \frac{1}{4}\right)\right],
\end{equation}
where a constant has been added to each constant of motion in order to eliminate the zero-point energy and have $Q_i \ket{0} = 0$, with $\ket{0} = \ket{\downarrow \dots \downarrow}$. The eigenvalue equation now simplifies to
\begin{equation}
Q_i \ket{v_1 \dots v_N} = -\frac{g}{2} \left[\sum_{a=1}^N Z(\epsilon_i,v_a)\right]\ket{v_1 \dots v_N},
\end{equation}
with rapidities satisfying the Bethe equations
\begin{equation}\label{EVB:EVG:RGeq}
\frac{1}{g}+ \frac{1}{2}\sum_{i=1}^L Z(\epsilon_i, v_a)-\sum_{b \neq a}^N Z(v_b,v_a) = 0, \qquad \forall a=1 \dots N.
\end{equation}
In this way, only the terms explicitly dependent on the rapidities have been kept in the eigenvalue, where $-g Z(\epsilon_i,v_a)/2$ can generally be interpreted as the excitation energy of $S^+(v_a)$ for a single conserved charge $Q_i$. The expressions for the eigenvalues then suggest the use of a set of `eigenvalue-based' variables as 
\begin{equation}\label{EVB:EVB:DefLam}
\Lambda_i \equiv \Lambda(\epsilon_i) = \sum_{a=1}^N Z(\epsilon_i,v_a), \qquad \textrm{with} \qquad \Lambda(u) \equiv \sum_{a=1}^N Z(u,v_a).
\end{equation}
Remarkably, it is possible to obtain the eigenvalues of the conserved charges in a numerically efficient and straightforward manner without explicitly obtaining the rapidities. This presents a method where an alternative set of equations is derived in terms of the eigenvalue-based variables. As will be shown, these can be seen either as a rewriting of the Bethe equations, or as a set of operator identities expressed at the level of the eigenvalues. In this way, this approach has several similarities with Baxter's $T-Q$ relation, also allowing the eigenvalues of conserved charges to be obtained without any knowledge of the rapidities \cite{baxter_exactly_2007}. Historically, this methods consists of a generalization of the numerical method first proposed by Faribault \emph{et al.} \cite{faribault_gaudin_2011} for a set of non-degenerate rational models, which was later extended to degenerate rational models by El Araby, Gritsev and Faribault \cite{el_araby_bethe_2012}. The equations for the Dicke model were independently presented by Babelon and Talalaev \cite{babelon_bethe_2007}, and the relevant variables follow naturally from a semi-classical limit of the Heisenberg model \cite{bettelheim_semi-classical_2014}. While the proposed framework holds for arbitrary spin representations, it is particularly convenient for doubly-degenerate spin-$1/2$ models and we will assume this to be the case unless mentioned otherwise (see Section \ref{subsec:EVB:deg}). 

In the non-interacting $g \to 0$ limit, the spin-$1/2$ conserved charges satisfy
\begin{equation}
\left(Q_i\right)^2 = Q_i, \qquad \forall i=1 \dots L,
\end{equation}
since the eigenvalues of $\left(S_i^z+\frac{1}{2}\right)$ will return either $0$ or $1$, and the eigenvalue-based variables can be obtained by solving a simple quadratic equation. This can be extended to interacting models using only the Bethe equations and the properties of the Gaudin algebra, leading to
\begin{align}
\left(Q_i\right)^2 &= Q_i -\frac{g}{2} \sum_{j \neq i}^L Z(\epsilon_i,\epsilon_j)\left(Q_i-Q_j\right) + \frac{g^2}{4} N (L-N) \Gamma , \qquad \forall i=1 \dots L,
\end{align}
or, re-expressed in the eigenvalue-based variables,
\begin{equation}\label{EVB:EVB:EVBeq}
\Lambda_i^2=-\frac{2}{g}\Lambda_i+\sum_{j \neq i}^L Z(\epsilon_i,\epsilon_j)(\Lambda_i-\Lambda_j)+ N (L-N) \Gamma, \qquad  \forall i=1 \dots L.
\end{equation}
The full derivation can be found in Appendix \ref{EVB:App:Eq}. These equations are also known as the substituted or quadratic Bethe equations. In principle, it is now possible to directly solve these equations and obtain the eigenvalues of the conserved charges. However, when solving these equations numerically, it immediately becomes clear that the total number of solutions exceeds the dimension of the Hilbert space for $N$ excitations distributed over $L$ levels. Therefore, these equations necessarily support unphysical solutions not corresponding to any eigenstate, implying that this new set of equations is not yet equivalent to the original set of Bethe equations (\ref{EVB:EVG:RGeq}). In order to obtain a set of equations equivalent to the original equations, additional constraints for the $\Lambda_i$ are needed.

It is clear from the $\Gamma=0$ case (XXX) that the new set of equations cannot distinguish between the different excitation sectors $N$. This can be imposed by noting that the sum of all constants of motion is given by the operator counting the number of excitations
\begin{equation}
\sum_{i=1}^L Q_i=\sum_{i=1}^L \left(S_i^z + \frac{1}{2}\right) = N.
\end{equation}
Writing out the eigenvalues of the constants of motion in the new variables results in
\begin{equation}\label{EVB:EVB:sumLam}
-\frac{g}{2}\sum_{i=1}^L \Lambda_i=N.
\end{equation}
This can also be obtained by summing the Bethe equations (\ref{EVB:EVG:RGeq}) for $a=1 \dots N$. In the following subsection it will be shown that the full set of equations (\ref{EVB:EVB:EVBeq}) and (\ref{EVB:EVB:sumLam}) supports as many solutions as the dimension of the Hilbert space, so introducing this additional equation leads to a system of equations fully equivalent to the original set of Bethe equations (\ref{EVB:EVG:RGeq}).

\begin{figure}
\begin{center}
\includegraphics{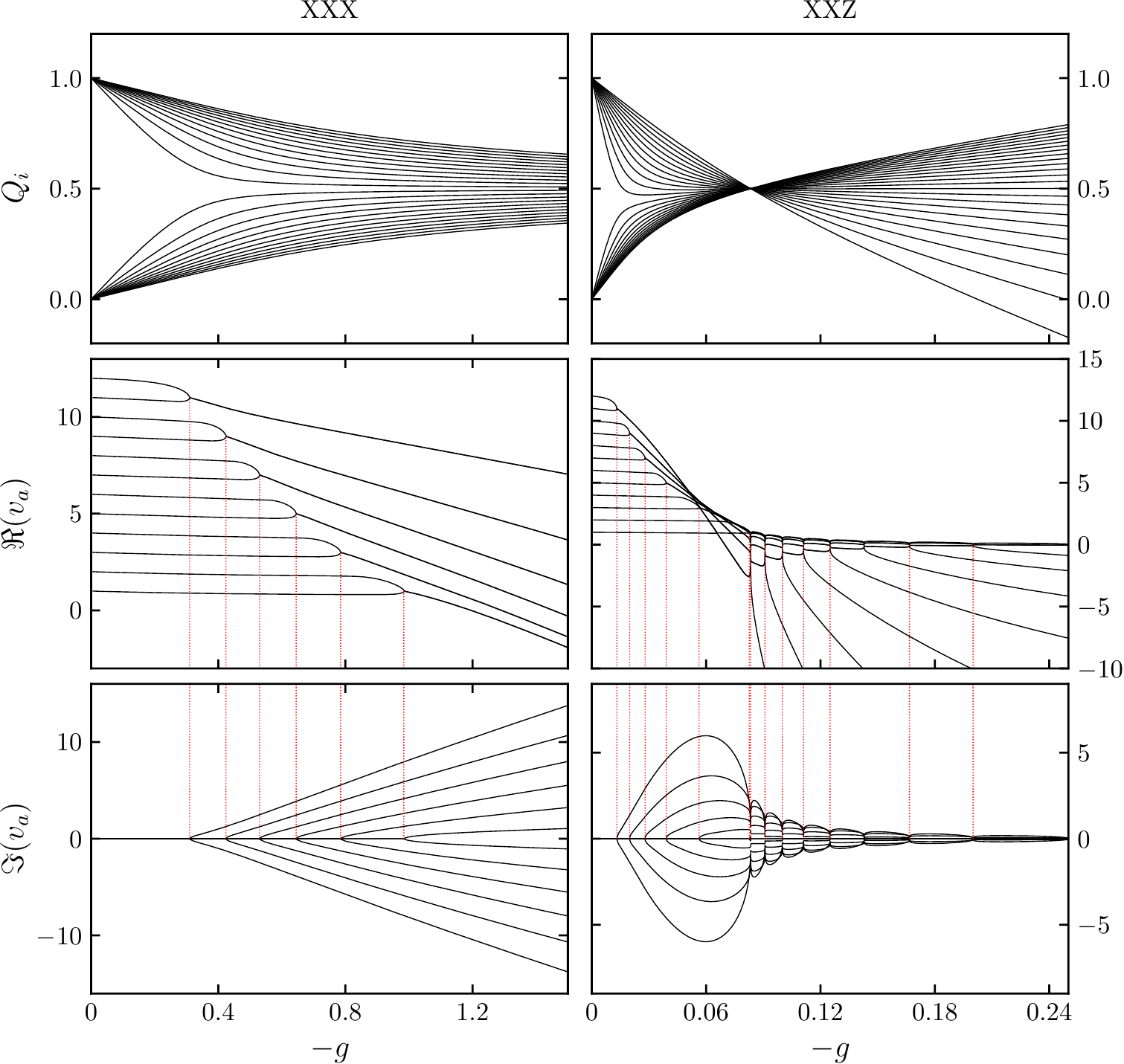}
\caption{Eigenvalues of the conserved charges $Q_i$ and real and imaginary part of the rapidities $\{v_a\}=\{v_1 \dots v_N\}$ for the XXX (left column) and XXZ (right column) model. Both models represent the ground state of $\sum_{i=1}^L \epsilon_i Q_i $ with a picket-fence parametrization ($\epsilon_i=i,i=1 \dots L$) at half-filling with $L=24$ and $N=12$. The red dotted lines mark the singular points where multiple rapidities coincide. Despite the seemingly singular behaviour of the rapidities, the eigenvalues behave smoothly when varying the interaction strength $g$.\label{fig:EVB:XXXvsXXZ}}
\end{center}
\end{figure}

The eigenvalue-based equations (\ref{EVB:EVB:EVBeq}) and (\ref{EVB:EVB:sumLam}) have several advantages compared to the regular Bethe equations (\ref{EVB:EVG:RGeq}). Firstly, and most importantly, they do not exhibit singular behaviour and can be straightforwardly solved numerically. Secondly, the variables to be solved for in the eigenvalue-based equations are necessarily real, since they represent eigenvalues of Hermitian operators. This is illustrated in Figure \ref{fig:EVB:XXXvsXXZ} for picket-fence XXX and XXZ models. Furthermore, the completeness of Bethe ansatz states and Bethe equations generally remains a major challenge, often treated on a case-by-case basis. However, since the eigenvalue-based equations follow from a set of operator identities, they are necessarily complete, as pointed out in Ref. \cite{links_completeness_2016}. Lastly, no physical observables can be associated with the rapidities, whereas the eigenvalue-based equations can be connected to the conserved charges and to the occupations of specific spins $\braket{S_i^z}$ (as will be shown in Section \ref{subsec:EVB:HellFeyn}). As such, they also exhibit the symmetries associated with physical operators, which remain hidden in the rapidities \cite{pogosov_electron-hole_2013,bork_particlehole_2015}.

\subsection{The weak-coupling limit}
In order to obtain some insight in the behaviour of the solutions of these equations, an approximate solution can be found in the weak-coupling limit ($|g| \ll 1 $). This can be done by proposing a series expansion in $g$ for the eigenvalue-based variables and solving the equations at each order or, equivalently, by applying perturbation theory to the eigenvalues of the conserved charges starting from the non-interacting limit.
For small $g$, a series expansion of $\Lambda_i$ in $g$ can be proposed up to $\mathcal{O}(g)$, keeping only the two dominant terms
\begin{equation}
\Lambda_i = \lambda_i^{(-1)}g^{-1}+\lambda_i^{(0)} g^0+\mathcal{O}(g), \qquad \forall i=1 \dots L.
\end{equation}
Plugging this expansion in Eq. (\ref{EVB:EVB:EVBeq}) and performing the expansion results in
\begin{align}
&g^{-2}\left[\lambda_i^{(-1)}\left(\lambda_i^{(-1)}+2\right)\right] + g^{-1} \left[2\lambda_i^{(0)}(1+\lambda_i^{(-1)})-\sum_{j \neq i}^L Z(\epsilon_i,\epsilon_j)\left(\lambda_i^{(-1)}-\lambda_j^{(-1)}\right)\right] \nonumber\\
&\qquad \qquad \qquad \qquad \qquad \qquad \qquad \qquad \qquad \qquad \qquad +\mathcal{O}(g^0)=0, \qquad \forall i=1 \dots L.
\end{align}
This can immediately be solved at each order to return
\begin{align}
&\lambda_i^{(-1)} = 0 \ \  \textrm{or} \  -2, \\
&\lambda_i^{(0)} = \frac{1}{2} \left(1+\lambda_i^{(-1)}\right)\sum_{j \neq i}^L Z(\epsilon_i,\epsilon_j)\left(\lambda_i^{(-1)}-\lambda_j^{(-1)}\right).
\end{align}
As expected, the first term returns eigenvalues $-\lambda_i^{(-1)}/2=0$ or $1$ in the non-interacting limit, whereas the second term contains the first-order perturbative corrections due to $\sum_{j \neq i}^L Z(\epsilon_i,\epsilon_j)S_i^z S_j^z$. This can also be observed in Figure \ref{fig:EVB:XXXvsXXZ}. In order to satisfy Eq. (\ref{EVB:EVB:sumLam}), the number of dominant terms different from $0$ has to equal the number of excitations $N$, resulting in a total number of $\binom{L}{N}$ solutions. The total number of solutions then equals the dimension of the Hilbert space for $N$ excitations distributed over $L$ doubly-degenerate levels. Any solution in the weak-coupling limit can be adiabatically connected to a solution for arbitrary coupling, indicating that all possible solutions are always found and no spurious solutions are present.

Another way of interpreting this perturbative expansion is by considering a similar series expansion for the rapidities \cite{ortiz_exactly-solvable_2005}. In the limit $g \to 0$ the rapidities $v_a$ converge to the parameters $\epsilon_i$, depending on the corresponding distribution of excitations over energy levels. A rapidity converging to $\epsilon_i$ then corresponds to an excited level $i$ in the noninteracting limit, where $\braket{S_i^z} \to \frac{1}{2}$. For finite but small $g$ the dominant corrections on the rapidities are of $\mathcal{O}(g)$ and can be shown to be proportional to the roots of orthogonal polynomials via a Heine-Stieltjes connection \cite{stieltjes_theoreme_1885,sriram_shastry_solution_2001}. For $v_a$ converging to $\epsilon_i$, this results in $Z(\epsilon_i,v_a)$ diverging as $g^{-1}$ in the weak-coupling limit, where the proportionality factor can be found to be $-2$ from the Heine-Stieltjes connection for $s_i=1/2$. This behaviour can be generalized through a connection of $\Lambda_i$ to occupation numbers, as will be shown in later subsections. Note that the divergence in $g \to 0$ can pose numerical problems, so $g\Lambda_i$ are commonly used as numerical variables instead of $\Lambda_i$.

\subsection{Solving the equations}
In general, no analytic solutions can be found for the eigenvalue-based equations, and it is necessary to resort to numerical methods. Starting from an initial guess, sets of nonlinear equations are commonly solved by an iterative approach such as the Newton-Raphson method \cite{press_numerical_2007}. This method converges quadratically if the initial guess lies in the basin of attraction, so an efficient numerical approach can be implemented once we have access to a sufficiently good initial guess for the solution. 

The known solution in the non-interacting limit can first be used as an initial guess for a solution in the weak-coupling limit ($|g|\ll 1$), and solutions at arbitrary values of the coupling constant can be obtained by adiabatically varying $g$ starting from the weak-coupling limit and using the solution at the previous step as the starting point for an iterative solution at the current step. From the solution at interaction strength $g$, an initial guess at $g+\delta g$ is given by a Taylor expansion
\begin{equation}
\Lambda_i(g+\delta g) \approx \Lambda_i(g)+\left.\frac{\partial \Lambda_i}{\partial g}\right\rvert_{g} \delta g , \qquad \forall i=1 \dots L.
\end{equation}
Taking the derivative of all involved equations w.r.t. $g$, their quadratic nature results in a linear system of equations for the derivatives
\begin{align}\label{EVB:EVB:eqDerLam_1}
&\frac{\partial \Lambda_i}{\partial g} \left[1+g \Lambda_i-\frac{g}{2}\sum_{j \neq i}^L Z(\epsilon_i,
\epsilon_j)\right] + \frac{g}{2} \sum_{j \neq i}^L Z(\epsilon_i,\epsilon_j) \frac{\partial \Lambda_j}{\partial g} = \frac{\Lambda_i}{g}, \qquad \forall i=1 \dots L, \\\label{EVB:EVB:eqDerLam_2}
&\frac{g^2}{2}\sum_{i=1}^L \frac{\partial \Lambda_i}{\partial g} = N.
\end{align}
All necessary derivatives of the eigenvalue-based variables at fixed interaction strength $g$ can be found by solving this linear system of equations depending on the eigenvalue-based variables at the same interaction strength. By taking higher-order derivatives of the original set of equations, linear equations can be found for higher-order derivatives of $\Lambda_i$, which can be used to construct the Taylor expansion up to arbitrary order. A further advantage of the quadratic form is that the matrix that needs to be (pseudo-)inverted when solving this system of equations is independent of the order of the derivative, so the inversion step only needs to be performed once. For an efficient numerical implementation, combining the Newton-Raphson method with a Taylor approximation up to first order already offers a remarkable increase in speed. 

\subsection{The Hellmann-Feynman theorem}
\label{subsec:EVB:HellFeyn}
\begin{figure}
\begin{center}
\includegraphics{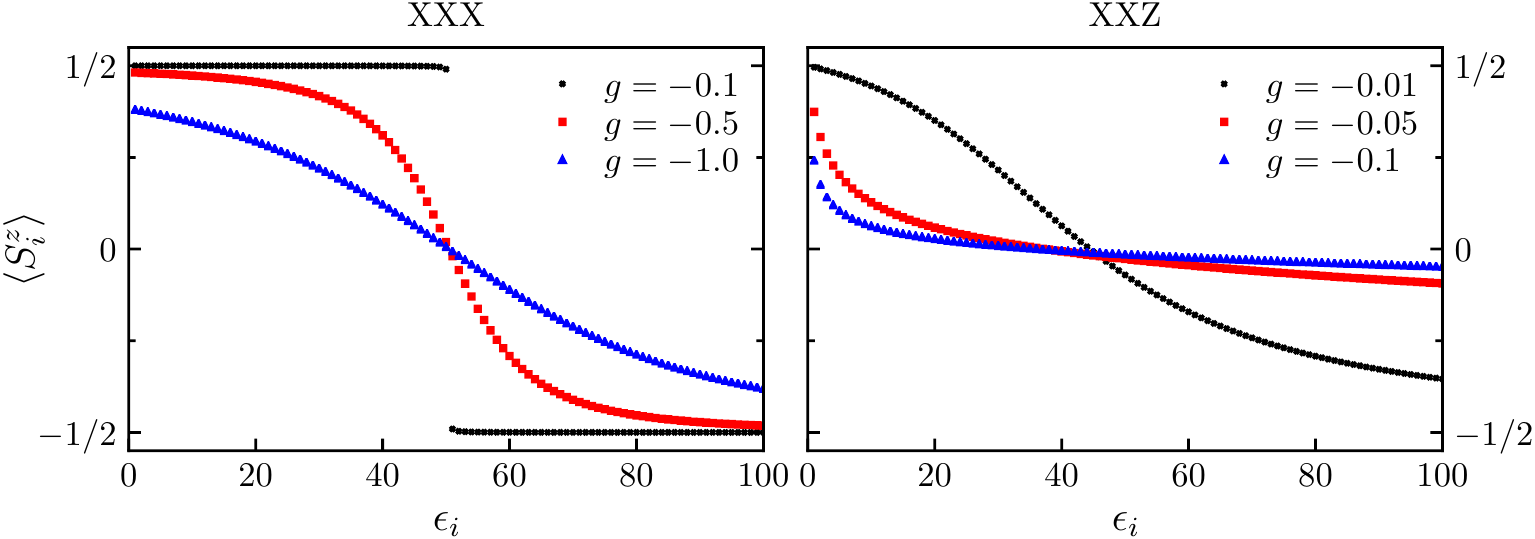}
\caption{Expectation values $\braket{S_i^z}$ for the XXX (left column) and XXZ (right column) model. Both models represent the ground state of $\sum_{i=1}^L \epsilon_i Q_i $ with a picket-fence parametrization ($\epsilon_i=i,i=1 \dots L$) at half-filling with $L=100$ and $N=50$. \label{fig:EVB:occ}}
\end{center}
\end{figure}

Through the Hellmann-Feynman theorem a clean interpretation can be obtained for these derivatives, connecting them to the occupation of the separate levels for a given eigenstate as
\begin{equation}
\left(S_i^z+\frac{1}{2}\right) = Q_i-g \frac{\partial Q_i}{\partial g} \qquad \rightarrow \qquad \braket{S_i^z} = -\frac{1}{2}+g^2 \frac{\partial \Lambda_i}{\partial g}.
\end{equation}
Knowledge of the evolution of $\Lambda_i$ with a changing coupling constant is then equivalent to knowing how the $N$ excitations are distributed over the $L$ levels. This can also be connected to Eq. (\ref{EVB:EVB:eqDerLam_2}), denoting conservation of total spin projection. In the non-interacting limit, this shows how $\braket{S_i^z} = -\frac{1}{2} - \frac{1}{2}\lambda_i^{(-1)}=\pm \frac{1}{2}$. This has been illustrated in Figure \ref{fig:EVB:occ} for the `picket-fence' models from Figure \ref{fig:EVB:XXXvsXXZ}.

\subsection{Inverting the transformation}
\label{subsec:EVB:invert}
The current scheme presents two alternative ways of solving for the eigenstates of RG integrable models: either by solving the Bethe equations for the rapidities, or by solving the quadratic equations for a set of eigenvalue-based variables. If the rapidities are known, these eigenvalues can immediately be determined from the definition (\ref{EVB:EVB:DefLam}). However, the inverse is more cumbersome, since the rapidities cannot be immediately obtained from the eigenvalues. This can be done by connecting the Bethe equations to a differential equation, a recurring technique in the theory of integrability \cite{dorey_ode/im_2007}. Following Eq.\ (\ref{GGA:XXZ:GaudinZ}), it is possible to reparametrize the Gaudin algebra as
\begin{equation}\label{EVB:EVB:Zia_1}
Z(\epsilon_i,v_a)=\frac{\Gamma+Z(u,\epsilon_i)Z(u,v_a)}{Z(u,\epsilon_i)-Z(u,v_a)}, \qquad \forall u \in \mathbb{C}.
\end{equation}
Alternatively, by taking $\Gamma = -\gamma^2$ and defining $z(v) = (\gamma-1)/(\gamma+Z(u,v))$ (at fixed $u$), this reduces to
\begin{equation}\label{EVB:EVB:Zia_2}
Z(\epsilon_i,v_a) = \frac{(1-\gamma)+\gamma (z(\epsilon_i)+z(v_a))}{z(\epsilon_i)-z(v_a)},
\end{equation}
which nicely interpolates between the rational ($\gamma=0$) and hyperbolic ($\gamma=1$) model if $\gamma \in \mathbb{R}$. Instead of directly solving for the rapidities $\{v_1 \dots v_N\}$, it is now possible to find a transformation that allows us to determine the set of Gaudin algebra elements $\{z(v_1) \dots z(v_N)\}$. Once these are known, the rapidities can generally be found by inverting $z(v)$. Eq. (\ref{EVB:EVB:Zia_2}) can now be used to show that
\begin{align}\label{inv:lam}
\Lambda_i = \sum_{a=1}^N \frac{(1-\gamma)+\gamma(e_i+z_a)}{e_i-z_a} = \sum_{a=1}^N \left[ \frac{(1-\gamma)+2 \gamma e_i }{e_i-z_a}-\gamma\right] 
\end{align}
in which $e_i = z(\epsilon_i)$ and $z_a = z(v_a)$. The main object of interest is the polynomial with the full set of roots $\{z_1 \dots z_N\}$
\begin{equation}
P(z)=\prod_{a=1}^N(z-z_a).
\end{equation}
Once a representation for this polynomial is known, its roots can be determined in order to obtain the rapidities\footnote{In this way, the rapidities are also sometimes referred to as \emph{Bethe roots}.}. This method arises naturally for different problems in the theory of integrable systems, such as the Heine-Stieltjes connection \cite{stieltjes_theoreme_1885,sriram_shastry_solution_2001}, the numerical methods by Guan \emph{et al.} \cite{guan_heine-stieltjes_2012} and Rombouts \emph{et al.} \cite{rombouts_solving_2004}, and the weak-coupling limit in RG models \cite{ortiz_exactly-solvable_2005}.

This can either be done in a straightforward way by setting $P(z)=\sum_{m=0}^N P_{N-m}z^m$ (with $P_0=1$), evaluating
\begin{equation}
\frac{P'(e_i)}{P(e_i)} = \frac{\sum_{m=0}^N m P_{N-m}e_i^{m-1}}{\sum_{m=0}^N P_{N-m}e_i^m} = \sum_{a=1}^N \frac{1}{e_i-z_a} = \frac{\Lambda_i + \gamma N}{(1-\gamma)+2\gamma e_i},  \qquad \forall i=1 \dots L,
\end{equation}
and solving this set of linear equations for the coefficients $P_{m}$. This provides a representation of the polynomial from which the roots can be obtained using a root-finding algorithm such as Laguerre's method. Although easy to implement, this method has the disadvantage of being prone to numerical errors. Indeed, it is well-known that the roots of a polynomial are highly sensitive to changes in the coefficients, sometimes even changing the solutions at the qualitative level (a set of complex conjugate roots can be found numerically instead of two separate real roots) \cite{wilkinson_evaluation_1959}. These considerations are not pressing for a limited number of excitations, but become more and more important for an increasing number of excitations.  

In order to circumvent this, it can be noted that the Bethe equations can be recast as a differential equation for $P(z)$ through the Heine-Stieltjes connection \cite{stieltjes_theoreme_1885,sriram_shastry_solution_2001} 
\begin{align}\label{EVB:EVB:BetheDiffEq}
&\left[2 g^{-1}-2\gamma(N-1)+\sum_{i=1}^N \frac{(1-\gamma)+\gamma(e_i+z)}{e_i-z}\right]P'(z)  \nonumber \\
&\qquad \qquad \qquad + \left[(1-\gamma)+2 \gamma z\right] P''(z) = \left[\sum_{i=1}^L \frac{\Lambda_i+\gamma}{e_i-z}\right]P(z).
\end{align}
If the eigenvalue-based variables are known, this reduces to an ordinary differential equation (ODE). This differential equation can now be solved for any arbitrary way of representing the polynomial (e.g. in Lagrange representation \cite{el_araby_bethe_2012}), circumventing the problem of expressing the polynomial through its roots.

The polynomial representation in Eq. (\ref{EVB:EVB:BetheDiffEq}) also suggests a hybrid way of solving the combined Bethe and eigenvalue-based equations,  solving the differential equation in a self-consistent way by expressing $\Lambda_i$ as a function of the roots of the polynomial. In this method it would not be necessary to start from the weak-coupling limit when numerically solving the equations \cite{guan_heine-stieltjes_2012,guan_numerical_2014,qi_exact_2015}. This method does not allow for the targeting of specific states, such as the ground state, but is useful when all eigenstates need to be efficiently determined for an arbitrary coupling constant, such as for the evaluation of partition functions.

\subsection{Degenerate models}
\label{subsec:EVB:deg}
The discussion has been limited to $s_i=1/2$ models only thus far. While this is sufficient for the majority of interacting (quasi-)spin systems, in situations with higher symmetries arbitrary degeneracies $2 s_i+1 > 2$ may occur. Following results for the rational model \cite{faribault_gaudin_2011,el_araby_bethe_2012}, the set of equations for the eigenvalue-based variables can be related to a functional equation for $\Lambda(u)$, which can be discretized in order to generate a closed set of equations.

\begin{figure}
\begin{center}
\includegraphics{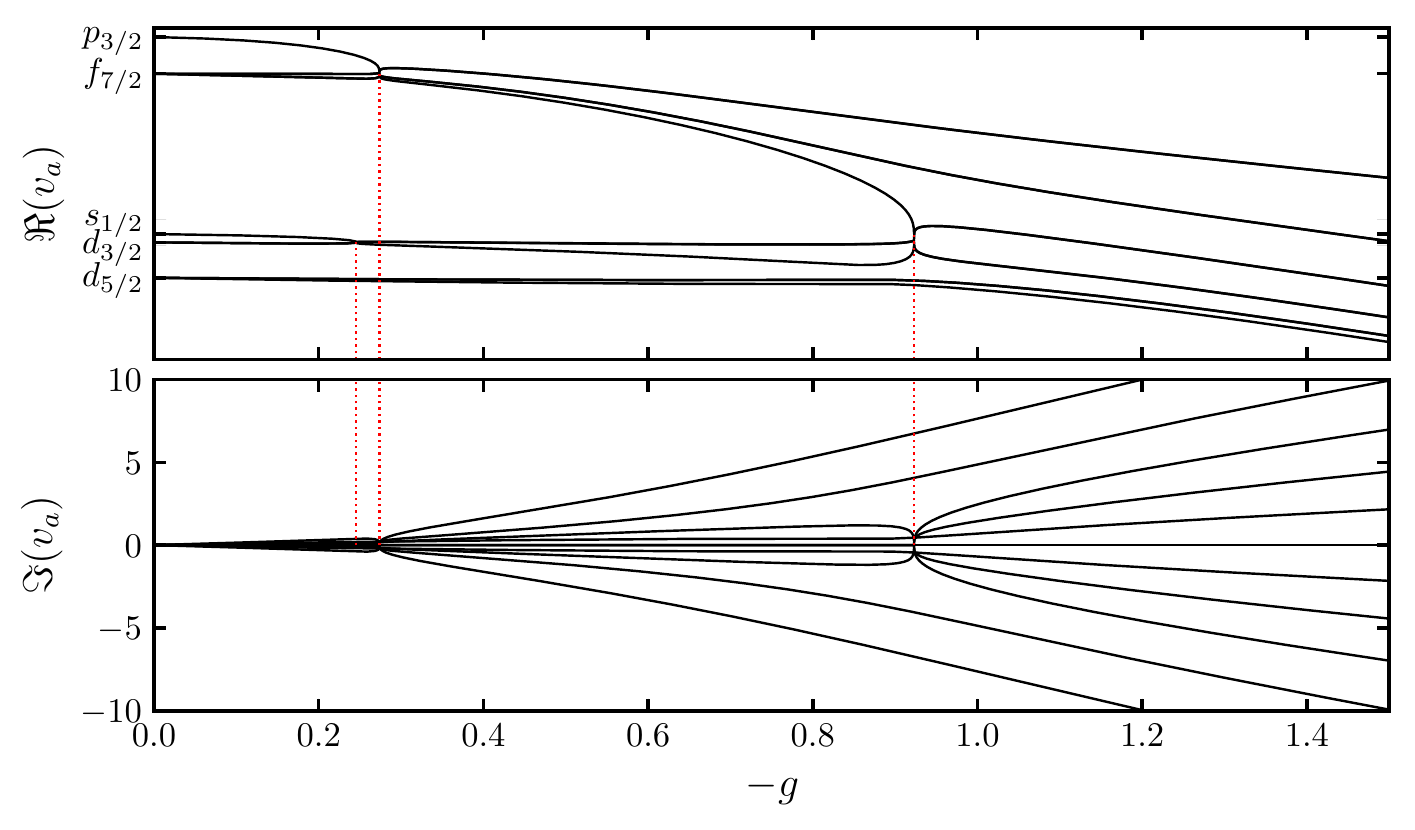}
\caption{Real and imaginary part of the rapidities $\{v_a\}=\{v_1 \dots v_N\}$ for the ground state of a reduced BCS model describing neutrons in $^{56}$Fe following the parametrization of Ref. \cite{rombouts_thermodynamical_1998}.\label{fig:EVB:degenerate_rap}}
\end{center}
\end{figure}

\begin{figure}
\begin{center}
\includegraphics{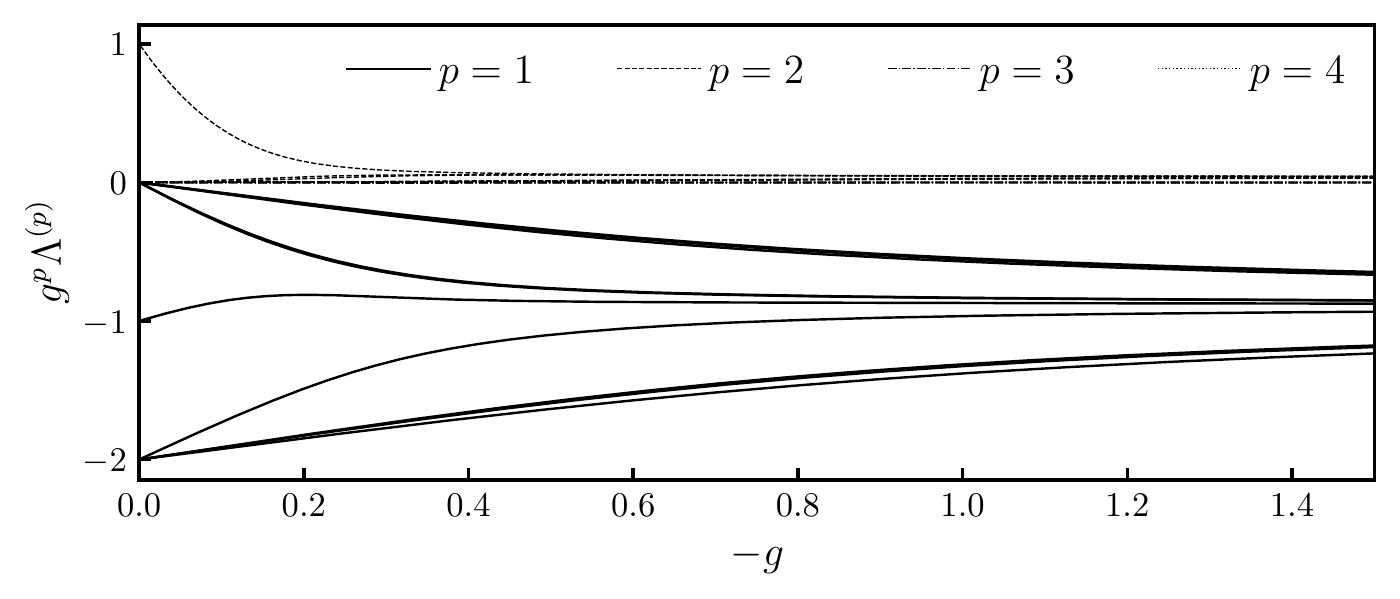}
\caption{Eigenvalue-based variables for the ground state of a reduced BCS model describing neutrons in $^{56}$Fe following the parametrization of Ref. \cite{rombouts_thermodynamical_1998}.\vspace{-\baselineskip}\label{fig:EVB:degenerate_EVB}}
\end{center}
\end{figure}

While the outlined procedure is slightly more involved for the class of XXZ RG models, it is analogous to the rational case, and remains therefore tractable. The main idea is identical: starting from a continuous representation of the equations it is possible to obtain any number of equations for the total set of variables. The continuous representation of the variables is given by
\begin{equation}
\Lambda(u)=\sum_{a=1}^N Z(u,v_a),
\end{equation}
where $\Lambda(\epsilon_i)\equiv\Lambda_i$, and the set of variables needs to be extended with
\begin{equation}
\Lambda_{i}^{(p)}=\Lambda_{i}^{(p)}(\epsilon_i) \qquad \textrm{with} \qquad \Lambda_i^{(p)}\equiv\sum_{a=1}^N  Z(u,v_a)^p,
\end{equation}
where $\Lambda_i \equiv \Lambda_i^{(1)}$. These are highly reminiscent of the variables introduced by Rombouts \emph{et al.} as a way of circumventing the singular points \cite{rombouts_solving_2004}. A similar derivation as for the $s_i=1/2$ model results in a continuous equation
\begin{align}\label{EVB:EVB:DiffEqDegen}
[\Lambda(u)]^2=&-\frac{2}{g}\Lambda(u)+ 2\sum_{j \neq i}^L s_j Z(u,\epsilon_j)\left(\Lambda(u)-\Lambda(\epsilon_j)\right) \nonumber\\
&+\sum_{a=1}^N Z(u,v_a)\left[Z(u,v_a)-2 Z(\epsilon_i,v_a) s_i \right]+\Gamma N\left(1-N+2\sum_{j \neq i}s_j\right),
\end{align}
which holds if $u \neq \epsilon_j$ for all $j\neq i$. The evaluation of these equations at $u=\epsilon_i$ results in the known set of equations if $s_i=1/2$, but for larger degeneracies the set of equations also depends on $\Lambda_i^{(2)}$. By taking the derivative of Eq. (\ref{EVB:EVB:DiffEqDegen}), additional equations can be obtained linking these higher-order variables to the original variables. These derivatives can be evaluated without any specific knowledge of the representation of the Gaudin algebra by noting from Eq. (\ref{EVB:EVB:Zia_1}) that
\begin{equation}
-X(z,u)^2 \frac{\partial Z(u,v_a)}{\partial Z(z,u)} = Z(u,v_a)^2 + \Gamma,
\end{equation}
where $z$ is fixed, explicitly relating the derivative of $\Lambda(u)$ to the higher-order variables by
\begin{equation}
-X(z,u)^2 \left.\frac{\partial \Lambda^{(p)}(u)}{\partial Z(z,u)}\right\rvert_{u=\epsilon_i}  = p \left(\Lambda_i^{(p+1)} + \Gamma \Lambda_i^{(p-1)} \right).
\end{equation}
Taking e.g. $s_i=1$, evaluating Eq. (\ref{EVB:EVB:DiffEqDegen}) and its first derivative at $u=\epsilon_i$ results in
\begin{align} 
&\Lambda_i^2=-\frac{2}{g}\Lambda_i+ 2\sum_{j \neq i}^L s_j Z(\epsilon_i,\epsilon_j)\left(\Lambda_i-\Lambda_j\right) 
+\Gamma N\left(1-N+2\sum_{j \neq i}^L s_j\right)-\Lambda_i^{(2)}, \\
&2\Lambda_i \left[\Lambda_i^{(2)}+\Gamma N\right] = -\frac{2}{g} \left[\Lambda_i^{(2)}+\Gamma N\right]+ 2 \sum_{j \neq i}^L s_j Z(\epsilon_i,\epsilon_j) \left(\Lambda_i-\Lambda_j\right),
\end{align}
leading to a closed set of equations independent of $\Lambda_i^{(p)}, p >2$ . For arbitrary $s_i$, the first $2s_i-1$ derivatives lead to a closed set of equations for the variables $\{\Lambda_i^{(p)},p=1 \dots 2s_i-1\}$. An additional equation for the total number of excitations can easily be determined as
\begin{equation}
N=-\sum_{i=1}^L g s_i \Lambda_i.
\end{equation}
For the rational model ($\Gamma=0$), a recursive expression for the higher-order equations can be found and equations can be constructed for arbitrary degeneracies \cite{el_araby_bethe_2012}. However, for general XXZ models no such results have been obtained and the resulting equations become quite burdensome. As such, this method is best suited to the treatment of  spin-$1/2$ models. This is illustrated in Figures \ref{fig:EVB:degenerate_rap} and \ref{fig:EVB:degenerate_EVB} for the parametrization referenced in Table \ref{table:degenerate}.

\setlength{\tabcolsep}{7pt}
\begin{table}
\begin{center}
\begin{tabular}{ l || c c c c c}
& $1 d _{5/2}$ & $1 d_{3/2}$ &  $2 s_{1/2}$ & $1 f_{7/2}$ &  $2p_{3/2}$ \\ \hline
 $\epsilon_i$ & -21.5607 & -19.6359 & -19.1840 & -10.4576 & -8.4804   \\
$\Omega_i$ & 4 & 3 & 2 & 5 & 3 \\ \hline \hline 
&  $1 f_{5/2}$  &  $2 p_{1/2}$ &  $3 s_{1/2}$ &  $2 d_{5/2}$  &$1 g_{9/2}$ \\  \hline
 $\epsilon_i$ & -7.7003 & -7.6512 & -0.3861 & 0.2225 & 0.5631 \\
 $\Omega_i$ & 4 & 2 & 2 & 4 & 6 \\ \hline
\end{tabular}
\caption{Single-particle levels following from a Woods-Saxon potential for $^{56}$Fe as used in Figures \ref{fig:EVB:degenerate_rap} and \ref{fig:EVB:degenerate_EVB} following Ref. \cite{rombouts_thermodynamical_1998}. All energies are measured in MeV.\label{table:degenerate}\vspace{-\baselineskip}}
\end{center}

\end{table}

\section{Expanding the Bethe state}
\label{RG:sec:expand}

When performing calculations with Bethe ansatz states, it is customary to expand them in a basis set of uncorrelated wave functions. As will be shown in this section, these expansion coefficients (or overlaps) are typically given by permanents of matrices in the theory of Richardson-Gaudin integrability. However, such expressions are not practical for computational purposes. Evaluating the permanent of an arbitrary matrix scales exponentially with the matrix size and is known to be an \#P-hard problem \cite{valiant_complexity_1979}. Compared to the determinant, which can be evaluated in a polynomial time, permanents require a factorial scaling computational time and therefore severely limit the size of the matrices that can be considered. Luckily, the underlying integrability again allows for an efficient way of calculating these permanents by connecting these to determinants of matrices. This fits within the general philosophy of integrability -- in a similar way that it circumvents the exponential scaling of the Hilbert space, it can be used to avoid the exponential scaling when calculating overlaps. Two dual ways of evaluating this permanent will again be presented, related to the dual ways of solving the Bethe equations.

The uncorrelated wave functions are typically given by direct product states
\begin{equation}
\ket{i(1) \dots i(N)} = \prod_{a=1}^N S^+_{i(a)}\ket{0},
\end{equation}
where $\{i(a)\}=\{i(1) \dots i(N)\}$ denote the occupied levels. For fermion pairing models, these are simply Slater determinants. Given a Bethe ansatz state, it can be expanded in the (Fock) basis spanned by these states as
\begin{align}
\ket{v_1 \dots v_N} &= \prod_{a=1}^N \left(\sum_{i=1}^L X(\epsilon_i,v_a)S_i^+\right)\ket{0} \\
&= \sum_{i(1), \dots ,i(N) }\ket{i(1) \dots i(N)}\braket{i(1) \dots i(N)|v_1 \dots v_N},
\end{align}
where expanding the product structure of the Bethe state results in 
\begin{equation}\label{EVB:exp:exp_term}
\braket{i(1) \dots i(N)|v_1 \dots v_N} = \sum_{\sigma \in S_N} \prod_{a=1}^N X(\epsilon_{i(a)},v_{\sigma(a)}),
\end{equation}
with $S_N$ the set of all permutations of $\{1,2, \dots, N\}$. This is a formula independent of the integrability or the explicit expression for $X(u,v)$ and is a direct result of the structure of the wave function \cite{percus_combinatorial_1971}. The right-hand side of Eq. (\ref{EVB:exp:exp_term}) is exactly the definition of the permanent of an $N \times N$ matrix $X_N$ with matrix elements $\left(X_N\right)_{ab} = X(\epsilon_{i(a)},v_b)$. The definition of a permanent can be contrasted with that of the determinant
\begin{align}
\per \left(C\right) &= \sum_{\sigma \in S_N} \prod_{a=1}^N C_{a,\sigma(a)}, \\
\det \left(C\right) &= \sum_{\sigma \in S_N} (-1)^{\sigma} \prod_{a=1}^N C_{a,\sigma(a)}.
\end{align}
Despite the apparent simplicity of the expression for the permanent, the absence of minus signs proves problematic when calculating the permanent\footnote{The absence of a minus sign can be related to the bosonic symmetry when exchanging two spin-raising operators $[S_i^+,S_j^+]=0$, whereas Slater determinants obtain a minus sign under exchange of fermionic creation operators $\{a_i^{\dagger},a_j^{\dagger}\}_+=0$.}. The determinant can be evaluated by Gauss elimination, making use of elementary row and column operations, whereas the permanent is not invariant under such operations and the full summation generally needs to be evaluated. However, by imposing a specific structure on the matrix this scaling may be circumvented. Following a famous result by Borchardt relating the permanent of Cauchy matrices to a ratio of determinants \cite{borchardt_bestimmung_1857,singer_bijective_2004}, the Gaudin algebra can be used to reduce the calculation of the permanent to
\begin{equation}\label{EVB:exp:detXX}
\per\left(X_N\right) = \frac{\det \left(X_N*X_N\right)}{\det \left(X_N\right)} = \frac{\prod_{a>b} X(\epsilon_{i(a)},\epsilon_{i(b)}) \prod_{b>a} X(v_a,v_b) }{\prod_{a,b} X(\epsilon_{i(a)},v_{b})}\det\left(X_N*X_N\right),
\end{equation}
where the Hadamard product of two matrices is introduced, defined as $(A*B)_{ab}=A_{ab}B_{ab}$, and the determinant of $X_N$ is explicitly known as a generalization of the determinant of a Cauchy matrix. Within the literature on integrability, such determinant expression are commonly known as Izergin determinants \cite{izergin_determinant_1992,korepin_quantum_1993}, and we will refer to these as Izergin-Borchardt determinants.

However, Eq. (\ref{EVB:exp:detXX}) is only one of the possible determinant expressions for this overlap. In Ref. \cite{faribault_determinant_2012}, a case was made for a theory of Richardson-Gaudin integrability that would require only the eigenvalue-based variables instead of the rapidities. This would be desirable from a numerical point of view as the eigenvalue-based variables are free of singularities, opposed to the singularity-prone rapidities. In addition, this would enable us to skip the inversion step, the main bottleneck of this method.

Multiple formulae exist for the rational model linking the permanents to determinants  \cite{faribault_determinant_2012,tschirhart_algebraic_2014,faribault_determinant_2016,faribault_common_2017,claeys_inner_2017}, and these can be generalized to the XXZ case. Starting from these expressions, determinant expressions for systems with arbitrary degeneracy can also be obtained as a limiting case where several levels $\epsilon_i$ coincide, but this will not be discussed here.

As shown in Appendix \ref{EVB:App:det} and discussed in detail in the next chapter, the ratio of determinants can be reworked in order to obtain a determinant expression which only depends on the rapidities through the eigenvalue-based variables. This results in
\begin{equation}\label{EVB:exp:detJ}
\left(\prod_{a=1}^N X(u,v_a)\right)^{-1}\per\left(X_N\right) = \left( \prod_{a=1}^N X(u, \epsilon_{i(a)})\right)^{-1}  \det\left(J_N(u)\right), \qquad \forall u \in \mathbb{C},
\end{equation}
with $J_N(u)$ an $N \times N$ matrix defined as
\begin{equation}
J_N(u)_{ab}=\begin{cases}
\sum_{c=1}^N Z(\epsilon_{i(a)},v_c)-\sum_{c \neq a}^N Z(\epsilon_{i(a)},\epsilon_{i(c)}) - Z(\epsilon_{i(a)},u)  &\text{if } a=b,\\
-X(\epsilon_{i(a)},\epsilon_{i(b)})  &\text{if } a \neq b.
\end{cases}
\end{equation}
The left-hand side contains a prefactor which can be absorbed in the Bethe states, whereas the only dependence on the rapidities in the right-hand side can be found in the diagonal elements, which can be re-expressed as
\begin{equation}
J_N(u)_{ab}=\begin{cases}
\Lambda_{i(a)}-\sum_{c \neq a}^N Z(\epsilon_{i(a)},\epsilon_{i(c)}) - Z(\epsilon_{i(a)},u)  &\text{if } a=b,\\
-X(\epsilon_{i(a)},\epsilon_{i(b)})  &\text{if } a \neq b.
\end{cases}
\end{equation}
In these expressions, $u$ is a gauge freedom which can generally be eliminated once a specific representation for the Gaudin algebra is introduced.

The advantage of this representation is that it only depends on the (real) eigenvalue-based elements, and as such exhibits no singular behaviour at the singular points. This can be compared with the Izergin-Borchardt determinants, where the divergence of the determinant is compensated by a vanishing prefactor and is as such highly prone to numerical noise. The connection between the different determinant representations and their connection to either the rapidities or eigenvalues will be discussed in more detail in the following chapter by introducing specific representations for the XXX and XXZ models, which allows for clearer determinant expressions.

\begin{subappendices}
\makeatletter
\@openrightfalse
\makeatother
\chapter*{Appendices}
\section{Deriving the equations}
\label{EVB:App:Eq}
In this Appendix the equations for the eigenvalue-based variables are derived by making use of the Bethe equations and the Gaudin algebra. All summations with  indices $i,j,k, \dots$ are taken to run over the $L$ levels $\epsilon_i$, while for indices labeled $a,b,c, \dots$ the summations run over the $N$ rapidities $v_a$. Starting from $\Lambda(u) = \sum_{a}Z(u,v_a)$, it is possible to write down
\begin{align}
\left[\Lambda(u)\right]^2 &= \sum_{a,b} Z(u,v_a) Z(u,v_b) = -\sum_{a} \sum_{b \neq a} Z(v_a,u) Z(u,v_b) + \sum_{a} Z(u,v_a)^2 \nonumber\\
&=-\sum_{a} \sum_{b \neq a} \left[Z(v_a,v_b)(Z(v_a,u)+Z(u,v_b))+\Gamma\right] + \sum_{a} Z(u,v_a)^2 \nonumber \\
&= - \Gamma N(N-1) - \sum_{a} \sum_{b \neq a} Z(v_a,v_b)(Z(v_a,u)+Z(u,v_b))+ \sum_{a} Z(u,v_a)^2,
\end{align}
where only the Gaudin equations for $Z(u,v)$ have been used. The dummy-indices $a$ and $b$ can now be exchanged in the second term of the summation, leading to
\begin{align}
\left[\Lambda(u)\right]^2 &= - \Gamma N(N-1) - 2 \sum_{a}Z(v_a,u) \sum_{b \neq a} Z(v_a,v_b)+ \sum_{a} Z(u,v_a)^2.
\end{align}
Now the Bethe equations can be introduced in order to rewrite the summation $b \neq a$,
\begin{align}
\left[\Lambda(u)\right]^2 &= - \Gamma N(N-1) + 2 \sum_{a}Z(v_a,u) \left[\frac{1}{g}+ \sum_{j}  Z(\epsilon_j, v_a)s_j\right]+ \sum_{a} Z(u,v_a)^2 \nonumber \\
&= -\Gamma N(N-1) -\frac{2}{g}\Lambda(u) + 2\sum_{a}\sum_{j \neq i}s_j\left[Z(\epsilon_j,u)(Z(\epsilon_j,v_a)+Z(v_a,u))+\Gamma\right] \nonumber \\
& \qquad \qquad \qquad \qquad + \sum_a Z(u,v_a) \left[Z(u,v_a)-2s_i Z(\epsilon_i,v_a)\right].
\end{align}
This then results in
\begin{align}
[\Lambda(u)]^2=&\Gamma N\left(2\sum_{j \neq i}s_j+1-N\right)-\frac{2}{g}\Lambda(u)+ 2\sum_{j \neq i}^L s_j Z(u,\epsilon_j)\left(\Lambda(u)-\Lambda(\epsilon_j)\right) \nonumber\\
&\qquad \qquad +\sum_{a=1}^N Z(u,v_a)\left[Z(u,v_a)-2 Z(\epsilon_i,v_a) s_i \right].
\end{align}
Evaluating this equation at $u=\epsilon_i$ for $s_i=1/2$ returns the set of eigenvalue-based equations (\ref{EVB:EVB:EVBeq}). Since these present a set of equations for the eigenvalues of commuting operators, this immediately results in an identical set of equations on the operator level.

\section{Eigenvalue-based determinant expressions}
\label{EVB:App:det}
The Gaudin equations can be used to rewrite the permanent in the XXZ model in a structure similar to a Cauchy matrix. Again making use of the reparametrization
\begin{equation}
X(\epsilon_i,v_a)=\frac{X(u,\epsilon_i)X(u,v_a)}{Z(u,\epsilon_i)-Z(u,v_a)}, \qquad Z(\epsilon_i,v_a)=\frac{\Gamma+Z(u,\epsilon_i)Z(u,v_a)}{Z(u,\epsilon_i)-Z(u,v_a)},
\end{equation}
the permanent can be rewritten as
\begin{align}
\per \left[X(\epsilon_{i(a)},v_b)\right] &= \per \left[\frac{X(u,\epsilon_{i(a)})X(u,v_b)}{Z(u,\epsilon_{i(a)})-Z(u,v_b)}\right] \nonumber \\
&= \prod_{a,b=1}^N \left(X(u,\epsilon_{i(a)})X(u,v_b)\right) \per \left[\frac{1}{Z(u,\epsilon_{i(a)})-Z(u,v_b)}\right],
\end{align}
where the matrix has been denoted by its matrix elements and all rows and columns have been multiplied by a constant absorbed in the prefactor. This results in the structure of a Cauchy matrix, for which it was shown by Faribault and Schuricht \cite{faribault_determinant_2012} that this could be reduced to a single determinant
\begin{equation}
\per \left[\frac{1}{Z(u,\epsilon_{i(a)})-Z(u,v_b)}\right]  = \det \left(J_N\right),
\end{equation}
with $J_N$ redefined as
\begin{equation}
J_N(u)_{ab}=
\begin{cases}
\sum_{c=1}^N\frac{1}{Z(u,\epsilon_{i(a)})-Z(u,v_c)}-\sum_{c \neq a}^N \frac{1}{Z(u,\epsilon_{i(a)})-Z(u,\epsilon_{i(c)})}  &\text{if } a=b,\\
\frac{1}{Z(u,\epsilon_{i(a)})-Z(u,\epsilon_{i(b)})}  &\text{if } a \neq b.\\
\end{cases}
\end{equation}
Multiplying each row and column $c$ with $X(u,\epsilon_{i(c)})$ and compensating for these factors in the prefactor, this can be written as
\begin{equation}
\per \left[X(\epsilon_{i(a)},v_b)\right]=\frac{\prod_{a=1}^NX(u,v_a)}{\prod_{a=1}^N X(u,\epsilon_{i(a)})} \det J_N(u),
\end{equation}
with
\begin{equation}
J_N(u)_{ab}=\begin{cases}\sum_{c=1}^N\frac{X(u,\epsilon_{i(a)})^2}{Z(u,\epsilon_{i(a)})-Z(u,v_c)}-\sum_{c \neq a}^N \frac{X(u,\epsilon_{i(a)})^2}{Z(u,\epsilon_{i(a)})-Z(u,\epsilon_{i(c)})}  &\text{if } a=b,\\
\frac{X(u,\epsilon_{i(a)})X(u,\epsilon_{i(b)})}{Z(u,\epsilon_{i(a)})-Z(u,\epsilon_{i(b)})}  &\text{if } a \neq b,
\end{cases}
\end{equation}
where $X(\epsilon_{i(a)},\epsilon_{i(b)})$ can be recognized in the off-diagonal elements and the diagonal elements can be rewritten using the Gaudin algebra until
\begin{equation}
J_N(u)_{ab}=\begin{cases}
\sum_{c=1}^N Z(\epsilon_{i(a)},v_c)-\sum_{c \neq a}^N Z(\epsilon_{i(a)},\epsilon_{i(c)}) - Z(\epsilon_{i(a)},u)  &\text{if } a=b,\\
-X(\epsilon_{i(a)},\epsilon_{i(b)})  &\text{if } a \neq b.
\end{cases}
\end{equation}

\end{subappendices}
\makeatletter
\@openrighttrue
\makeatother



\chapter{Inner products}
\label{chap:innerproducts}


\setlength\epigraphwidth{.4\textwidth}
\epigraph{\emph{Algebra is generous: she often gives more than is asked for.}}{{Jean D'Alembert}}


Having found an efficient way of solving the Bethe equations, the next step consists of extracting physics from the Bethe state. This is commonly done by calculating correlation functions, providing insight in the behaviour of physical observables and making the connection with experiments. Such correlation functions can be calculated in a surprisingly efficient way in integrable models. Here, one of the key results is the well-known Slavnov formula for the inner product between two Bethe states, one with rapidities satisfying the Bethe equations (an on-shell state), and one with arbitrary rapidities (an off-shell state) \cite{slavnov_calculation_1989}. In integrable models the exponentially hard calculation of such inner products can again be reduced to the evaluation of a determinant, a routine task in most calculations. Such determinant expressions provide a basic building block for the calculation of form factors and correlation coefficients from the Bethe states, which has allowed for massive simplifications in the calculations of correlation coefficients in these models \cite{slavnov_calculation_1989,von_delft_algebraic_2002,amico_exact_2002, zhou_superconducting_2002,links_algebraic_2003,faribault_exact_2008,gorohovsky_exact_2011,amico_bethe_2012} and led to applications in various physical contexts, where e.g. their appearance in quantum quenches has attracted a lot of attention (see Ref. \cite{jstatmech_outofequilibrium_2016} and references therein.).

In the previous chapter, it was shown how a Bethe state can be characterized in two different ways -- either in terms of the rapidities or in terms of the eigenvalues of the conserved charges. This dual description is extended in this chapter, showing how inner products can also be calculated in two distinct ways, either best suited to the rapidities or to these eigenvalues. The requirement that one of the states is on-shell arises naturally by demanding that a state has a dual representation. By implicitly combining these different representations, inner products can be recast as domain wall boundary partition functions (DWPFs). This is first done for the rational model, where the structure of all involved matrices in terms of Cauchy matrices is made explicit and used to present a purely algebraic derivation of the Slavnov determinant formula. Whereas all previous results were independent of the realization of the GGA, the results in this chapter are specific to the XXX model, and extensions to XXZ models are presented in the last section. The present chapter is largely based on Ref. \cite{claeys_inner_2017}.

\section{Inner products}
\label{ip:sec:results}
The main object of interest is now the inner product of Bethe states in spin-$1/2$ rational models, where Bethe states are given by
\begin{equation}\label{ip:ov:BetheState}
\ket{v_1 \dots v_N} = \prod_{a=1}^N \left(\sum_{i=1}^L \frac{S_i^+}{\epsilon_i-v_a} \right)\ket{\downarrow \dots \downarrow}.
\end{equation}
In this section, a series of determinant expressions for such inner products will be presented without proof, after which the mathematical identities and underlying reasoning will be presented in later sections. These will prove their use when expanding eigenstates of integrable models in terms of eigenstates of integrable models with different interaction strengths $g$ (e.g. quantum quenches and Floquet dynamics), and when calculating correlation coefficients (e.g. when using the Bethe ansatz as a variational ansatz). 

Suppose we have two sets of rapidities, one ($\{v_a\} = \{v_1 \dots v_N\}$) obtained by solving the Bethe equations, and another ($\{w_b\} = \{w_1 \dots w_N\}$) given by arbitrary complex variables. If the rapidities satisfy the Bethe equations, the resulting Bethe state is also known as an \emph{on-shell} state, whereas a Bethe state defined by arbitrary variables is known as an \emph{off-shell} state.  The inner product between an on-shell and an off-shell state is famously given by Slavnov's determinant formula \cite{slavnov_calculation_1989, korepin_quantum_1993,kitanine_form_1999,zhou_superconducting_2002}
\begin{equation}\label{ip:ov:Slavnov}
\braket{v_1 \dots v_N|w_1 \dots w_N} = \frac{\prod_{b} \prod_{a \neq b} (v_a-w_b)}{\prod_{a<b}(v_b-v_a) \prod_{b<a} (w_b-w_a) } \det S_N(\{v_a\},\{w_b\}),
\end{equation}
with $S_N(\{v_a\},\{w_b\})$ an $N \times N$ matrix defined as
\begin{equation}
S_N(\{v_a\},\{w_b\})_{ab} = \frac{v_b-w_b}{v_a-w_b}\left(\sum_{i=1}^L \frac{1}{(v_a-\epsilon_i)(w_b-\epsilon_i)}-2\sum_{c \neq a}^N \frac{1}{(v_a-v_c)(w_b-v_c)}  \right).
\end{equation}
Such inner products can now be rewritten in two distinct ways, where the connection with both the Bethe equations and the eigenvalue-based equations can be made clear. 
\newline
First, these can be re-expressed as
\begin{equation}
\braket{v_1 \dots v_N|w_1 \dots w_N} =(-1)^N \left(\frac{g}{2}\right)^{L-2N}\det{J_L(\{v_a\},\{w_b\})},
\end{equation}
with $J_L(\{v_a\},\{w_b\})$ an $L \times L$ matrix defined as 
\begin{equation}\label{ip:over:detJ_1}
 J_L(\{v_a\},\{w_b\})_{ij} =
  \begin{cases}
   \frac{2}{g}+\Lambda_i(\{v_a\})+\Lambda_i(\{w_b\})-\sum_{k \neq i}^{L} \frac{1}{\epsilon_i-\epsilon_k} &\text{if}\ i=j, \\
   -\frac{1}{\epsilon_i-\epsilon_j}       & \text{if}\ i \neq j,
  \end{cases}
\end{equation}
in which the dependence on the rapidities can be made explicit as
\begin{equation}\label{ip:over:detJ}
 J_L(\{v_a\},\{w_b\})_{ij} =
  \begin{cases}
   \frac{2}{g}+\sum_{a=1}^N \frac{1}{\epsilon_i-v_a}+\sum_{b=1}^N \frac{1}{\epsilon_i-w_b}-\sum_{k \neq i}^{L} \frac{1}{\epsilon_i-\epsilon_k} &\text{if}\ i=j, \\
   -\frac{1}{\epsilon_i-\epsilon_j}       & \text{if}\ i \neq j.
  \end{cases}
\end{equation}
This matrix only depends on the rapidities through the eigenvalue-based variables in the diagonal elements, so it is not necessary to explicitly know the rapidities in order to evaluate the matrix elements. Furthermore, the diagonal elements have a structure similar to the eigenvalue-based equations
\begin{align}\label{ip:RG:EVBeq}
 \left( \frac{2}{g} + \Lambda_i \right) \Lambda_i-\sum_{k \neq i}^L \frac{\Lambda_i-\Lambda_k}{\epsilon_i-\epsilon_k} = 0, \qquad \textrm{with} \qquad
\Lambda_i \equiv \Lambda_i(\{v_a\})=\sum_{a=1}^N \frac{1}{\epsilon_i-v_a}.
\end{align}
Second, this is equivalent to the related determinant expression
\begin{equation}\label{ip:over:detK}
\braket{v_1 \dots v_N|w_1 \dots w_N} =(-1)^N \det{K_{2N}(\{v_a\},\{w_b\})},
\end{equation}
with $K_{2N}(\{v_a\},\{w_b\})$ a $2N \times 2N$ matrix defined in terms of $\{x_1 \dots x_{2N}\}=$ $\{v_1 \dots v_N\}$ $\cup$ $\{w_1 \dots w_N\}$ as
\begin{equation}
 K_{2N}(\{v_a\},\{w_b\})_{ab} =
  \begin{cases}
  \frac{2}{g}-\sum_{i=1}^L \frac{1}{x_a-\epsilon_i}+\sum_{c \neq a}^{2N} \frac{1}{x_a-x_c} &\text{if}\ a=b, \\
   -\frac{1}{x_a-x_b}       & \text{if}\ a \neq b.
  \end{cases}
\end{equation}
The diagonal elements can now be compared to the Bethe equations
\begin{align}\label{ip:RG:BAE}
\frac{2}{g}-\sum_{i=1}^L \frac{1}{v_a-\epsilon_i} + 2 \sum_{c \neq a}^N \frac{1}{v_a-v_c} = 0, \qquad a=1 \dots N.
\end{align}
These two expressions are clearly related by exchanging the role of the rapidities $\{x_1 \dots x_{2N}\}$ and the inhomogeneities $\{\epsilon_1 \dots \epsilon_L\}$. The orthogonality of two different eigenstates can then easily be shown by exploiting the similarity of the diagonal elements to the eigenvalue-based/Bethe equations, as detailed in Appendix \ref{ip:app:ortho}. These matrices also exhibit the same structure as the Gaudin matrix for the normalization of an on-shell state \cite{gaudin_bethe_2014}, hence the denomination of Gaudin-like matrices. For this kind of matrices, the off-diagonal elements only depend on the difference of two rapidities (inhomogeneities). The diagonal elements then contain two summations, one over all but one rapidities (inhomogeneities), and one over all inhomogeneities (rapidities). Remarkably, these Gaudin-like structures are not limited to Richardson-Gaudin models and have been observed in general integrable models such as the XXZ spin chain \cite{izergin_spontaneous_1999,kitanine_algebraic_2009,kozlowski_surface_2012,brockmann_gaudin-like_2014} and the Lieb-Liniger model \cite{de_nardis_solution_2014,brockmann_overlaps_2014}. Such determinant expressions can then be used to obtain similar expressions for expectation values and form factors, as shown in Appendix \ref{innerproducts:app:formfactors}.

The second main result is the identification of the Cauchy matrix as the fundamental building block underneath all matrix expressions, being directly responsible for the Gaudin-like structure. In the following sections, it will also be shown that the Slavnov determinant (\ref{ip:ov:Slavnov}) can be derived starting from Eq. (\ref{ip:over:detK}), similarly exposing its structure in terms of Cauchy matrices. Two well-known specific cases of inner products have special use in calculations, leading to the Gaudin determinant and the Izergin-Borchardt determinant, and it will also be shown how they fit within this scheme. 

\subsection{Gaudin determinant}
If the two sets of rapidities coincide, Slavnov's determinant expression (\ref{ip:ov:Slavnov}) reduces to the Gaudin determinant for the normalization of Bethe states \cite{gaudin_bethe_2014}, given by
\begin{equation}
\braket{v_1 \dots v_N|v_1 \dots v_N} =\det{G_N(v_1 \dots v_N)},
\end{equation}
with $G_N(v_1 \dots v_N)$ an $N \times N$ matrix defined as 
\begin{equation}
 G_N(v_1 \dots v_N)_{ab} =
  \begin{cases}
   \sum_{i=1}^L\frac{1}{(\epsilon_i-v_a)^2} -2 \sum_{c \neq a}^N \frac{1}{(v_c-v_a)^2}&\text{if}\ a=b, \\
   \frac{2}{(v_a-v_b)^2}       & \text{if}\ a \neq b.
  \end{cases}
\end{equation}
The Gaudin matrix is also obtained by taking the limit of coinciding rapidities for Eq. (\ref{ip:over:detK}). Using elementary row and column operations, the determinant of the $2N \times 2N$ matrix from Eq. (\ref{ip:over:detK}) can in this limit be reduced to that of the $N \times N$ Gaudin matrix. The eigenvalue-based expression for the inner product (\ref{ip:over:detJ}) now leads to an alternative expression for the normalization as
\begin{equation}
\braket{v_1 \dots v_N|v_1 \dots v_N} =(-1)^N \left(\frac{g}{2}\right)^{L-2N}\det{J_L(\{v_a\})},
\end{equation}
with $J_L(\{v_a\})$ an $L \times L$ matrix defined as 
\begin{align}\label{ip:over:detJGaudin}
 J_L(\{v_a\})_{ij} & =
  \begin{cases}
   \frac{2}{g}+2\Lambda_i(\{v_a\})-\sum_{k \neq i}^{L} \frac{1}{\epsilon_i-\epsilon_k} &\text{if}\ i=j, \\
   -\frac{1}{\epsilon_i-\epsilon_j}       & \text{if}\ i \neq j,
  \end{cases} \\ 
  & =
  \begin{cases}
   \frac{2}{g}+2\sum_{a=1}^N \frac{1}{\epsilon_i-v_a}-\sum_{k \neq i}^{L} \frac{1}{\epsilon_i-\epsilon_k} &\text{if}\ i=j, \\
   -\frac{1}{\epsilon_i-\epsilon_j}       & \text{if}\ i \neq j.
  \end{cases}
\end{align}
It has already been mentioned how the structure of this matrix resembles that of the presented determinant expressions (\ref{ip:over:detJ}) and (\ref{ip:over:detK}). This similarity and the related use of both determinant expressions can be even further established, since the Gaudin matrix is identical to the Jacobian of the Bethe equations (\ref{ip:RG:BAE}), a common property of integrable systems \cite{reshetikhin_calculation_1989,gohmann_hubbard_1999,mukhin_norm_2005,slavnov_scalar_2015,basso_asymptotic_2017,hutsalyuk_norm_2017}, while the alternative matrix (\ref{ip:over:detJGaudin}) is exactly the Jacobian of the eigenvalue-based equations (\ref{ip:RG:EVBeq}).

\subsection{Izergin-Borchardt determinant}
\begin{figure}
\begin{center}
\includegraphics[width=\textwidth]{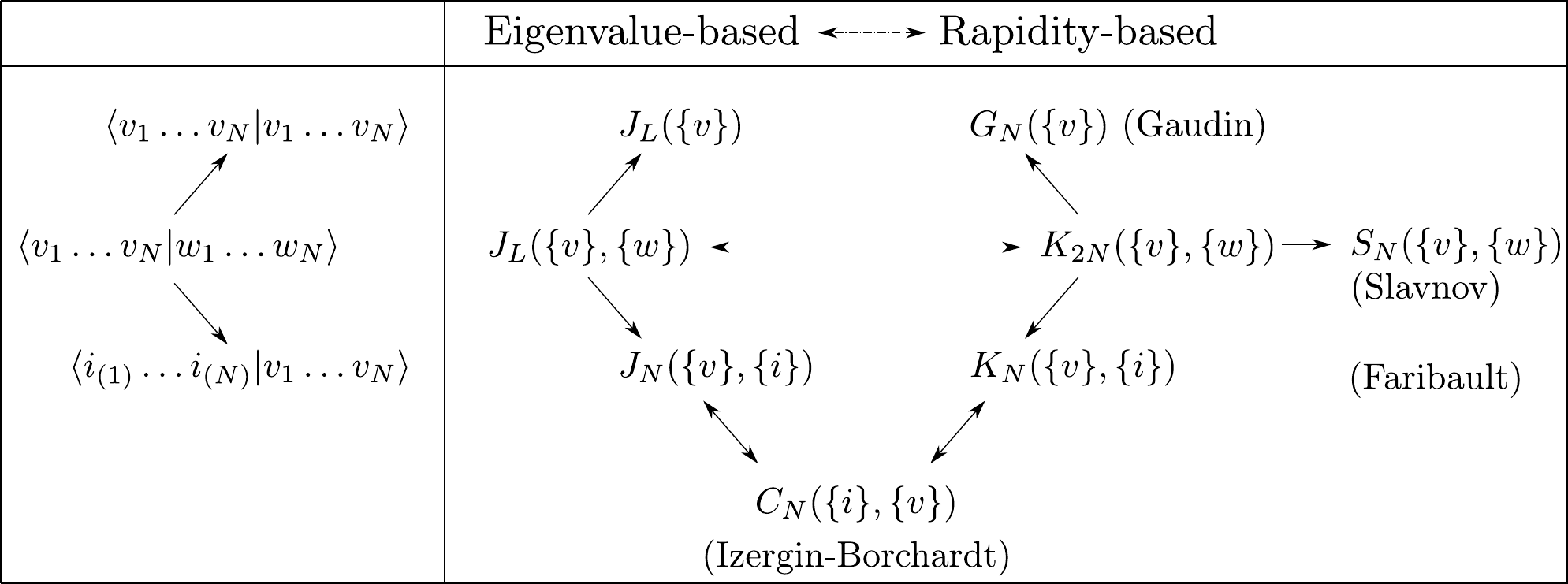}
\caption{Overview of all relevant matrices and their interconnections for inner products in Richardson-Gaudin models.\label{ip:fig:overview}}
\end{center}
\end{figure}

Another frequently encountered case is where the off-shell state simplifies to a simple product state $\ket{i(1) \dots i(N)}=\prod_{b=1}^N S^{+}_{i(b)}\ket{\downarrow \dots \downarrow}$ and Slavnov's determinant expression results in an Izergin-Borchardt determinant \cite{borchardt_bestimmung_1857,izergin_partition_1987,izergin_determinant_1992}
\begin{equation}
\braket{i(1) \dots i(N) | v_1 \dots v_N}=\frac{\prod_{a,b}(\epsilon_{i(b)}-v_a)}{\prod_{c>b}(\epsilon_{i(b)}-\epsilon_{i(c)})\prod_{c<a}(v_a-v_c)} \det \left[\frac{1}{(\epsilon_{i(b)}-v_a)^2}\right],
\end{equation}
which contains the determinant of an $N \times N$ matrix with matrix elements $\frac{1}{(\epsilon_{i(b)}-v_a)^2}$. The alternative matrix following from Eq. (\ref{ip:over:detJ}) corresponding to the Izergin-Borchardt determinant then reads
\begin{equation}
\braket{\{i(b)\}|\{v_a\}}=\det J_N(\{v_a\},\{i(b)\}),
\end{equation}
with $J_N(\{v_a\},\{i(b)\})$ an $N \times N$ matrix defined as 
\begin{equation}
 J_N(\{v_a\},\{i(b)\})_{ab}=
  \begin{cases}
   \sum_{c=1}^N \frac{1}{\epsilon_{i(a)}-v_c}-\sum_{c \neq a}^{N} \frac{1}{\epsilon_{i(a)}-\epsilon_{i(c)}} &\text{if}\ a=b, \\
   -\frac{1}{\epsilon_{i(a)}-\epsilon_{i(b)}}       & \text{if}\ a \neq b,
  \end{cases}
\end{equation}
while Eq. (\ref{ip:over:detK}) gives rise to
\begin{equation}
\braket{\{i(b)\}|\{v_a\}}=\det K_N(\{v_a\},\{i(b)\}),
\end{equation}
with $K_N(\{v_a\},\{i(b)\})$ an $N \times N$ matrix defined as 
\begin{equation}
 K_N(\{v_a\},\{i(b)\})_{ab}=
  \begin{cases}
   -\sum_{c=1}^N \frac{1}{v_a-\epsilon_{i(c)}}+\sum_{c \neq a}^{N} \frac{1}{v_a-v_c} &\text{if}\ a=b, \\
   -\frac{1}{v_a-v_b}       & \text{if}\ a \neq b.
  \end{cases}
\end{equation}
In fact, these results were originally obtained in Ref. \cite{faribault_determinant_2012}, and served as the building blocks for this work. The matrices in Eqs. (\ref{ip:over:detJ}) and  (\ref{ip:over:detJGaudin}) also appeared in Ref. \cite{faribault_determinant_2012}, where it was shown that inner products and normalizations are \emph{proportional} to the determinants of these respective matrices. It is here shown that the proportionality factor can be immediately evaluated as a single constant, independent of both the rapidities and the inhomogeneities of the model. This will be proven in section \ref{ip:sec:connect}, and an overview of all relevant matrices and their counterparts in both frameworks is given in Figure \ref{ip:fig:overview}.

\section{Properties of Cauchy matrices}
\label{ip:sec:cauchy}
In order to derive the previously-presented determinant expressions, the Cauchy structure of all matrices needs to be made clear. Several properties of Cauchy matrices will then be used in order to navigate between different determinant expressions, which will be presented in this section. Necessary proofs can be found in Ref. \cite{claeys_inner_2017}. 

To set the stage, assume we have two sets of variables $\{\epsilon_1 \dots \epsilon_N\}=\{\epsilon_i\}$ and $\{x_1 \dots x_N\}=\{x_{\alpha}\}$. Given two such sets, an $N \times N$ Cauchy matrix $C$ is defined by the matrix elements
\begin{equation}
C_{i\alpha} = \frac{1}{\epsilon_i-x_{\alpha}}.
\end{equation}
The inverse of this matrix is related to the transposed Cauchy matrix through two diagonal matrices \cite{schechter_inversion_1959}, which are defined in terms of two polynomials $p(x)=\prod_{i}(x-\epsilon_i)$ and $q(x)=\prod_{\alpha}(x-x_{\alpha})$ as
\begin{equation}
(D_\epsilon)_{ii} = \frac{q(\epsilon_i)}{p'(\epsilon_i)}, \qquad (D_x)_{\alpha\alpha} = \frac{p(x_{\alpha})}{q'(x_{\alpha})}, 
\end{equation}
such that 
\begin{equation}\label{ip:cauchy:inv}
C^{-1} = -D_{x} C^T D_{\epsilon}.
\end{equation}
Permanents of Cauchy matrices are ubiquitous in the theory of Richardson-Gaudin integrability, and an important result by Borchardt \cite{borchardt_bestimmung_1857} showed how the permanent of a Cauchy matrix can be evaluated as a ratio of determinants given by
\begin{equation}
\per \left[C\right] = \frac{\det \left[C *C \right]}{\det \left[C\right]},
\end{equation}
with $*$ the Hadamard product, defined as $(A*B)_{ij}=A_{ij}B_{ij}$. The denominator can be explicitly evaluated as
\begin{equation}
\det\left[C\right] = \frac{\prod_{j<i}(\epsilon_i-\epsilon_j)\prod_{\alpha < \beta}(x_{\alpha}-x_{\beta})}{\prod_{i, \alpha} (\epsilon_i-x_{\alpha})}.
\end{equation}
However, instead of directly evaluating these determinants, it is also possible to rewrite this as
\begin{equation}
\per \left[C\right] = \det \left[C^{-1}(C*C)\right] = \det\left[(C*C)C^{-1}\right],
\end{equation}
as noted in Ref. \cite{faribault_determinant_2016}. Because of the known structure of the inverse of a Cauchy matrix, these matrices can be explicitly calculated as
\begin{equation}
 C^{-1} (C * C)=D_x J_{x} D_x^{-1}, \qquad (C * C) C^{-1} = D_{\epsilon}^{-1} J_{\epsilon} D_{\epsilon},
\end{equation}
with the $J$-matrices given by 
\begin{align}\label{ip:cauchy:JxJe}
 (J_x)_{\alpha\beta} &=
  \begin{cases}
 - \sum_{i} \frac{1}{x_{\alpha}-\epsilon_i} + \sum_{\kappa \neq \alpha} \frac{1}{x_{\alpha}-x_{\kappa}} &\text{if}\ \alpha=\beta, \\
   -\frac{1}{x_{\alpha}-x_{\beta}}  &\text{if}\ \alpha \neq \beta,
  \end{cases}
  \\
 (J_\epsilon)_{ij} &=
  \begin{cases}
  \sum_{\alpha} \frac{1}{\epsilon_i-x_{\alpha}} - \sum_{k \neq i} \frac{1}{\epsilon_i-\epsilon_k} &\text{if}\ i=j, \\
   -\frac{1}{\epsilon_i-\epsilon_j}  &\text{if}\ i \neq j,
  \end{cases}
\end{align}
and the $D$-matrices are those arising in the equation for the inverse (\ref{ip:cauchy:inv}). This now presents two different ways of evaluating the permanent as a determinant
\begin{equation}\label{ip:cauchy:CvsJ}
\per\left[C\right] = \det\left[J_{\epsilon}\right] =  \det\left[J_{x}\right].
\end{equation}
Because of the product structure of the $J$-matrices, an additional identity holds as
\begin{equation}\label{ip:cauchy:sylv}
\det\left[\mathbbm{1}_N+J_{x}\right] = \det\left[\mathbbm{1}_N+C^{-1} (C * C) \right]=\det\left[\mathbbm{1}_N+(C*C)C^{-1} \right]=\det\left[\mathbbm{1}_N+J_{\epsilon}\right].
\end{equation}
Note that so far all involved matrices were $N \times N$ matrices defined in terms of two sets of variables $\{\epsilon_1 \dots \epsilon_N\}$ and $\{x_1 \dots x_N\}$. Remarkably, this final expression (\ref{ip:cauchy:sylv}) can immediately be generalized towards matrices of different dimensions, defined in terms of sets with a different number of variables, playing an important role in connecting the inner products to DWPFs. Given two such sets $\{\epsilon_1 \dots \epsilon_L\}$ and $\{x_1 \dots x_N\}$, with $L \neq N$, the following also holds
\begin{equation}\label{ip:cauchy:sylv2}
\det\left[\mathbbm{1}_N+J_{x}\right] =\det\left[\mathbbm{1}_L+J_{\epsilon}\right],
\end{equation}
with $J_x$ and $N \times N$ matrix and $J_{\epsilon}$ an $L \times L$ matrix defined as in Eq. (\ref{ip:cauchy:JxJe}), but now in terms of sets with a different number of variables. This result can be obtained by generalizing Eq. (\ref{ip:cauchy:sylv}) towards non-square matrices and applying Sylvester's determinant identity
\begin{equation}
\det\left[\mathbbm{1}_N+AB\right] =\det\left[\mathbbm{1}_L+BA\right],
\end{equation}
with $A$ an arbitrary $N \times L$ matrix and $B$ an arbitrary $L \times N$ matrix.

Returning to $N \times N$ matrices, one final property of Cauchy matrices which will be key in relating the eigenvalue-based determinant to Slavnov's determinant, is that
\begin{equation}\label{ip:cauchy:had3}
2 (C*C)^{-1}(C*C*C)  = D_x \left[J_x+ C^T J_{\epsilon}^{-1} C \right] D_x^{-1}.
\end{equation}
This can be seen as a higher-order extension of the previous equality
\begin{equation}
 C^{-1} (C * C)=D_x J_{x} D_x^{-1},
\end{equation}
and leads to the determinant identity
\begin{equation}
\frac{\det\left[2 \left(C*C*C\right)\right]}{\det \left[C*C\right]} = \det \left[J_x+ C^T J_{\epsilon}^{-1} C \right].
\end{equation}

\section{From eigenvalue-based to Slavnov through dual states}
\label{ip:sec:connect}
\subsection{Dual states}
The crucial element in all eigenvalue-based expressions is the (implicit) existence of dual states for on-shell Bethe states. Any eigenstate of the RG integrable models can be constructed in two different ways: either by creating excitations on top of a vacuum, or by annihilating excitations from a dual vacuum state \cite{faribault_determinant_2012,claeys_eigenvalue-based_2015}. The former approach was used so far, where eigenstates were constructed as
\begin{equation}
\ket{v_1 \dots v_N} = \prod_{a=1}^N \left(\sum_{i=1}^L\frac{S^{+}_i}{\epsilon_i-v_a}\right)\ket{\downarrow \dots \downarrow},
\end{equation}
with rapidities $\{v_1 \dots v_N\}$ satisfying the usual equations
\begin{align}
\frac{1}{g}+\frac{1}{2}\sum_{i=1}^L \frac{1}{\epsilon_i-v_a} - \sum_{b \neq a}^N \frac{1}{v_b-v_a} = 0, \qquad a=1 \dots N.
\end{align}
However, the latter approach states that eigenstates also have a \emph{dual} representation given by
\begin{equation}
\ket{v_1' \dots v_{L-N}'} = \prod_{a=1}^{L-N} \left(\sum_{i=1}^L\frac{S^{-}_i}{\epsilon_i-v_a'}\right)\ket{\uparrow \dots \uparrow},
\end{equation}
with the rapidities of this dual state satisfying
\begin{equation}
-\frac{1}{g}+\frac{1}{2}\sum_{i=1}^L \frac{1}{\epsilon_i-v_a'} -\sum_{b \neq a}^{L-N} \frac{1}{v_b'-v_a'} = 0, \qquad a=1 \dots L-N.
\end{equation}
Note that these equations are related to the previous ones by changing the sign of $g$, which can be seen as a consequence of the spin-flip symmetry of the conserved charges (\ref{RG:spin:Qi}). Because total spin-projection $S^z = \sum_{i=1}^L S_i^z$ is a symmetry of the system, this also implies that the same state will in general be described by a different number of rapidities in both representations.

Two such states are eigenstates of a given integrable Hamiltonian by construction, and these can be made to represent the same eigenstate by demanding that their eigenvalues coincide, leading to
\begin{equation}\label{dual:corr}
\sum_{a=1}^N \frac{1}{\epsilon_i-v_a} = \sum_{a=1}^{L-N} \frac{1}{\epsilon_i-v_a'}-\frac{2}{g}, \qquad i=1 \dots L.
\end{equation}
While the correspondence between the rapidities of both representations is far from intuitive, the correspondence between the eigenvalue-based variables is simple and given by
\begin{equation}
\Lambda_i(\{v_a\}) =\Lambda_i(\{v_a'\})-\frac{2}{g}, \qquad i=1 \dots L.
\end{equation}
A special case is the $g \to \infty$ limit, where both sets of variables are equal up to a number of diverging rapidities going to infinity. In this limit, the conserved charges have an additional $su(2)$ total spin symmetry, which is reflected in $\lim _{v\to \infty} S^+(v) \sim \sum_i S_i^+$, as also observed in Ref. \cite{kostov_inner_2012}.

These two different representations of a single eigenstate obviously have different normalizations, but it is shown in Appendix \ref{ip:app:rat} that the ratio of these normalizations is simply given by
\begin{equation}\label{ip:dual:ratio}
\ket{v_1 \dots v_N} = (-1)^N \left(\frac{g}{2}\right)^{L-2N} \ket{v_1' \dots v_{L-N}'}.
\end{equation}
It is this ratio that is directly responsible for the appearance of the prefactor in Eq. (\ref{ip:over:detJ}), as will now be shown.

\subsection{Inner products as DWPFs}
\begin{figure}
    \centering
    \begin{subfigure}[b]{0.45\textwidth}
        \includegraphics[width=\textwidth]{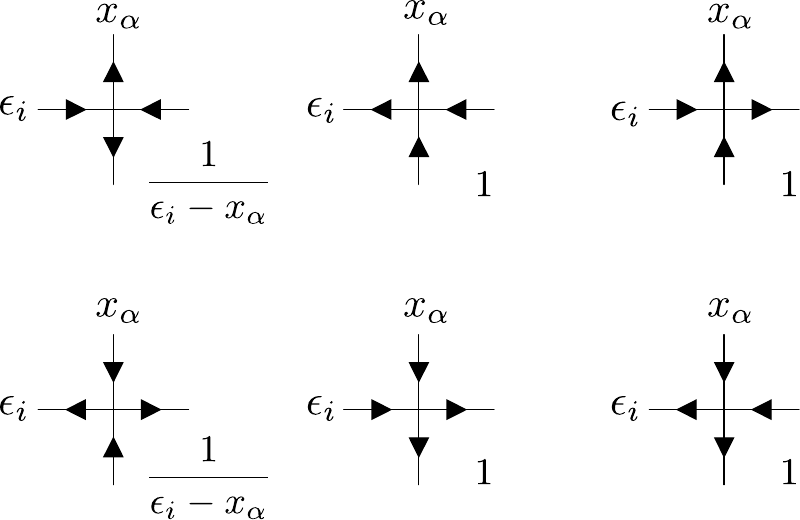}
        \caption{Vertices with non-zero weight.}
    \end{subfigure}\hspace{0.15\textwidth}
    \begin{subfigure}[b]{0.3\textwidth}
        \includegraphics[width=\textwidth]{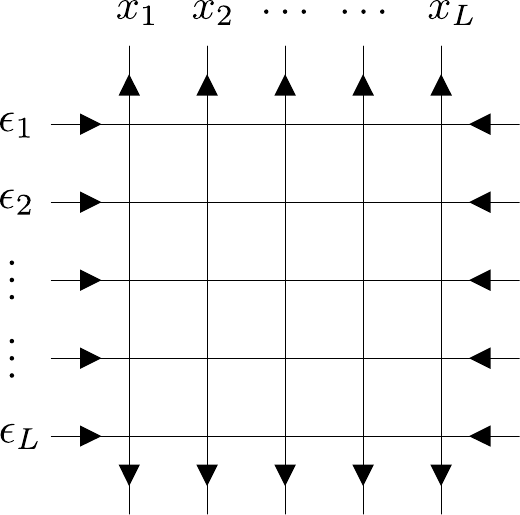}
        \caption{Boundary conditions.}
    \end{subfigure}
    \caption{Graphical representation of the DWPF associated with the Bethe states. The partition function sums over all configurations of vertices in the bulk consistent with the boundary conditions, with the total weight of each configuration given by the product of the weights of the vertices. The weights in the first column follow from $\bra{\uparrow} S^{+}(x_{\alpha}) \ket{\downarrow}_i$.\label{ip:fig:DWPF}}
\end{figure}

Since the existence of dual states is guaranteed for on-shell Bethe states, it is always possible to express inner products in terms of the dual state. The overlap between a (possibly off-shell) state $\ket{w_1 \dots w_N}$ and an on-shell state $\ket{v_1 \dots v_N}$ can now be written as
\begin{equation}
\braket{v_1 \dots v_N | w_1 \dots w_N} = (-1)^N \left(\frac{g}{2}\right)^{L-2N} \braket{v_1' \dots v_{L-N}' | w_1 \dots w_N}.
\end{equation}
The overlap between a dual state and a normal state can be written out as
\begin{align}
\braket{v_1' \dots v_{L-N}' | w_1 \dots w_N} &= \bra{\uparrow \dots \uparrow} \prod_{a=1}^{L-N} \left(\sum_{i=1}^L \frac{S_i^{+}}{\epsilon_i-v'_a} \right)  \prod_{b=1}^{N} \left(\sum_{i=1}^L \frac{S_i^{+}}{\epsilon_i-w_b} \right) \ket{\downarrow \dots \downarrow}\\
&=\bra{\uparrow \dots \uparrow} \prod_{\alpha=1}^L \left(\sum_{i=1}^L \frac{S_i^{+}}{\epsilon_i-x_{\alpha}} \right) \ket{\downarrow \dots \downarrow},
\end{align}
which can be interpreted as the overlap of a Bethe state defined by $L$ rapidities $\{x_{\alpha}\} = \{x_1 \dots x_L\} = \{v'\} \cup \{w\}$ with the dual vacuum $\ket{\uparrow \dots \uparrow}$, 
\begin{equation}\label{ip:DWPF:per}
\bra{\uparrow \dots \uparrow} \prod_{\alpha=1}^L S^{+}(x_{\alpha}) \ket{\downarrow \dots \downarrow} = \sum_{\sigma \in S_L} \prod_{i=1}^L \frac{1}{\epsilon_i-x_{\sigma(i)}},
\end{equation}
which is known as a domain wall boundary partition function (DWPF), as introduced by Korepin \cite{korepin_calculation_1982}. This partition function also has a graphical interpretation, as illustrated in Figure \ref{ip:fig:DWPF}. 

For the models at hand and following Section \ref{RG:sec:expand}, the expression (\ref{ip:DWPF:per}) is exactly the definition of the permanent of a Cauchy matrix
\begin{equation}
\bra{\uparrow \dots \uparrow} \prod_{\alpha=1}^L S^{+}(x_{\alpha}) \ket{\downarrow \dots \downarrow} = \per \left[ C\right],
\end{equation}
with $C$ the Cauchy matrix defined as
\begin{equation}
C_{i\alpha} = \frac{1}{\epsilon_i-x_{\alpha}}.
\end{equation}
Using the results on Cauchy matrices from the previous section (\ref{ip:cauchy:CvsJ}), this permanent can be rewritten as a determinant 
\begin{equation}
 \per \left[ C\right] = \det \left[J_{\epsilon}\right],
\end{equation}
with $J_{\epsilon}$ an $L \times L$ matrix defined as
\begin{equation}
 \left(J_{\epsilon}\right)_{ij}=
  \begin{cases}
   \sum_{\alpha=1}^L \frac{1}{\epsilon_{i}-x_{\alpha}}-\sum_{k \neq i}^{L} \frac{1}{\epsilon_{i}-\epsilon_{k}} &\text{if}\ i=j, \\
   -\frac{1}{\epsilon_{i}-\epsilon_{j}}       & \text{if}\ i \neq j.
  \end{cases}
\end{equation}
Reintroducing the rapidities $\{x_1 \dots x_L\} = \{v'\} \cup \{w\}$ in the diagonal elements leads to
\begin{equation}
 \left(J_{\epsilon}\right)_{ij}=
  \begin{cases}
   \sum_{a=1}^{L-N} \frac{1}{\epsilon_{i}-v_a'}+ \sum_{b=1}^{N} \frac{1}{\epsilon_{i}-w_b}-\sum_{k \neq i}^{L} \frac{1}{\epsilon_{i}-\epsilon_{k}} &\text{if}\ i=j, \\
   -\frac{1}{\epsilon_{i}-\epsilon_{j}}       & \text{if}\ i \neq j,
  \end{cases}
\end{equation}
where, because of the correspondence between the eigenvalues of the original state and the dual state (\ref{dual:corr}), this can be written entirely in terms of the original rapidities
\begin{equation}\label{ip:DWPF:resJ}
 \left(J_{\epsilon}\right)_{ij}=
  \begin{cases}
   \frac{2}{g}+\sum_{a=1}^{N} \frac{1}{\epsilon_{i}-v_a}+ \sum_{b=1}^{N} \frac{1}{\epsilon_{i}-w_b}-\sum_{k \neq i}^{L} \frac{1}{\epsilon_{i}-\epsilon_{k}} &\text{if}\ i=j, \\
   -\frac{1}{\epsilon_{i}-\epsilon_{j}}       & \text{if}\ i \neq j,
  \end{cases}
\end{equation}
resulting in the proposed determinant expression (\ref{ip:over:detJ}) for the inner product if the ratio of normalizations for the original state and the dual state (\ref{ip:dual:ratio}) is taken into account. Note that the existence of the dual state was necessary to derive these determinant expressions, but is only implicit in the final results due to the demand that one of the two states in the inner product must be on-shell. 

At this point, the first proposed expression (\ref{ip:over:detJ}) has been derived, which is expressed in the eigenvalue-based variables. In order to derive the equivalent expression (\ref{ip:over:detK}) for the rapidities, the crucial observation is that the matrix (\ref{ip:DWPF:resJ}) has the structure $(\mathbbm{1}_{L}+\cdots)$ after multiplying all matrix elements with $g/2$. This leads to the exact same structure obtained when applying Sylvester's determinant identity in Eq. (\ref{ip:cauchy:sylv2}). The involved matrices are now defined in terms of two sets of variables of unequal size, being the $L$ inhomogeneities $\{\epsilon_1 \dots \epsilon_L\}$ and the $2N$ combined rapidities $\{v_1 \dots v_N\} \cup \{w_1 \dots w_N\}$. Applying the relation (\ref{ip:cauchy:sylv2}) then connects the determinant of the $L \times L$ matrix $(\mathbbm{1}_{L}+\cdots)$ to that of an $2N \times 2N$ matrix $(\mathbbm{1}_{2N}+\cdots)$, equaling the matrix proposed in (\ref{ip:over:detK}) after correcting for the missing prefactors of $(g/2)$ in the matrix elements by changing the prefactor of the determinant. Afterwards, the related Gaudin and Izergin-Borchardt determinants follow immediately by taking the appropriate limits.

\subsection{Reduction to Slavnov's determinant}
Starting from the eigenvalue-based inner product (\ref{ip:over:detJ}), it is possible to rederive Slavnov's determinant expression, obtaining other determinant expressions in the process and shedding some light on the structure of all involved identities.

First, by making using of the Bethe equations (\ref{ip:RG:BAE}), the well-known Slavnov's determinant expression (\ref{ip:ov:Slavnov}) can be straightforwardly rewritten as
\begin{equation}\label{ip:slavnov:slavnov}
\braket{v_1 \dots v_N|w_1 \dots w_N} = \frac{\prod_{b} \prod_{a} (v_a-w_b)}{\prod_{b<a} (w_b-w_a) \prod_{a<b}(v_b-v_a) } \det S_N(\{v_a\},\{w_b\}),
\end{equation}
\begin{equation}
S_N(\{v_a\},\{w_b\})_{ab} = \frac{1}{(v_a-w_b)^2}\left(\sum_{i=1}^L \frac{1}{w_b-\epsilon_i}-2\sum_{c \neq a}^N \frac{1}{w_b-v_c} -\frac{2}{g} \right).
\end{equation}
Note that, by rewriting the matrix elements in this way, the only explicit dependence on the inhomogeneities is through $\sum_{i}\frac{1}{w_b-\epsilon_i}$, bringing to mind the eigenvalue-based determinants and Eq. (\ref{ip:over:detK}). The prefactor is the inverse of the determinant of the Cauchy matrix $U$ defined by 
\begin{equation}\label{ip:slavnov:U}
U_{ab}=\frac{1}{v_a-w_b}.
\end{equation}
Defining a diagonal $N \times N$ matrix $\Lambda$ in order to absorb the dependency on the inhomogeneities as 
\begin{equation}
\Lambda_{aa}=\sum_{i=1}^L \frac{1}{w_a-\epsilon_i}-2\sum_{c=1 }^N \frac{1}{w_a-v_c}-\frac{2}{g},
\end{equation}
the matrix in Slavnov's determinant can be decomposed as
\begin{equation}
S_N(\{v_a\},\{w_b\})  =  (U*U)  \Lambda - 2 (U*U*U).
\end{equation}
This can be summarized in
\begin{equation}
\braket{v_1 \dots v_N|w_1 \dots w_N} = \det \left[ U \right]^{-1} \det\left[  (U*U)  \Lambda - 2 (U*U*U)\right].
\end{equation} 

It is now possible to work towards this expression starting from the eigenvalue-based expressions using the presented properties of Cauchy matrices. The inner product as given in Eq. (\ref{ip:over:detK}) depends on a $2N \times 2N$ matrix, which can be interpreted as a $2 \times 2$ block matrix of $N \times N$ matrices as
\begin{equation}
\braket{v_1 \dots v_N|w_1 \dots w_N} =(-1)^N \det{K_{2N}(\{v_a\},\{w_b\})},
\end{equation}
with 
\begin{equation}
K_{2N}(\{v_a\},\{w_b\})=\begin{bmatrix}
J_v & -U \\
U^T & J_w-\Lambda
\end{bmatrix}
\end{equation}
and the diagonal matrices are given by 
\begin{align}
 \left(J_v\right)_{ab} &=
  \begin{cases}
   \frac{2}{g}-\sum_{i=1}^L \frac{1}{v_a-\epsilon_i}+\sum_{c \neq a}^N\frac{1}{v_a-v_c}+\sum_{c=1}^N\frac{1}{v_a-w_c} &\text{if}\ a=b, \\
   -\frac{1}{v_a-v_b}       & \text{if}\ a \neq b,
  \end{cases}   \\
\left(J_w\right)_{ab} &=
  \begin{cases}
   -\sum_{c=1}^N\frac{1}{w_a-v_c} + \sum_{c \neq a}^N\frac{1}{w_{a}-w_{c}} &\text{if}\ a=b, \\
   -\frac{1}{w_a-w_b}       & \text{if}\  a\neq b.
  \end{cases}
\end{align}
The dependence of $J_w$ on the inhomogeneities can be absorbed in the same diagonal matrix $\Lambda$ as defined for Slavnov's determinant and the off-diagonal matrices are then determined by the same Cauchy matrix (\ref{ip:slavnov:U}) defined previously. Using the Bethe equations (\ref{ip:RG:BAE}) for $\{v_a\}$ in the diagonal elements of $J_v$ then leads to
\begin{equation}
 \left(J_v\right)_{ab} =
  \begin{cases}
   -\sum_{c \neq a}^N\frac{1}{v_a-v_c}+\sum_{c=1}^N\frac{1}{v_a-w_c} &\text{if}\ a=b, \\
   -\frac{1}{v_a-v_b}       & \text{if}\ a \neq b.
  \end{cases}
\end{equation}
By rewriting the matrix in this way, all submatrices exhibit the structures previously introduced since $J_v \sim  U^{-1} (U*U)$ and $J_w \sim (U*U) U^{-1}$. One final step now consists of relating the determinant of a $2N \times 2N$ matrix to that of an $N \times N$ one. For this, it is possible to evaluate the determinant of a $2 \times 2$ block matrix as
\begin{equation}
\det \begin{bmatrix}
A & B \\
C & D
\end{bmatrix} = \det \left(A\right)\det\left(D-C A^{-1}B\right).
\end{equation}
Applying this to the equation for the inner product yields
\begin{align}(-1)^N\det \begin{bmatrix}
J_v & -U \\
U^T & J_w-\Lambda
\end{bmatrix} 
&=(-1)^N \det(J_v)\det(J_w-\Lambda+U^TJ_v^{-1}U) \nonumber\\
&=\det(U)^{-1} \det(U*U) \det\left[\Lambda-2 (U*U)^{-1} (U*U*U)\right]\nonumber\\
&=\det(U)^{-1} \det\left[(U*U) \Lambda -2(U*U*U)\right] ,
\end{align}
where $\det(J_v) = \det(U*U)/\det(U)$ and Eq. (\ref{ip:cauchy:had3}) have been used to evaluate the inverse of $J_v$ as
\begin{align}
-D_w(J_w-\Lambda+U^TJ_v^{-1}U) D_w^{-1} = \Lambda-2 (U*U)^{-1} (U*U*U).
\end{align}
This corresponds exactly to Slavnov's determinant expression (\ref{ip:slavnov:slavnov}). Alternative ways of calculating the determinant of the block matrix would result in alternative $N \times N$ matrices, but their structure is more involved than that of the Slavnov determinant. Again, the on-shell requirement for one of the two states is crucial, arising here due to the implied existence of a dual state.

In conclusion, there are three ways of evaluating the inner product - by calculating the determinants of $L \times L$, $2N \times 2N$, or $N \times N$ matrices. The structure of all these matrices is intricately related to Cauchy matrices, where the $2N \times 2N$ matrix can be reduced to the $N \times N$ Slavnov determinant, while the $L \times L$ matrix follows from the eigenvalue-based framework. Similar results were also obtained in Refs. \cite{kostov_inner_2012} and \cite{foda_variations_2012} for the rational six-vertex model, connecting DWPFs with Slavnov determinants, however without invoking the dual representation of Bethe states. Alternatively, similar results were obtained by Kitanine \emph{et al.} for the integrable XXX Heisenberg chain through both a separation of variables and an ABA approach \cite{kitanine_determinant_2016}.

\section{Extension to hyperbolic models}
\label{ip:sec:hyp}
The presentation thus far was focused on the rational (XXX) Richardson-Gaudin models, because of their clear-cut connection with Cauchy matrices. However, it is possible to extend all results to the hyperbolic (XXZ) Richardson-Gaudin case, for which we will only provide an overview and refer to Ref. \cite{claeys_inner_2017} for a detailed derivation. Because of their richer phase diagram and various symmetries, these expressions are slightly more involved. \cite{ibanez_exactly_2009,skrypnyk_spin_2009,skrypnyk_non-skew-symmetric_2009,dunning_exact_2010,rombouts_quantum_2010,van_raemdonck_exact_2014,lukyanenko_boundaries_2014}. This class is now defined by a similar set of commuting operators\footnote{We follow the presentation from Refs.\ \cite{lukyanenko_boundaries_2014,lukyanenko_integrable_2016}, which differs from the usual one in the asymmetry in the last term. However, this allows for a clearer presentation and can be connected to the fact that these models can also be constructed from a non-skew-symmetric $r$-matrix \cite{skrypnyk_non-skew-symmetric_2009,skrypnyk_non-skew-symmetric_2009-1,skrypnyk_z_2016}.} given by
\begin{equation}\label{hyp:com}
Q_i = \left(S_i^z+\frac{1}{2}\right) + g \sum_{j\neq i}^L \frac{1}{\epsilon_i-\epsilon_j}\left[\sqrt{\epsilon_i \epsilon_j} \left(S_i^{+}S_j^{-}+S_i^{-}S_j^{+}\right) +2 \epsilon_i \left(S_i^zS_j^z-\frac{1}{4}\right)\right],
\end{equation}
and this section concerns the inner products of Bethe states
\begin{equation}
\ket{v_1 \dots v_N} = \prod_{a=1}^N \left(\sum_{i=1}^L\frac{\sqrt{\epsilon_i}}{\epsilon_i-v_a}S^{+}_i\right)\ket{\downarrow \dots \downarrow}.
\end{equation}
The normalizations of the original and the dual state are now related through (for $L-2N > 0$)
\begin{equation}\label{hyp:rat}
\ket{v_1' \dots v_{L-N}'}  = (-1)^N \frac{\prod_{a=1}^N v_a}{\prod_{i=1}^L \sqrt{\epsilon_i}} \left[\prod_{k=1}^{L-2N}\left(g^{-1}+1-k\right)\right]\ket{v_1 \dots v_N}.
\end{equation}
This is a more involved expression compared to the rational model (\ref{ip:dual:ratio}), which can be seen as a consequence of the various symmetries and dualities present in this model  -- for specific values $g^{-1} \in \mathbb{Z}$ the dual wave function vanishes because some of the dual rapidities diverge \cite{links_exact_2015}. Similar terms also arise in the investigation of the phase diagram of these models \cite{ibanez_exactly_2009,rombouts_quantum_2010,van_raemdonck_exact_2014}. 

From this, it is possible to obtain a similar set of determinant expressions for inner products in the hyperbolic model, leading to (for on-shell $\{v_a\}$)
\begin{equation}\label{hyp:detJ}
\braket{v_1 \dots v_N|w_1 \dots w_N} = (-1)^N \left(\frac{\prod_{i=1}^L \epsilon_i}{\prod_{a=1}^N v_a}\right) \left[\prod_{k=1}^{L-2N}\left(g^{-1}+1-k\right)\right]^{-1} \det J_L(\{v_a\},\{w_b\}),
\end{equation}
with $J_L(\{v_a\},\{w_b\})$ an $L \times L$ matrix given by
\begin{equation}
 J_L(\{v_a\},\{w_b\})_{ij} =
  \begin{cases}
   \frac{g^{-1}}{\epsilon_i}+\sum_{a=1}^N \frac{1}{\epsilon_i-v_a}+\sum_{b=1}^N \frac{1}{\epsilon_i-w_b}-\sum_{k \neq i}^{L} \frac{1}{\epsilon_i-\epsilon_k} &\text{if}\ i=j, \\
   -\frac{1}{\epsilon_i-\epsilon_j}       & \text{if}\ i \neq j.
  \end{cases}
\end{equation}
The diagonal elements of these matrices again reflect the structure of the eigenvalue-based equations
\begin{equation}\label{ip:hyp:EVBeq}
\epsilon_i \Lambda_i \left(\frac{g^{-1}}{\epsilon_i}+ \Lambda_i  \right) - \sum_{j \neq i}^L \frac{\epsilon_i \Lambda_i - \epsilon_j \Lambda_j}{\epsilon_i-\epsilon_j} = 0, \qquad \Lambda_i \equiv \Lambda_i(\{v_a\}) = \sum_{a=1}^N \frac{1}{\epsilon_i-v_a},
\end{equation}
which can also be used to show the orthogonality of different eigenstates. When calculating the normalization of an on-shell state, this reduces to the Jacobian of the eigenvalue-based equations (\ref{ip:hyp:EVBeq}), where all derivatives are now taken w.r.t. $\epsilon_i \Lambda_i$ instead of $\Lambda_i$. Alternatively, there is again a direct equivalence with the determinant of a $2N \times 2N$ matrix defined in terms of $\{x_1 \dots x_{2N}\}=\{v_1 \dots v_N\} \cup \{w_1 \dots w_N\}$ as
\begin{equation}
\braket{v_1 \dots v_N|w_1 \dots w_N} = (-1)^N \left(\prod_{b=1}^N w_b\right) \det K_{2N}(\{v_a\},\{w_b\}),
\end{equation}
with $K_{2N}(\{v_a\},\{w_b\})$ a $2N \times 2N$ matrix given by
\begin{equation}\label{ip:hyp:detK}
 K_{2N}(\{v_a\},\{w_b\})_{ab} =
  \begin{cases}
  \frac{1+g^{-1}}{x_a}-\sum_{i=1}^L \frac{1}{x_a-\epsilon_i}+\sum_{c \neq a}^{2N} \frac{1}{x_a-x_c}  &\text{if}\ a=b, \\
   -\frac{1}{x_a-x_b}       & \text{if}\ a \neq b,
  \end{cases}
\end{equation}
where the diagonal elements now clearly reflect the Bethe equations
\begin{equation}\label{ip:hyp:RGeq}
\frac{1+g^{-1}}{v_a}-\sum_{i=1}^L \frac{1}{v_a-\epsilon_i} + 2\sum_{c \neq a}^N \frac{1}{v_a-v_c}=0, \qquad \forall a=1 \dots N.
\end{equation}
The Slavnov determinant for the inner product of an on-shell and an off-shell state in the XXZ model \cite{dunning_exact_2010} then follows from evaluating this matrix as a block matrix, where all steps in the derivation are fundamentally the same as for the rational XXX model. The resulting expression is given by
\begin{equation}\label{hyp:detS}
\braket{v_1 \dots v_N|w_1 \dots w_N} = \frac{\prod_{a} \prod_{b} (v_a-w_b)}{ \prod_{a<b}(v_b-v_a) \prod_{b<a} (w_b-w_a)} \det S_N(\{v_a\},\{w_b\}),
\end{equation}
with $S_N(\{v_a\},\{w_b\})$ an $N \times N$ matrix given by
\begin{equation}
S_N(\{v_a\},\{w_b\})_{ab} = \frac{w_b}{(v_a-w_b)^2}\left(\sum_{i=1}^L \frac{1}{w_b-\epsilon_i}-2\sum_{c \neq a}^N \frac{1}{w_b-v_c} - \frac{g^{-1}+1}{w_b} \right).
\end{equation}

\begin{subappendices}
\makeatletter
\@openrightfalse
\makeatother
\chapter*{Appendices}
\section{Orthogonality of Bethe states}
\label{ip:app:ortho}
Starting from the determinant expressions of Eq. (\ref{ip:over:detJ_1}) or Eq. (\ref{ip:over:detK}), the orthogonality of different Bethe states can easily be shown. Starting from two non-equal on-shell states $\{v_a\}$ and $\{w_b\}$, it is straightforward to show that the rows of (\ref{ip:over:detJ}) are linearly dependent as
\begin{align}
\sum_{i=1}^L &\left[\Lambda_{i}(\{v_a\}) - \Lambda_{i}(\{w_b\})\right] J_L(\{v_a\},\{w_b\})_{ij}   \nonumber \\
=& \left[\Lambda_j(\{v_a\})^2 + \frac{2}{g} \Lambda_j(\{v_a\}) -\sum_{i \neq j}^L \frac{\Lambda_j(\{v_a\})-\Lambda_i(\{v_a\})}{\epsilon_j-\epsilon_i}\right] \nonumber \\
&-\left[\Lambda_j(\{w_b\})^2 + \frac{2}{g} \Lambda_j(\{w_b\}) -\sum_{i \neq j}^L \frac{\Lambda_j(\{w_b\})-\Lambda_i(\{w_b\})}{\epsilon_j-\epsilon_i}\right] = 0,
\end{align}
which vanish since these are exactly the eigenvalue-based equations satisfied by on-shell states (\ref{ip:RG:EVBeq}). The linear dependence of the rows implies that the determinant of this matrix also vanishes, proving the orthogonality of different eigenstates

The orthogonality also follows from the similarity of the diagonal elements of (\ref{ip:over:detK}) to the Bethe equations (\ref{ip:RG:BAE}). Taking $\{x_1 \dots x_N\} = \{v_1 \dots v_N\}$ and $\{x_{N+1} \dots x_{2N}\} = \{w_1 \dots w_N\}$, this results in
\begin{align}
\sum_{c=1}^N &K_{2N}(\{v_a\},\{w_b\})_{cb} - \sum_{c=N+1}^{2N} K_{2N}(\{v_a\},\{w_b\})_{cb}  \nonumber \\
  &= \begin{cases}
   \frac{2}{g}+\sum_{i=1}^L \frac{1}{\epsilon_i-v_b} - 2 \sum_{c \neq b}^N \frac{1}{v_c-v_b} = 0 
 &\text{if}\ b=1 \dots N, \\
   \frac{2}{g}+\sum_{i=1}^L \frac{1}{\epsilon_i-w_{b}} - 2 \sum_{c \neq b}^N \frac{1}{w_c-w_b} = 0    & \text{if}\ b=N+1 \dots 2N,
  \end{cases}
\end{align}
where we have identified $w_b$ and $w_{N+b}$ in the second line. Since these are exactly the Bethe equations (\ref{ip:RG:BAE}), the rows are again linearly dependent, leading to a vanishing determinant and orthogonal eigenstates.

\section{From inner products to form factors}
\label{innerproducts:app:formfactors}
The method of obtaining form factors starting from inner products will be illustrated on the off-diagonal matrix elements $\braket{v_1 \dots v_N | S_k^z| w_1 \dots w_N} = \braket{\{v_a\}| S_k^z| \{w_b\}}$, where both states are assumed to be on-shell. The first step consists of writing this matrix element as a summation over inner products, for which various determinant expressions have been presented. This can be done using the standard commutation relations, where the action of $S_k^z$ on $\ket{\{w_b\}}$ can be found as
\begin{align}
&S_k^z  \ket{\{w_b\}} = S_k^z \prod_{b=1}^N S^+(w_b) \ket{\downarrow \dots \downarrow} \nonumber \\
&\qquad \qquad = \sum_{\tilde{b}=1}^N \left( \prod_{c \neq \tilde{b}}^N S^+(w_c) \right)[S_k^z, S^+(w_{\tilde{b}})]\ket{\downarrow \dots \downarrow} + \left(\prod_{b=1}^N S^+(w_b)\right) S_k^z \ket{\downarrow \dots \downarrow},
\end{align}
which can be related to the Gaudin algebra elements as
\begin{equation}
[S_k^z,S^+(w_{\tilde{b}})] = \frac{S_k^+}{\epsilon_k-w_{\tilde{b}}} = \lim_{\lambda \to \epsilon_k} \frac{\epsilon_k-\lambda}{\epsilon_k-w_{\tilde{b}}} S^+(\lambda).
\end{equation}
Taking the inner product with the on-shell state $\ket{\{v_a\}}$, this matrix element can be written as
\begin{align}\label{ip:expansionSz}
\braket{\{v_a\}| S_k^z| \{w_b\}} =&  \lim_{\lambda \to \epsilon_k} \sum_{\tilde{b}=1}^N   \frac{\epsilon_k-\lambda}{\epsilon_k-w_{\tilde{b}}} \braket{v_1 \dots v_N| w_1 \dots w_{\tilde{b}} \to \lambda \dots w_N}-\frac{1}{2}\braket{\{v_a\}| \{w_b\}},
\end{align}
leading to a summation of inner products between an on-shell state $\ket{v_1 \dots v_N}$ and a set of off-shell states $\ket{ w_1 \dots w_{\tilde{b}} \to \lambda \dots w_N}$. Since the two states were assumed to be orthogonal, the term $-\frac{1}{2}\braket{\{v_a\}| \{w_b\}}$ generally vanishes, although this can easily be taken into account otherwise.

\subsection{Rapidity-based form factors}
This can be rewritten as the determinant of a single matrix starting from Slavnov's determinant expression. The matrix representation from Eq. (\ref{ip:slavnov:slavnov}) has the advantage that the only dependence on the rapidity $w_{b}$ is in column $b$, so each term in the summation (\ref{ip:expansionSz}) returns a matrix where only a single column $\tilde{b}$ depends on $\lambda$. The limit $\lambda \to \epsilon_k$ can be evaluated in this column, returning
\begin{equation}
\lim_{\lambda \to \epsilon_k} (\epsilon_k-\lambda) S_N(\{v_a \},\{w_b\}, w_{\tilde{b}} \to \lambda )_{a\tilde{b}} = -\frac{1}{(v_a-\epsilon_k)^2}.
\end{equation}
Defining this column as $Q$ and denoting the separate columns of the matrix $S_N(\{v_a\},\{w_b\}) $ $=$  $ \left(S_1| S_2 | \dots |S_N \right)$, this gives rise to
\begin{align}
\braket{\{v_a\}| S_k^z| \{w_b\}} = - \sum_{\tilde{b}=1}^N  \frac{\prod_{a=1}^N(v_a-\epsilon_k)}{\prod_{a=1}^N (w_a - \epsilon_k)} \frac{\prod_{a,c \neq \tilde{b}} (v_a-w_c) \prod_{c \neq \tilde{b}}(w_c-w_{\tilde{b}})}{\prod_{a<c}(v_c-v_a)\prod_{a>c} (w_c-w_a)} \nonumber \\
\times  \det \left(S_1| \dots | S_{\tilde{b}-1}|Q| S_{\tilde{b}+1}| \dots | S_N\right).
\end{align}
All dependence on the summation index $\tilde{b}$ can be absorbed in the determinant by defining
\begin{align}
\left(T\right)_{ab} =  S_N(\{v_a\},\{w_b\})_{ab} \cdot \prod_{c=1}^N (v_c-w_b), \qquad \tilde{Q}_{ab} = \frac{1}{(v_a-\epsilon_k)^2}\cdot \prod_{c \neq b}^N(w_c-w_b).
\end{align}
Writing out both matrices as $T = (T_1 | T_2 | \dots| T_N )$ and $\tilde{Q} = (\tilde{Q}_1 | \tilde{Q}_2 | \dots| \tilde{Q}_N )$ leads to
\begin{align}
\braket{\{v_a\}| S_k^z| \{w_b\}} = \sum_{\tilde{b}=1}^N\frac{\prod_{a=1}^N(v_a-\epsilon_k)}{\prod_{a=1}^N (w_a - \epsilon_k)} \frac{\det \left(
T_1 | \dots | T_{\tilde{b}-1}| \tilde{Q}_{\tilde{b}}| T_{\tilde{b}+1}| \dots |T_N \right)}{\prod_{a<c}(v_c-v_a)\prod_{a>c} (w_c-w_a)}.
\end{align}
The determinant is a multilinear function in its columns and the determinant of a matrix containing multiple linearly-dependent columns vanishes by definition, so this summation can be seen as an expansion for the determinant of $T + \tilde{Q}=(T_1+\tilde{Q}_1 | T_2+\tilde{Q}_2 | \dots| T_N+\tilde{Q}_N )$. Expanding the determinant using the multilinearity in the columns, any term in the expansion containing multiple columns of $\tilde{Q}$ will vanish because of their linear dependence. This only leaves terms $\det (T_1 | T_2 | \dots| T_N )= \det(T) = 0$, also vanishing because of orthogonality, and terms $\det \left(
T_1 | \dots | T_{\tilde{b}-1}| \tilde{Q}_{\tilde{b}}| T_{\tilde{b}+1}| \dots |T_N \right)$. The full expression then results in a single determinant
\begin{align}\label{ip:vSzw:rap}
\braket{\{v_a\}| S_k^z| \{w_b\}} = \frac{\prod_{a=1}^N(v_a-\epsilon_k)}{\prod_{a=1}^N (w_a - \epsilon_k)} \frac{\det\left[T + \tilde{Q}\right]}{\prod_{a<c}(v_c-v_a)\prod_{a>c} (w_c-w_a)}.
\end{align}

\subsection{Eigenvalue-based form factors}
Alternatively, the eigenvalue-based framework returns 
\begin{align}\label{ip:vSzw:EVB}
\braket{\{v_a\}| S_k^z| \{w_b\}} =& (-1)^N \left(\frac{g}{2}\right)^{L-2N} \nonumber \\
&\ \ \times \Big(\left[1+\Lambda_k(\{w_b\})\right] \det\left[{J^k_{L-1}}\right]  -  \det \left[{J^k_{L-1}+F^k_{L-1}}\right]\Big),
\end{align}
where two $(L-1) \times (L-1)$ matrices have been introduced. $J^k_{L-1} = J^k_{L-1}(\{v_a\},\{w_b\})$ has the common structure of the $J$-matrices, but the row and column associated with the level $k$ are removed\footnote{The indices $i,j=1 \dots k-1, k+1, \dots L$ now function as labels for $1 \dots L-1$ when denoting matrix elements of a $(L-1) \times (L-1)$ matrix.} and $\epsilon_k$ has been excluded from the diagonal elements as
\begin{align}\label{ip:defJk}
 J^k_{L-1}(\{v_a\},\{w_b\})_{ij} =
  \begin{cases}
   \frac{2}{g}+\Lambda_i(\{v_a\})+\Lambda_i(\{w_b\})-\sum_{l \neq i,k}^{L} \frac{1}{\epsilon_i-\epsilon_l} &\text{if}\ i=j, \\
   -\frac{1}{\epsilon_i-\epsilon_j}       & \text{if}\ i \neq j,
  \end{cases}
\end{align}
and $i,j \neq k$. The second matrix $F^k_{L-1} = F^k_{L-1}(\{w_b\})$ is a matrix where all rows are identical and given by
\begin{equation}\label{ip:defFk}
F^k_{L-1}(\{w_b\})_{ij} = -\frac{\Lambda_k(\{w_b\})-\Lambda_j(\{w_b\})}{\epsilon_k-\epsilon_j},
\end{equation}
where again $i,j \neq k$. The proof of this determinant expression is surprisingly similar to that of Eq. (\ref{ip:vSzw:rap}). From Eq.\ (\ref{ip:expansionSz}), it immediately follows that 
\begin{align}\label{ip:expandEVB}
\braket{\{v_a\}| S_k^z| \{w_b\}}  &=  \sum_{\tilde{b}=1}^N \frac{1}{\epsilon_k-w_{\tilde{b}}} \braket{v_1 \dots v_N | S_k^+| w_1 \dots w_{\tilde{b}-1} w_{\tilde{b}+1} \dots w_N} \nonumber \\ &=(-1)^N \left(\frac{g}{2}\right)^{L-2N} \sum_{\tilde{b}=1}^N \frac{1}{\epsilon_k-w_{\tilde{b}}} \det \left[J^k_{L-1}-D_{L-1}(w_{\tilde{b}})\right],
\end{align}
with $J_{L-1}^k$ defined as in Eq. (\ref{ip:defJk}) and $D_{L-1}(w_{\tilde{b}})$ an $(L-1) \times (L-1)$ diagonal matrix with $D_{L-1}(w_{\tilde{b}})_{ii} = \frac{1}{\epsilon_i-w_{\tilde{b}}}$. This diagonal matrix effectively removes the rapidity $w_{\tilde{b}}$ from the summations in the diagonal elements of $J_{L-1}^k$.

The crucial step is to note that this equality remains valid if $D_{L-1}(w_{\tilde{b}})$ is extended from a diagonal matrix to a matrix with the same diagonal elements and matrix elements
\begin{equation}
D_{L-1}(w_{\tilde{b}})_{ij} = \frac{1}{\epsilon_j-w_{\tilde{b}}},
\end{equation}
since this modifies the off-diagonal elements of $\left(J^k_{L-1}-D_{L-1}(w_{\tilde{b}})\right)$ as
\begin{equation}
\left(J^k_{L-1}-D_{L-1}(w_{\tilde{b}})\right)_{ij} = -\frac{1}{\epsilon_i-\epsilon_j}-\frac{1}{\epsilon_j-w_{\tilde{b}}} = -\frac{1}{\epsilon_i-\epsilon_j} \frac{\epsilon_i-w_{\tilde{b}}}{\epsilon_j-w_{\tilde{b}}},
\end{equation}
corresponding to multiplying the row $i$ by $(\epsilon_i-w_{\tilde{b}})$ and dividing column $i$ by the same factor, which leaves both the diagonal elements and the determinant invariant \cite{faribault_private_2017}. The separate rows of both matrices can now be written down as
\begin{equation}
J^k_{L-1} = \left(J_1 | J_2 | \dots |J_{L-1}\right)^T, \qquad D_{L-1}(w_{\tilde{b}}) = \left(D(w_{\tilde{b}}) | D(w_{\tilde{b}}) | \dots |D(w_{\tilde{b}})\right)^T, 
\end{equation}
where the determinant of $J^k_{L-1}-D_{L-1}(w_{\tilde{b}})$ can be expanded as
\begin{align}
&\det \left( J^k_{L-1}-D_{L-1}(w_{\tilde{b}}) \right)^T = \det \left( J_1 -D(w_{\tilde{b}}) | J_2 -D(w_{\tilde{b}}) | \dots |J_{L-1} -D(w_{\tilde{b}}) \right)^T \nonumber \\
&\qquad \qquad = \det \left( J_1 | J_2 | \dots |J_{L-1} \right)^T - \sum_{i=1}^{L-1} \det \left(J_1 | \dots |J_{i-1} | D(w_{\tilde{b}}) |J_{i+1}| \dots |J_{L-1} \right)^T,
\end{align}
since the determinant is a multilinear function in its rows and the determinant of a matrix containing multiple equal $D$-rows vanishes by definition. The first matrix returns exactly $J^k_{L-1}$, while all other matrices reduce to $J^k_{L-1}$ where a single row $i$ has been replaced by the row $D$. Plugging this into Eq. (\ref{ip:expandEVB}) leads to
\begin{align}
 \braket{\{v_a\}| S_k^z| \{w_b\}}  \propto & \sum_{\tilde{b}=1}^N \frac{1}{\epsilon_k-w_{\tilde{b}}} \det \left(J^k_{L-1}\right) \nonumber \\
& \qquad  - \sum_{i=1}^{L-1} \sum_{\tilde{b}=1}^N\frac{1}{\epsilon_k-w_{\tilde{b}}} \det \left(J_1 | \dots |J_{i-1} | D(w_{\tilde{b}}) |J_{i+1}| \dots |J_{L-1} \right)^T,
\end{align}
with the usual proportionality factor $(-1)^N \left(g/2\right)^{L-2N}$. Performing the summation over rapidities and absorbing the prefactor in the definition of the row $F =  \sum_{\tilde{b}=1}^N \frac{1}{\epsilon_k-w_{\tilde{b}}} D(w_{\tilde{b}})$, the same arguments can be used to reduce the summation to
\begin{align}
\braket{\{v_a\}| S_k^z| \{w_b\}}  &\propto \sum_{\tilde{b}=1}^N \frac{1}{\epsilon_k-w_{\tilde{b}}} \det \left(J^k_{L-1}\right) - \sum_{i=1}^L \det \left(J_1 | \dots |J_{i-1} | F |J_{i+1}| \dots |J_{L-1} \right) \nonumber \\
&\propto \left(1+\sum_{\tilde{b}=1}^N \frac{1}{\epsilon_k-w_{\tilde{b}}}\right) \det \left(J^k_{L-1}\right) - \det \left(J_1+F | J_2+F| \dots |J_{L-1} +F \right).
\end{align}
Defining $F^k_{L-1}$ as the matrix where all rows equal $F$, which can be evaluated as Eq. (\ref{ip:defFk}), this returns the proposed expression (\ref{ip:vSzw:EVB}). Alternatively, this can be obtained from the matrix determinant lemma \cite{faribault_private_2017}.

\section{Normalizations of the normal and the dual state}
\label{ip:app:rat}
From the identities presented in Section \ref{ip:sec:cauchy}, it is possible to derive the ratio of the normalizations of a dual Bethe state and a regular Bethe state. Since these are both eigenstates with identical eigenvalues, they describe the same state but with different normalizations. It is now possible to calculate the ratio of the normalization of these states by taking the overlap of both states with an (arbitrary) reference state, and to show how it leads to the proposed expression (\ref{ip:dual:ratio}). To reiterate, a Bethe ansatz eigenstate for the Richardson-Gaudin models can be written in two representations as
\begin{equation}
\ket{\{v_a\}} = \prod_{a=1}^N \left(\sum_{i=1}^L\frac{S^{+}_i}{\epsilon_i-v_a}\right)\ket{\downarrow \dots \downarrow}, \qquad \ket{\{v_a'\}}= \prod_{a=1}^{L-N} \left(\sum_{i=1}^L\frac{S^{-}_i}{\epsilon_i-v'_a}\right)\ket{\uparrow \dots \uparrow},
\end{equation}
with the (dual) rapidities satisfying the Richardson-Gaudin equations
\begin{align}\label{ip:rat:RGeq}
\frac{1}{g}+\frac{1}{2}\sum_{i=1}^L\frac{1}{\epsilon_i-v_a}- \sum_{b \neq a}^N\frac{1}{v_b-v_a}&=0, \qquad a=1 \dots N, \\
-\frac{1}{g}+\frac{1}{2}\sum_{i=1}^L\frac{1}{\epsilon_i-v_a'}- \sum_{b \neq a}^{L-N}\frac{1}{v_b'-v_a'}&=0, \qquad a=1 \dots L-N,
\end{align}
describing the same eigenstate if these variables are coupled through
\begin{equation}\label{ip:rat:corrEVB}
\sum_{a=1}^N \frac{1}{\epsilon_i-v_a}+\frac{2}{g}=\sum_{a=1}^{L-N}\frac{1}{\epsilon_i-v_a'}, \qquad  i=1 \dots L.
\end{equation}
The crucial element in this proof is the existence of eigenvalue-based expressions for the inner product of a Bethe state and a reference state (see also Section \ref{RG:sec:expand}). Defining a reference state from a set of occupied levels can again be done in two ways
\begin{equation}
\ket{\{i_{occ}\}} = \prod_{i \in \{i_{occ}\}} S^{+}_i \ket{\downarrow \dots \downarrow} = \prod_{i \notin \{i_{occ}\}}S_i^{-}\ket{\uparrow \dots \uparrow},
\end{equation}
where both states are already normalized. Then the overlap of the original Bethe state with this reference state is given by \cite{faribault_determinant_2012}
\begin{equation}
\braket{\{i_{occ}\}|\{v_a\}} = \det J_N(\{v_a\},\{i_{occ}\}),
\end{equation}
with $J_N(\{v_a\},\{i_{occ}\})$ an $N \times N$ matrix defined as
\begin{equation}
 J_N(\{v_a\},\{i_{occ}\})_{ij} =
  \begin{cases}
   \sum_{a=1}^N \frac{1}{\epsilon_i-v_a}-\sum_{\substack{k \in \{i_{occ}\} \\ k \neq i }} \frac{1}{\epsilon_i-\epsilon_k} &\text{if}\ i=j, \\
   -\frac{1}{\epsilon_i-\epsilon_j}       & \text{if}\ i \neq j,
  \end{cases} \qquad i,j \in \{i_{occ}\}.
\end{equation}
Alternatively, the overlap of the dual state with the same reference state can be written as
\begin{align}
\braket{\{i_{occ}\}|\{v_a'\}}  = \bra{\downarrow \dots \downarrow} \left(\prod_{a=1}^{L-N} S^{-}(v_a')\right) \left(  \prod_{i \notin \{i_{occ}\}} S_i^+ \right)\ket{\downarrow \dots \downarrow} .
\end{align}
This final expression is the complex conjugate of the inner product of a regular Bethe state defined by the dual rapidities and a different reference state. Because all eigenvalue-based variables are real, these inner products are always real, and the same determinant expressions can be invoked to write
\begin{equation}
\braket{\{i_{occ}\}|\{v_a'\}} = \det J_{L-N}(\{v_a'\},\{i | i \notin \{i_{occ}\} \}),
\end{equation}
with $J_{L-N}(\{v_a'\},\{i | i \notin \{i_{occ}\} \})$ an $(L-N) \times (L-N)$ matrix defined as
\begin{equation}
 J_{L-N}(\{v_a'\},\{i | i \notin \{i_{occ}\} \})_{ij} =
  \begin{cases}
   \sum_{a=1}^{L-N} \frac{1}{\epsilon_i-v_a'}-\sum_{\substack{k \notin \{i_{occ}\} \\ k \neq i }} \frac{1}{\epsilon_i-\epsilon_k} &\text{if}\ i=j, \\
   -\frac{1}{\epsilon_i-\epsilon_j}       & \text{if}\ i \neq j,
  \end{cases} \qquad i,j \notin \{i_{occ}\}.
\end{equation}
Because this only depends on the dual rapidities through the eigenvalue-based variables in the diagonal elements, the correspondence (\ref{ip:rat:corrEVB}) can be used to express this determinant in the rapidities of the original state as
\begin{align}
\braket{\{i_{occ}\}|\{v_a'\}} &= \det \left[\frac{2}{g} + J_{L-N}(\{v_a\},\{i | i \notin \{i_{occ}\} \})\right] \nonumber \\
&=\left(\frac{2}{g}\right)^{L-N} \det \left[\mathbbm{1} + \frac{g}{2} J_{L-N}(\{v_a\},\{i | i \notin \{i_{occ}\} \})\right].
\end{align} 
Now Sylvester's determinant identity can be used, as explained in Section \ref{ip:sec:cauchy}, to exchange the role of rapidities and inhomogeneities and obtain
\begin{align}
\braket{\{i_{occ}\}|\{v_a'\}} &= \left(\frac{2}{g}\right)^{L-N} \det \left[\mathbbm{1} + \frac{g}{2} K_{N}(\{v_a\},\{i | i \notin \{i_{occ}\} \})\right],
\end{align} 
with $K_{N}(\{v_a\},\{i | i \notin \{i_{occ}\} \})$ an $N \times N$ matrix defined as
\vspace{-0.4\baselineskip}
\begin{equation}
 K_N(\{v_a\},\{i | i \notin \{i_{occ}\} \})_{ab}=
  \begin{cases}
   -\sum_{i \notin \{i_{occ}\}} \frac{1}{v_a-\epsilon_{i}}+\sum_{c \neq a}^{N} \frac{1}{v_a-v_c} &\text{if}\ a=b, \\
   -\frac{1}{v_a-v_b}       & \text{if}\ a \neq b.
  \end{cases}
\end{equation}
Here, it is important to note the similarity between the diagonal elements of this matrix and the Bethe equations (\ref{ip:rat:RGeq}). By slightly rewriting these equations, the diagonal elements can be rewritten as
\begin{equation}
1-\frac{g}{2}\sum_{i \notin \{i_{occ}\}} \frac{1}{v_a-\epsilon_i}+\frac{g}{2}\sum_{c \neq a}^{N} \frac{1}{v_a-v_c} = \frac{g}{2}\sum_{i \in \{i_{occ}\}} \frac{1}{v_a-\epsilon_i}-\frac{g}{2}\sum_{c \neq a}^{N} \frac{1}{v_a-v_c},
\end{equation}
resulting in
\begin{equation}
\braket{\{i_{occ}\}|\{v_a'\}} = (-1)^N\left(\frac{2}{g}\right)^{L-2N} \det  K_{N}(\{v_a\},\{i_{occ}\}).
\end{equation}
Both $g/2$ and the minus sign in the diagonal elements have been absorbed in a prefactor, since multiplying all rows with $-1$ and taking the transpose of the matrix returns the original matrix but with the sign of the diagonal elements exchanged. The roles of both variables can again be exchanged in this final expression to obtain the original equality
\begin{equation}
\braket{\{i_{occ}\}|\{v_a'\}} =  (-1)^N\left(\frac{2}{g}\right)^{L-2N}\braket{\{i_{occ}\}|\{v_a\}}.
\end{equation}
Since the set of reference states forms a complete basis and this equality holds for all such states, this yields the proposed expression (\ref{ip:dual:ratio}) as
\begin{equation}
\ket{\{v_a'\}} = (-1)^N\left(\frac{2}{g}\right)^{L-2N} \ket{\{v_a\}}.
\end{equation}

\end{subappendices}
\makeatletter
\@openrighttrue
\makeatother



\chapter{Integrability in the contraction limit of the quasispin}
\label{chap:contraction}

\setlength\epigraphwidth{.4\textwidth}
\epigraph{\emph{I like the speed \\
And the incomparably blurred \\
Sensation of being deformed \\
Into being and about to begin.}}{{Frederick Seidel}}

So far, the Richardson-Gaudin models were formulated in terms of $su(2)$-algebras describing spin states or fermion pairs, but these are far from exhaustive. The GGA construction does not depend explicitly on the $su(2)$ realization, and various integrable models are known which contain e.g. a bosonic degree of freedom. Unfortunately, the direct connection between a solution to the Gaudin equations and the conserved charges and Bethe ansatz is somewhat muddled in these models. One possible way of retaining the clear structure of the spin models is by starting from a spin model and introducing a bosonic degree of freedom by means of a limiting procedure, deforming one of the spin degrees of freedom into a bosonic mode \cite{gaudin_diagonalisation_1976,dukelsky_exactly_2004,lerma_h._integrable_2011}. 

Two main classes of systems can be obtained in this way. Representative models for these separate classes are the Dicke model \cite{jaynes_comparison_1963,tavis_exact_1968,dicke_coherence_1954}, part of a family of Dicke-Jaynes-Cummings-Gaudin (DJCG) models describing two-level systems interacting with a single photonic mode of an electromagnetic field, and the two-channel extension of the $p_x+ip_y$ model interacting through a $p$-wave Feshbach resonance \cite{lerma_h._integrable_2011}. Originally, the class of DJCG models were obtained from the trigonometric RG model, which includes the Jaynes-Cummings \cite{jaynes_comparison_1963}, the Tavis-Cummings \cite{tavis_exact_1968} and the inhomogeneous Dicke model \cite{dicke_coherence_1954}, all describing the interaction between a (set of) two-level system(s) and a single bosonic electromagnetic mode. The second class of systems considers the coupling of an integrable $p_x+ip_y$-wave superfluid to a bosonic mode, which can be obtained as the limiting case of a hyperbolic RG model \cite{lerma_h._integrable_2011}. This model was initially introduced by Dunning \emph{et al.}  \cite{dunning_becbcs_2011} and can be shown to be equivalent to a model which couples Cooper pairs to condensed molecular bosons \cite{hibberd_bethe_2006}. These two classes were later obtained as two distinct cases of integrable Hamiltonians containing a bosonic degree of freedom, starting from a variational approach \cite{birrell_variational_2012}. 

In this chapter, we show how these models can be constructed through a \emph{pseudo-deformation} of the (quasi-)spin algebra \cite{de_baerdemacker_richardson-gaudin_2012}, providing a controlled way of transforming the hard-core bosonic $su(2)$ algebra into a genuinely bosonic Heisenberg-Weyl algebra $hw(1)$. The pseudo-deformation scheme  was originally proposed as a way to shed light on the singularities arising in the RG equations by connecting the spin models to purely bosonic models. However, here this method is used as a way to obtain a bosonic algebra as the contraction limit of a single $su(2)$ quasispin algebra. The connection can be made adiabatically by means of a continuous pseudo-deformation parameter, allowing for a systematic extension of the $su(2)$-based integrable systems towards those containing a bosonic degree of freedom \cite{claeys_dicke_2015}.

This can also be used to find eigenvalue-based determinant expressions for overlaps in these models. The dependency on the $hw(1)$ algebra makes the generalization of results for the $su(2)$-models towards models containing a bosonic degree of freedom far from straightforward, because this algebra is non-compact and therefore lacks a highest weight state. Tschirhart and Faribault recently showed how determinant expressions could be found for the form factors of the DJCG models by means of the introduction of an intricate alternative Algebraic Bethe Ansatz (ABA) \cite{tschirhart_algebraic_2014}. In this chapter, these expressions are obtained as a limiting case of a renormalized pseudo-deformed spin model. The purpose of this chapter is twofold. It is shown how the pseudo-deformation scheme provides a convenient and relatively straightforward way of deriving all properties (eigenstates, eigenvalue-based variables, form factors, ...) of both the Dicke and extended $p_x+ip_y$ model, and how an eigenvalue-based framework can be obtained for the two classes of integrable RG models containing a bosonic degree of freedom. This chapter is largely based on Refs. \cite{claeys_dicke_2015,claeys_eigenvalue-based-bosonic_2015}.

\section{Pseudo-deformation of the quasispin}
\begin{figure}                  
 \begin{center}
 \includegraphics{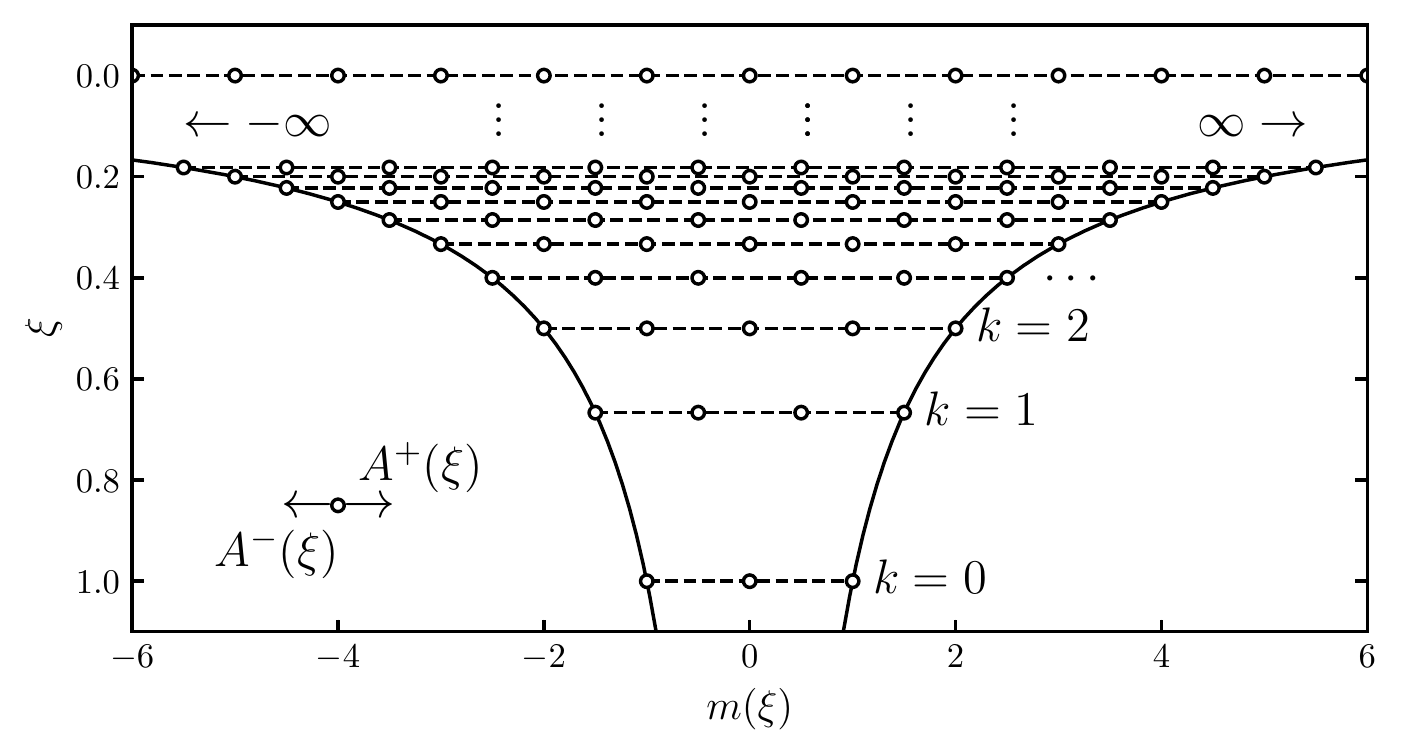}
 \caption{An illustration of the pseudo-deformed quasispin irreps $\ket{s(\xi),m(\xi)}$ of $su(2)_{\xi}$. The unitary irreps, at $\xi=\frac{2s}{2s+k}$ with $k\in\mathbb{N}$, are denoted by open circles, connected by dashed lines. The lines connecting the lowest and highest weights with $m(\xi)=\mp s(\xi)$ are drawn in full. For $\xi=1$, the $s(1)=1$ irrep is retained. Based on a similar figure in Ref. \cite{de_baerdemacker_richardson-gaudin_2012}. \label{fig:cont:irreps}}
 \end{center}
\end{figure}

There are multiple possible approaches for the process of bosonization, of which the Holstein-Primakoff transformation is arguably the best-known \cite{holstein_field_1940}. In this chapter, the recently-proposed pseudo-deformation scheme \cite{de_baerdemacker_richardson-gaudin_2012} is presented as a similar way of obtaining bosonic commutation relations. This then provides an adiabatic, and therefore controlled, mapping from $su(2)$ to $hw(1)$. A pseudo-deformed algebra can be defined as
\begin{align}
&[S^z(\xi),S^+(\xi)]=S^+(\xi),\quad[S^z(\xi),S^-(\xi)]=-S^-(\xi), \nonumber \\
&[S^+(\xi),S^-(\xi)]=2\left(\xi S^z(\xi) +(\xi-1)s\right)\label{cont:pseudo:def}
\end{align}
with $\xi \in [0,1]$ the pseudo-deformation parameter and $s$ the original ($\xi=1$) $su(2)$ irrep label. This definition can be interpreted as providing a linear interpolation between two known limits: $\xi=1$ gives rise to the original $su(2)$ quasispin algebra, while $\xi=0$ results in a (unnormalized) bosonic $hw(1)$ algebra. This latter limit was also termed the contraction limit of the algebra \cite{gilmore_lie_2008}. The nomenclature \emph{pseudo}-deformation was originally proposed because this algebra can be reduced to a canonical $su(2)_{\xi}$ algebra
\begin{align}\label{cont:pseudo:su2}
&[A^z(\xi),A^+(\xi)]=A^+(\xi),\qquad[A^z(\xi),A^-(\xi)]=-A^-(\xi), \nonumber \\
&[A^+(\xi),A^-(\xi)]=2A^z(\xi),
\end{align}
by defining
\begin{equation}
A^+(\xi)=\frac{1}{\sqrt{\xi}}S^+(\xi),\quad A^-(\xi)=\frac{1}{\sqrt{\xi}}S^-(\xi),\quad A^z(\xi)=S^z(\xi)+\left(1-\frac{1}{\xi}\right)s.
\end{equation}
This holds for all values of $\xi$ except for the contraction limit $\xi=0$. In this limit, the following operators
\begin{equation}
b^\dag=\sqrt{\frac{1}{2s}}S^+(0),\quad b=\sqrt{\frac{1}{2s}}S^-(0),\quad b^\dag b=S^z(0)+s,
\end{equation}
close the bosonic $hw(1)$ algebra
\begin{equation}
[b^\dag b,b^\dag]=b^\dag,\quad [b^\dag b,b]=-b,\quad [b,b^\dag]=1.
\end{equation}
The irreps of the $\{A^+(\xi),A^-(\xi),A^z(\xi)\}$ algebra are labeled by $s(\xi)\equiv s/\xi$. The interpretation behind this is a gradual increase of the effective multiplicity $\Omega(\xi)=(2s(\xi)+1)$ of the $su(2)$ irrep with decreasing $\xi$, in the contraction limit leading to an effective unlimited occupation consistent with the bosonic character. This is illustrated in Figure \ref{fig:cont:irreps}. It should be noted that only discrete values of $\xi_n=\frac{2s}{n}$ (with $n=2s,2s+1,\dots$) give rise to unitary irreps.  Nevertheless, this is not problematic because the theory of RG integrability is only based on commutation relations, so the parameter $\xi$ can be regarded as a continuous variable.

Recently, it has been shown how the Bethe ansatz states in the RG solution \cite{richardson_restricted_1963,richardson_exact_1964} of the reduced BCS Hamiltonian for conventional superconductivity \cite{bardeen_microscopic_1957,bardeen_theory_1957} can be connected adiabatically to a product state of generalized bosons \cite{de_baerdemacker_richardson-gaudin_2012} employing a pseudo deformation of the quasispin algebra.  The product structure of the Bethe ansatz state is reminiscent of the projected BCS approximation, in which the superconducting state is approximated by a condensate of collective Cooper pairs \cite{bardeen_microscopic_1957,bardeen_theory_1957,cooper_bound_1956}. However, in general the rapidities parametrizing the generalized quasispin creation operators in the Bethe ansatz state differ from each other, which is not reconcilable with the concept of a true condensate. The pseudo deformation can then be used to continuously deform the $su(2)$ quasispin algebra of the BCS pairing Hamiltonian into a genuinely bosonic $hw(1)$. The reduced BCS Hamiltonian remains integrable along the path of deformation, allowing for an adiabatic and injective mapping of the exact Bethe ansatz states into a true condensate of (orthogonal) bosonic modes \cite{de_baerdemacker_tamm-dancoff_2011}.  Accordingly, the coupled set of RG equations of integrability reduce to a single decoupled equation, equivalent to the secular equation of the particle-particle Tamm-Dancoff Approximation ($pp$-TDA) for the elementary pairing modes for the reduced BCS Hamiltonian \cite{ring_nuclear_2004}. Reversely, the pseudo deformation enables one to numerically reconstruct the solution of the coupled set of non-linear RG equations from the simpler decoupled secular $pp$-TDA equation, by adiabatically reintroducing the Pauli principle in the integrable model \cite{de_baerdemacker_richardson-gaudin_2012,van_raemdonck_exact_2014}.

\section{The Dicke model}
\label{cont:dicke}

The interaction of a single quantized mode of electromagnetic radiation (photons) with a two-state system, such as a nuclear spin or two-level atom, can be modeled by means of the Rabi Hamiltonian \cite{grynberg_introduction_2010}.  Although simple and physically transparent in its formulation, the structure of the eigenstates of the Rabi Hamiltonian are quite involved due to the `counter-rotating' interaction terms in the Hamiltonian \cite{batchelor_integrability_2014}. In the resonance regime however, these fast-frequency counter-rotating terms average out with respect to the collective two-level oscillations, and can be neglected in first-order perturbation theory (the so-called Rotating Wave Approximation (RWA)).  This gives rise to the Jaynes-Cummings (JC) Hamiltonian of quantum electrodynamics \cite{jaynes_comparison_1963}.  Due to the RWA, the JC Hamiltonian exhibits an additional symmetry, which allows for the decomposition of the Hilbert space into irreps conserving the total number of (bosonic and atomic) excitations.  For the two-level ($s=1/2$ quasispin) formulation of the JC model, these irreps remain two-dimensional and therefore reduce to two-level mixing models.  One of the successes of theoretical and experimental quantum mechanics is that such simple theoretical models have been validated in sophisticated experiments situated in cavity quantum electrodynamics \cite{brune_quantum_1996}, circuit quantum electrodynamics \cite{wallraff_strong_2004},...

A direct consequence of the additional symmetry is that the two-level JC model exhibits two linearly independent conserved operators, \emph{i.e.}\ the Hamiltonian and the operator counting the total number of excitations.  As the model has two degrees of freedom (the electromagnetic photon mode and the two-level system), the model is integrable by definition, and supports a Bethe ansatz wave function \cite{gaudin_bethe_2014}. This result can be extended to general $(2s+1)$-level systems, also referred to as Tavis-Cummings models \cite{tavis_exact_1968}, or a collection of inequivalent $(2s+1)$-level systems interacting via a single mediating bosonic mode, called the Dicke model \cite{dicke_coherence_1954}. The integrability of the Dicke model was first established by Gaudin \cite{gaudin_bethe_2014}, by means of an infinite-dimensional Schwinger representation of the central spin in the Gaudin magnet.  Later, the full set of conserved charges of the Dicke model was derived by Dukelsky \emph{et al.} \cite{dukelsky_exactly_2004} by mapping one of the $su(2)$ quasispin copies of the Richardson-Gaudin conserved charges onto the bosonic mode.  As an integrable model, the conserved charges and Bethe ansatz state of the Dicke model can also be obtained from the ABA \cite{babelon_bethe_2007}, however a direct and simplified derivation of the Bethe ansatz state with the corresponding Bethe equations solution is feasible using a commutator scheme \cite{tsyplyatyev_simplified_2010}.

The Dicke Hamiltonian \cite{dicke_coherence_1954} is given by
\begin{equation}
H=\epsilon_0 b^{\dagger}b+\sum_{i=1}^L \epsilon_i S^z_i +G \sum_{i=1}^L \left(S^+_i b +S^-_i b^{\dagger}\right),
\end{equation}
and describes a set of $L$ $(2s_i+1)$-level systems with energy differences $\epsilon_i$ interacting with a single mode of the bosonic field, represented by a photon with energy $\epsilon_0$. The interaction strength is tunable through the single coupling constant $G$.

This Hamiltonian can now be related to the models from Chapter \ref{Chap_RGmodels} through the pseudo-deformation scheme. Starting from the constants of motion for a set of $L+1$ interacting spins (\ref{RG:spin:Qi}), these are labeled $i=0,1, \dots, L$, and the $su(2)$ algebra labeled $i=0$ is replaced by a pseudo-deformed $su(2)_{\xi}$ algebra. The trigonometric Gaudin algebra (\ref{RG:RG:trig}) can be mapped to
\begin{equation}
X(\eta_i,\eta_j)=\frac{\sqrt{(1+\eta_i^2)(1+\eta_j^2)}}{\eta_i-\eta_j}, \qquad Z(\eta_i,\eta_j)=\frac{1+\eta_i \eta_j}{\eta_i-\eta_j},
\end{equation}
and taking the limit $\eta_0 \to \infty$ leads to a Gaudin algebra determined by (for $i,j \neq 0$)
\begin{align}
&X(\eta_0,\eta_j)=\sqrt{1+\eta_j^2}, \qquad Z(\eta_0,\eta_j)=\eta_j, \\
&X(\eta_i,\eta_j)=\frac{\sqrt{(1+\eta_i^2)(1+\eta_j^2)}}{\eta_i-\eta_j}, \qquad Z(\eta_i,\eta_j)=\frac{1+\eta_i \eta_j}{\eta_i-\eta_j}.
\end{align}
The conserved charge associated with the deformed algebra is given by 
\begin{align}
Q_0(\xi)&=A^z(\xi)+g\sum_{j=1}^L \left[\frac{1}{2}X(\eta_0,\eta_j)\left(A^+(\xi)S^-_j+A(\xi)^-S^{+}_j\right)+Z(\eta_0,\eta_j) A^z(\xi)S_j^z\right]\nonumber\\
&=A^z(\xi)+g\sum_{j=1}^L \left[\frac{1}{2}\sqrt{1+\eta_j^2}\left(A^+(\xi) S^-_j + A^-(\xi)S^+_j\right)+\eta_j A^z(\xi)S_j^z\right]. 
\end{align}
It is now possible to define a $\xi$-dependent coupling constant as $g=\sqrt{\frac{2\xi}{s_0 G^2}}\frac{G^2}{\epsilon_0}$ and rescale the variables $\eta_j=-\sqrt{\frac{\xi}{2s_0G^2}}\epsilon_j$. Near the contraction limit ($\xi \ll 1$) the Gaudin algebra reduces to
\begin{equation}\label{cont:dicke:param}
X(\eta_0,\eta_j)=1+\frac{\xi}{4s_0 G^2}\epsilon_j^2+\mathcal{O}(\xi^2), \qquad Z(\eta_0,\eta_j)=-\sqrt{\frac{\xi}{2s_0G^2}}\epsilon_j,
\end{equation}
which is related to the parametrization proposed by Dukelsky \emph{et al.} \cite{dukelsky_exactly_2004}. By making use of this parametrization, the Dicke Hamiltonian can be obtained (up to a diverging constant) in the contraction limit as
\begin{equation}
\epsilon_0 Q_0(\xi \to 0)=\epsilon_0 b^{\dagger}b+\sum_{i=1}^L \epsilon_i S^z_i +G \sum_{i=1}^L \left(S^+_i b + S^-_i b^{\dagger}\right),
\end{equation}
A similar procedure leads to the other properties of the Dicke model.
\begin{enumerate}
\item The constants of motion
\begin{align}\label{cont:dicke:com}
\epsilon_0 Q_i(\xi \to 0) =&(\epsilon_0-\epsilon_i)S^z_i-G(S^+_i b+S^-_i b^{\dagger}) \nonumber \\
&-2 G^2 \sum_{j \neq i}^L \frac{1}{\epsilon_i-\epsilon_j}\left[\frac{1}{2}(S^+_i S^-_j+S^-_i S^+_j)+S^z_iS^z_j\right].
\end{align}
\item The underlying GGA, where the rescaling of the inhomogeneities results in a rational GGA instead of the original trigonometric one,
\begin{align}
&S^{+}(u) = b^{\dagger}+G \sum_{i=1}^L \frac{S_i^+}{u-\epsilon_i}, \qquad S^{-}(u) = b+G \sum_{i=1}^L \frac{S_i^-}{u-\epsilon_i}, \nonumber \\
&S^z(u)= \frac{\epsilon_0-u}{2G}-G \sum_{i=1}^L \frac{S_i^z}{u-\epsilon_i}.
\end{align}
\item The Bethe ansatz for the Dicke model
\begin{equation}\label{cont:dicke:ba}
\ket{v_1 \dots v_N}=\left(\frac{2s_0}{\xi}\right)^{\frac{N}{2}}\prod_{a=1}^N\left(b^{\dagger}-G\sum_{i=1}^L \frac{S^+_i}{\epsilon_i-v_a}\right)\ket{0},
\end{equation}
where the prefactor can be absorbed in the normalization and the vacuum state is given by $\ket{0} = \ket{0}\otimes \ket{s_1,-s_1}\otimes\cdots\otimes\ket{s_L, -s_L}$.
\item The Bethe equations
\begin{equation}
(\epsilon_0-v_a)-2 G^2 \sum_{i=1}^L \frac{s_i}{\epsilon_i-v_a}+2 G^2 \sum_{b \neq a}^N \frac{1}{v_b-v_a}=0, \qquad \forall a=1 \dots N.
\end{equation}
\end{enumerate}
Remarkably, it remains possible to transform the equations for the rapidities to an equivalent set of equations for the set of eigenvalue-based variables provided $s_i=1/2, \forall i=1,\dots,L$ \cite{babelon_bethe_2007,faribault_gaudin_2011}. For the Dicke model, the relevant variables are given by
\begin{equation}
\Lambda_0 = \sum_{a=1}^N v_a, \qquad \Lambda_i \equiv \Lambda_i(\{v_a\})=\sum_{a=1}^N \frac{1}{\epsilon_i-v_a}, \qquad i=1 \dots L,
\end{equation}
and these determine the eigenvalues of the constants of motion (\ref{cont:dicke:com}) as 
\begin{align*}
\epsilon_0 Q_i \ket{v_1 \dots v_N}&=\frac{1}{2}\left[(\epsilon_i-\epsilon_0)+2G^2\sum_{a=1}^N \frac{1}{\epsilon_i-v_a}-G^2 \sum_{j \neq i}^L \frac{1}{\epsilon_i-\epsilon_j}\right]\ket{v_1 \dots v_N},
\end{align*}
while satisfying the coupled quadratic equations
\begin{align}
&G^2\Lambda_i^2=N-\Lambda_i(\epsilon_i-\epsilon_0)+G^2\sum_{j\neq i}^L\frac{\Lambda_i-\Lambda_j}{\epsilon_i-\epsilon_j}, \qquad \forall i=1 \dots L,\nonumber\\
&\Lambda_0 + G^2 \sum_{i=1}^L \Lambda_i  = \epsilon_0 N.
\end{align}
These equations can either be determined starting from the Bethe equations for the Dicke model \cite{babelon_bethe_2007,faribault_gaudin_2011,tschirhart_algebraic_2014}, or by taking the contraction limit of the eigenvalue-based equations for the trigonometric RG model. Similar to the $su(2)$-based RG models, it is possible to express the overlap and form factors of the Dicke model in the eigenvalue-based variables only, again circumventing the need to calculate the singularity-prone rapidities \cite{tschirhart_algebraic_2014} (see Section \ref{cont:EVB:det}).

\section{The extended $p_x+ip_y$-wave pairing model}
\label{cont:pip}
A similar procedure can be used to prove the integrability of the $p_x+ip_y$-wave pairing model coupled to a bosonic degree of freedom. This model was introduced by Dunning \emph{et al.} \cite{dunning_becbcs_2011} as an extension of the integrable fermionic $p_x+ip_y$-pairing model \cite{ibanez_exactly_2009,rombouts_quantum_2010,dunning_exact_2010}. Lerma \emph{et al.} consequently showed how this model is given by the limit of a hyperbolic RG model \cite{lerma_h._integrable_2011}, which here will be reformulated by making use of the pseudo-deformation. The Hamiltonian is given by 
\begin{align}
H=&\delta\, b^{\dagger}b+\sum_{\mathbf{k}}\frac{\mathbf{k}^2}{2m}c^{\dagger}_{\mathbf{k}}c_\mathbf{k}-\frac{G}{4}\sum_{\mathbf{k} \neq \pm \mathbf{\mathbf{k}}'}(k_x-ik_y)(k_x'+ik_y')c^{\dagger}_{\mathbf{k}}c^{\dagger}_{-\mathbf{k}}c_{\mathbf{k}'}c_{-\mathbf{k}'} \nonumber \\
& \qquad \qquad  -\frac{K}{2}\sum_{\mathbf{k}}\left((k_x-ik_y)c^{\dagger}_{\mathbf{k}}c^{\dagger}_{-\mathbf{k}}b+\textrm{h.c.}\right),
\end{align}
and was shown to be integrable if
\begin{equation}\label{cont:pip:intcond}
\delta=-F^2G, \qquad K=FG.
\end{equation}%
By making use of the quasispin formalism \cite{talmi_simple_1993} and again absorbing a phase in the quasispin operators, the Hamiltonian can be rewritten as
\begin{equation}\label{cont:pip:ham}
H=\delta\, b^{\dagger}b+\sum_{k=1}^L\epsilon_k S_k^z - G\sum_{k,k'=1}^L\sqrt{\epsilon_k \epsilon_k'}S_k^{+}S_{k'}^{-}-K\sum_{k=1}^L \sqrt{\epsilon_k}\left(S_k^+ b+S_k^- b^{\dagger}\right),
\end{equation}
which can again be related to a $su(2)$-based RG model. This suggests starting from the hyperbolic Gaudin algebra (\ref{RG:RG:rat}),
\begin{equation}
X(\epsilon_i,\epsilon_j)=2\frac{\sqrt{\epsilon_i \epsilon_j}}{\epsilon_i-\epsilon_j}, \qquad Z(\epsilon_i,\epsilon_j)=\frac{\epsilon_i+\epsilon_j}{\epsilon_i-\epsilon_j}.
\end{equation}
Taking the conserved charges for $(L+1)$ levels and exchanging a single $su(2)$ algebra with a pseudo-deformed $su(2)_{\xi}$-algebra, the Hamiltonian can be obtained as a linear combination of the conserved charges in the contraction limit. Renormalizing the coupling constant and bosonic energy level $\epsilon_0$ as
\begin{equation}
g=\frac{\xi}{s_0+\xi \kappa}, \qquad \epsilon_0=\frac{\xi}{2s_0}\eta_0^2,
\end{equation}
the conserved charges reduce to
\begin{align}\label{cont:pip:number}
&\lim_{\xi \to 0} Q_0(\xi) =b^{\dagger}b+\sum_{i=1}^L S_i^z  \equiv N, \\
&\lim_{\xi \to 0 } \frac{s_0}{\xi} Q_i(\xi ) = \sum_{j \neq i}^L \left[\frac{\sqrt{\epsilon_i\epsilon_j}}{\epsilon_i-\epsilon_j}\left(S^{+}_i S^-_j+S^-_i S^{+}_j \right)+\frac{\epsilon_i+\epsilon_j}{\epsilon_i-\epsilon_j}S_i^zS_j^z\right] \nonumber\\
&\qquad \qquad \qquad \qquad \qquad  +\frac{\eta_0}{\sqrt{\epsilon_i}}\left(S^+_i b+S^-_i b^{\dagger}\right)+S_i^z \left(\kappa+b^{\dagger}b-\frac{\eta_0^2}{\epsilon_i}\right),
\end{align}
with $N$ the number operator counting the number of excitations. Note that $Q_i(\xi \to 0)=0$, but $Q_i(\xi)/\xi$ remains finite for the whole range of $\xi$ and results in a non-zero conserved operator in the contraction limit. These operators are the building blocks for the Hamiltonian (\ref{cont:pip:ham}), similar to the results presented for the fermionic $p_x+ip_y$ pairing model \cite{rombouts_quantum_2010}. The integrability condition (\ref{cont:pip:intcond}) arises naturally from the parametrization of the Gaudin algebra. The related Hamiltonian studied in Ref. \cite{hibberd_bethe_2006} can be obtained by taking the linear combination
\begin{equation}
H=\sum_{i=1}^L Q_i = \sum_{i=1}^L \frac{\eta_0}{\sqrt{\epsilon_i}}\left(S^+_{i}b+S^-_i b^{\dagger}\right)-\sum_{i=1}^L \frac{\eta_0^2}{\epsilon_i}S_i^z +\sum_{i=1}^L S_i^z(b^{\dagger}b+\kappa),
\end{equation}
which also equals the two-channel $p_x+ip_y$-wave superfluid Hamiltonian introduced by Lerma \emph{et al.} \cite{lerma_h._integrable_2011} up to a rescaling, a redefinition of the energy levels as $1/\epsilon_i$, and by expressing $\sum_{i=1}^L S_i^z=N-b^{\dag}b$. All other properties of these models can similarly be obtained from the contraction limit.

\begin{enumerate}
\item The underlying GGA remains the hyperbolic model, with generators given by
\begin{align}
&S^+(u) = 2\frac{\eta_0}{\sqrt{u}} b^{\dagger} +2 \sum_{i=1}^L \frac{\sqrt{u\epsilon_i}}{u-\epsilon_i}S_i^+, \qquad S^-(u) = 2\frac{\eta_0}{\sqrt{u}} b+2 \sum_{i=1}^L \frac{\sqrt{u\epsilon_i}}{u-\epsilon_i}S_i^-, \nonumber \\
&S^z(u) = \frac{\eta_0^2}{u}-b^{\dagger}b-\kappa \ \sum_{i=1}^L \frac{u+\epsilon_i}{u-\epsilon_i}S_i^z.
\end{align}

\item The Bethe states for these models can similarly be found from the contraction limit as
\begin{equation}
\ket{v_1 \dots v_N}=\prod_{a=1}^N\left(b^{\dagger}-\sum_{i=1}^L \frac{\sqrt{\epsilon_i} v_a}{\epsilon_i-v_a}\frac{S^{+}_i}{\eta_0}\right)\ket{0}.
\end{equation}
\item The resulting Bethe equations are now given by
\begin{equation}
\kappa-\frac{\eta_0^2}{v_a}+\sum_{i=1}^L s_i \frac{\epsilon_i+v_a}{\epsilon_i-v_a}-\sum_{b \neq a}^N\frac{v_b+v_a}{v_b-v_a}=0, \qquad \forall a=1 \dots N.
\end{equation}
\end{enumerate}
Taking again $s_i=1/2, \forall i=1 \dots L$, these now result in a set of quadratic equations in the variables
\begin{align}\label{cont:pip:deflam}
\Lambda_0=\sum_{a=1}^N\frac{\eta_0^2}{v_a}, \qquad \Lambda_i&=\sum_{a=1}^N\frac{\epsilon_i+v_a}{\epsilon_i-v_a}, \qquad \forall i=1 \dots L,
\end{align}
which have to satisfy
\begin{align}\label{cont:pip:eqlam}
&\Lambda_i^2 =-N(L-N)-2\kappa \Lambda_i +2 \eta_0^2\left(\Lambda_0+\frac{\Lambda_i+L}{\epsilon_i}\right)+\sum_{j \neq i}^L \frac{\epsilon_i+\epsilon_j}{\epsilon_i-\epsilon_j}(\Lambda_i-\Lambda_j), \qquad \forall i=1 \dots L, \nonumber \\
&2 \Lambda_0=\sum_{i=1}^L \Lambda_i+2\kappa L.
\end{align}
These again fully determine the eigenvalues of the conserved charges as
\begin{align}
Q_i \ket{v_1 \dots v_N} &= \frac{1}{2}\left[-\kappa-\sum_{a=1}^N \frac{\epsilon_i+v_a}{\epsilon_i-v_a}+\frac{\eta_0^2}{\epsilon_i}+\sum_{j \neq i}^L \frac{\epsilon_i+\epsilon_j}{\epsilon_i-\epsilon_j}\right]\ket{v_1 \dots v_N}.
\end{align}

\section{Eigenvalue-based determinant expressions}
\label{cont:EVB:det}
In the previous section, it was shown how many results for integrable systems containing a bosonic degree of freedom can be obtained in the contraction limit. However, when deriving expressions for normalizations and form factors starting from the $su(2)$ spin models, we face the problem that the bosonic level has no state of maximum occupation number. A dual vacuum now cannot be defined since this would contain an infinite number of bosonic excitations. This problem was first envisioned for the Dicke model by Tschirhart and Faribault \cite{tschirhart_algebraic_2014}, who devised an alternative formulation of the ABA, introducing a pseudovacuum, which allowed for a description in terms of the eigenvalue-based variables for the Dicke model. The pseudo-deformation scheme now allows for a simpler derivation of these results by means of a renormalization procedure. Starting from the eigenvalue-based determinant expressions for the XXZ RG models, the dual state can still be defined for any Bethe ansatz eigenstate. Determinant expressions can then be obtained while keeping the dual state implicit, leading to finite expressions in the contraction limit. This approach can then be extended to the $p_x+ip_y$-wave pairing model. In a similar manner as presented in the previous chapter, it can be expected that these results can be recast using the underlying Cauchy matrix structure to return the Slavnov determinant \cite{strater_nonequilibrum_2012}. 

A determinant expression for the overlap of an arbitrary Bethe ansatz state in the Dicke model with an uncorrelated product state will first be derived. The expressions for these overlaps do not depend on the existence of a dual state, nor a dual vacuum state, so the inclusion of a bosonic level into the existing fermionic $su(2)$ models in the pseudo-deformation is well defined. The main difference with the previously-considered spin-$1/2$ models is the occurrence of the bosonic level with arbitrary occupation number. The expansion with permanents as expansion coefficients still holds, but since multiple columns of the permanent can be equal, these can not immediately be rewritten as determinants. This problem can be avoided by introducing a limiting procedure. The permanent of a coefficient matrix with distinct columns can be rewritten as a determinant, so we will introduce a matrix with different columns and consider the limit where multiple columns become equal. The overlaps can then be found by a two-step limiting procedure: first the permanent with multiple equal columns can be reduced to a determinant, after which the contraction limit can be taken. The expressions for inner products can be derived in a similar way with only minor modifications, as shown in Ref. \cite{claeys_eigenvalue-based-bosonic_2015}.

Assume a model where all spins $s_i=1/2$, except for one level (again labeled $0$) with an arbitrary large degeneracy. The Bethe state can then be expanded in a set of basis states as
\begin{align}
&\prod_{a=1}^N\left(X(\eta_0,\mu_a)S_0^{+}+\sum_{i=1}^L X(\eta_i,\mu_a)S^+_i\right)\ket{0} \nonumber \\
&\qquad \qquad \qquad \qquad =\sum_{[\{i(a)\}]}  \phi_{[\{i(a)\}]}  N_0! \left(S_0^+\right)^{N_0} \left(\prod_{a=1}^{N-N_0}S^+_{i(a)} \right)\ket{0},
\end{align}
with the summation running over all subsets $[\{i(a)\}] = [i(1) \dots i(M)]$ of the set $[1,2 \dots L]$ containing $M=N-N_0$ elements. The expansion coefficients can now be written down as
\begin{equation}
\phi_{[\{i(a)\}]}=\per \left(
\begin{array}{cccccc}
X(\eta_0,\mu_1) & \dots & X(\eta_0,\mu_1) & X(\eta_{i(1)},\mu_1) & \dots & X(\eta_{i(M)},\mu_1) \\
X(\eta_0,\mu_2) & \dots & X(\eta_0,\mu_2) & X(\eta_{i(1)},\mu_2) & \dots & X(\eta_{i(M)},\mu_2) \\
\vdots &  & \vdots & \vdots &  & \vdots \\
\undermat{N_0}{X(\eta_0,\mu_N) & \dots & X(\eta_0,\mu_N)} & X(\eta_{i(1)},\mu_N) & \dots & X(\eta_{i(M)},\mu_N) \\
\end{array}
\right).
\end{equation}
\vspace{\baselineskip}

\noindent 
In the following, it will be shown how this permanent can be written as the determinant of an $M \times M$-matrix for the trigonometric realization of the Gaudin algebra leading to the Dicke model. The Gaudin algebra elements (\ref{cont:dicke:param}) associated with the bosonic level are given by
\begin{equation}
X(\eta_0,\mu_a)=\sqrt{1+\mu_a^2}=\lim_{\eta_0 \to \infty}\frac{\sqrt{(1+\eta_0^2)(1+\mu_a^2)}}{\eta_0-\mu_{a}}.
\end{equation}
Instead of immediately taking the limit $\eta_0 \to \infty$, it is possible to first evaluate the permanent for arbitrary $\eta_0$ and later take this limit. Instead of using the same parameter $\eta_0$ for each column and taking the limit for each column simultaneously, it is also possible to introduce a different variable $\lambda_{i}$ (replacing $\eta_0$) for each column $i$, and taking consecutive limits to infinity of these parameters. This reduces the problem to the evaluation of
\vspace{\baselineskip}
\begin{equation}
\lim_{\lambda_{1}, \dots \lambda_{N_0} \to \infty} \per \left(
\begin{array}{cccccc}
X(\lambda_1,\mu_1) & \dots & X(\lambda_{N_0},\mu_1) & X(\eta_{i(1)},\mu_1) & \dots & X(\eta_{i(M)},\mu_1) \\
X(\lambda_1,\mu_2) & \dots & X(\lambda_{N_0},\mu_2) & X(\eta_{i(1)},\mu_2) & \dots & X(\eta_{i(M)},\mu_2) \\
\vdots &  & \vdots & \vdots &  & \vdots \\
\undermat{N_0}{X(\lambda_1,\mu_N) & \dots & X(\lambda_{N_0},\mu_N)} & X(\eta_{i(1)},v_N) & \dots & X(\eta_{i(M)},\mu_N) \\
\end{array}
\right).
\end{equation}
\vspace{\baselineskip}

\noindent 
This is the permanent of a matrix where each matrix element satisfies the Gaudin algebra (\ref{RG:RG:trig}), for which a determinant representation can be obtained using the results from Section \ref{RG:sec:expand}. Extending this set of variables as $\{\lambda_1 \dots \lambda_N\}=\{\lambda_1 \dots \lambda_{N_0}\} \cup \{ \eta_{i(1)} \dots \eta_{i(M)}\}$
\begin{equation}
\phi_{[\{i(a)\}]}=\lim_{\lambda_{1} \to \infty} \dots \lim_{\lambda_{{N_0}} \to \infty} \frac{\prod_{a=1}^N\sqrt{1+\mu_a^2}}{\prod_{a=1}^{N}\sqrt{1+\lambda_a^2}} \det J_N,
\end{equation}
with $J_N$ defined as
\begin{equation}
 \left(J_N\right)_{ab} =
  \begin{cases}
   \lambda_a+\sum_{c=1}^N \frac{1+\lambda_a \mu_c}{\lambda_a-\mu_c}-\sum_{c \neq a}^N\frac{1+\lambda_a \lambda_c}{\lambda_a-\lambda_c} &\text{if}\ a=b, \\
   \frac{\sqrt{1+\lambda_a^2}\sqrt{1+\lambda_b^2}}{\lambda_a-\lambda_b}       &\text{if}\ a \neq b.
  \end{cases}
\end{equation}
This is simply the application of Eq. (\ref{EVB:exp:detJ}) with the gauge $u$ tending to infinity. Absorbing the factors $(1+\lambda_a^2)^{1/4}\approx \sqrt{\lambda_{a}}, a=1 \dots N_0$, from the prefactor into the first $N_0$ columns $i$ and the first $N_0$ rows $i$ and taking the subsequent limits to infinity as $\lambda_{1}\gg \lambda_{2}\gg \dots \gg \lambda_{N_0}$, the overlap returns
\begin{align}
\phi_{[\{i(a)\}]}=\frac{\prod_{a=1}^N\sqrt{1+\mu_a^2}}{\prod_{a=1}^M\sqrt{1+\eta_{i(a)}^2}} \det
\left[
\begin{array}{ccc|c}
1 & \dots & 0 & 0  \\
\vdots &  & \vdots &   \\
0 & \dots & N_0 & 0 \\\hline
0 & \dots &  & J_M \\
\end{array}
\right]=N_0!\frac{\prod_{a=1}^M\sqrt{1+\mu_a^2}}{\prod_{a=1}^M\sqrt{1+\eta_{i(a)}^2}} \det [J_M],
\end{align}
with the $M \times M$ matrix $J_M$ defined as 
\vspace{-0.1\baselineskip}
\begin{equation}
 \left(J_M\right)_{ab} =
  \begin{cases}
  (N_0+1) \eta_{i(a)}+\sum_{c=1}^N \frac{1+\eta_{i(a)} \mu_c}{\eta_{i(a)}-\mu_c}-\sum_{c \neq a}^N\frac{1+\eta_{i(a)} \eta_{i(c)}}{\eta_{i(a)}-\eta_{i(c)}} &\text{if}\ a=b, \\
   \frac{\sqrt{1+\eta_{i(a)}^2}\sqrt{1+\eta_{i(b)}^2}}{\eta_{i(a)}-\eta_{i(b)}}       &\text{if}\ a \neq b.
  \end{cases}
\end{equation}
The contraction limit can now be taken without any problems, leading to the overlap between a basis state $\ket{N_0;\{i(a)\}} = \left(b^{\dagger}\right)^{N_0} \prod_{a=1}^M S^+_{i(a)}\ket{0}$ and a Bethe state in the Dicke model as
\begin{equation}
\braket{N_0;\{i(a)\}|v_1 \dots v_N} = \sqrt{N_0!} (-G)^M \det J_M,
\end{equation}
with $J_M$ exhibiting the usual structure of
\vspace{-0.2\baselineskip}
\begin{equation}
 \left(J_M\right)_{ab} =
  \begin{cases}
   \sum_{c=1}^N\frac{1}{\epsilon_{i(a)}-v_c}-\sum_{c \neq a}^{M}\frac{1}{\epsilon_{i(a)}-\epsilon_{i(c)}} &\text{if}\ a=b, \\
   \frac{1}{\epsilon_{i(a)}-\epsilon_{i(b)}}       & \text{if}\ a \neq b.
  \end{cases}
\end{equation}
A similar procedure can be used to obtain determinant expressions for the overlaps between original states and dual states, for which this final expression provides the building block. Repeating this for the extended $p_x+ip_y$ model then returns an expansion of the Bethe state
\begin{equation}
\ket{v_1 \dots v_N}=\prod_{a=1}^N\left(b^{\dagger}-\sum_{i=1}^L \frac{\sqrt{\epsilon_i}v_a}{\epsilon_i-v_a}\frac{S^{+}_i}{\eta_0}\right)\ket{0},
\end{equation}
with expansion coefficients
\begin{equation}
\braket{N_0; \{i(a)\}|v_1 \dots v_N}=\frac{\sqrt{N_0!}}{\sqrt{\prod_a \epsilon_{i(a)}}} \frac{\det J_M}{\eta_0^{M}},
\end{equation}
where $J_M$ is now defined as 
\vspace{-0.2\baselineskip}
\begin{equation}
\left(J_M\right)_{ab}=\begin{cases}
\frac{1}{2}\sum_{c=1}^N \frac{\epsilon_{i(a)}+v_c}{\epsilon_{i(a)}-v_c}-\sum_{c \neq a}^{M}\frac{1}{2}\frac{\epsilon_{i(a)}+\epsilon_{i(c)}}{\epsilon_{i(a)}-\epsilon_{i(c)}}-\frac{N_0+1}{2} &\text{if}\ a=b,\\
\frac{\sqrt{\epsilon_{i(a)}\epsilon_{i(b)}}}{\epsilon_{i(a)}-\epsilon_{i(b)}}  &\text{if}\ a \neq b.
\end{cases}
\end{equation}
These results can be extended towards normalizations and inner products by following the same route. Here, the results for XXZ models with spin-$1/2$ containing a single spin with arbitrary degeneracy can be immediately extended towards models with a bosonic degree of freedom. While no dual state exists in the contraction limit, a dual state can be found for the XXZ models at arbitrary degeneracies and determinant expressions can again be obtained in which the dual state is implicit. These will not be repeated due to the largely similar derivation, but can be found in Ref. \cite{claeys_eigenvalue-based-bosonic_2015}.

\part{Applications}

\chapter[Read-Green resonances]{Read-Green resonances in a topological superconductor \\ coupled to a bath}
\label{chap:readgreen}

\setlength\epigraphwidth{.6\textwidth}
\epigraph{\emph{Topology is the property of something that doesn't change when you bend it or stretch it as long as you don't break anything.}}{{Edward Witten}}

The coming chapters now involve applications of the framework presented in the first half of this thesis to various physical domains. In this chapter, a topological $p_x+ip_y$ superconductor capable of exchanging particles with an environment is studied. This additional interaction breaks particle-number symmetry, but can still be modeled by means of an integrable Hamiltonian building on the class of XXZ Richardson-Gaudin pairing models. The isolated system supports zero-energy modes at a topological phase transition, which disappear when allowing for particle exchange with an environment. However, it is shown from the exact solution that these still play an important role in system-environment particle exchange, which can be observed through resonances in low-energy and -momentum level occupations. These fluctuations signal topologically protected Read-Green points and cannot be observed within traditional mean-field theory. 

The first successful description of superconductivity was given by Bardeen, Cooper, and Schrieffer (BCS) by introducing a collective ground state consisting of condensed Cooper pairs, explicitly violating conservation of particle number \cite{bardeen_theory_1957}. This original mean-field theory was formulated for an $s$-wave pairing interaction, where the gap function is assumed isotropic. As mentioned in Section \ref{RG:subsec:pip}, it was later shown how topological superconductivity arose from a chiral $p_x+ip_y$-wave interaction by breaking time-reversal symmetry \cite{read_paired_2000,ryu_topological_2010}. Most of the theoretical insights into topological superconductivity are based on mean-field Bogoliubov-de Gennes theory, as initiated by Read and Green  \cite{read_paired_2000}. One of their crucial results was the uncovering of a phase transition between a (topologically nontrivial) weak-pairing and a (topologically trivial) strong-pairing state. At this transition, the chemical potential vanishes and the Bogoliubov quasiparticle spectrum becomes gapless. The main goal is now to put the robustness of the topological phase transition in a finite system to test when exchange of particles with an environment is allowed. For this, a $p_x+ip_y$ superconductor coupled to a bath system is modeled by means of a recently-proposed integrable model \cite{lukyanenko_integrable_2016}. The Hamiltonian can be regarded as a hybrid between an integrable number-conserving Hamiltonian describing superconductivity and a mean-field term coupling this system to an infinite bath, allowing for particle-number fluctuations. 

First, the topological nature of the isolated system is discussed both from the mean-field and the Bethe ansatz wave function, after which the integrable Hamiltonian is extended by coupling it to a bath. The exact Bethe eigenstates are presented together with the Bethe and eigenvalue-based equations. As can be expected from a particle-number non-conserving Hamiltonian, the exact eigenstates mix separate $U(1)$-gauge (particle-number) symmetry sectors. Interestingly, they retain the common factorized form of Bethe states. This exact solution is found to be essential to observe the effects of particle-exchange on the topological phase transition. It is shown how zero-energy excitations associated with the phase transition govern the particle-exchange with the bath, resulting in avoided crossings between states from different $U(1)$-symmetry sectors at the Read-Green points. These can be observed from strong fluctuations in the single-particle level occupations. This connects the physics of zero-energy excitations, arising from topological phase transitions, with the physics of open quantum systems. Furthermore, it is shown that while the breaking of particle-number symmetry in mean-field Bogoliubov-de Gennes theory has known great success in the description of number-conserving superconductors, it fails in describing the fluctuations observed in the exact solution.  The goal of this chapter is then twofold: to show how Bethe states can still be obtained even if particle-number $U(1)$ symmetry is broken and to investigate the physical effects of particle-exchange. The present chapter is largely based on Ref. \cite{claeys_read-green_2016}.

\section{Topological superconductivity}
\subsection{Mean-field theory}
\begin{figure}                    
 \begin{center}
 \includegraphics{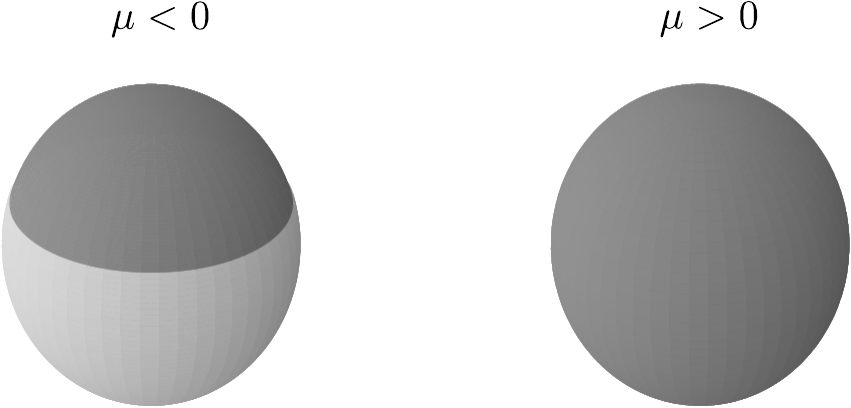}
 \caption{Graphical representation of the different topological behaviours in the strong-pairing ($\mu <0$) and the weak-pairing ($\mu >0$) phase. The dark gray area is the area covered by $\hat{\mathbf{h}}(\mathbf{k})$. Based on a similar figure in \cite{alicea_new_2012}.\label{fig:pip:spheres}\vspace{-\baselineskip}}
 \end{center}
\end{figure}

The $p_x+ip_y$ pairing Hamiltonian as introduced in Chapter \ref{Chap_RGmodels} is given by 
\begin{align}\label{pip:top:hamiltoniannobath}
&H_{p+ip} = \sum_{\mathbf{k}}\frac{|\mathbf{k}|^2}{2m}c^{\dagger}_{\mathbf{k}}c_{\mathbf{k}} -\frac{G}{4m}\sum_{\substack{\mathbf{k},\textbf{k}' \\ \mathbf{k} \neq \pm \textbf{k}'}}(k_x+ik_y)(k_x'-ik_y')c^{\dagger}_{\mathbf{k}}c^{\dagger}_{-\mathbf{k}}c_{-\mathbf{k}'}c_{\mathbf{k}'},
\end{align}
in which $c_{\mathbf{k}}$ and $c^{\dagger}_{\mathbf{k}}$ denote annihilation and creation operators respectively for spinless fermions of mass $m$ with two-dimensional momentum $\mathbf{k}=(k_x,k_y)$, and a dimensionless coupling constant $G$ has been introduced (\ref{RG:pip:ham}). Within mean-field theory, the Hamiltonian (\ref{pip:top:hamiltoniannobath}) can be approximated by the quadratic Hamiltonian
\begin{equation}\label{pip:top:hammf}
H_{mf} = \sum_{\mathbf{k}} \tilde{\epsilon}(\mathbf{k})c^{\dagger}_{\mathbf{k}}c_{\mathbf{k}} - \left[\frac{\Delta}{2} \sum_{\mathbf{k}}(k_x+ik_y)c^{\dagger}_{\mathbf{k}}c^{\dagger}_{-\mathbf{k}} + \textrm{h.c.} \right] + \frac{2m}{G}|\Delta|^2+\mu N,
\end{equation}
where a BCS pairing gap/order parameter $\Delta$ and mean-field single-particle energies have been defined as
\begin{equation}
\Delta = \frac{G}{2m}\sum_{\mathbf{k}}(k_x-i k_y)\braket{c_{-\mathbf{k}}c_{\mathbf{k}}}, \qquad \tilde{\epsilon}(\mathbf{k}) = (1+G)\frac{|\mathbf{k}|^2}{2m}-\mu.
\end{equation}
The interaction terms in this Hamiltonian have been approximated as $AB = \braket{A}B+A\braket{B}-\braket{A}\braket{B}$. The resulting Hamiltonian explicitly breaks particle-number $U(1)$ symmetry, and a chemical potential $\mu$ has been added, acting as a Lagrange multiplier when fixing the total particle number. Due to its quadratic nature, it can be diagonalized using a Bogoliubov transformation \cite{bogoliubov_new_1958}, and the chemical potential and gap follow from a self-consistent treatment of the mean-field Hamiltonian (\ref{pip:top:hammf}). The quadratic part of the Hamiltonian can be recast as
\begin{equation}
H_{mf} \propto \sum_{\mathbf{k}} 
\left(
\begin{matrix}
 c^{\dagger}_{\mathbf{k}} & c_{-\mathbf{k}}
\end{matrix}
\right)
\left(
\begin{matrix}
\tilde{\epsilon}(\mathbf{k}) & -\Delta (k_x+ik_y) \\
-\Delta^* (k_x-ik_y) & -\tilde{\epsilon}(\mathbf{k})
\end{matrix}
\right)
\left(
\begin{matrix}
 c_{\mathbf{k}} \\ c^{\dagger}_{-\mathbf{k}}
\end{matrix}
\right) + \textrm{Cst}.
\end{equation}
The topological character is already contained in this Hamiltonian after a self-consistent determination of $\mu$, which will depend on both the interaction strength and the total particle number. The $2 \times 2$ Hamiltonian matrix $\mathcal{H}(\mathbf{k})$ appearing in this expression can be expanded as $
\mathcal{H}(\mathbf{k}) = \mathbf{h}(\mathbf{k})\cdot \vec{\sigma}$, with $\vec{\sigma}$ the vector of Pauli matrices and $\mathbf{h}(\mathbf{k})$ defined as
\begin{equation}
{h}_x(\mathbf{k}) = -\Re\left[\Delta (k_x+ik_y) \right], \qquad {h}_y(\mathbf{k}) = \Im\left[\Delta (k_x+ik_y) \right], \qquad {h}_z(\mathbf{k}) =\tilde{\epsilon}(\mathbf{k}).
\end{equation}
The topological information is now encoded in the behaviour of this vector as the momentum is varied over all possible momenta $\mathbf{k}=(k_x,k_y)$ \cite{alicea_new_2012}. Taking the normalized unit vector $\hat{\mathbf{{h}}}(\mathbf{k}) = \mathbf{h}(\mathbf{k}) / | \mathbf{h}(\mathbf{k}) |$, this vector maps momenta to the unit sphere (see Figure \ref{fig:pip:spheres}). For momenta with fixed $|\mathbf{k}|$, $\hat{h}_x$ and $\hat{h}_y$ sweep out a circle on the sphere at fixed height $\hat{h}_z$. The height at the initial and final point are given by
\begin{equation}
\hat{h}_z(|\mathbf{k}| \to 0) = -\sgn (\mu), \qquad \hat{h}_z(|\mathbf{k}| \to \infty) = 1.
\end{equation}
The global behaviour is now only dependent on the sign of the chemical potential, so it is possible to distinguish two phases. If $\mu < 0$, $\hat{\mathbf{h}}(\mathbf{k})$ starts at the north pole and initially sweeps out the shaded region in the northern hemisphere of Figure \ref{fig:pip:spheres}, but then `unsweeps' the same area before ending up at the north pole again. In contrast, for $\mu >0$, $\hat{\mathbf{h}}(\mathbf{k})$ transitions from the north to the south pole, fully covering the unit sphere. 

Topology then underlies the fact that these regimes constitute distinct phases that cannot be smoothly connected without closing the bulk gap $|\mathbf{h}(\mathbf{k})| = \sqrt{\tilde{\epsilon}(\mathbf{k})^2+|\Delta|^2|\mathbf{k}|^2}$. This bulk gap fully determines the Bogoliubov quasi-particle spectrum and only vanishes if both $|\mathbf{k}|=0$ and $\mu=0$, exactly at the phase transition. The underlying topological invariant is the Chern number $C$ (winding number), with either $|C|=0$ (for the topologically trivial `strong-pairing' phase with $\mu < 0$) or  $|C|=1$ (for the topologically non-trivial `weak-pairing' phase with $\mu > 0$)  \cite{volovik_analog_1988,foster_quantum_2013}.
Physically, these can be distinguished by the real-space behaviour of the mean-field wave function $\ket{\psi_{mf}}$, following from the Hamiltonian \ref{pip:top:hammf} as
\begin{equation}
\ket{\psi_{mf}} \propto \prod_{\mathbf{k} > 0 }\left[1+\varphi(\mathbf{k})c^{\dagger}_{\mathbf{k}}c^{\dagger}_{\mathbf{-k}}\right] \ket{0}, \qquad \textrm{with} \qquad \varphi(\mathbf{k}) = \frac{\sqrt{\tilde{\epsilon}(\mathbf{k})^2+|\Delta|^2|\mathbf{k}|^2}-\tilde{\epsilon}(\mathbf{k})}{\Delta^*(k_x-ik_y)}.
\end{equation}
The qualitative real-space behaviour follows from
\begin{equation}
\lim_{\mathbf{k} \to 0}\braket{\psi_{mf}|c^{\dagger}_{\mathbf{k}}c_{\mathbf{k}}|\psi_{mf}} = \frac{1}{2}+\frac{1}{2}\sgn(\mu).
\end{equation}
As a result, the strong-pairing phase exhibits `BEC'-like behaviour where the Cooper pairs are strongly confined in real space, whereas the weak-pairing phase is a `BCS'-like phase corresponding to weakly-bound Cooper pairs \cite{dunning_exact_2010,rombouts_quantum_2010}.

\subsection{Bethe ansatz}
Alternatively, the theory of Richardson-Gaudin integrability also provides insights into topological superconductivity by means of the exact solution in finite-size systems \cite{dukelsky_colloquium_2004}. Richardson's original solution for the reduced $s$-wave pairing Hamiltonian was generalized to $p_x+ip_y$ interactions through a variety of means \cite{ibanez_exactly_2009,skrypnyk_non-skew-symmetric_2009,dunning_exact_2010,rombouts_quantum_2010,van_raemdonck_exact_2014,lukyanenko_boundaries_2014}, allowing for an exact calculation of spectral properties and correlation coefficients \cite{dunning_exact_2010,claeys_eigenvalue-based_2015}. From this exact solution, it was shown how the topological phase transition is reflected in the \emph{Read-Green} points for finite systems. At these points, it is possible to reach excited states through a fixed number of zero-energy pair excitations. When a single zero-energy excitation is allowed, this corresponds to a vanishing chemical potential and the topological phase transition is recovered in the thermodynamic limit of the Richardson-Gaudin solution \cite{ibanez_exactly_2009,rombouts_quantum_2010}.  Again applying the quasispin formalism \cite{talmi_simple_1993}
\begin{align}
S^{+}_{\mathbf{k}}=\frac{k_x+ik_{y}}{|\mathbf{k}|}c^{\dagger}_{\mathbf{k}}c^{\dagger}_{-\mathbf{k}}, \qquad S^{-}_{\mathbf{k}}=\frac{k_x-ik_{y}}{|\mathbf{k}|}c_{-\mathbf{k}}c_{\mathbf{k}}, \qquad S_{\mathbf{k}}^z = \frac{1}{2}(c^{\dagger}_{\mathbf{k}}c_{\mathbf{k}} + c^{\dagger}_{-\mathbf{k}}c_{-\mathbf{k}}-1),
\end{align} 
defining $\epsilon_k=|\mathbf{k}|/\sqrt{2m}$ and labeling $L$ different states $\pm|\mathbf{k}|$ with integers\footnote{We will restrict ourselves to spin-$1/2$. For the model at hand, this does not allow $|\mathbf{k}|=|\mathbf{k'}|$ if $\mathbf{k} \neq \pm \mathbf{k'}$. The $\epsilon_{\mathbf{k}}^2$ in this chapter also correspond to the $\epsilon_{\mathbf{k}}$ in Chapter \ref{Chap_RGmodels}.}, the exact eigenstates containing $N$ fermion pairs (or $2N$ fermions) are given by (\ref{RG:pip:BA}) as
\begin{equation}\label{wavefunctionnointeraction}
\ket{v_1 \dots v_N}_N=\prod_{a=1}^NS^+(v_a)\ket{0}=\prod_{a=1}^N\left(\sum_{k=1}^L \frac{\epsilon_k }{\epsilon_k^2-v_{a}^2}S^{+}_k\right)\ket{0},
\end{equation}
with $\ket{0}$ denoting the particle vacuum state and the rapidities $\{v_1 \dots v_N\}$ coupled through a set of Bethe equations \cite{ibanez_exactly_2009}
\begin{equation}
(1+G)-G\sum_{j=1}^L \frac{\epsilon_j^2}{\epsilon_j^2-v_{a}^2}+2G\sum_{b \neq a}^L\frac{v_{b}^2}{v_{b}^2-v_{a}^2} = 0, \qquad \forall a=1 \dots N.
\end{equation}
The total energy of this state is (up to a constant) given by $(1+G)\sum_{a=1}^N v_{a}^2$, so each rapidity $v_{a}^2$ can be loosely interpreted as the energy of a single Cooper pair. These Bethe equations have the remarkable property that zero-energy fermion pair excitations are supported at specific fractional values of the coupling constant $G^{-1}=L-2N-p$, $p \in \mathbb{N}$, the so-called Read-Green points \cite{ibanez_exactly_2009,rombouts_quantum_2010,links_exact_2015}. At $G^{-1}=L-2N-p$, if the set of $N$ rapidities $\{v_1,\dots, v_N\}$ is a solution to the Bethe equations, then the set of $N+p$ rapidities $\{v_1,\dots, v_N,0,\dots,0\}$ is another solution to the Bethe equations. These zero-solutions do not contribute to the energy, so the states defined by these variables are degenerate. This can also be found through operator identities, since $[H_{p+ip},S^+(0)^p]\propto G^{-1}-L+2\hat{N}+p$ \cite{links_exact_2015}.

\begin{figure}                    
 \begin{center}
 \includegraphics{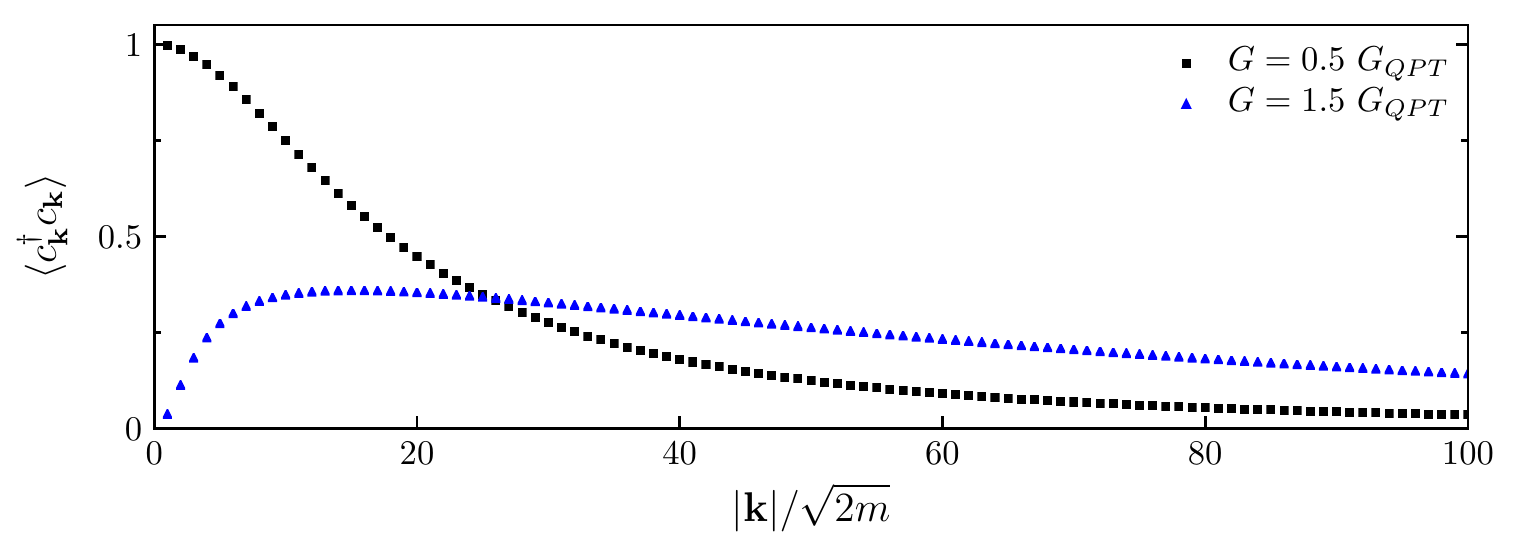}
 \caption{Occupation numbers $\braket{c^{\dagger}_{\mathbf{k}}c_{\mathbf{k}}}$ for a system with 25 fermion pairs in 100 levels, where $G_{QPT}^{-1}= L-2N-1$.\label{fig:pip:occ} \vspace{-\baselineskip}}
 \end{center}
\end{figure}

The $p=1$ case corresponds to a vanishing chemical potential in mean-field theory, since a fermion pair can then be added without changing the energy. Indeed, it has been shown that in the thermodynamic limit a third-order topological phase transition occurs precisely at this point, accompanied by nonanalytic behaviour of the ground-state energy \cite{rombouts_quantum_2010,ortiz_many-body_2014}. Still, these zero-solutions should be contrasted to the gapless Bogoliubov quasiparticles found in mean-field theory: while both correspond to zero-energy excitations, the latter is a single-fermion quasiparticle excitation, while the former corresponds to a collective fermion pair excitation, as can be seen from Eq. (\ref{wavefunctionnointeraction}). The value of the coupling constant $G^{-1}=L-2N-1$ where the phase transition occurs depends on the pair density $N/L$ but not on the single-particle energies $\epsilon_k$, reflecting its topological nature. In finite-size systems it is no longer possible to explicitly define a Chern number, but an equivalent topological invariant has been suggested as the occupation of the $|\mathbf{k}|=0$ zero-momentum level $\lim_{\mathbf{k} \to 0} \braket{c^{\dagger}_{\mathbf{k}}c_{\mathbf{k}}}$ \cite{ortiz_many-body_2014,ortiz_what_2016}. This is illustrated in Figure \ref{fig:pip:occ}, where this occupation exactly corresponds to the proposed topological invariant.

\section{Interaction with a bath}

The topological phase transition is closely related to a degeneracy between states with different numbers of fermions. When the Hamiltonian conserves particle number, these states are symmetry-protected and do not interact. However, once this symmetry is broken, e.g. by coupling to a bath, it's possible for these degenerate states to interact strongly. Such an interaction with a bath can be modeled by introducing a coupling term
\begin{equation}\label{pip:bath:ham}
H =H_{p+ip}+\frac{\gamma}{\sqrt{2m}} \sum_{\mathbf{k}}\left[(k_x+ik_y)c^{\dagger}_{\mathbf{k}}c^{\dagger}_{-\mathbf{k}}+\text{h.c.}\right].
\end{equation}
This coupling term makes abstraction of the exact nature of the bath, and allows for particle exchange tunable by a single parameter $\gamma$. Alternatively, this additional term can be seen as a partial mean-field approximation of a more general $p_x+ip_y$ interaction Hamiltonian, allowing for particle-number fluctuations\footnote{Note that this interaction conserves fermion parity, so we will not consider the effect of Majorana zero modes associated with the spontaneous breaking of this symmetry.}. As such, the gap of this model is given by $\Delta=\frac{G}{2m}\sum_{\mathbf{k}}(k_x-ik_y)\braket{c_{-\mathbf{k}}c_{\mathbf{k}}}-\frac{\gamma}{\sqrt{2m}}$, which is the pairing gap for an isolated system shifted by the bath-coupling $\gamma/\sqrt{2m}$ \cite{ibanez_exactly_2009,rombouts_quantum_2010}.

The set of conserved quantities\footnote{We again follow the presentation from \cite{lukyanenko_boundaries_2014,lukyanenko_integrable_2016}, differing from the usual one in the asymmetry in the last term. This already incorporates the particle-number dependent redefinition of the coupling constant and simplifies the resulting operator identities. } associated with the integrable Hamiltonian (\ref{pip:bath:ham}) are given by \cite{lukyanenko_integrable_2016},
\begin{align}\label{pip:bath:com}
Q_{k}=&\left(S_k^z+\frac{1}{2}\right) +\gamma \epsilon_k^{-1} \left(S^{+}_k + S^-_k \right) \nonumber \\
&\qquad -G \sum_{k' \neq k}^L \frac{\epsilon_k \epsilon_{k'}}{\epsilon_k^2-\epsilon_{k'}^2}\left(S^{+}_{k'} S^-_k+S^-_{k'} S^{+}_k\right) -2G\sum_{k' \neq k}^L \frac{\epsilon_{k'}^2}{\epsilon_k^2-\epsilon_{k'}^2}\left(S_k^zS_{k'}^z-\frac{1}{4}\right),
\end{align}
where $[H,Q_{k}]=[Q_{k},Q_{k'}]=0$ and $H=\sum_k \epsilon_k^2 Q_{k}$. From direct calculation it can be shown that the relations
\begin{equation}\label{pip:bath:evbeq}
Q_{k}^2 = Q_{k} +\gamma^2 \epsilon_k^{-2}+ G\sum_{k' \neq k}^L \epsilon_{k'}^2\frac{Q_{k}-Q_{k'}}{\epsilon_k^2-\epsilon_{k'}^2},\qquad \forall k=1 \dots L,
\end{equation}
hold at the operator level provided all spins are spin-$1/2$. These present a direct generalization of the previously-presented identities (\ref{EVB:EVB:EVBeq}) and immediately lead to a new set of eigenvalue-based equations, which can be solved following the methods of Chapter \ref{chap:EVB}.

\section{Signatures of the topological phase transition}

\begin{figure}                   
 \begin{center}
 \includegraphics{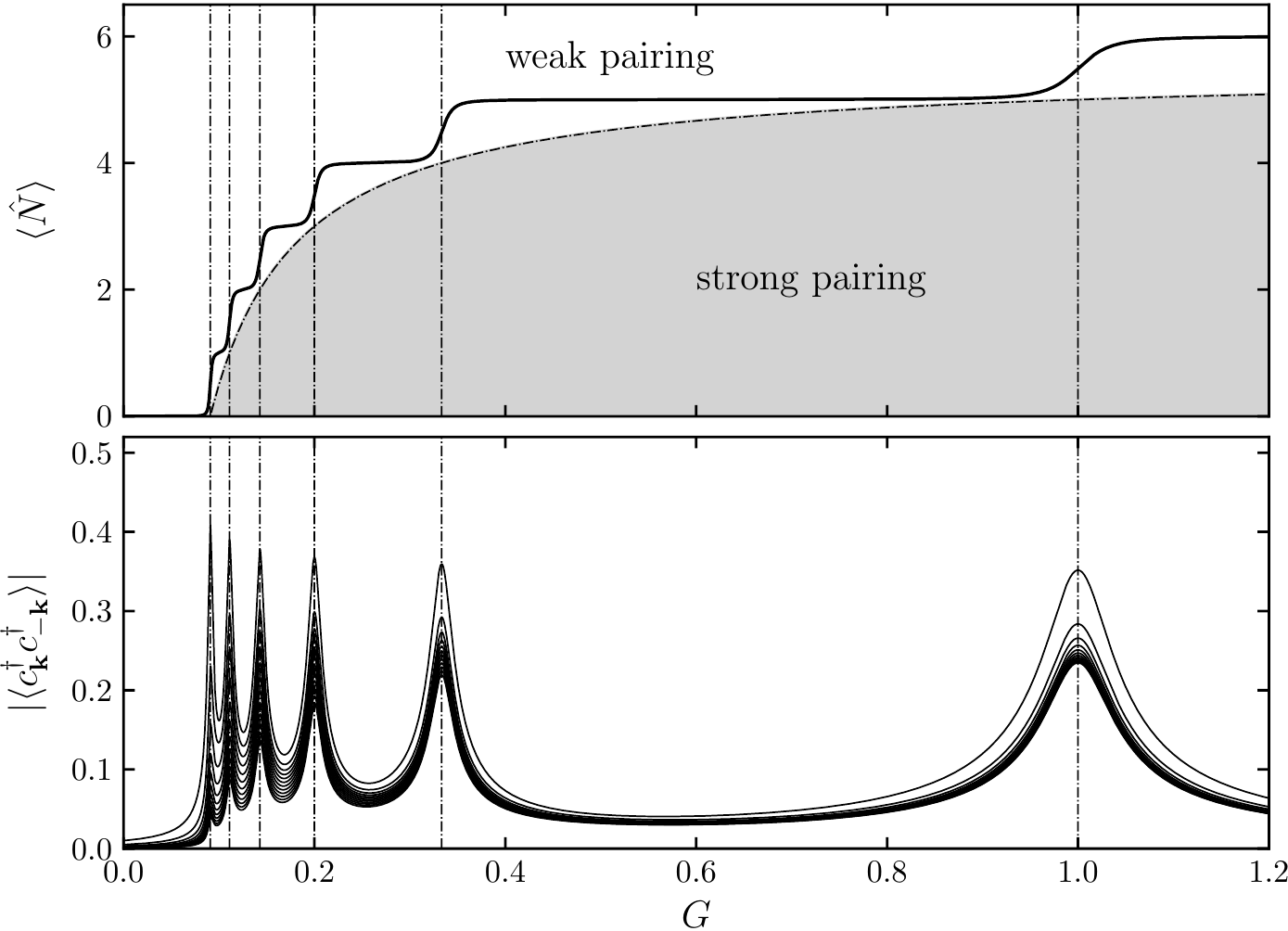}
 \caption{Average pair number $\braket{\hat{N}}$ and expectation values $\braket{c^{\dag}_{\mathbf{k}}c^{\dag}_{-\mathbf{k}}}$ for the ground state of  a picket-fence model ($L=12$) with $\gamma=10^{-2}$. At the Read-Green point $G^{-1}=L-2N-1$ the average pair number changes from $N$ to $N+1$, accompanied by sharp peaks in $\braket{c^{\dag}_{\mathbf{k}}c^{\dag}_{-\mathbf{k}}}, \forall \mathbf{k}$. The line $\braket{\hat{N}} = (L-G^{-1}-1)/2$ separating the two phases in the isolated system ($\gamma=0$) has also been given in the top figure. \vspace{-\baselineskip}\label{groundstate}}
 \end{center}
\end{figure}

\begin{figure}                 
 \begin{center}
 \includegraphics{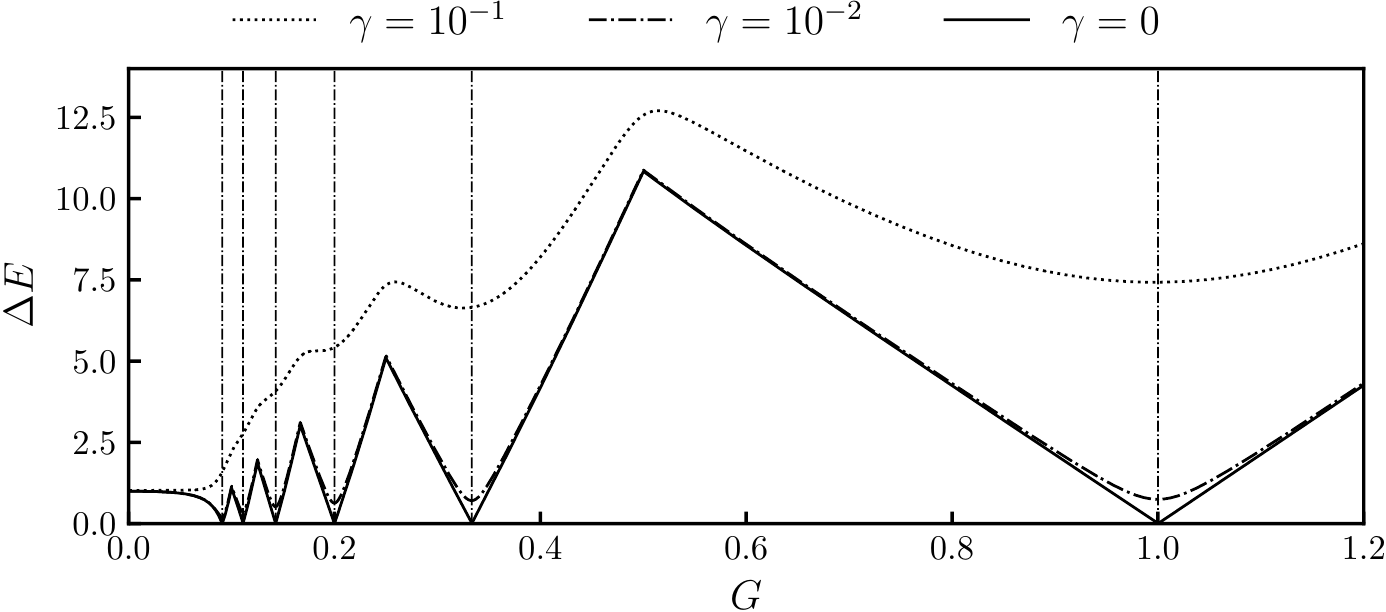}
 \caption{Energy gap $\Delta E$ between the first excited state and the ground state of a picket-fence model ($L=12$) for different values of $\gamma$. The Read-Green points $G^{-1}=L-2N-1$ with $N=0 \dots L/2-1$ are marked by vertical lines. \vspace{-\baselineskip}\label{gap}}
 \end{center}
\end{figure}

It's now a natural question to ask if (and how) the phase transition is modified by this additional interaction term. This will be done through the calculation of expectation values and energy spectra. The expectation values in this section can all be obtained from the Hellmann-Feynman theorem, so no explicit knowledge of the Bethe state is necessary yet. Consider the population of the ground state $\braket{\hat{N}}$ for a small system, with $\hat{N}$ the pair number operator, together with the expectation values $\braket{c^{\dag}_{\mathbf{k}}c^{\dag}_{-\mathbf{k}}}$ in Figure \ref{groundstate}. It can be seen that the average number of Cooper pairs in the ground state increases by one unit near a Read-Green point $G^{-1}=L-2N-1, N \in \mathbb{N}$. The Read-Green points mark the phase transitions of isolated systems with different (fixed) densities, and the fluctuations in the density here lead to a series of Read-Green points, each associated with a different density.
The mechanism underlying this particle exchange with the environment can be understood from Figure \ref{gap}. At the Read-Green points, the isolated system ($\gamma=0$) is gapless by definition, and a gap opens up for increasing $|\gamma|$. Where a phase transition is expected in the canonical regime with fixed pair density, an (avoided) level crossing occurs instead when the exchange of particles with an environment is allowed. Instead of crossing the Read-Green line with increasing $G$, the system absorbs a fermion pair from the environment near the Read-Green point, remaining in the weak-pairing phase at each interaction strength.

At the phase transition, the isolated system is gapless and the degenerate states are symmetry-protected due to the differing particle number. However, the interaction with the environment couples states with different particle number, and for small coupling $|\gamma|$, the states at the Read-Green points are (up to a perturbative correction) given by the superposition $\frac{1}{\sqrt{2}}\ket{v_1 \dots v_N}_N-\frac{1}{\sqrt{2}}\ket{v_1 \dots v_N \ 0}_{N+1}$, as follows immediately from perturbation theory. This strong deviation from the symmetry-protected states can be inferred from Figure \ref{groundstate}, where the expectation values $\braket{c^{\dag}_{\mathbf{k}}c^{\dag}_{-\mathbf{k}}}$ are exactly zero when particle-number symmetry is conserved, but here exhibit sharp resonances exactly at the Read-Green points.

\begin{figure}                    
 \begin{center}
 \includegraphics{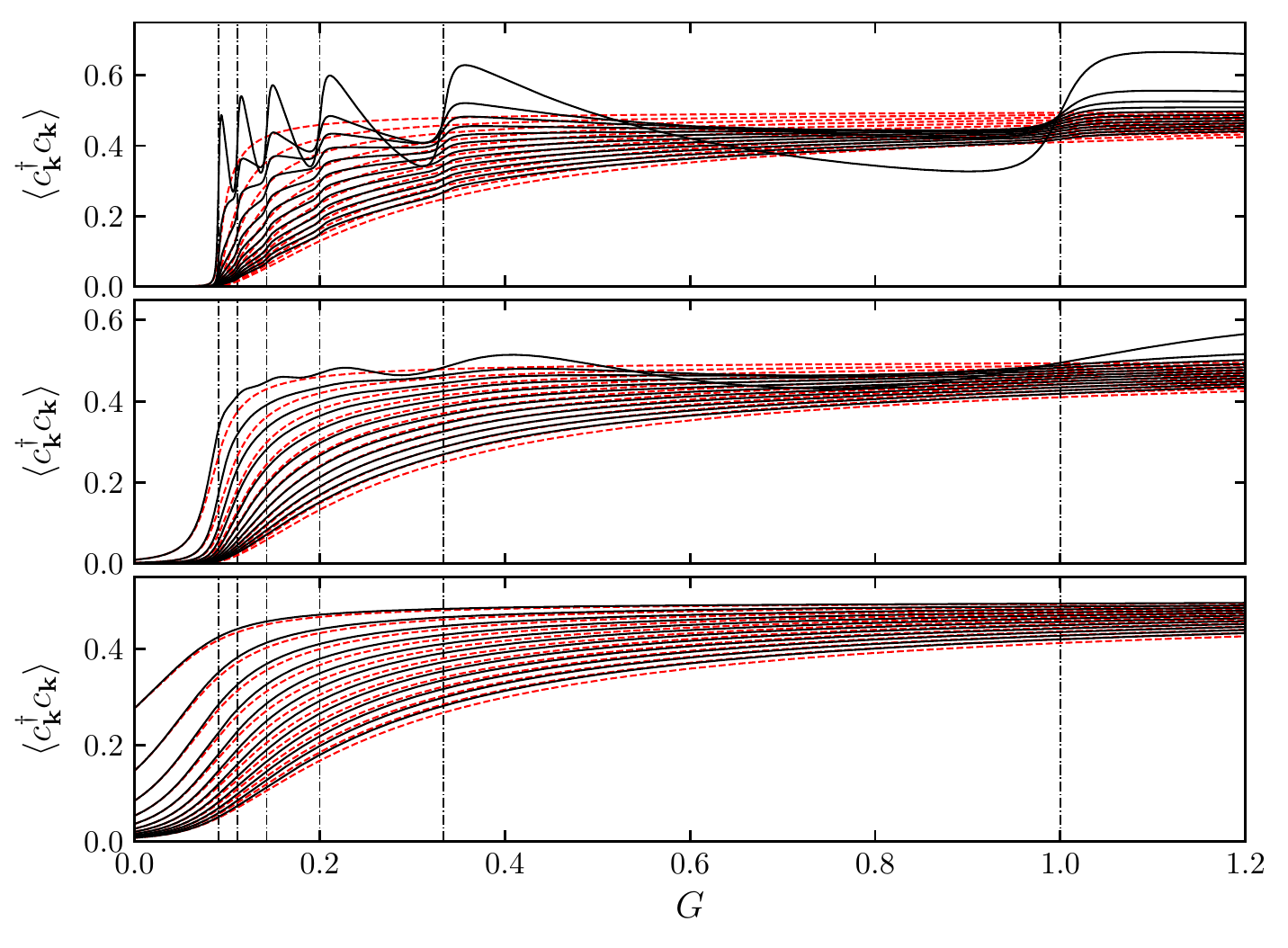}
 \caption{Occupation numbers of the ground state for varying $G$ at three different values of $\gamma=10^{-2},10^{-1}, 10^0$ (top to bottom). The exact solution is marked by black lines, while the mean-field solution is marked by red dashed lines. In the top figure, the lowest-momentum state exhibits peaks near the Read-Green points, which are again marked by vertical lines.\vspace{-\baselineskip}\label{expectationvalues}}
 \end{center}
\end{figure}

While the zero-energy excitations at the phase transition are not allowed for non-zero system-bath coupling, they can still be observed in the level occupations of the ground state, as shown in Figure \ref{expectationvalues}. For small $|\gamma|$, strong fluctuations are observed in the occupancy of the lowest-energy and momentum states. These can be seen as the signatures of the zero-energy modes existing at each Read-Green point. For small interaction strengths the zero modes result in large fluctuations of the occupation of the lowest-energy and momentum states, which exhibit sharp resonances near the Read-Green points, and can as such be termed \emph{Read-Green resonances}. This can again be connected to the topological invariant in finite-size systems -- if the systems were to undergo a phase transition, the lowest-momentum states would become unoccupied. However, it now becomes energetically favourable for the system to absorb a low-momentum fermion pair from the environment, leading to occupied low-momentum states indicated by these resonances.

A comparison with the mean-field treatment, on which the majority of theoretical insights for topological superconductors are based \cite{read_paired_2000}, is also given in Figure \ref{expectationvalues}. These mean-field results are based on the standard derivation for the $p_x+ip_y$-Hamiltonian \cite{read_paired_2000,botelho_quantum_2005,ibanez_exactly_2009,rombouts_quantum_2010}, where in Eq. (\ref{pip:top:hammf}) the chemical potential $\mu$ is now set to zero and the gap $\Delta$ instead contains a contribution from $\gamma$. For small $|\gamma|$ it can be seen that mean-field theory fails in capturing the fluctuations in the low-energy states, while for larger $|\gamma|$ mean-field theory becomes increasingly more accurate. While it is known that mean-field methods are not adequate in finite-size systems, the extent to which they fail at detecting these resonances is remarkable. For large system-bath coupling the mean-field term in the Hamiltonian becomes dominant, so it is expected that mean-field theory will provide reliable results in this regime. From this, it is clear the regime with small system-bath coupling is the most physically interesting.

Although all calculations were performed on an integrable model, these results do not depend on integrability. They originate purely from the topological phase transition and the related zero-energy excitations, coupling states within different particle-number symmetry sectors. Apart from the ground state, Read-Green points are also spread throughout the entire spectrum, connecting these results with recent work on strong zero modes \cite{alicea_topological_2016}.

\section{Rapidities}

The structure of the eigenstates can also be used to shed light on the particle-exchange mechanism and its relation with zero-energy excitations. Remarkably, the Hamiltonian (\ref{pip:bath:ham}) can be solved with a Bethe ansatz state
\begin{equation}\label{wavefunctionwithinteraction}
\ket{v_1 \dots v_L}_{\gamma}=\prod_{a=1}^L\left(\frac{\gamma}{ v_{a}^2}+G\sum_{k=1}^L \frac{\epsilon_k }{\epsilon_k^2-v_{a}^2}S^{+}_k\right)\ket{0},
\end{equation}
as a generalization of (\ref{wavefunctionnointeraction}), where the rapidities are now coupled through
\begin{equation}\label{pip:rap:bae}
(1+G)-G\sum_{j=1}^L \frac{\epsilon_j^2}{\epsilon_j^2-v_{a}^2}+2G\sum_{b \neq a}^L\frac{v_{b}^2}{v_{b}^2-v_{a}^2} = -\frac{\gamma^2}{G} \frac{\prod_{j=1}^L(v_{a}^{-2}-\epsilon_j^{-2})}{\prod_{b \neq a}^L (v_{a}^{-2}-v_{b}^{-2})}.
\end{equation}
These Bethe equations were originally presented in Ref. \cite{lukyanenko_integrable_2016} and the Bethe state in Ref. \cite{claeys_read-green_2016}, and are derived in Appendix \ref{app:pip:BAE} using a commutator scheme.  
\begin{figure}                    
 \begin{center}
 \includegraphics{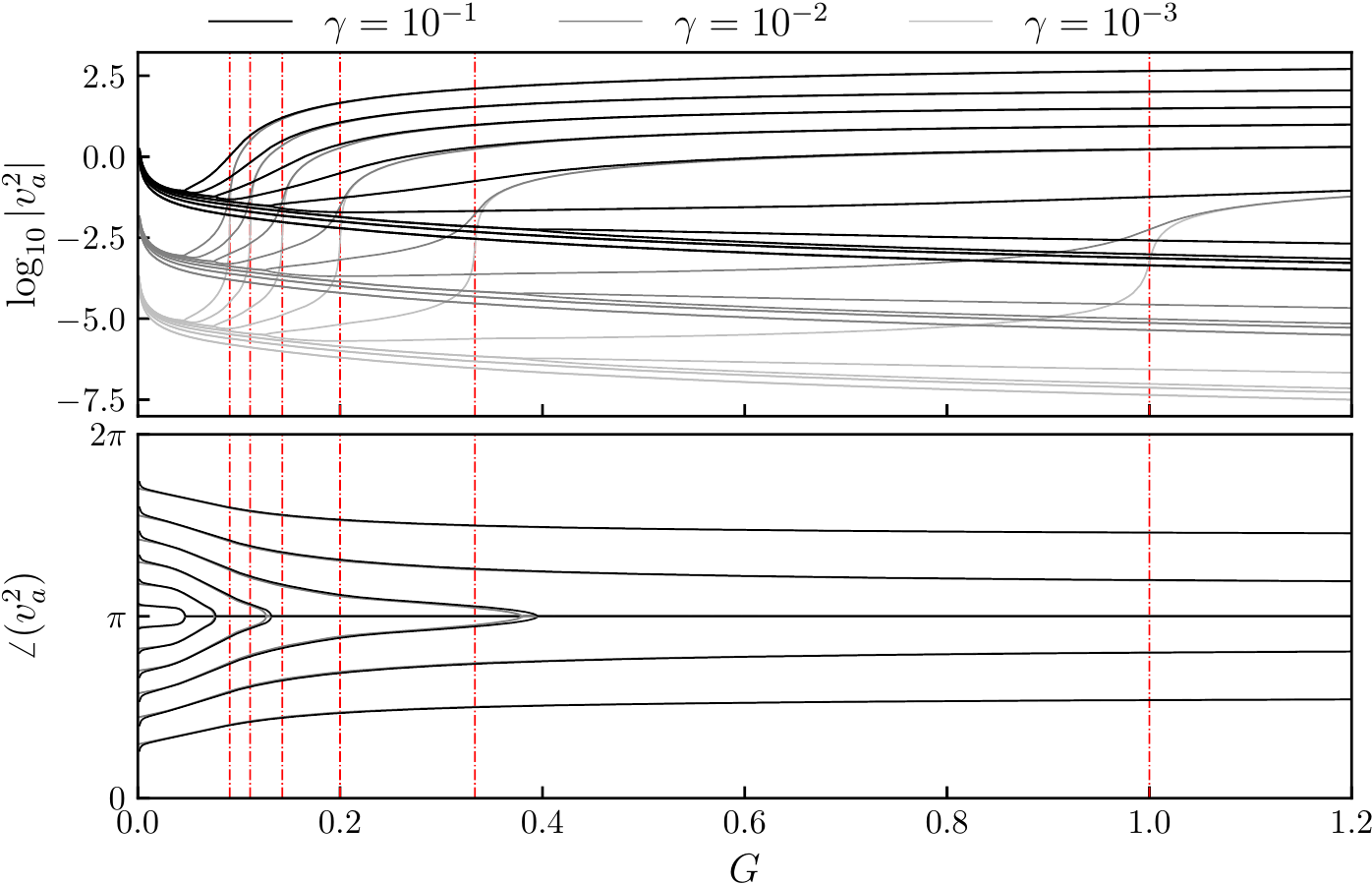}
 \caption{Modulus and phase ($\angle$) of the rapidities for the model in Figure \ref{groundstate} at three different values of $\gamma$. Note the logarithmic scale for the modulus. For decreasing $|\gamma|$ the transitions at the Read-Green points become steeper and the subset of rapidities below $|\gamma|$ decrease in magnitude, while the finite rapidities (above $|\gamma|$) remain approximately unchanged.\vspace{-\baselineskip}\label{rapidities}}
 \end{center}
\end{figure}
This parametrization in terms of a set of rapidities allows for additional insight in the particle-exchange mechanism. From the factorized expression in Eq. (\ref{wavefunctionwithinteraction}) it can be seen that the factors (excitations) for which $|v_{a}^2| \ll |\gamma|$ only rescale the wavefunction (up to a small correction term) and do not lead to particle creation. Subsequently, if $L-N$ rapidities are small compared to $|\gamma|$, the average number of Cooper pairs will be approximately $N$. Furthermore, the energy contribution of a single rapidity is proportional to $v_{a}^{2}$, so these also do not contribute to the energy and can be associated with zero modes. A clear separation of scales can be seen in Figure \ref{rapidities}, where at each Read-Green point a single rapidity quickly increases in magnitude, entailing a change by one in the average pair number which can be understood as the activation of a single dormant zero-energy rapidity.

\section{Conclusions}
In this chapter, it was shown how particle-exchange with an external environment influences the phase transition in a topological superconductor. When a topological phase transition is expected, a single zero-energy excitation is instead created from the environment, increasing the average number of Cooper pairs in the superconductor, and the topological phase transition is changed to an avoided level crossing between topologically non-trivial states with different particle number. As such, the system remains in the topologically non-trivial weak-pairing phase. Each of these crossings is accompanied by a resonance in the level occupations of the lowest-energy single-particle states as a remainder of the zero-energy excitations and the topological invariant, and cannot be observed within traditional mean-field theory. Changing the interaction strength, the occupation of the ground state corresponds to the occupation of the isolated system along the phase transition, allowing the phase diagram to be mapped out by coupling the system to the bath.

\begin{subappendices}
\makeatletter
\@openrightfalse
\makeatother
\chapter*{Appendices}

\section{Bethe equations}
\label{app:pip:BAE}

In this Appendix, the Bethe equations are derived for the commuting operators
\begin{align}
Q_{k} =&(S_k^z+\frac{1}{2})+\gamma \epsilon_k^{-1} S^{+}_{k} -\lambda \epsilon_k^{-1} S_k^- \nonumber \\
&\qquad - G \sum_{j \neq k}^L \left[\frac{\epsilon_k \epsilon_j}{\epsilon_k^2-\epsilon_j^2}\left(S^{+}_{k} S^-_{j}+S^-_{k}S^{+}_j\right)+\frac{2 \epsilon_j^2}{\epsilon_k^2-\epsilon_j^2}\left(S_{k}^z S_{j}^z-\frac{1}{4}\right)\right],
\end{align}
which reduce to the conserved charges of the $p_x+ip_y$ model in the limiting case $\gamma=-\lambda$, but commute for arbitrary values of $\gamma$ and $\lambda$, leading to possible non-Hermitian conserved charges. The Bethe states are again product wave functions
\begin{align}\label{pip:app:eigenstate}
\ket{v_1 \dots v_L}=\prod_{a=1}^L S^{+}(v_{a})\ket{0}, \qquad S^{+}(v_{a})=-\frac{\lambda}{v_{a}^2}+G\sum_{j=1}^L \frac{\epsilon_j}{\epsilon_j^2-v_{a}^2} S^{+}_{j},
\end{align}
with the set of rapidities $\{v_{a},a=1 \dots L\}$ to be determined. The reference state is again given by $\ket{0}=\ket{\downarrow \dots \downarrow}$, despite it not being an eigenstate of $Q_k$ for non-zero $\gamma$. This wave function does not contain a definite particle number due to the presence of the constant factor $\lambda/v_{a}^2$ in the generalized creation operator $S^{+}(v_{a})$. Alternatively, these can also be obtained from a non-Hermitian and non-symmetric GGA defined by the operators\footnote{The GGA from Ref. \cite{faribault_common_2017} can here be recovered in the rational limit.}
\begin{align}
&S^{+}(u) = -\frac{\lambda}{u^2}+G\sum_{j=1}^L \frac{\epsilon_j}{\epsilon_j^2-u^2} S^{+}_{j}, \qquad S^{-}(u) = \frac{\gamma}{u^2}+G\sum_{j=1}^L \frac{\epsilon_j}{\epsilon_j^2-u^2} S^{-}_{j} \\
&S^z(u) = G \sum_{j=1}^L  \frac{\epsilon_j}{\epsilon_j^2-u^2} S_j^z.
\end{align}
Because of the commutativity, all conserved operators again share a common set of eigenstates, and the conditions for the product state (\ref{pip:app:eigenstate}) to be an eigenstate of $Q_{k}$ can be derived. The action on the wave function can be determined through a Richardson-Gaudin commutator scheme as
\begin{align}
Q_{k}\prod_{a=1}^L S^{+}(v_{a})\ket{0}=&\sum_{a=1}^L\sum_{b=a+1}^L\left(\prod_{c \neq a,b}^LS^{\dagger}(v_{c})\right)[[Q_{k},S^+(v_{a})],S^+(v_{b})]\ket{0} \nonumber\\
&+\sum_{a=1}^L\left(\prod_{b \neq a}^L S^{+}(v_b)\right)[Q_{k},S^+(v_{a})]\ket{0}+\left(\prod_{a=1}^L S^{+}(v_{a})\right)Q_{k}\ket{0}.
\end{align}
The necessary commutation relations are given by
\begin{align}
&[Q_k, S^{+}(v_{a})]=2G^2 \frac{\epsilon_k}{\epsilon_k^2-v_{a}^2}S^{+}_k\left(\sum_{j=1}^L \frac{v_{a}^2}{\epsilon_j^2-v_{a}^2}S_j^z\right)+G\frac{\epsilon_k}{\epsilon_k^2-v_{a}^2}S^{+}_k\left(1+2G\sum_{j \neq k}^L S_j^z\right) \nonumber\\
&\qquad \qquad \qquad \qquad -2 G \frac{v_{a}^2}{\epsilon_k^2-v_{a}^2}S^{+}(v_{a})S_k^z , \nonumber\\
&[[Q_k, S^{+}(v_{a})],S^{+}(v_{b})]=2G^2\frac{ \epsilon_k }{v_{a}^2-v_{b}^2}S^{+}_k\left[\frac{v_{a}^2 }{\epsilon_k^2-v_{b}^2}S^{+}(v_{a})-\frac{v_{b}^2}{\epsilon_k^2-v_{a}^2}S^{+}(v_{b})\right].
\end{align}
Taking these results together, the action of a single conserved operator on the product wave function can be written as
\begin{align}
&Q_k \prod_{a=1}^LS^{+}(v_{a})\ket{0} = \left[G\sum_{a=1}^L \frac{v_{a}^2}{\epsilon_k^2-v_{a}^2}\right]\prod_{a=1}^LS^{+}(v_{a})\ket{0} + \gamma \epsilon_k^{-1}S^{+}_k \prod_{a=1}^LS^{+}(v_{a})\ket{0}\nonumber\\
&\qquad +G\sum_{a=1}^L \frac{\epsilon_k}{\epsilon_k^2-v_{a}^2}\left[(1+G)-G\sum_{j=1}^L \frac{\epsilon_j^2}{\epsilon_j^2-v_{a}^2}+2G\sum_{b \neq a}^L\frac{v_{b}^2}{v_{b}^2-v_{a}^2}\right]S^{+}_k\prod_{b \neq a}^LS^+(v_b)\ket{0}.
\end{align}
In order to be an eigenstate, the two unwanted (non-diagonal) terms should cancel. It can be shown that the contributions from different excitation sectors cancel termwise provided the rapidities $\{v_1 \dots v_L\}$ satisfy a set of Bethe equations. First, the contribution from both terms to the state $S^{+}_k\ket{0}$ containing a single excitation is checked, which vanishes provided the set of equations
\begin{equation}\label{pip:app:matvecmult}
\sum_{a=1}^L \frac{\epsilon_k^2v_{a}^2}{\epsilon_k^2-v_{a}^2}\left[(1+G)-G\sum_{j=1}^L \frac{\epsilon_j^2}{\epsilon_j^2-v_{a}^2}+2G\sum_{b \neq a}^L\frac{v_{b}^2}{v_b^2-v_{a}^2}\right] = \frac{\gamma \lambda}{G}, 
\end{equation}
are satisfied. These can be brought in the previously-obtained form \cite{lukyanenko_integrable_2016}
\begin{equation}\label{pip:app:bae}
(1+G)-G\sum_{j=1}^L \frac{\epsilon_j^2}{\epsilon_j^2-v_{a}^2}+2G\sum_{b \neq a}^L\frac{v_{b}^2}{v_{b}^2-v_{a}^2} = \frac{\gamma \lambda}{G} \frac{\prod_{j=1}^L(v_{a}^{-2}-\epsilon_j^{-2})}{\prod_{b \neq a}^L (v_{a}^{-2}-v_{b}^{-2})},
\end{equation}
by interpreting (\ref{pip:app:matvecmult}) as a matrix-vector multiplication and multiplying with the well-known inverse of a Cauchy matrix (see Chapter \ref{chap:innerproducts}). However, this requirement only cancels a single contribution from both unwanted terms, when all other contributions should also cancel exactly in order for (\ref{pip:app:eigenstate}) to be an eigenstate. Remarkably, this leads to a set of equivalent equations. The coefficient in front of a state with $N+1$ excitations where the set of spins labeled $\{k,i(1), \dots, i(N)\}$ are flipped up is proportional to

\begin{align}\label{pip:app:unwantedterms}
C^N_{\{k,i(1), \dots, i(N)\}}=\sum_{a=1}^L \frac{F_{a}}{\epsilon_k^{-2}-v_{a}^{-2}}&\left[\sum_{\hat{A}\in S^N_{\hat{a}}}\frac{1}{\prod_{c=1}^N(\epsilon_{i(c)}^{-2}-\hat{\nu}^{-2}_{c})}\right]+\frac{\gamma\lambda}{G} \left[\sum_{A \in S^N}\frac{1}{\prod_{c=1}^N(\epsilon_{i(c)}^{-2}-\nu_c^{-2})}\right],
\end{align}
with
\begin{equation}
F_{a}=(1+G)-G\sum_{j=1}^L \frac{\epsilon_j^2}{\epsilon_j^2-v_{a}^2}+2G\sum_{b \neq a}^L\frac{v_{b}^2}{v_{b}^2-v_{a}^2}
\end{equation}
and $S^N_{\hat{a}}$ the set of $N$-tuples built out of $N$ non-repeated elements $\{v_1, \dots v_{a-1},v_{a+1}, \dots, v_L\}$ and $S_N$ the set of $N$-tuples built from $\{v_1, \dots, v_L\}$. The elements of these sets are denoted $\hat{A}=\{\hat{\nu}_1, \dots, \hat{\nu}_{N}\}$ and $A=\{\nu_1, \dots, \nu_N\}$. A highly similar expression was obtained in the study of the DJCG models, where it was proven that these expressions vanish provided the Bethe equations are satisfied \cite{tschirhart_algebraic_2014}. The full proof that (\ref{pip:app:unwantedterms}) equals zero is completely analogous and does not depend on the explicit form of the Bethe equations, rather on the Cauchy-matrix structure when evaluating $F_a$ as in Eq. (\ref{pip:app:matvecmult}) and plugging this in Eq. (\ref{pip:app:unwantedterms}). Following Ref. \cite{tschirhart_algebraic_2014}, this then reduces to a purely combinatorial problem.

\end{subappendices}
\makeatletter
\@openrighttrue
\makeatother


\chapter{Variational application of the Bethe ansatz}
\label{chap:VarRG}

\setlength\epigraphwidth{.5\textwidth}
\epigraph{\emph{All models are wrong, but some are useful.}}{George E.P. Box}

While integrable models can be solved exactly in a polynomial time, this advantage comes at a price -- for a model to be integrable, and thus exactly solvable, all parameters and interactions of the system need extraordinary fine-tuning. Even slight perturbations to the Hamiltonian break integrability, and it is still an open question how much of the features of integrability are retained for systems `close to integrability'. Theoretically, we immediately lose the full underlying framework, and it is in general no longer possible to solve such systems using Bethe ansatz techniques.

It is then a natural question to ask how well the wave functions of systems close to integrability can be approximated using exact eigenstates of integrable systems. In this chapter, we propose to perform a variational optimization within the set of eigenstates of integrable Richardson-Gaudin models in order to obtain the optimal approximation to the ground state of a given Hamiltonian. While this method cannot be expected to return accurate approximations to ground states of arbitrary Hamiltonians, it is tailor-made for models which are in some sense `close to' Richardson-Gaudin integrable models. In such models, the approximations can be made clearer and the advantages and limitations of the integrable wave functions can be better understood. 

The main requirement for any variational method to be feasible is being able to efficiently and accurately calculate and minimize the energy functional 
\begin{equation}\label{var:int:en}
E\left[\psi\right] = \frac{\braket{\psi|{H}|\psi}}{\braket{\psi|\psi}},
\end{equation}
for any given Hamiltonian ${H}$ and any given trial state $\ket{\psi}$ \cite{sakurai_modern_2010}. As shown in the first half of this thesis, the theoretical and numerical toolbox of integrability provides us with exactly this. Following Chapter \ref{chap:innerproducts}, the expectation value of a given Hamiltonian can be evaluated as a sum of determinants. These can be evaluated at a computationally favorable (polynomial) scaling, and a gradient descent method can be applied to minimize the energy functional \cite{press_numerical_2007}.

Here, this method is applied to spin systems consisting of an integrable model plus an integrability-breaking perturbation term. If no integrability-breaking terms are present, the proposed method leads to the exact ground state by construction. Furthermore, thanks to the use of eigenstates of unperturbed integrable models as trial states, the variational energy is also guaranteed to be an improvement upon the energy obtained from first-order perturbation theory, serving as further motivation for the choice of trial states. The specific class of Richardson-Gaudin integrable models then provides a large amount of variational parameters and a wide variety of physical systems \cite{ortiz_exactly-solvable_2005}. 

It is shown that this method is also able to return accurate approximations in the region where perturbation theory is not expected to hold, provided the perturbative interactions do not influence the qualitative physics of the model. In this case, the bulk of the correlations in the ground state of the non-integrable system is captured by the ground state of the integrable system, and the variational optimization returns an accurate approximation. If this is not the case, it is shown that a more accurate description can be obtained by variationally optimizing an excited state of an integrable model. This is illustrated by comparing overlaps and correlation functions, and can be understood as (avoided) level crossings in the spectrum of the non-integrable Hamiltonian \cite{dalessio_quantum_2016}, when the perturbative energy of an excited state drops below that of the original ground state.

In a broader context, this research fits within the general development of wave function-based methods (as compared to density-based methods) for the description of strongly-correlated models. In this aspect, this approach is also motivated by recent developments in the theory of Antisymmetric Product of Geminals (APG) in molecular physics and quantum chemistry \cite{surjan_introduction_1999,johnson_size-consistent_2013,limacher_new_2013,tecmer_assessing_2014}. Composed as a generalized valence-bond wave function, APG wavefunctions are tailor-made for the description of resonating electron-pair configurations, and tie in directly with the Lewis picture of molecular bonding. Recently, it has been realized that the Richardson-Gaudin eigenstates fit within the class of geminal wave functions. This has given rise to various computationally tractable versions of APG, including a variational formulation based on the Richardson-Gaudin wave functions \cite{johnson_model_2015,tecmer_assessing_2014}. However, initial calculations for simple molecular systems \cite{tecmer_assessing_2014} showed that the variational method was surpassed in accuracy and efficiency by Coupled-Cluster-based APG methods \cite{limacher_new_2013,boguslawski_efficient_2014}. These preliminary results then naturally shifted the research focus to the Coupled-Cluster variant of APG theory in recent years \cite{boguslawski_efficient_2014,boguslawski_nonvariational_2014,limacher_simple_2014,henderson_seniority-based_2014, stein_seniority_2014,degroote_polynomial_2016}. However, it is presently becoming clear that further developments in APG theory will benefit from a well-defined Hilbert space, which is conveniently obtained through the connection with a variational Richardson-Gaudin APG state and the associated integrable Hamiltonian. This then paves the way for an extension of the variational method by including excitations on top of the variational approximation, as discussed in the second half of this chapter. The present chapter is largely based on Refs. \cite{claeys_variational_2017,de_baerdemacker_richardson-gaudin_2017}.



\section{Moving away from integrability}
\label{sec:var:bi}
A rich variety of methods has been developed for the approximation of the ground state of general non-integrable systems. Here, the distinction can be made between wave function-based methods such as mean-field theory \cite{chaikin_principles_2000}, the related coupled cluster and configuration interaction theories \cite{helgaker_molecular_2014}, tensor networks \cite{orus_practical_2014} and variational quantum Monte Carlo methods \cite{sorella_wave_2005}, as compared to density-based density functional theory \cite{parr_density-functional_1994}. Within the wave function-based methods, the common approach is that a specific structure is imposed on a wave function, which is then optimized (often variationally) in order to approximate the ground state of a given system \cite{sakurai_modern_2010}. The success of any approach is then judged by how well the proposed structure of the wave function matches that of the exact ground state.

In this chapter, the Bethe states $\ket{\psi_{RG}}$ from rational Richardson-Gaudin models (\ref{RG:phys:BA_BCS}) will be investigated as variational ansatz, where
\begin{equation}\label{var:bi:RGwf}
\ket{\psi_{RG}} = \prod_{a=1}^N\left(\sum_{i=1}^L \frac{S^{+}_i}{\epsilon_i - v_a}\right)\ket{\downarrow \dots \downarrow},
\end{equation}
the different spins in the system are labeled $i=1, \dots, L$ and the spin operators constitute an $su(2)$ algebra (see Chapter \ref{Chap_RGmodels}). Although the parameters $\{\epsilon_1 \dots \epsilon_L \}$ and $\{v_1 \dots v_N\}$ have a clear physical interpretation within integrability, these can be thought of as arbitrary parameters when regarding (\ref{var:bi:RGwf}) as a variational ansatz. In order to obtain the computationally favourable scaling from integrability following Chapter \ref{chap:innerproducts}, the state must be on-shell and the variables are coupled through the Bethe (or Richardson-Gaudin) equations (\ref{ip:RG:BAE}). It is worth stressing that although all variables in the wave function can be connected to the physics of integrable systems, it is not strictly necessary to interpret them as such. They can equally be treated as variational parameters. Given the Hamiltonian ${H}$ of a strongly-correlated system, the goal is now to find the Bethe state minimizing the energy (\ref{var:int:en}), resulting in a variational energy
\begin{equation}\label{var:bi:Evar}
E_{\textrm{Var.}} = \min E\left[\psi_{RG}\right].
\end{equation}
For such a generalized product state to be computationally tractable, it needs to be dressed with additional structure, which is here provided by integrability and the on-shell requirement. It is worth noting that the projected BCS method can be reinterpreted as a special case of variational RG integrability, providing a connection between the variational wave function (\ref{var:bi:RGwf}) and the BCS mean-field wave function \cite{roman_large-n_2002}. Systems successfully described by mean-field theory, where the particles can be treated as non-interacting particles, also arise as a particular limit of the Bethe ansatz. In fact, a crucial feature of the wave function (\ref{var:bi:RGwf}) is that it exhibits a similar product structure as the Hartree-Fock wave function \cite{sakurai_modern_2010}, returning a Slater determinant in the non-interacting limit $g \to 0$ (see Chapter \ref{Chap_RGmodels}). The variational method can thus already be expected to return accurate results for weakly-correlated systems.

The key question is then if the on-shell condition restricts the physics that can be captured by this ansatz. While integrable Hamiltonians are necessarily quite schematic, they have shown remarkable success in the description of general physical phenomena. Richardson's original solution to the (reduced) BCS Hamiltonian \cite{richardson_restricted_1963,richardson_exact_1964} already succeeded in qualitatively describing regular superconductivity \cite{bardeen_theory_1957,von_delft_spectroscopy_2001}, and only afterwards was it recognized that this Hamiltonian is integrable \cite{cambiaggio_integrability_1997}. Furthermore, form factors in integrable theories are exactly known, and can be used to build a perturbation theory for non-integrable models \cite{delfino_non-integrableqft_1996, controzzi_mass_2005,delfino_non-integrable_1998,pozsgay_characterization_2006,delfino_decay_2006,groha_spinon_2017}. Approximate scattering matrices for low-lying excited states of non-integrable systems have also been constructed from approximate (coordinate) Bethe ansatz techniques \cite{krauth_bethe_1991,kiwata_hirohito_bethe-ansatz_1994,okunishi_magnetization_1999,vanderstraeten_scattering_2015}. Despite these models being non-integrable, accurate results could still be obtained by applying techniques from integrability. Integrability-based methods have also been proposed in the description of time evolution governed by an integrable Hamiltonian plus a perturbation, both in the description of the initial behaviour \cite{brandino_relaxation_2013,van_den_berg_competing_2014} and the infinite-time behaviour \cite{lange_pumping_2017} of observables. Such problems have also been tackled using a numerical renormalization group expressed in the basis of eigenstates of the integrable model \cite{caux_constructing_2012}.

The majority of these results essentially build on the same idea as our proposed method -- integrability can be used to describe the bulk of the correlations, on which corrections can be added. While using the same technical toolbox as these methods, our results are mainly similar in spirit to the use of perturbation theory for non-integrable system, where the important distinction is that the variational optimization guarantees a more accurate approximation of the ground state wave function than perturbation theory.


\section{A variational method}
\label{sec:met}
The procedure is conceptually straightforward. For a given Hamiltonian
\begin{equation}
{H}={H}_{int}+{V},
\end{equation}
with ${H}_{int}$ an integrable (Richardson-Gaudin) Hamiltonian, and ${V}$ containing additional interactions breaking the integrability, we wish to minimize
\begin{equation}
E\left[\psi_{RG}\right] = \frac{{\braket{\psi_{RG}|{H}|\psi_{RG}}}}{\braket{\psi_{RG}|\psi_{RG}}}.
\end{equation}
It is important to note that the variables in the wave function are independent from those in the integrable Hamiltonian $H_{int}$, since the former are the degrees of freedom over which we optimize, while the latter are a characteristic of the unperturbed system. Obviously, in the limit of a vanishing perturbation ${V}=0$ the Hamiltonian ${H}=H_{int}$ becomes integrable, and the variational optimization should return the variables in the Hamiltonian as variational parameters, since this wave function is then the exact ground state of the integrable Hamiltonian. While these states explicitly depend on $L+N$ variables $\{\epsilon_1 \dots \epsilon_L\}$ and $\{v_1 \dots v_N\}$, the demand that these states are on-shell (\ref{ip:RG:BAE}) leaves $L$ degrees of freedom over which to optimize, which can be chosen as $\{\epsilon_1 \dots \epsilon_L\}$, and we denote $E[\psi_{RG}] \equiv E[\epsilon_1 \dots \epsilon_L] \equiv E[\vec{\epsilon}]$, with the implicit assumption that all rapidities $\{v_1 \dots v_N\}$ uniquely follow through the Bethe equations, resulting in a manifold of states only determined by the variables $\vec{\epsilon}$. However, an additional discrete degree of freedom exists -- the choice of eigenstate. Each eigenstate of an integrable Hamiltonian defined by a set of variables $\vec{\epsilon}$ can be written as (\ref{var:bi:RGwf}), so the targeted eigenstate needs further specification. This degree of freedom will initially be disregarded, and we will restrict ourselves to the state that is adiabatically connected to the ground state of the integrable Hamiltonian in the limit of a vanishing perturbation. For small perturbations, it is expected that this state will be the most relevant. Later, it will be shown that this choice is not guaranteed to be optimal for large perturbations, and the excited states will prove to be important. 


\section{Results}
\label{sec:res}
\subsection{Perturbing the central spin model}
The results will first be illustrated on the central spin model (see Section \ref{RG:subsec:cs}) with perturbations restricted to operators acting on one or two spins. The central spin Hamiltonian is given by
\begin{equation}\label{res:csHam}
{H}_{cs} = B_z S_1^z + g \sum_{i \neq 1}^L \frac{\vec{S}_1 \cdot \vec{S}_i}{\epsilon_{0,1}-\epsilon_{0,i}} ,
\end{equation}
describing the interaction of a single spin on which a magnetic field $B$ is applied, with a bath of surrounding spins. This model is integrable for any choice of the interaction modulated by $\epsilon_{0,i}$, which has been written in this way in order to make this connection explicit. In this model, the bath spins do not interact among themselves and do not experience the magnetic field applied to the central spin. However, such interactions may be added in a perturbative way by introducing terms of the form $S_i^z$ and $\vec{S}_i \cdot \vec{S}_j$ in the Hamiltonian. The basic physics in this model can be easily understood -- $B_z$ determines the orientation of the central spin $\braket{S_1^z}$, either parallel or anti-parallel to the quantization axis, while the signs of $g/(\epsilon_{0,1}-\epsilon_{0,i})$ determine the relative orientation of the bath spin $\vec{S}_i$ with the central spin $\braket{\vec{S}_1 \cdot \vec{S}_i}$. 

In the following, system sizes $L=12$ are considered for which exact diagonalization methods can still be used as benchmark and the Hamiltonian is parametrized with $\epsilon_{0,i} = L- i$ as a picket-fence model \cite{hirsch_fully_2002}. This leads to algebraically decaying interactions between the central spin and the bath spin, where the strength of the interaction is fixed by setting $B_z=1$ and $g=-2$, intermediate between strong- and weak-coupling \cite{de_baerdemacker_richardson-gaudin_2012}. For this choice of parametrization, the central spin and all surrounding spins tend to align, while also being restricted by conservation of spin projection $S^z = \sum_i S_i^z$. In the following, we always choose $S^z=0$ (or $L=2N$), since this is the sector where the dimension of the Hilbert space is maximal.

\begin{figure}
\begin{center}
\includegraphics{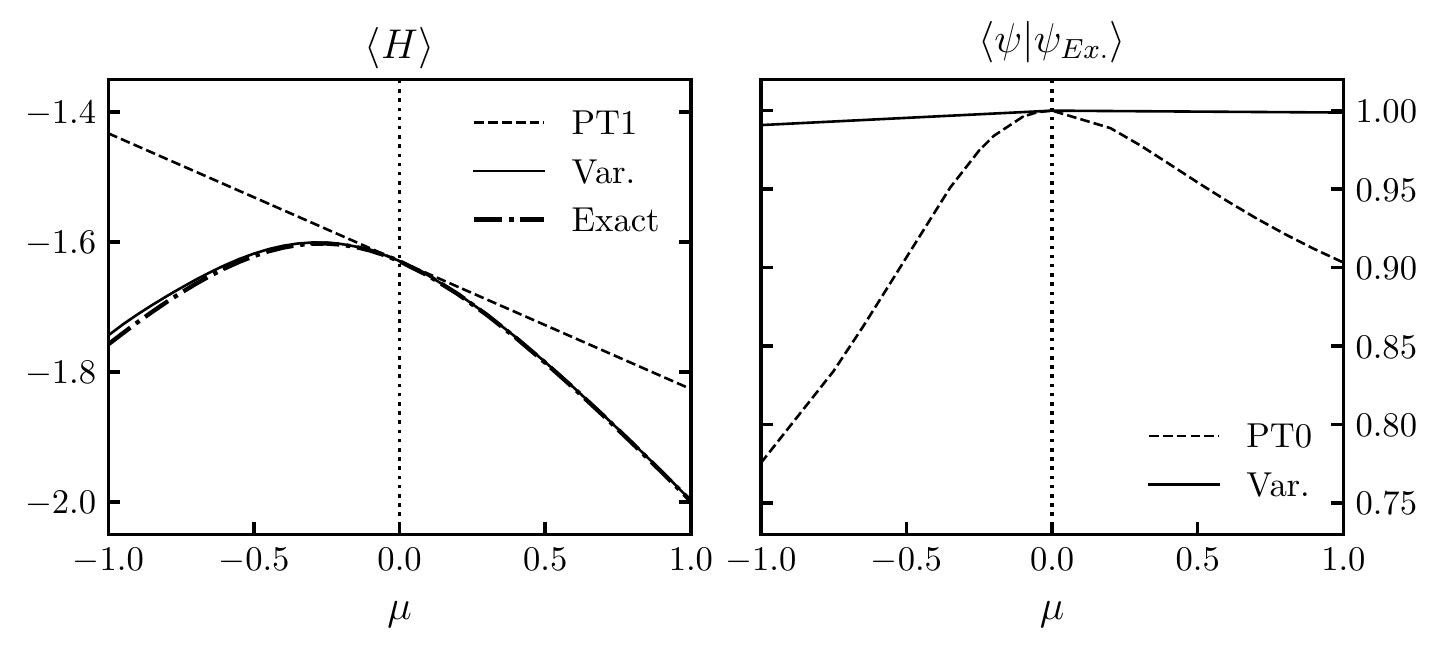}
\caption{Results for the central spin model with perturbation $\mu S_i^z$. \textbf{Left}: Variational energy (Var.), exact ground state energy (Exact), and first-order perturbation theory (PT1) energy for different values of the perturbation strength. \textbf{Right}: Overlap of the exact ground state with the variational ground state (Var.) and the ground state of the unperturbed model (PT0).
 \label{fig:cs:ener_ov_Ji}}
\end{center}
\end{figure}

\textbf{Single-spin perturbation.} Firstly, calculations are performed for a Hamiltonian 
\begin{equation}\label{res:HamJi}
{H} = {H}_{cs}+\mu S_i^z,
\end{equation}
applying a magnetic field of size $\mu$ to one of the spins in the bath (here labeled $i$). Such a model has previously also been investigated in the context of integrability-breaking \cite{erbe_different_2010,schliemann_spins_2010}. The variational energy is calculated and compared with the ground-state energy obtained by exact diagonalization. In Figure \ref{fig:cs:ener_ov_Ji}, we plot the variational energy (Var.), the exact ground-state energy (Exact), and the energy obtained by first-order perturbation theory (PT1) for varying perturbation strengths $\mu$, with $i=2$ chosen to maximize the deviation from the integrable model, since the central spin interacts most strongly with this bath spin. Since the ground-state energy deviation is intimately connected to the overlap between the approximate ground state and the exact ground state, this is also given in Figure \ref{fig:cs:ener_ov_Ji}. As the error in the energy is generally quadratic in the error in the overlap, the latter can be seen as a more sensitive measure for the accuracy of the proposed method.

Labeling the parameters of the unperturbed integrable model as $\vec{\epsilon}_0$, the relevant energies can be contrasted as
\begin{align}
E_{\textrm{Var.}} = \min_{\vec{\epsilon}} E\left[\vec{\epsilon}\right], \qquad E_{\textrm{PT}1} = E\left[\vec{\epsilon}_0\right],
\end{align}
making clear why the variational method provides a guaranteed improvement on first-order perturbation theory. In the chosen model, perturbation theory is guaranteed to provide a good approximation to the exact ground state energy only when $|\mu \braket{S_i^z}| \ll\Delta E$, with $\Delta E$ the energy difference between the ground state and the first excited state. In the following, this roughly corresponds to $|\mu| \ll 1$, which we will consider to be a small perturbation.

The overlaps given are those between the variationally obtained wave function (Var.) and the exact ground state, together with the overlap between the ground state of the unperturbed model and the exact ground state (PT0)\footnote{The first-order correction to the energy (PT1) follows from the unperturbed wave function (PT0), hence the discrepancy in order.}. The variational wave function is able to accurately model the ground state for a wide range of the perturbation strength, even going up to the limit where the size of the perturbation interaction equals that of the unperturbed central spin interaction ($|\mu|=1$), providing a substantial improvement over first-order perturbation theory. Here, the variational optimization plays a crucial role, as can be seen by comparing the overlap of the exact ground state with the ground state of the unperturbed Hamiltonian to the overlap with the variationally optimized wave function, which is improved by several orders of magnitude (from an overlap of 0.7754 to 0.9908 for $\mu=-1$). However, since the perturbation only acts on a single spin site, it is not expected that this will fundamentally influence the correlations in the model, and more intrusive perturbations may be more physical.
\begin{figure}             
 \begin{center}
 \includegraphics{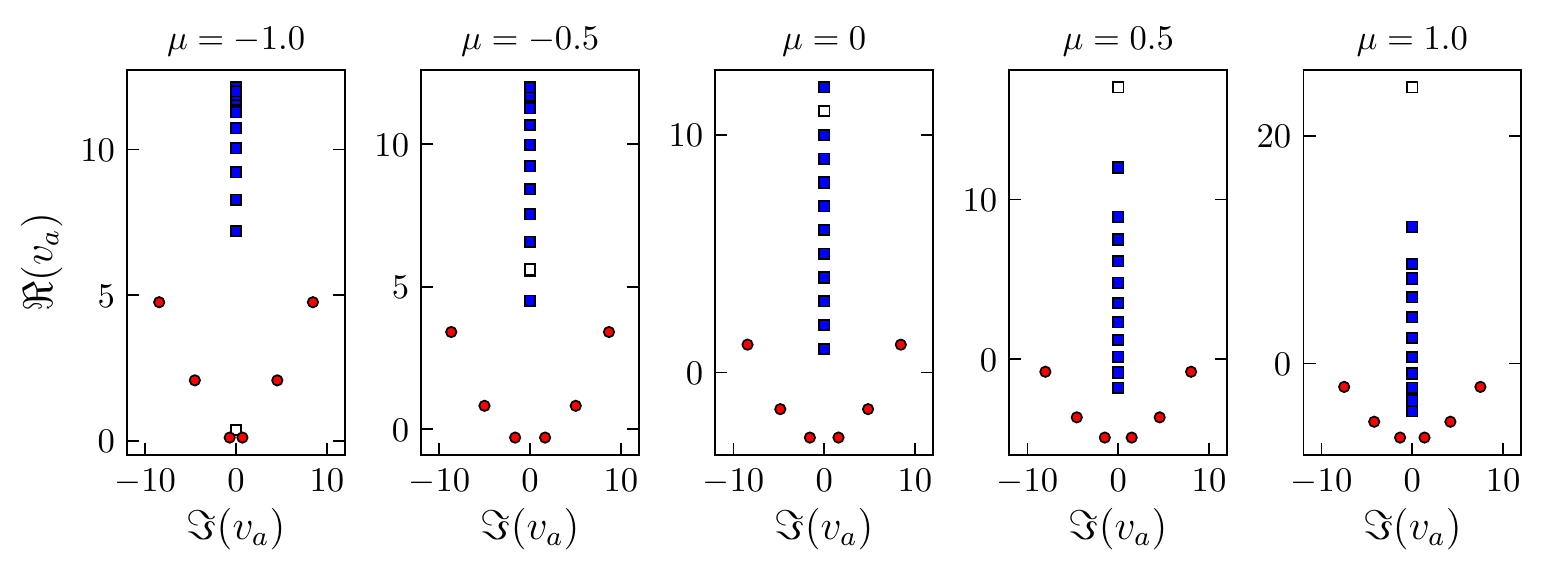} 
 \caption{Variational parameters for the Hamiltonian (\ref{res:HamJi}). Position of $\vec{\epsilon}$ (squares) and $\vec{v}$ (dots) for the variationally optimized wave function in the complex plane at different values of the perturbation strength. The white square denotes the variable $\epsilon_i$ associated with the level on which the perturbation is applied following Eq. (\ref{res:HamJi}).
\label{fig:rap_Ji}}
 \end{center}
\end{figure}

Some more insight in the role of the optimization and the structure of the wave function can be obtained by considering the evolution of the variables $\vec{\epsilon}$ and $\vec{v}$ in the wave function. These are given in Figure \ref{fig:rap_Ji} for different values of the perturbation strength. The variables $\vec{\epsilon}$ are restricted to be real, while the rapidities $\vec{v}$ are either real or arise as complex conjugate pairs. The single-spin character of the perturbation is clear from these figures. Only the variable $\epsilon_i$ ($i=2$), associated with the perturbed level, is significantly sensitive to the perturbation, whereas all other variables are largely unaffected. While the on-shell condition still connects both sets of variables, it can be seen that the variables $\vec{v}$ are quite robust against perturbations. This also motivates the use of $\vec{\epsilon}$ as variational parameters.

\begin{figure}
\begin{center}
\includegraphics{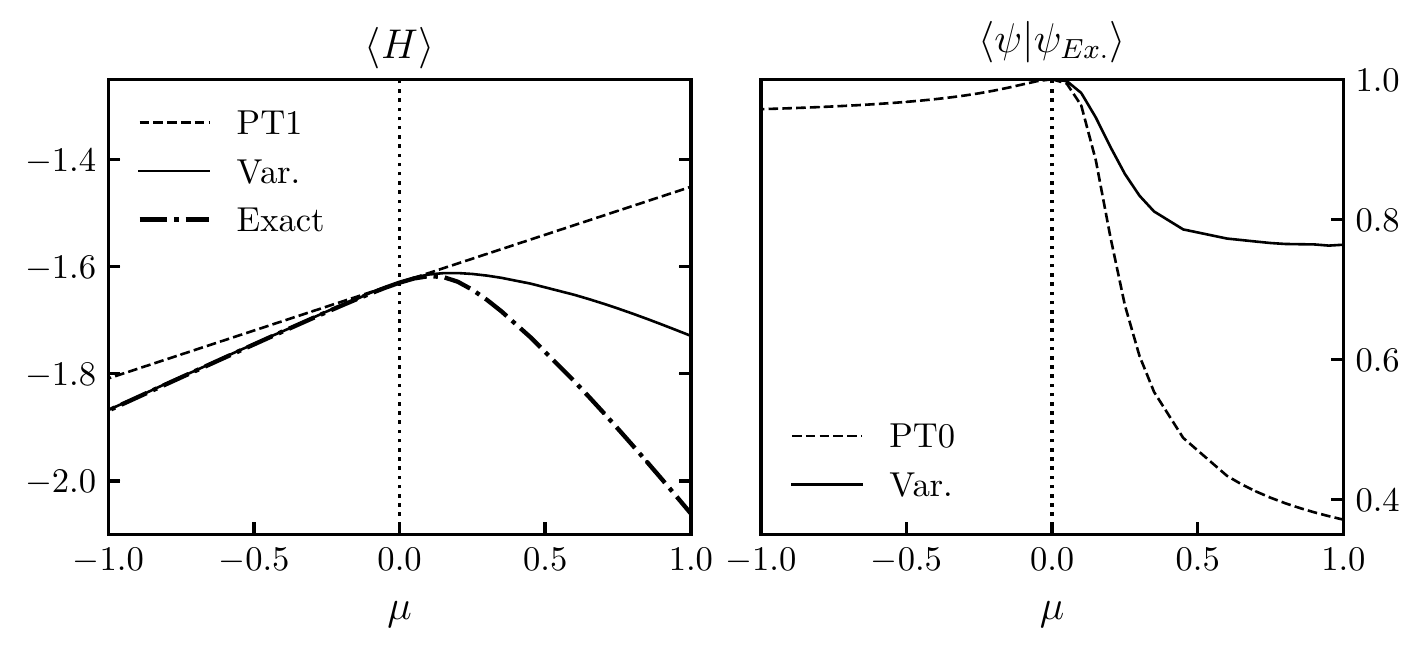}
\caption{Results for the central spin model with perturbation $\mu \vec{S}_i \cdot \vec{S}_j$, with $i,j=2,L-1$. \textbf{Left}: Variational energy (Var.), exact ground state energy (Exact), and first-order perturbation theory energy (PT1) for different values of the perturbation strength. \textbf{Right}: Overlap of the exact ground state with the variational ground state (Var.) and the ground state of the unperturbed model (PT0). \label{fig:ener_ov_Jij}}
\end{center}
\end{figure}

\textbf{Double-spin perturbation.}  Secondly, and more interestingly, this method is applied to a non-integrable Hamiltonian
\begin{equation}\label{res:HamJij}
{H} = {H}_{cs} + \mu \vec{S}_i \cdot \vec{S}_j,
\end{equation}
where $\mu$ again determines the perturbation strength, and the same calculations are repeated, where the bath spins are chosen as $i,j=2,L-1$ for similar reasons as before (while also coupling states with different occupation number in the weak-interaction limit). The results for the energy and overlap are given in Figure \ref{fig:ener_ov_Jij}.

It can be seen that the variational method still provides an accurate description for negative $\mu$, but interestingly fails to model the behaviour of the wave function for large positive $\mu$. The method holds in the limit where we can interpret the additional term as a perturbation ($|\mu \braket{\vec{S}_i \cdot \vec{S}_j} | \ll \Delta E$), but moving away from this limit the method quickly breaks down. The reason for this can be inferred from perturbation theory for the two different regimes (positive and negative $\mu$). In the ground state of the unperturbed model $\braket{\vec{S}_i \cdot \vec{S}_j} > 0$, since all spins tend to align. So, the perturbation will lower the ground state energy if $\mu<0$ and increase the energy if $\mu>0$. In the former case, the perturbation does not qualitatively change the physics in the model, whereas the latter introduces a counteracting interaction, lowering the energy if the two spins are anti-parallel. For larger $\mu$ ($\mu 
\gtrsim 0.2$),  the energy then again lowers, pointing to a change in qualitative character of the ground state. The sudden drop in overlap with the exact ground state in Figure \ref{fig:ener_ov_Jij} then hints at an avoided crossing between the ground state and an excited state for increasing $\mu$, where if $\mu$ is increased the ground state would resemble an excited state of the original system rather than the ground state. The variational optimization is still capable of increasing the overlap by more than a factor 2, but is ultimately unable to obtain an accurate description for large $\mu$. This can be understood since, while the perturbation increases the energy of the unperturbed ground state, it simultaneously lowers the energy of selected excited states of the unperturbed model.
\begin{figure}
\begin{center}
\includegraphics{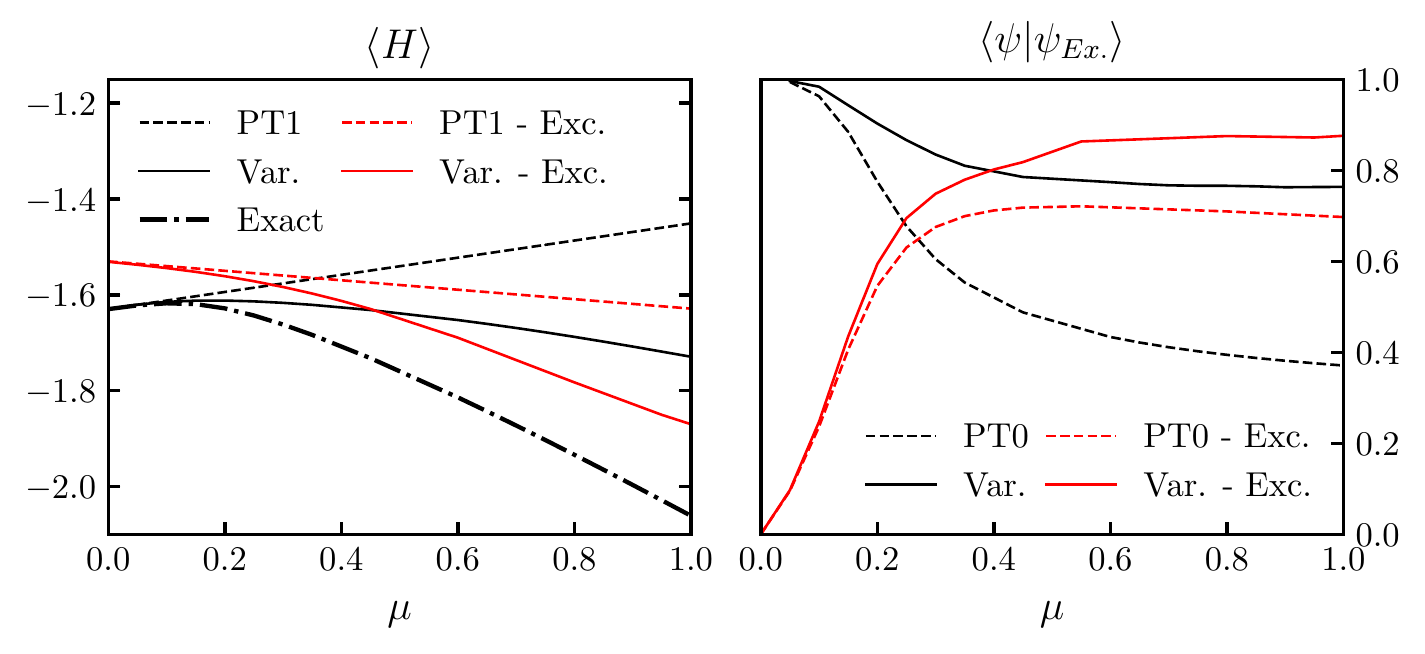}
\caption{Results for the central spin model with perturbation $\mu \vec{S}_i \cdot \vec{S}_j$. \textbf{Left}: Variational energy (Var.), exact ground state energy (Exact), and first-order perturbation theory energy (PT1) starting from the ground and excited state of the integrable model for different values of the perturbation strength. \textbf{Right}: Overlap of the exact ground state with the variational ground state (Var.) and the ground state of the unperturbed model starting from both the ground and excited state of the integrable model (PT0). \label{fig:ener_ov_Jij_exc}}
\end{center}
\end{figure}

For relatively simple perturbations, the relevant excited state can be gathered from the limit $|\mu| \to \infty$, where the perturbation becomes dominant, and the state which is adiabatically connected to this excited state in the limit $\mu \to 0$ can be variationally optimized. For positive $\mu$, the results for a variational optimization starting from both the ground state and this excited state are presented in Figure \ref{fig:ener_ov_Jij_exc}. At small $\mu$, the unperturbed ground state is the energetically favourable one, while for increasing perturbation strength the energy of the unperturbed excited state drops below that of the unperturbed ground state. Such crossings are observed both in perturbation theory and in the variational method, albeit occurring for smaller values of the perturbation in the variational method. This behaviour can also be observed from the overlaps, where a similar crossing occurs in the same region. The variational optimization again plays an important role in lowering the energy and increasing the overlap, both for the variational state obtained from the unperturbed ground- and excited state, resulting in an improved approximation to the ground state. Note that, while this results in a much improved description, there is still a part of the wave function that cannot be captured by the variational method, and for which either perturbation theory would need to be applied on the optimized wave function, or a multi-reference approach should be implemented with multiple Bethe ansatz wave functions in the variational optimization.

\begin{figure}            
 \begin{center}
 \includegraphics{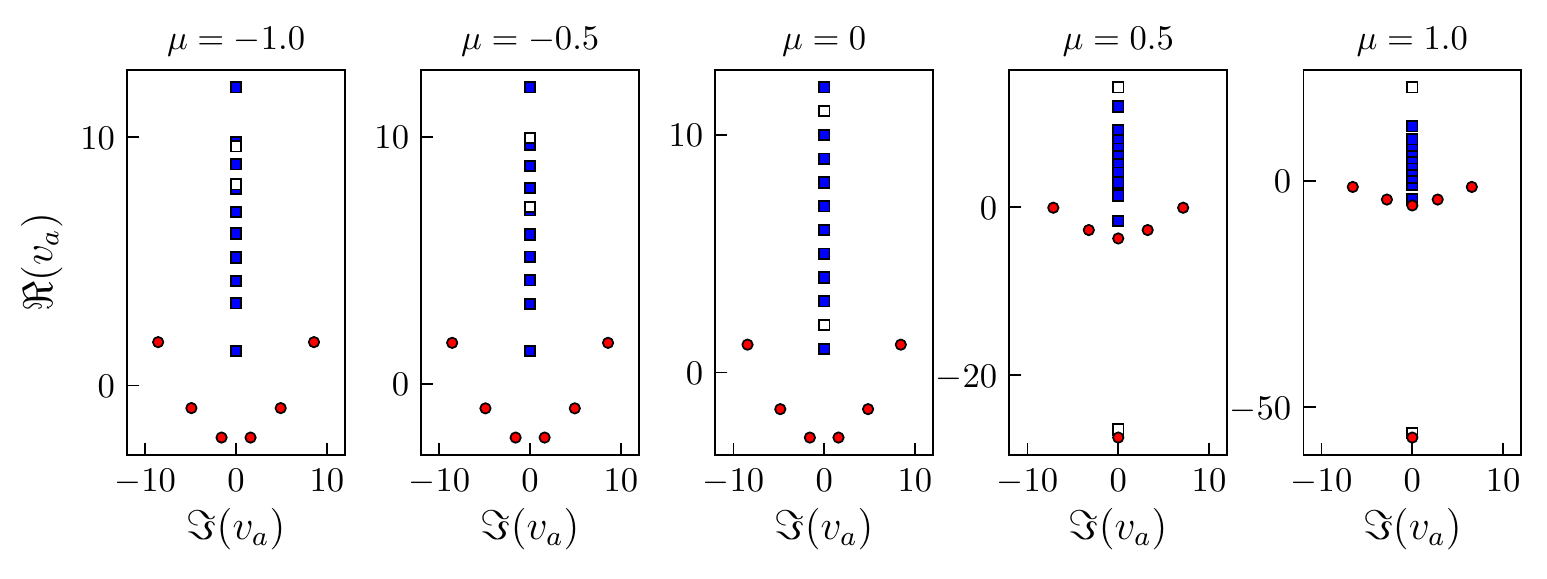} 
 \caption{Variational parameters for the Hamiltonian (\ref{res:HamJij}). Position of $\vec{\epsilon}$ (squares) and $\vec{v}$ (dots) for the variationally optimized wave function in the complex plane at different values of the perturbation strength. The white squares denote the variables $\epsilon_i, \epsilon_j$ associated with the levels on which the perturbation is applied following Eq. (\ref{res:HamJij}). \label{fig:rap_Jij_gs}}
 \end{center}
\end{figure}

\begin{figure}
\begin{center}
\includegraphics{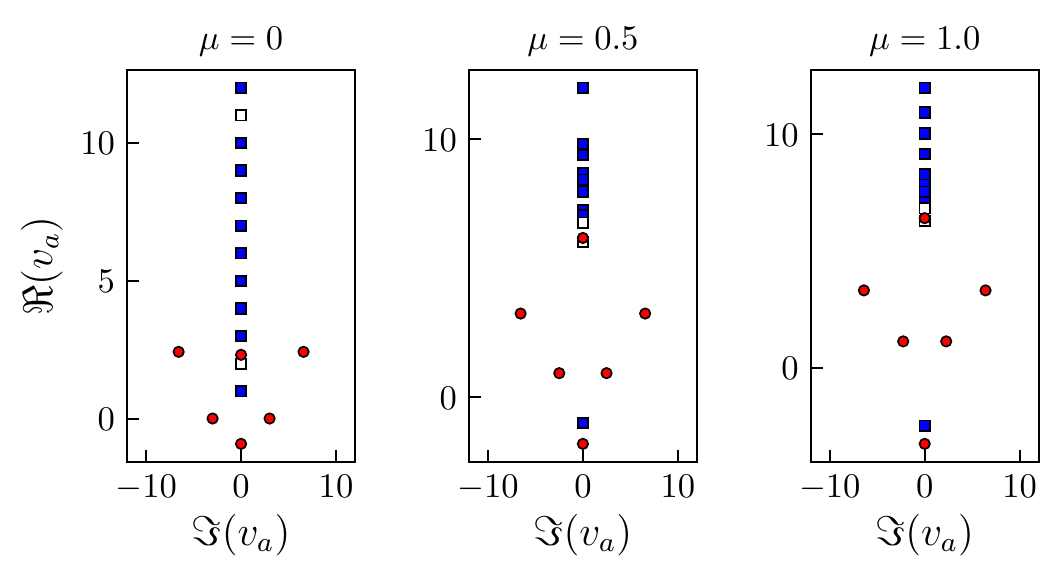}
\caption{Results for the central spin model with perturbation $\mu \vec{S}_i \cdot \vec{S}_j$. Position of $\vec{\epsilon}$ (squares) and $\vec{v}$ (dots) for the variationally optimized wave function starting from an excited state in the complex plane at different values of the perturbation strength. The white squares denote the variables $\epsilon_i, \epsilon_j$ associated with the levels on which the perturbation is applied. \label{fig:rap_Jij_exc}}
\end{center}
\end{figure}

The structure of the optimized variables $\vec{\epsilon}$ and $\vec{v}$ can again be analyzed (Figures \ref{fig:rap_Jij_gs} and \ref{fig:rap_Jij_exc}). The two-spin character of the interaction is clearly visible, where the optimization is mainly sensitive to the two variables $\epsilon_i, \epsilon_j$ $(i=2, j=11)$ in the region where the optimization performs well. When the optimization fails to provide an accurate wave function, the rapidities exhibit a qualitative change (complex conjugate variables become real) and quickly increase in absolute value, pointing out that they are qualitatively wrong. Starting from the excited state in the unperturbed model, it is observed that the rapidities already have the correct structure, and remain bounded during the optimization.

\begin{figure*}                 
 \begin{center}
 \includegraphics[scale=1.15]{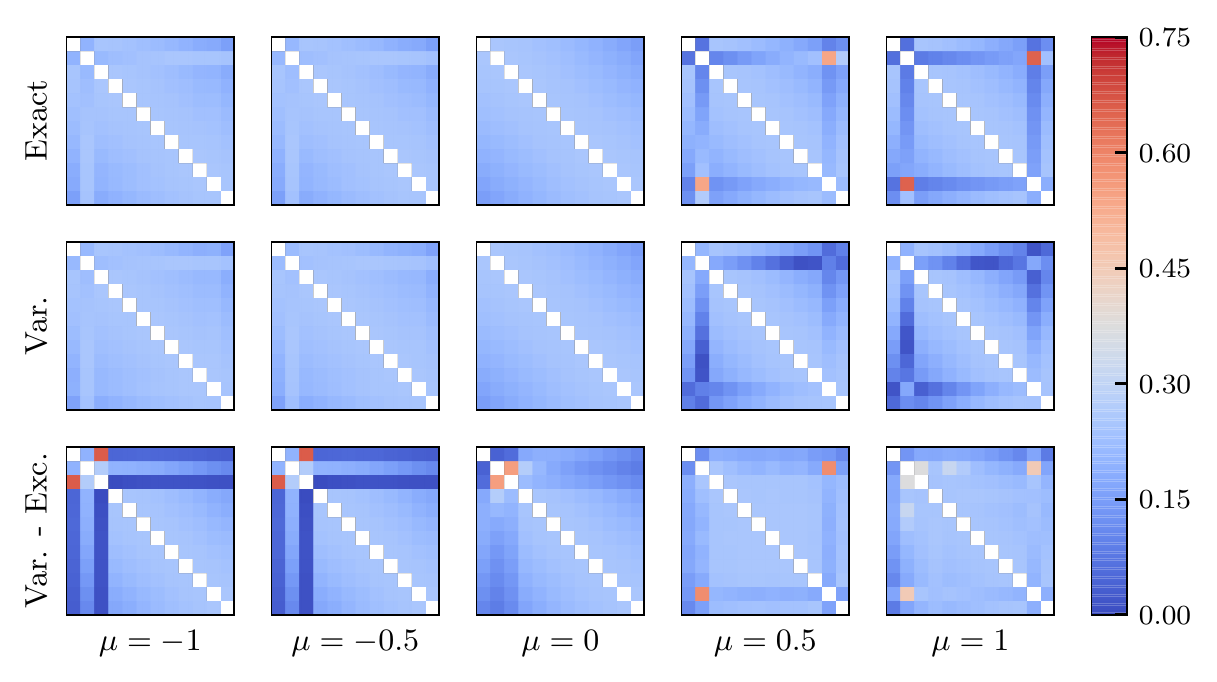} 
 \caption{Absolute value of the expectations values $\braket{\vec{S_i} \cdot \vec{S}_j}$ for the central spin model with perturbation $\mu \vec{S}_i \cdot \vec{S}_j$, with $i,j=2,L-1$. The expectation values are taken w.r.t. the exact ground state and the variational state obtained by starting from both the ground state and excited state of the unperturbed model. \label{fig:exp_Jij}}
 \end{center}
\end{figure*}

\begin{figure*}           
 \begin{center}
 \includegraphics[scale=1.15]{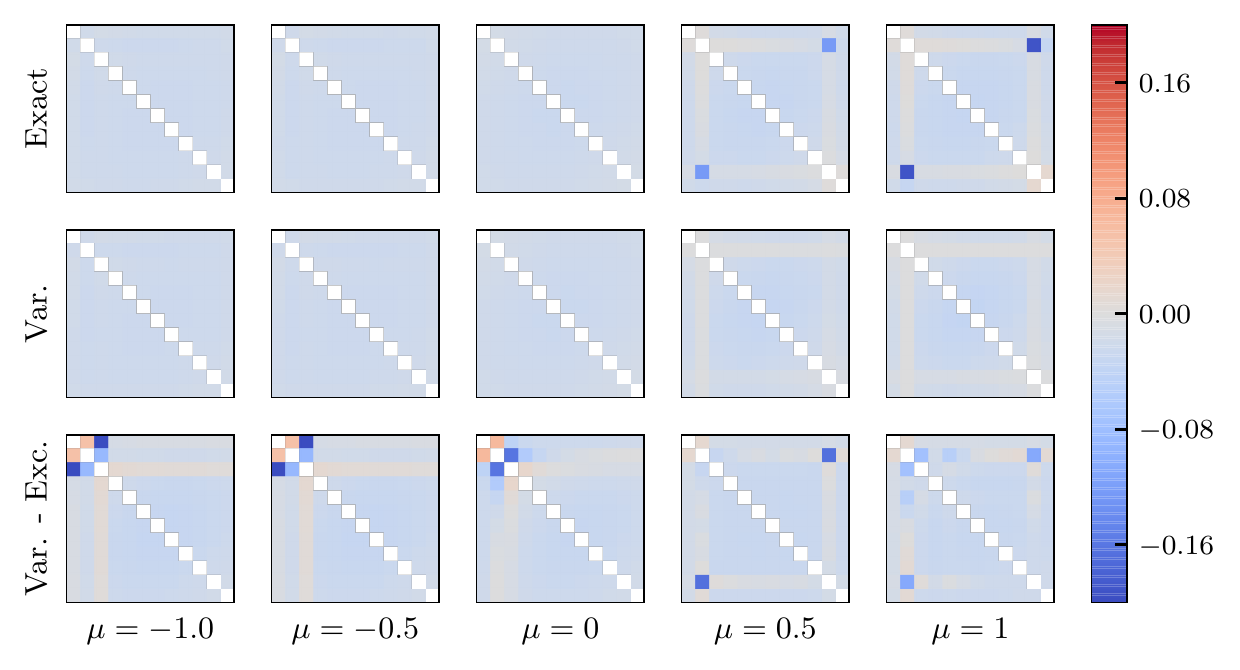} 
 \caption{Correlation coefficients $\braket{S_i^z S_j^z}-\braket{S_i^z}\braket{S_j^z}$ for the central spin model with perturbation $\mu \vec{S}_i \cdot \vec{S}_j$, with $i,j=2,L-1$. These correlation coefficients are calculated for the exact ground state and the variational state obtained by starting from both the ground state and excited state of the unperturbed model.  \label{fig:corr_Jij}}
 \end{center}
\end{figure*}

Some more physical insight can be gathered from expectation values and correlation coefficients calculated from both wave functions. In Figures \ref{fig:exp_Jij} and \ref{fig:corr_Jij}, we present the expectation values $\braket{\vec{S_i} \cdot \vec{S_j}}, \forall i,j$, motivated by the choice of perturbation interactions, and the unconnected correlation coefficients $\sigma_{ij} = \braket{S_i^z S_j^z}-\braket{S_i^z}\braket{S_j^z}, \forall i,j$ for both exact and variational wave functions at different values of $\mu$. It is clear that the correlations within the wave function only change slightly for negative $\mu$, and as such the wave function is able to easily adapt to the perturbation. Comparing the exact and the variational ground state for positive $\mu$, it is notable that the correlations between the two spins affected by the interactions have not been captured by the variational ground-state wave function. Comparing this with the results from the variational excited wave function, it can be seen that the missing correlations are reintroduced there, as was expected. 

For low perturbation strengths, the variational wave function is able to adapt to the correlation structure of the exact ground state  through the optimization. The change in correlation coefficients also points towards the failure of perturbation theory. In the region $\mu \gtrsim 0.2$, the unconnected correlation coefficients from the approximate wave function for the levels on which we apply the perturbation vanish, and this level effectively decouples from the many-body system. In the exact wave function, this decoupling does not occur and instead these coefficients change sign. From this, it can be concluded that the wave function can adapt to the perturbation for as long as the general structure of the correlations does not change. By starting the variational optimization from the excited state, the correct structure is again recovered, as can be seen in the bottom row of Figures \ref{fig:exp_Jij} and \ref{fig:corr_Jij}.

This can now also be related to the expected range of applicability of perturbation theory. For the given Hamiltonians ${H}={H}_{cs}+\mu {V}$, perturbation theory starting from the integrable $\mu=0$ limit can be expected to provide accurate results only if $|\mu \braket{{V}}| \ll \Delta E$, in the regime where the additional term can be considered a small perturbation on the integrable model. The variational optimization starting from the ground state results in a relatively accurate approximation for a larger range of $\mu$, even when the additional term can no longer be considered to be a small perturbation, provided there occur no avoided crossings between the ground- and excited states of the integrable Hamiltonian in the spectrum of the non-integrable Hamiltonian when the perturbation strength $\mu$ is adiabatically increased from $0$ to the given value. Because these Hamiltonians are non-integrable these are expected to be avoided crossings, but this reasoning should also hold for allowed level crossings.


It can be checked that the same behaviour is observed when introducing more involved perturbations, where small perturbations can be accurately described starting from the ground state of the integrable Hamiltonians, and for larger perturbations variational optimization starting from an excited state is necessary in order to obtain the optimal approximative state. However, at present it is not always clear which excited state should be chosen for arbitrary perturbations. In practice, this problem could be circumvented using a stochastic approach, since it was found that several excited states can lead to the same variationally optimized state.

\begin{figure}
\includegraphics[width=\columnwidth]{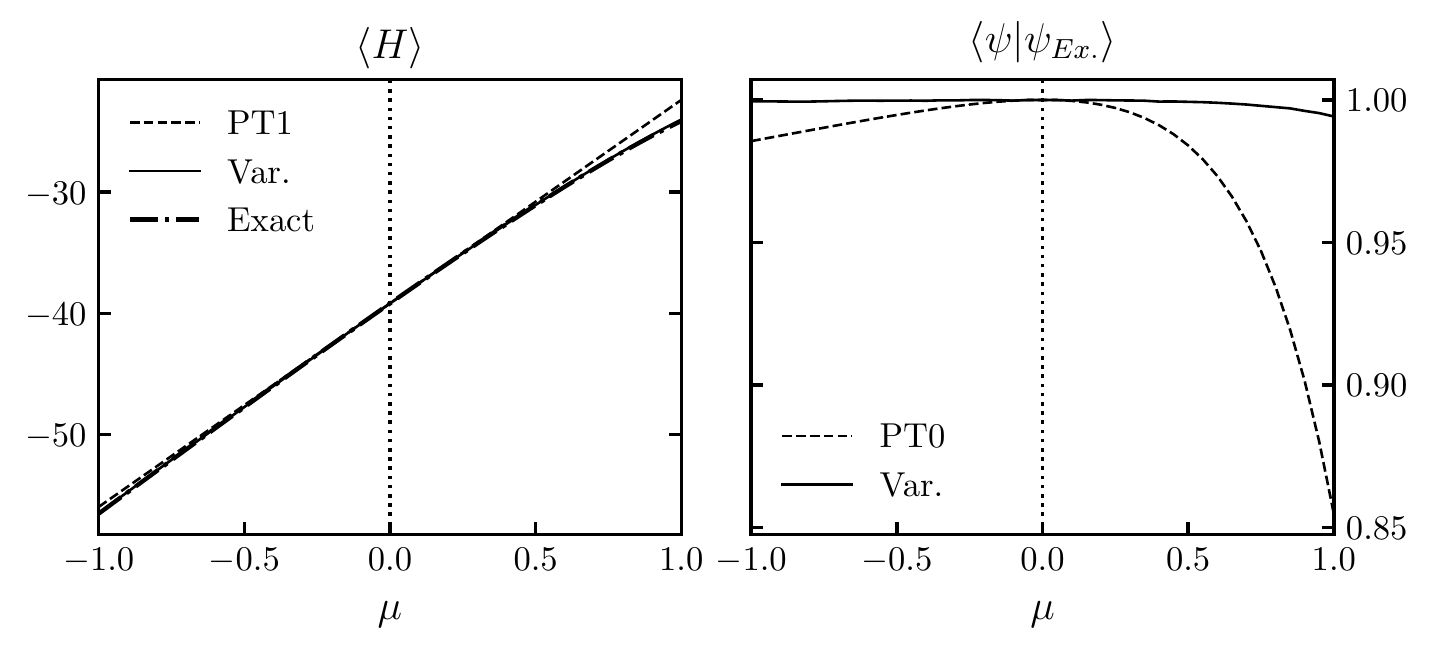}
\caption{Results for the inhomogeneous BCS model. \textbf{Left}: Variational energy (Var.), exact ground state energy (Exact), and first-order perturbation theory (PT1) for different values of the perturbation strength. \textbf{Right}: Overlap of the exact ground state with the variational ground state (Var.) and the ground state of the unperturbed model (PT0).  \label{fig:bcs:ener_ov}}
\end{figure}

\subsection{Perturbing the Richardson model}
The other emblematic example of Richardson-Gaudin models is the Richardson Hamiltonian (see Section \ref{RG:subsec:redBCS}) as given by 
\begin{equation}
{H}_{BCS} = \sum_{i=1}^L \epsilon_i S_i^z + g \sum_{i,j=1}^LS_i^{+}S_{j}^-.
\end{equation}
This Hamiltonian can be used to describe fermion pairing in e.g. nuclear pairing and superconductivity \cite{talmi_simple_1993}, and is exactly solvable under the key assumption that the pairing interactions are uniform and fully determined by a single pairing constant $g$ \cite{richardson_restricted_1963,richardson_exact_1964}. Because of this exact solvability, this model has recently become a testing ground for novel many-body methods focusing on pairing interactions \cite{sambataro_treatment_2013,degroote_polynomial_2016,ripoche_combining_2017,gomez_attenuated_2017}. It's worthwhile to stress that the proposed integrability-based method will return the exact ground state energy of this model by construction. Moving away from integrability, the restriction of uniform interactions can be relaxed by introducing non-uniformities in a perturbative way, resulting in a more physical model. The Hamiltonians under consideration are of the form
\begin{equation}
{H}_{BCS} = \sum_{i=1}^L \epsilon_i S_i^z +  \sum_{i,j=1}^L G_{ij}S_i^{+}S_{j}^-.
\end{equation}
While such models are solvable by $U(1)$-breaking BCS mean-field theory in the thermodynamic limit, it is important to obtain an accurate description for medium-size systems as well \cite{degroote_polynomial_2016,ripoche_combining_2017,gomez_attenuated_2017}. In fact, it has been shown that the Richardson-Gaudin equations are equivalent to the BCS mean-field equations for thermodynamically large systems, and as such the BCS wave function and the Bethe ansatz wave function coincide in this limit \cite{roman_large-n_2002}. The results are presented in Figure \ref{fig:bcs:ener_ov} for a single Richardson Hamiltonian with interaction matrix $G_{ij} = g + \mu g_{ij}$, with $g=-1$ and $g_{ij}$ random numbers uniformly distributed over the interval $[0,1]$. We again take system size $L=12$, parameters according to the picket-fence model, and take $L=2N$ corresponding to half-filling.

\begin{figure*}                 
 \begin{center}
 \includegraphics{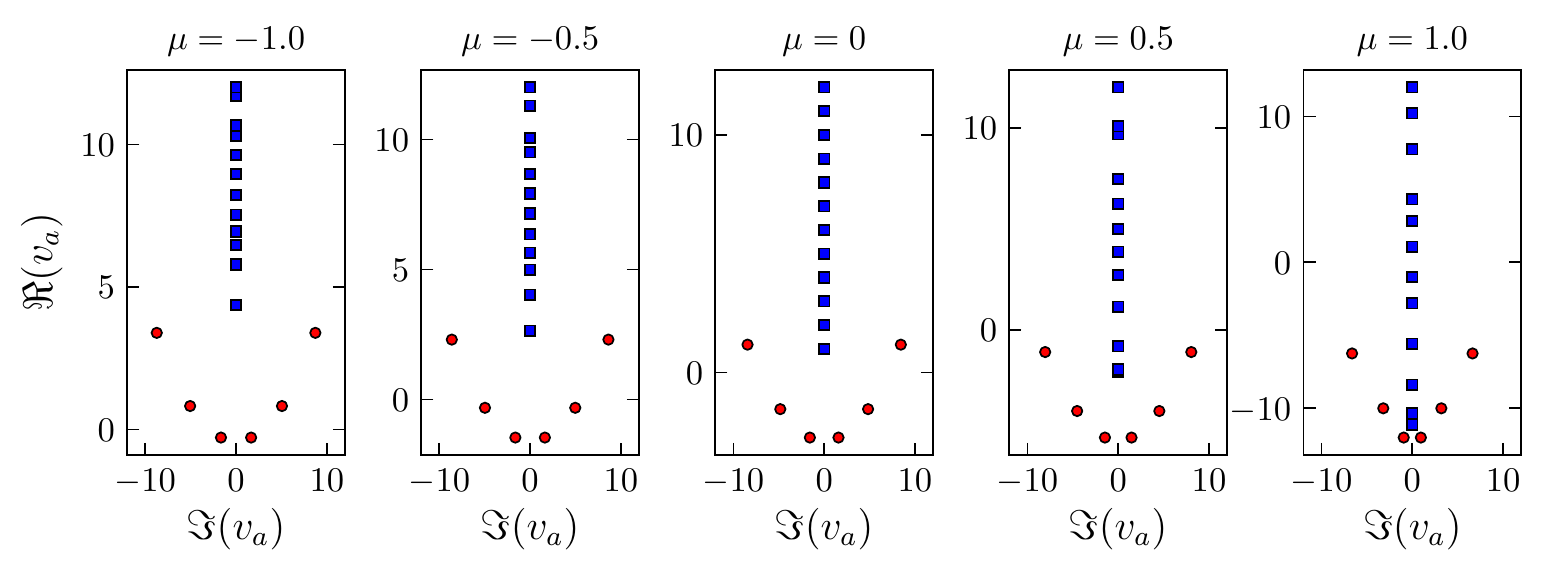} 
 \caption{Results for the inhomogeneous BCS model. Position of $\vec{\epsilon}$ (squares) and $\vec{v}$ (dots) for the variationally optimized wave function in the complex plane at different values of the perturbation strength.\label{fig:rap_BCS}}
 \end{center}
\end{figure*}

The same behaviour as for the central spin model can be observed, where it should be noted that the error on the energy and overlap is much smaller compared to the results for the central spin model. This implies that a general pairing Hamiltonian can already be efficiently approximated by taking the average pairing interaction as single parameter, consistent with the success of BCS mean-field theory in the description of such Hamiltonians.  From the structure of the optimized wave function  (Figure \ref{fig:rap_BCS}), it can be seen that only minor modifications are necessary in order for the wave function to provide an accurate description.

\subsection{Discussion}
\label{sec:concl}

At present, the selection of the proper excited state on which to perform the variational optimization is the main bottleneck in the procedure. One can envision several methods to cope with this problem. The method used in this chapter is to capitalize on the physical insight in the perturbation. Often, the integrability-breaking term in the Hamiltonian itself has a clear physical interpretation, and it is only the competition between the integrable and non-integrable part of the Hamiltonian which is the main cause for complications. Consequently, the correct choice of variational manifold among the excited states can be deduced from the ground state structure of the integrability-breaking term. Other approaches could make use of ideas of stochastic sampling, the correspondence with Coupled-Cluster approaches \cite{limacher_new_2013,boguslawski_efficient_2014,boguslawski_nonvariational_2014,limacher_simple_2014,henderson_seniority-based_2014, stein_seniority_2014,degroote_polynomial_2016} or the $pp$-TDA adiabatic connection \cite{de_baerdemacker_richardson-gaudin_2012,ring_nuclear_2004}. In practice, all relevant excited states were obtained as so-called 1p-1h or 2p-2h excitations of the ground state \cite{dukelsky_colloquium_2004}.

\section{Richardson-Gaudin Configuration Interaction}

Given the remarkable accuracy of the variational approach in the description of general pairing Hamiltonians, this method can be extended by including an increasing number of Bethe states in the ansatz. The exact ground states of all models have a clear multi-reference character in the basis of Slater determinants, and given the similarity of the Bethe state to a Slater determinant it is now possible to construct the multi-reference character of exact ground states in the basis of (variationally optimized) Bethe states. This would then lead to a more accurate description of both the ground state and the lowest-lying excited states, extending the variational scheme to a two-part process.
\begin{enumerate}
\item In a first step, a variational Richardson-Gaudin calculation is performed. Because of the integrability of the underlying model, the variationally obtained state not only gives an approximation to the ground state, but also a complete set of orthogonal basis states, leading to an optimized basis used in the consecutive step.  
\item In the second step, the actual Configuration Interaction (CI) step, the non-integrable Hamiltonian of interest is diagonalized in an increasingly large basis set until convergence is obtained. This step is very much related to other CI methods acting in a basis of on-shell integrable states, such as the Truncated Space Approach \cite{yurov_truncated_1990,yurov_truncated-fermionic-space_1991,james_non-perturbative_2017} which has been used to diagonalize perturbed integrable quantum field theories in one dimension. In this work, the use of an optimized Richardson-Gaudin basis is key.
\end{enumerate}
This then yields an approximation to both the ground state and the lowest-lying excited states, where both steps will be discussed in more detail in the following subsections.  This method can be seen as an adaptation of traditional HF+CI methods.  In these methods, an optimal single-particle Hartree-Fock (HF) product state is obtained first.  This state then defines a Fock Hilbert space in which residual interactions can be systematically included until convergence.  Again, the main difference in this work is that the HF state is replaced by a variational Richardson-Gaudin state, already incorporating pairing correlations in the initial step.

\subsection{Nuclear pairing models}

This scheme will now be applied to a more realistic nuclear-pairing Hamiltonian describing the Sn region. In such models, nuclear pairing correlations are traditionally incorporated from symmetry-broken Bardeen-Cooper-Schrieffer (BCS) theory.  However, as nuclear interactions are becoming better constrained and more accurate \cite{bogner_low-momentum_2010}, this mean-field description will no longer be sufficient, and many-body methods are urged to follow along.  Ripoche \emph{et al.} \cite{ripoche_combining_2017} recently argued for a method that combines symmetry projection and CI for pairing correlations, producing ground-state correlation energies with an accuracy of 0.1\% and better.  The core idea is to construct an optimized set of basis states built from the projected BCS state and selected quasi-particle excitations, which are subsequently used in a non-orthogonal CI method. This can be quite well understood physically, because the pairing correlations have already been optimized in the basis states, either at the BCS mean-field level in the strong interaction regime, or at the perturbative particle-hole level in the weak interaction regime. As such, the approach can be regarded as a natural generalization of the Polynomial Similarity Transformation method (PoST) \cite{degroote_polynomial_2016}, a many-body method that interpolates between projected BCS theory and pairs Coupled-Cluster Doubles (pCCD) \cite{henderson_seniority-based_2014}. The differences between Refs. \cite{ripoche_combining_2017} and \cite{degroote_polynomial_2016} is that the former employs a non-orthogonal CI method, whereas the latter is based on a Coupled Cluster formulation of projected BCS \cite{dukelsky_structure_2016}. Again, Bethe states from a variational Richardson-Gaudin calculation can also be used as an orthogonal set of basis states, further building on these ideas.

The nuclear pairing Hamiltonian can generally be written as
\begin{equation}\label{rgci:hamiltonian}
{H} = \sum_{i=1}^L \epsilon_i n_i+\sum_{i,j=1}^L G_{ij} S_i^+ S_j^-,
\end{equation}
where the effective pairing interaction can be obtained from a $G$-matrix construction for the Sn isotopes in the neutron valence shell $A=100-132$ ($g_{7/2},d_{5/2},s_{1/2},h_{11/2},d_{3/2}$) \cite{holt_effective_1998,zelevinsky_nuclear_2003}. The specific values of the pairing interaction can be found in Ref. \cite{zelevinsky_nuclear_2003}, and are also listed in Table \ref{table:interactionparameters} for quick reference.  
\setlength{\tabcolsep}{7pt}
\begin{center}
\begin{table}
\begin{center}
\begin{tabular}{l | c c c c c }
  & $g_{7/2}$ & $d_{5/2}$ & $s_{1/2}$ & $h_{11/2}$ & $d_{3/2}$ \\ \hline
$\Omega_i$ & $8$ & $6$ & $2$ & $12$ & $4$ \\ \hline
$\epsilon_i$  & $-6.121$ & $-5.508$ & $-3.891$ & $-3.778$ & $-3.749$ \\ \hline
   $g_{7/2}$  & $-0.2463$ & $-0.1649$ & $-0.1460$ & $-0.2338$ & $-0.1833$ \\
   $d_{5/2}$  &           & $-0.2354$ & $-0.1995$ & $-0.2250$ & $-0.3697$ \\
   $s_{1/2}$  &           &           & $-0.7244$ & $-0.1741$ & $-0.2486$ \\
   $h_{11/2}$ &           &           &           & $-0.1767$ & $-0.1762$ \\
   $d_{3/2}$  &           &           &           &           & $-0.2032$ \\ \hline
\end{tabular}
\caption{Pairing interaction parameters $\epsilon_i$ and $G_{ij}$ in the Hamiltonian (\ref{rgci:hamiltonian}) for Sn isotopes from a $G$-matrix formalism \cite{holt_effective_1998,zelevinsky_nuclear_2003}.  All energies are measured in MeV.}\label{table:interactionparameters}
\end{center}
\end{table}
\end{center}

\begin{figure}[!htb]
\begin{center}
	\includegraphics{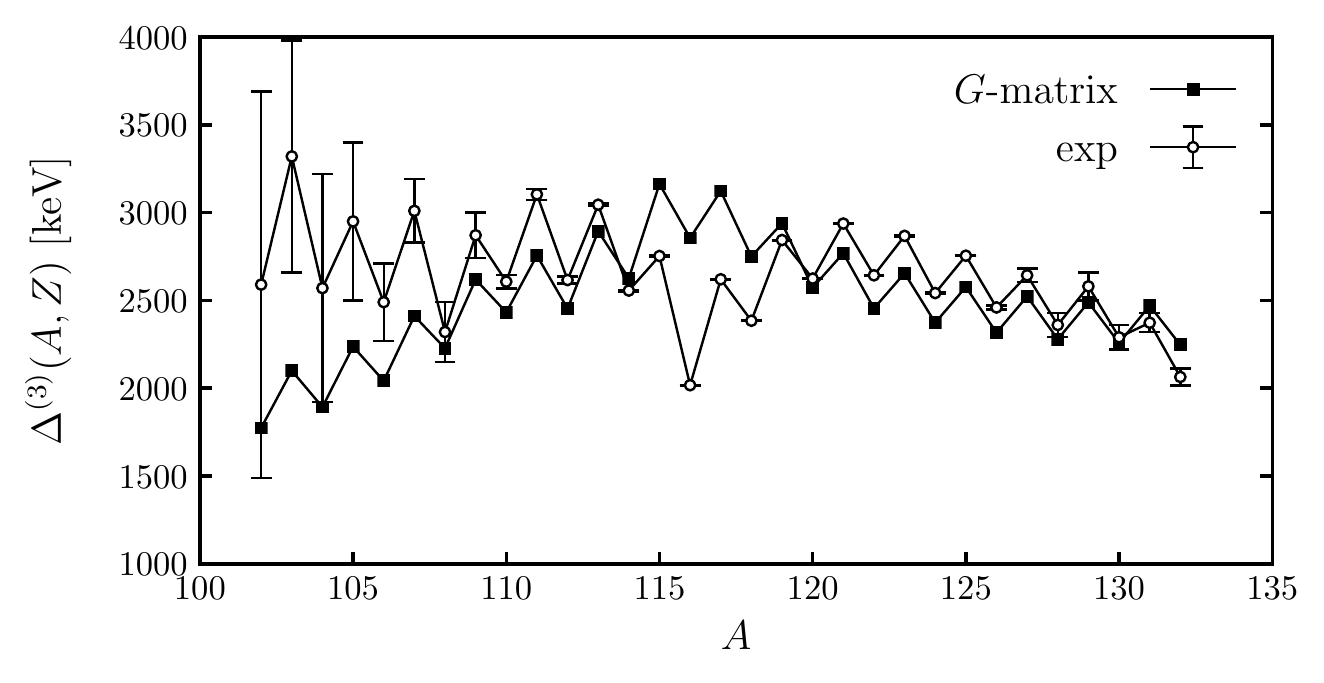}
	\caption{3-point neutron pairing gaps $\Delta^{(3)}$ (\ref{rgci:3pointgaps}) calculated from an effective $G$-matrix interaction (See Table \ref{table:interactionparameters} and Ref. \cite{zelevinsky_nuclear_2003}), and compared with experimental values \cite{audi_ame2003_2003}.}\label{figure:rgci:gmatrixvsexperiment}
\end{center}
\end{figure}

This is an ideal benchmark system for multiple reasons.  First, the pairing strength is known to be very stable in the Sn isotopes, with a slight experimentally-observed decrease around the neutron number $64$ subshell closure \cite{jungclaus_evidence_2011,morales_generalized_2011,de_baerdemacker_probing_2013}.  Second, the dimensions of the pairing Hamiltonian (\ref{rgci:hamiltonian}) are rather limited for this shell, so a comparison with exact results from conventional exact CI remains possible \cite{volya_exact_2001}.  To illustrate the performance of the effective interaction (Table \ref{table:interactionparameters}) with respect to experimental values, calculated 3-point neutron pairing gaps, derived from nuclear binding energies $BE(A,Z)$ \cite{bohr_nuclear_1998},
\begin{align}\label{rgci:3pointgaps}
\Delta^{(3)}(A,Z)=(-)^A[&BE(A,Z)-2BE(A-1,Z)+BE(A-2,Z)],
\end{align}
are compared to experimental values taken from \cite{audi_ame2003_2003} in Figure \ref{figure:rgci:gmatrixvsexperiment}. For the purpose of this work, it is sufficient to note the qualitative agreement between the $G$-matrix results and experimental values within their errorbars. As such, this interaction will be used solely to test the RGCI method, as is done before with other methods like \cite{sambataro_multipair_2012,sambataro_treatment_2013}, and we will refrain explicitly from making further comparison with experimental data.
\begin{figure}
    \centering
    \begin{subfigure}[b]{0.49\textwidth}
        \includegraphics[scale=0.92]{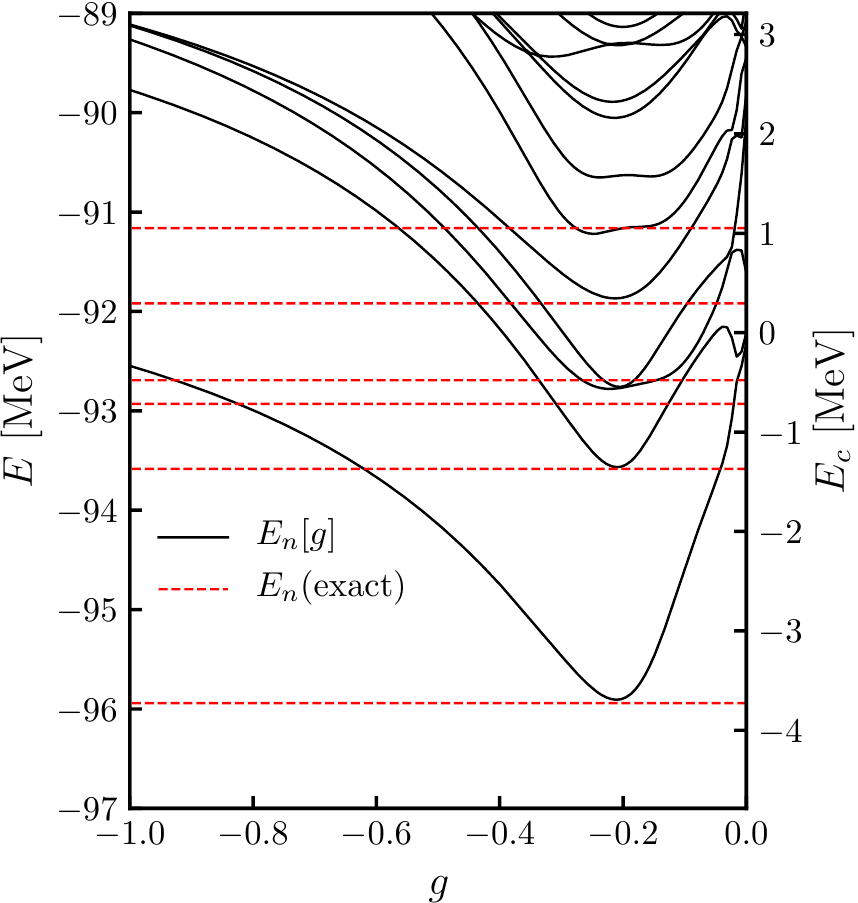}
        \caption{\label{fig:rgci:spectrum}}
    \end{subfigure}
    \begin{subfigure}[b]{0.49\textwidth}
        \includegraphics[scale=0.92]{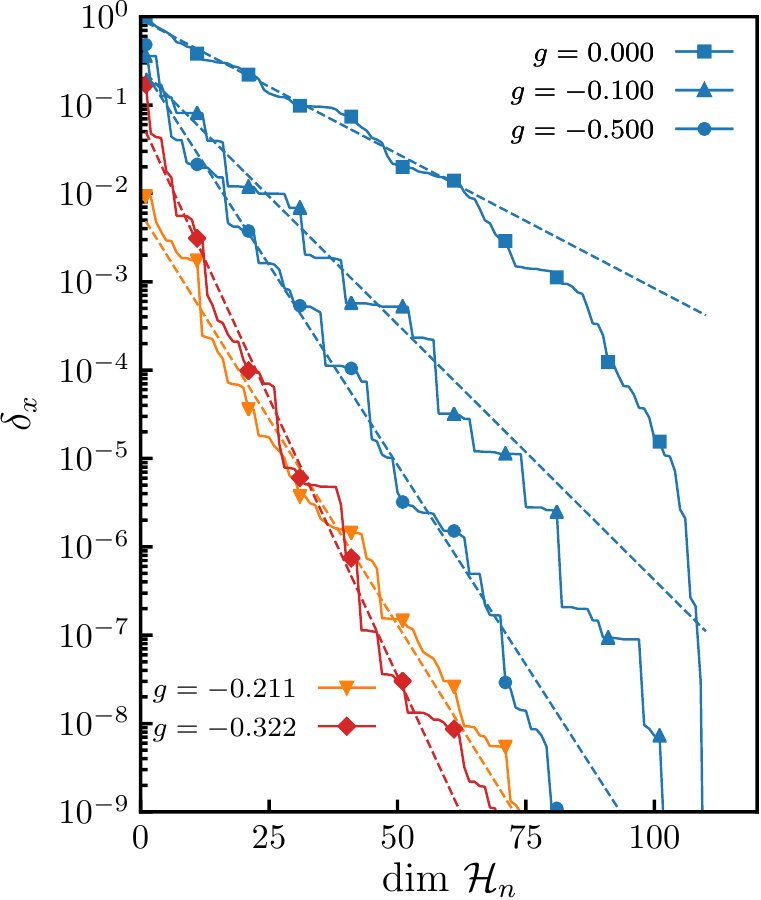}
        \caption{\label{fig:rgci:convergence}}
    \end{subfigure}
    \caption{\textbf{(a)} Full lines represent the energy function profiles $E_n[g]$ for different eigenstates $n$ of the corresponding RG Hamiltonian (\ref{rgci:intham}).  Exact eigenstate energies $E_n$(exact) of the effective Hamiltonian (\ref{rgci:hamiltonian}) are given in dashed lines ($n=0\dots 5$).  The energy scale shows total (left axis) and correlation energies (right axis). The calculations are performed for $N=8$ pairs (${}^{116}$Sn). \textbf{(b)} Convergence rate of missing correlation error $\delta_c$ for the RGCI method with $N=8$ (${}^{116}$Sn).  Convergence rates for different integrable bases are denoted by the corresponding value of $g$. Units of $g$ are given in MeV, and $\delta_c$ is dimensionless.}\label{figure:rgci:rgci:116sn}
\end{figure}

The two-part process is now presented in Figures \ref{fig:rgci:spectrum} and \ref{fig:rgci:convergence}. Because of the remarkable success of the variational approach for pairing models, the initial variational optimization can be restricted to the pairing strength $g$, while the single-particle energies $\epsilon_i$ are fixed and taken to be those from Table \ref{table:interactionparameters}. This has several advantages. First, the variational optimization of $E[\psi_{RG}]=E[g]$ is reduced to a line-search compared to a gradient descent method. Second, the connection with HF mean-field theory can be made more explicit, where the variationally-optimized pairing strength $g_0$ and the resulting reduced BCS Hamiltonian
\begin{equation}\label{rgci:intham}
{H} = \sum_{i=1}^L \epsilon_i n_i+g_0 \sum_{i,j=1}^L S_i^+ S_j^-
\end{equation}
can be seen as a mean-field pairing description of the original Hamiltonian (\ref{rgci:hamiltonian}). This first step is illustrated in Figure \ref{fig:rgci:spectrum}. The lowest full curve corresponds to the energy functional of the Richardson-Gaudin ground state, and gives the best approximation of the exact ground-state energy, as was expected.  The variationally obtained energy is reached at $g_0=-0.211$ MeV, giving rise to $E[g_0]=-95.907$ MeV, which is equivalent to $99.07\%$ of the exact correlation energy.  A more practical measure for gauging the quality of a method is given by 1 minus this correlation-energy ratio, being
\begin{equation}
\delta_c=1-\frac{E_c(\textrm{method})}{E_c(\textrm{exact})},
\end{equation}
which in the present example amounts to $0.93\%$. Apart from the ground-state energy curve $E[g] \equiv E_0[g]$, it is also interesting to investigate the performance of other Bethe eigenstates energy curves. These are also included in Figure \ref{fig:rgci:spectrum}.  It is clear that the low-lying excited energy curves $E_n[g]$ all approach an exact eigenstate energy $E_n$(exact) in the vicinity of the optimal $g_0=-0.211$ MeV, pointing out that the integrable Hamiltonian with $g=g_0=-0.211$ MeV is indeed a good approximation to the effective Hamiltonian (\ref{rgci:hamiltonian}). 

In Figure \ref{fig:rgci:convergence}, the convergence in this missing correlation energy error $\delta_c$ with increasing basis size is then presented for $N=8$ pairs (${}^{116}$Sn) for different values of $g$ (including $g_0$ in (red) diamonds). Obviously, $\delta_c$ is an adequate measure for the validation of the procedure, but only make sense when the exact ground state is known.  However, one can easily envision other suitable convergence measures in practical situations. The ordering of the on-shell basis states is fixed by means of the eigenstate energy spectrum of the integrable Hamiltonian (\ref{rgci:intham}), and the Hamiltonian (\ref{rgci:hamiltonian}) is diagonalized in an increasingly large active Hilbert space $\mathcal{H}_n$ ($n=1\dots \dim{\mathcal{H}}$) of on-shell states until convergence or the complete basis limit ($\mathcal{H}_{\dim\mathcal{H}}\equiv\mathcal{H}$) is reached. From Figure \ref{figure:rgci:rgci:116sn}, the following observations can be made, where the observed behaviour is generic for all isotopes in the Sn region. 
\begin{itemize}
\item Because of the variational principle, the error $\delta_c$ is monotonically decreasing with increasing size of the active Hilbert space $\mathcal{H}_n$, and vanishes by definition as soon as the complete basis set limit is reached, regardless of the value of $g$.  For $N=8$ (${}^{116}$Sn), the complete basis limit is reached for $\dim\mathcal{H}=110$.  
\item Different values of $g$ give rise to different convergence rates.  The $g=0$ curve (blue squares) corresponds to the traditional approach in which the Hamiltonian (\ref{rgci:hamiltonian}) is diagonalized in an uncorrelated Fock space with increasing dimension.  As can be expected, the convergence rate of $\delta_c$ is steady but slow.  From Figure \ref{fig:rgci:convergence}, it can be seen that approximately half of the Hilbert space is required to build up the necessary degree of collectivity to reach the desired $\delta_c\le 1\%$ accuracy.  
\item For non-zero values of $g$, the convergence is considerably improved (note the logarithmic scale). This is visible in both the intercept and the slope of the $g\neq0$ curves.  The values of the intercept correspond to the ground-state energy expectation value $E[g]$ (\ref{fig:rgci:spectrum}), so the more $g$ approaches the variational minimum $g_0$, the lower the value of the intercept.  The (orange) curve with triangles depicts exactly the RG basis constructed with the variationally optimized $g_0=-0.211$ MeV.  Not only is the intercept lowest of all possible $g$ values by definition, the slope of convergence is also among the steepest, pointing out again that this is a very suitable basis and hinting at an exponential convergence rate. The fastest convergence is obtained at $g_b=-0.322$ MeV, as evidenced by the (red) curve with diamonds.
\end{itemize}

\subsection{Pre-diagonalization and Similarity Renormalization Group}
\begin{figure}[!htb]
\begin{center}
	\includegraphics{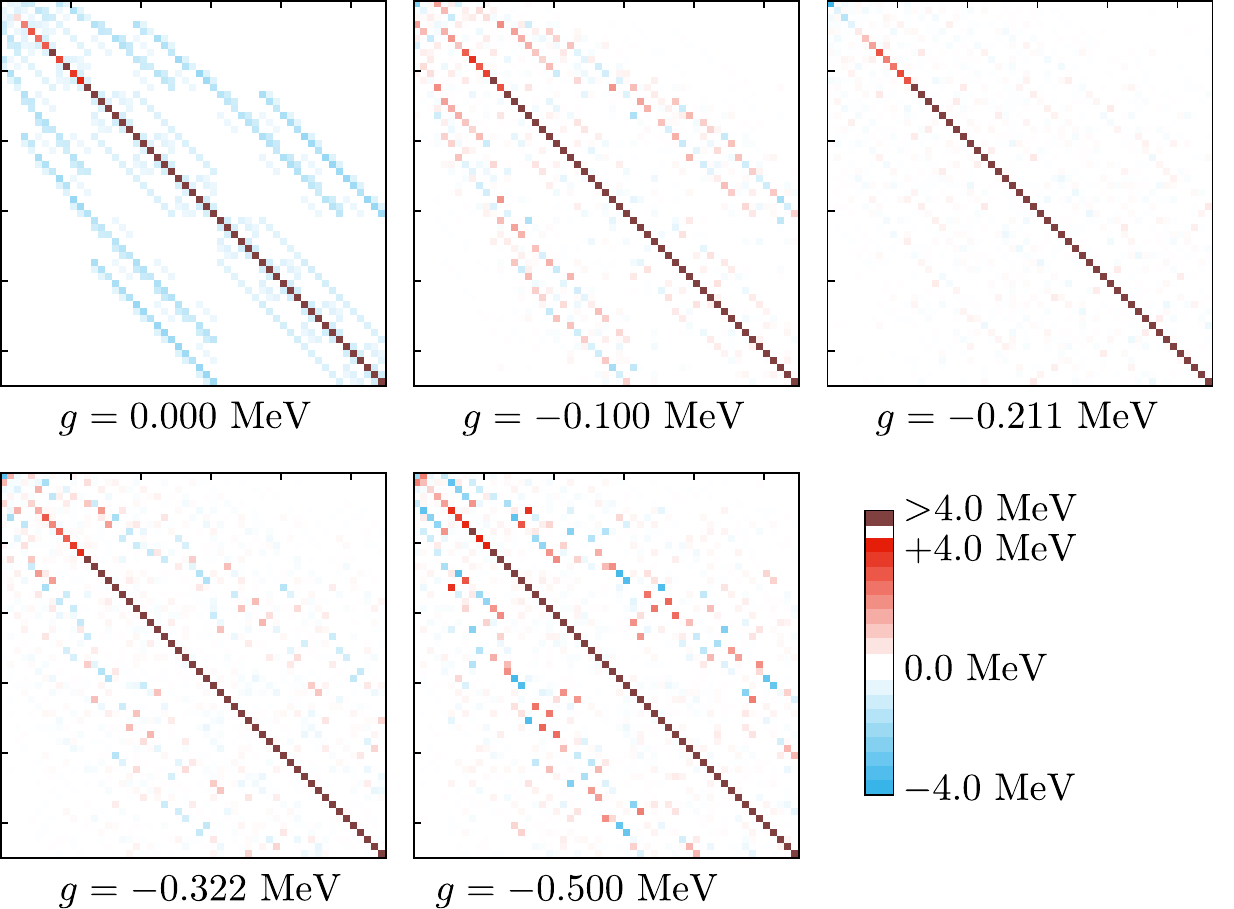}
	\caption{Visual representation of the exact Hamiltonian $\hat{H}$ (\ref{rgci:hamiltonian}) matrix in different normalized Richardson-Gaudin bases, labeled by $g$ (in MeV). Each square represents a matrix element (\ref{rgci:srg:matrixelement}), with the value of the shading denoting the magnitude of the matrix element.  The sign of the matrix element (blue/red color) is irrelevant for the discussion.  Only the lowest (quarter) part of the Hamiltonian matrix is shown. }\label{figure:rgci:srg:matrixelements}
\end{center}
\end{figure}

Although intuitive, the good convergence rate of the RGCI method at the variational minimum is by no means guaranteed from the variational principle.  For a better understanding of the convergence performance of RGCI, it is instructive to investigate the matrix elements of the non-integrable Hamiltonian (\ref{rgci:hamiltonian}) in the basis of on-shell Bethe states
\begin{equation}\label{rgci:srg:matrixelement}
\frac{\langle \psi_{RG}(g)_m| {H}| \psi_{RG}(g)_n \rangle}{\sqrt{\langle\psi_{RG}(g)_m|\psi_{RG}(g)_m\rangle\langle\psi_{RG}(g)_n|\psi_{RG}(g)_n\rangle }},
\end{equation}
as a function of $g$.  These matrix elements are visualized in Figure (\ref{figure:rgci:srg:matrixelements}) for $N=8$ (${}^{116}$Sn) with the same selected values of $g$ as in Figure \ref{figure:rgci:rgci:116sn}.  

For visual purposes, only the lowest part of the total matrix is given.  Each matrix element is represented by a colored dot, with the color saturation proportional to the magnitude of the matrix elements (\ref{rgci:srg:matrixelement}).  While the Figure distinguishes between positive and negative matrix elements,this distinction is irrelevant as particular non-diagonal matrix elements can be sign flipped by an appropriate phase similarity transformation.  More importantly, zero-valued matrix elements are represented by white dots.  The diagonal matrix elements are shifted such that the first matrix element (upper left dot in each panel) represents the correlation energy $E_c$ (see right axis of Figure \ref{fig:rgci:spectrum}). The $g=0.000$ MeV panel corresponds to the traditional Hamiltonian matrix in Fock space.  Accordingly, the upper-left matrix element is zero (white) in this panel, by definition.  Moving away from $g=0.000$ MeV, the off-diagonal elements of the matrix become suppressed, with the Hamiltonian matrix (\ref{rgci:srg:matrixelement}) approaching diagonality around the variationally optimal value $g_0=-0.211$MeV.  It is worth noting that the diagonality is again lost when further increasing $g$, even at the fastest converging point $g_b=-0.322$ MeV.  From this, it is easy to understand the fast convergence of the RGCI method at the optimal variational point, as the Hamiltonian matrix was already very close to diagonal from the start.

This observation appears to be in line with ideas from Similarity Renormalization Group (SRG) methods \cite{wegner_flow-equations_1994,bogner_low-momentum_2010,hergert_-medium_2017}. The SRG describes an isospectral flow of a Hamiltonian in such a way that it finds a representation (basis) in which part of the Hamiltonian matrix is suppressed. The variational Richardson-Gaudin method shares the characteristics of an isospectral flow because the Hamiltonian matrix (\ref{rgci:srg:matrixelement}) can be recast as a unitary similarity transformation
\begin{align}
\frac{\langle \psi_{RG}(g)_m| {H}| \psi_{RG}(g)_n \rangle}{\sqrt{\langle\psi_{RG}(g)_m|\psi_{RG}(g)_m\rangle\langle\psi_{RG}(g)_n|\psi_{RG}(g)_n\rangle }}
=\sum_{{i},{j}}\frac{\langle \psi_{RG}(g)_m|{i}\rangle\langle {i}|{H}|{j}\rangle\langle{j} | \psi_{RG}(g)_n\rangle}{\sqrt{\langle\psi_{RG}(g)_m|\psi_{RG}(g)_m\rangle\langle\psi_{RG}(g)_n|\psi_{RG}(g)_n\rangle }},
\end{align}
with $\{|{i}\rangle\}$ and $\{|{j}\rangle\}$ both a complete set of (normalized) basis states in Fock space.  In operator form, this can be clarified as
\begin{equation}
{H}(g)=U(g){H}(g=0)U(g)^\dagger,
\end{equation}
with ${H}(g=0)$ the matrix representation in Fock space, and $U(g)$ the unitary matrix with matrix elements
\begin{equation}\label{rgci:srg:unitary}
U(g)_{m,i}=\frac{\langle \psi_{RG}(g)_m|{i}\rangle}{\sqrt{\langle \psi_{RG}(g)_m|\psi_{RG}(g)_m\rangle}}.
\end{equation}
The variational method then shares the properties of isospectral flow with SRG because each value of $g$ not only characterizes a (variational) trial state, but also a complete basis of on-shell Bethe states, leading to a full-rank unitary matrix (\ref{rgci:srg:unitary}).  This is in contrast with other variational approaches, where typically only the trial state is properly defined.  Nevertheless, the main difference with SRG is that SRG generates a dynamical flow from local updates driven towards a suppression of unwanted off-diagonal matrix elements.  In the proposed variational approach, the suppression of the off-diagonal part of the Hamiltonian matrix appears to be a convenient byproduct of the variational approach leading to optimal convergence properties in the RGCI step.

\section{Conclusions}
In this chapter, it was investigated how the ground states of non-integrable Hamiltonians consisting of an integrable (Richardson-Gaudin) Hamiltonian and an integrability-breaking Hamiltonian can be approximated by modified eigenstates of related integrable Hamiltonians. Due to the inherent structure of these Bethe ansatz eigenstates, it is possible to efficiently calculate and minimize the expectation value of given Hamiltonians with respect to these states, and it was shown how such a variational approach can be implemented. This was then shown to provide accurate results for selected perturbed non-integrable Hamiltonians, where the accuracy of the variational approach is only limited by the appearance of avoided level crossings in the spectrum of non-integrable Hamiltonians. When the exact ground state can be considered a perturbation of the non-perturbed integrable Hamiltonian (i.e. there are no avoided crossings), the variational optimization starting from the non-perturbed ground state will provide accurate results. The effects of such crossings can then be taken into account by variationally optimizing excited states of the integrable Hamiltonian, instead of restricting the optimization to the ground state.

This could then be extended for the treatment of general pairing correlations, leading to a two-step approach. The first step is a variational optimization of an on-shell Richardson-Gaudin state. Here, the wave function is already qualitatively correct for the description of collective Cooper pair condensation. This eliminates the need for a sophisticated selection scheme to identify the correct manifold of on-shell states upon which to vary. The second step is to use the resulting set of excited states on top of the variationally optimized Richardson-Gaudin state as a basis in which to perform a Configuration Interaction calculation in an increasingly large active Hilbert space until convergence. Again, the integrability of the Richardson-Gaudin model is key for a computationally soft (polynomial) scaling. Interestingly, the convergence to the exact values is exponential, mainly due to a strong suppression of the off-diagonal matrix elements in the Hamiltonian when expressed in this optimized basis.

In all steps, the computationally favourable scaling of expectation values and inner products of on-shell integrable states is exploited. A first possible extension of this method would be to enlarge the Richardson-Gaudin basis set to include non-zero seniority states.  In the theory of Richardson-Gaudin integrability, this corresponds to the simple blocking of a given orbital, so all useful features of integrability for the variational/RGCI method are kept. A second approach would be to generalize the Slavnov-like theorems of integrability to higher-order algebras, including the isovector/scalar proton-neutron pairing algebras $so(5)$ and $so(8)$ \cite{dukelsky_integrable_2006,lerma_exactly_2006,lerma_exact_2007,lerma_su3_2007}.  However, much more mathematical results are needed for the efficient calculation of off-diagonal matrix elements \cite{johnson_size-consistent_2013,johnson_strategies_2017}, so the first suggestion seem much more straightforward in the short run.  From a physical point of view, it would be interesting to further investigate the connection between the variational Richardson-Gaudin method and Similarity Renormalization Group ideas \cite{hergert_-medium_2017}. The applicability of the RGCI in other domains of physics also seems worth exploring, suggesting the use of variationally-optimized basis sets for perturbed integrable quantum field theories in the Truncated Space Approach.

\chapter{Floquet dynamics from integrability}
\label{chap:floquet}

\setlength\epigraphwidth{.4\textwidth}
\epigraph{\emph{Everything takes time.  Bees have to move very fast to stay very still.}}{{David Foster Wallace}}

Integrability has recently found a major use in the field of quantum quenches \cite{jstatmech_outofequilibrium_2016}. Within a quantum quench, a system is prepared in the ground state of an initial Hamiltonian, after which this initial Hamiltonian is abruptly changed (quenched) and the state is left free to evolve in time under this final Hamiltonian. The availability of exact Bethe eigenstates combined with numerically efficient expressions for inner products have made integrable systems the perfect testing ground for the study of quantum quenches, since these are the necessary ingredients for time evolution. Although much more involved, the resulting dynamics can then be seen as the quantum equivalent of the exactly-solvable dynamics in classical integrability. In such a way, the dynamics of Bethe states have been extensively studied, where Richardson-Gaudin models presented one of the first applications of integrability in quantum quenches \cite{faribault_bethe_2009,faribault_quantum_2009,buccheri_structure_2011,foster_quantum_2013,barmettler_non-equilibrium_2013}.

A natural extension would then be to consider a system which is `periodically quenched', where the Hamiltonian is repeatedly changed. Such periodic quenches are a specific case of periodically-driven systems. Here, Floquet theory presents a convenient framework for the study of periodic dynamics \cite{shirley_solution_1965,sambe_steady_1973,goldman_periodically_2014,bukov_universal_2015}, allowing for the definition of a Floquet Hamiltonian encoding all periodic dynamics. However, the explicit construction of the Floquet Hamiltonian involves the full Hilbert space and it is generally impossible to obtain this Hamiltonian exactly. Furthermore, even if the system is being driven using integrable Hamiltonians, the resulting Floquet Hamiltonian will generally not be integrable \cite{prosen_time_1998,claeys_breaking_2017,seetharam_absence_2017}.

More generally, periodically driven systems are ubiquitous in both nature and experimental physics, with a rich history ranging from the simple kicked rotor to recent experimental progress on cold atoms in optical fields \cite{goldman_periodically_2014,bukov_universal_2015}. The dynamics in these systems has some remarkable features, such as the absence of a well-defined adiabatic limit \cite{hone_time-dependent_1997,weinberg_adiabatic_nodate} and the heating to infinite temperature which is expected to occur \cite{dalessio_long-time_2014,lazarides_equilibrium_2014, ponte_periodically_2015,else_pre-thermal_2017,moessner_equilibration_2017}. The same physical mechanism underlies these two phenomena -- in the presence of periodic driving it is possible for states to interact resonantly. By coupling to the driving, states whose energies are separated by an integer multiple of the driving frequency will interact strongly, leading to Floquet or many-body resonances (similar to multi-photon resonances) \cite{eckardt_avoided-level-crossing_2008,hone_statistical_2009,bukov_heating_2016,russomanno_floquet_2017}. While this is generally seen as a disadvantage because of the experimental problems posed by heating, there is hope that in large but finite systems such many-body resonances can be well understood and controlled. This could then be used to induce transitions between resonant eigenstates of the time-averaged Hamiltonian or engineer coherent superpositions of specific eigenstates by adiabatically tuning the driving frequency to resonance. Such ``driven driving'' protocols, if smartly conceived, could lead to states not necessarily realizable as eigenstates of physical (stationary) Hamiltonians, with properties beyond these associated with the driving Hamiltonians \cite{bendall_broadband_1995,hediger_adiabatic_1995,hwang_broadband_1997,takayoshi_laser-induced_2014,chen_floquet_2015,sato_laser-driven_2016}.

In this chapter, we show how Richardson-Gaudin integrability can be exploited in order to obtain a description of periodically-driven systems by explicitly constructing the (non-integrable) Floquet Hamiltonian in a restricted many-body basis consisting of Bethe states. By adiabatically varying the driving frequency of a periodically-driven central spin system it is possible to induce controlled transitions between resonant eigenstates of the time-averaged (integrable) Hamiltonian. This corresponds to adiabatic transitions in the Floquet Hamiltonian at quasi-degeneracies, and it is illustrated how such transitions can be used to construct a coherent superposition of the ground state and the highest excited state in a driven central spin model. Remarkably, Floquet resonances can here be used to construct pure spin states at even-order resonances, seemingly at odds with the inevitable interaction with the environment and the resulting decoherence effects. Alternatively, at odd-order resonances the magnetization is shown to vanish. This presents a first step toward applying the toolbox from integrability to driven interacting systems, where integrability is generally expected to lose its usefulness. The present chapter is based on Ref. \cite{claeys_spin_2017}.

\section{Floquet theory}

The starting point in the study of periodically driven systems is the Floquet theorem \cite{shirley_solution_1965,sambe_steady_1973,goldman_periodically_2014,bukov_universal_2015}, which allows the unitary evolution operator to be recast as
\begin{equation}\label{floq:evolop}
{U}(t) = {P}(t) e^{-i {H}_F t},
\end{equation}
with ${P}(t)$ a periodic unitary operator with the same period $T$ as the driving Hamiltonian $H(t+T)=H(t)$, and ${H}_F$ the so-called Floquet Hamiltonian. Furthermore, the fast-motion operator ${P}(t)$ reduces to the identity at stroboscopic times $t = n T, n \in \mathbb{N}$. Hence, at stroboscopic times the system behaves as if it evolves under the time-independent Floquet Hamiltonian, since $U(nT) = e^{-i H_F nT}$. Considering time evolution over one full cycle then leads to the Floquet operator $U_F$, from which the Floquet Hamiltonian follows as
\begin{equation}
{U}_{F} \equiv {U}(T) = e^{-i {H}_F T}.
\end{equation}
Simultaneously diagonalizing these operators leads to
\begin{align}
{H}_F = \sum_n \epsilon_n \ket{\phi_n}\bra{\phi_n}, \qquad 
{U}_F = \sum_n e^{-i \theta_n} \ket{\phi_n}\bra{\phi_n},
\end{align}
where the eigenvalues of the Floquet Hamiltonian are denoted quasi-energies, defined as $\epsilon_n=\theta_n/T$. Floquet theory in periodically-driven systems has several parallels with Bloch theory in periodic lattices, and the quasi-energies here provide the Floquet equivalent of quasi-momenta in Bloch waves. Similarly, they are only defined up to shifts $k \cdot 2 \pi/T, k \in \mathbb{N}$, since the phases $\theta_n$ are only defined up to similar shifts $k \cdot 2 \pi, k \in \mathbb{N}$. As such, they are commonly restricted to a Brillouin (Floquet) zone $[-\pi/T,\pi/T]$ and quasi-energies separated by shifts $k \cdot 2 \pi/T, k \in \mathbb{N}$ are said to be quasi-degenerate. Although not clear from the notation, the Floquet Hamiltonian itself is also strongly dependent on the driving period $T$, which will be crucial in the following. In periodic quenches, the driving protocol can generally be written as
\begin{equation}
{H}(t) = 
  \begin{cases}
 {H}_1 &\text{for}\qquad 0 < t < \eta T, \\
  {H}_2    & \text{for}\qquad \eta T < t < T,
  \end{cases}
\end{equation}
with ${H}(t+T) = {H}(t)$ and $\eta \in [0,1]$, leading to 
\begin{equation}
{U}_F \equiv e^{-i {H}_F T} = e^{-i (1-\eta) {H}_2 T} e^{-i \eta {H}_1 T}.
\end{equation}
\begin{figure}
\begin{center}
\includegraphics{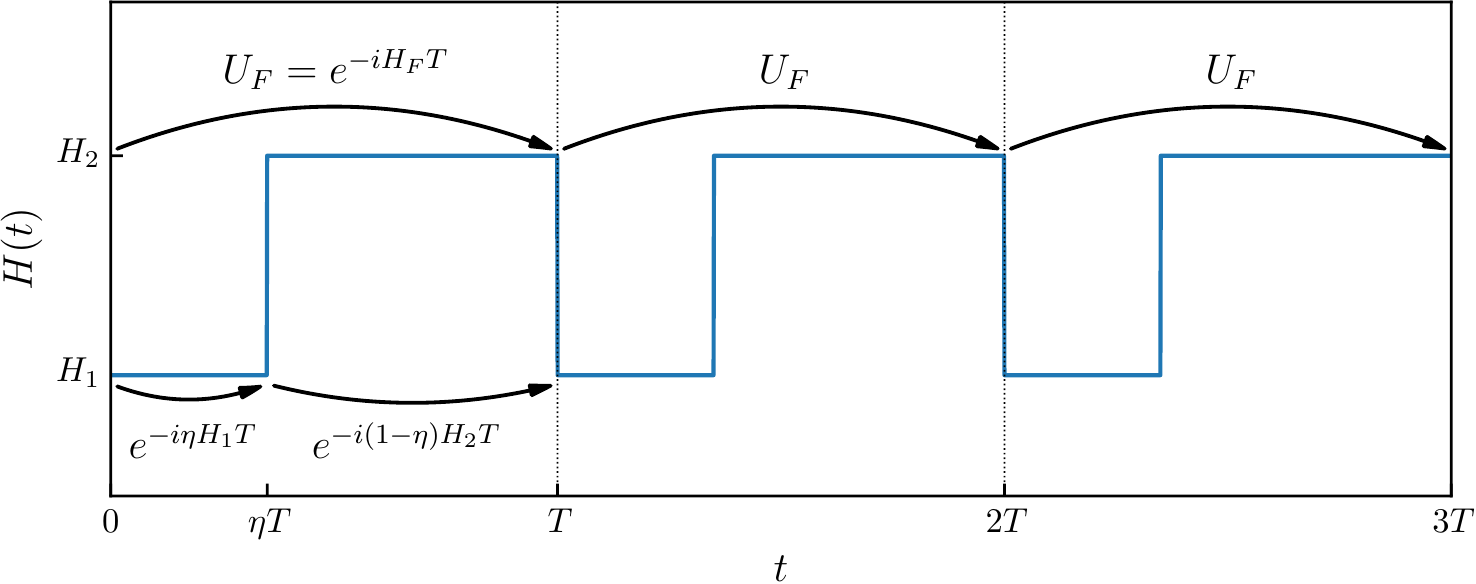}
\caption{Illustration of the double quench driving protocol. \label{fig:floq:driving}\vspace{-\baselineskip}}
\end{center}
\end{figure}
This is illustrated in Figure \ref{fig:floq:driving}. Obtaining the Floquet Hamiltonian from this expression is a non-trivial task, with exact results restricted to systems where there is a clear commutator structure in all involved Hamiltonians \cite{gritsev_integrable_2017} (e.g. non-interacting systems \cite{russomanno_periodic_2012,lazarides_periodic_2014,russomanno_entanglement_2016}) or small systems for which exact diagonalization is feasible \cite{weinberg_quspin:_2017}.

\subsection{The Floquet-Magnus expansion}
In the following, the most physically interesting protocol will be high-frequency driving. Here, the Magnus expansion \cite{klarsfeld_baker-campbell-hausdorff_1989,blanes_magnus_2009,kuwahara_floquetmagnus_2016} provides a series expansion of ${H}_F$ in $T$, allowing the Floquet Hamiltonian to be approximated as ${H}_F = \sum_{n=0}^{\infty}{H}_F^{(n)}$. For periodic quenches, this expansion reduces to the Baker-Campbell-Hausdorff expansion and the first terms are immediately given by
\begin{align}
{H}_F^{(0)} = {H}_{Avg}, \qquad {H}_F^{(1)} = -\frac{iT}{4}[{H}_{Avg},{V}], \qquad {H}_F^{(2)} =-\frac{T^2}{24}[[{H}_{Avg},{V}],{V}],
\end{align}
with
\begin{align}
{H}_{Avg}=\eta {H}_1 + (1-\eta) {H}_2, \qquad {V} = \eta {H}_1-(1-\eta) {H}_2,
\end{align}
in which $H_{Avg}$ corresponds to the driving Hamiltonian $H(t)$ averaged over a single driving cycle. These higher-order terms can now be given a physical interpretation by returning to the Floquet operator. The dynamics of the eigenstates of $H_F$ follows from Eq. (\ref{floq:evolop}) as
\begin{equation}
U(t) \ket{\phi_n} = e^{-i \epsilon_n t} \ket{\phi_n(t)}, \qquad \textrm{with} \qquad \ket{\phi_n(t)} = {P}(t) \ket{\phi_n},
\end{equation}
in which $\ket{\phi_n(t+T)} = \ket{\phi_n(t)}$ presents the periodic part of the time-evolved state. Plugging this in the time-dependent Schr\"odinger equation, the Floquet phases can be written as 
\begin{equation}
\theta_n = \int_{0}^T \braket{\phi_n(t) | {H}(t) | \phi_n(t)} \mathrm{d}t -i\int_{0}^T \braket{\phi_n(t) | \partial_t | \phi_n(t)}  \mathrm{d}t,
\end{equation}
where the first term is the average energy of the state during a single cycle, leading to a dynamical phase contribution, while the second term describes a nonadiabatic (i.e. generalized) Berry phase \cite{grifoni_driven_1998}. For periodic quenches, both contributions to the Floquet phases can be simplified to single expectation values as
\begin{align}\label{floq:contr_phases}
\frac{1}{T}\int_{0}^T \braket{\phi_n(t) | {H}(t) | \phi_n(t)} \mathrm{d}t &= \braket{\phi_n|{H}_{Avg}|\phi_n}, \\
-\frac{i}{T}\int_{0}^T \braket{\phi_n(t) |  \partial_t | \phi_n(t)} \mathrm{d}t &= \braket{\phi_n|{H}_F-{H}_{Avg}|\phi_n},
\end{align}
where it is precisely the higher-order terms ($n \neq 0$) in the Floquet-Magnus expansion that give rise to the Berry phase in the second expression. As similarly shown in Appendix \ref{app:floquetenergies}, these can then be connected to the Floquet phases as 
\begin{equation}
\frac{\theta_n}{T}=\bra{\phi_n} {H}_F \ket{\phi_n}, \qquad \frac{\partial \theta_n}{\partial T} = \bra{\phi_n}{H}_{Avg} \ket{\phi_n}.
\end{equation}

\subsection{Many-body resonances}
\label{sec:floq:mbr}
While this might make it seem like the Floquet Hamiltonian is (up to good approximation) the time-averaged Hamiltonian with additional perturbative terms, this is far from true. As mentioned in the introduction, the Floquet Hamiltonian is distinguished from stationary Hamiltonians due to the presence of \emph{many-body resonances} where quasi-degenerate states interact strongly and hybridize. 

\begin{figure}
\begin{center}
\includegraphics{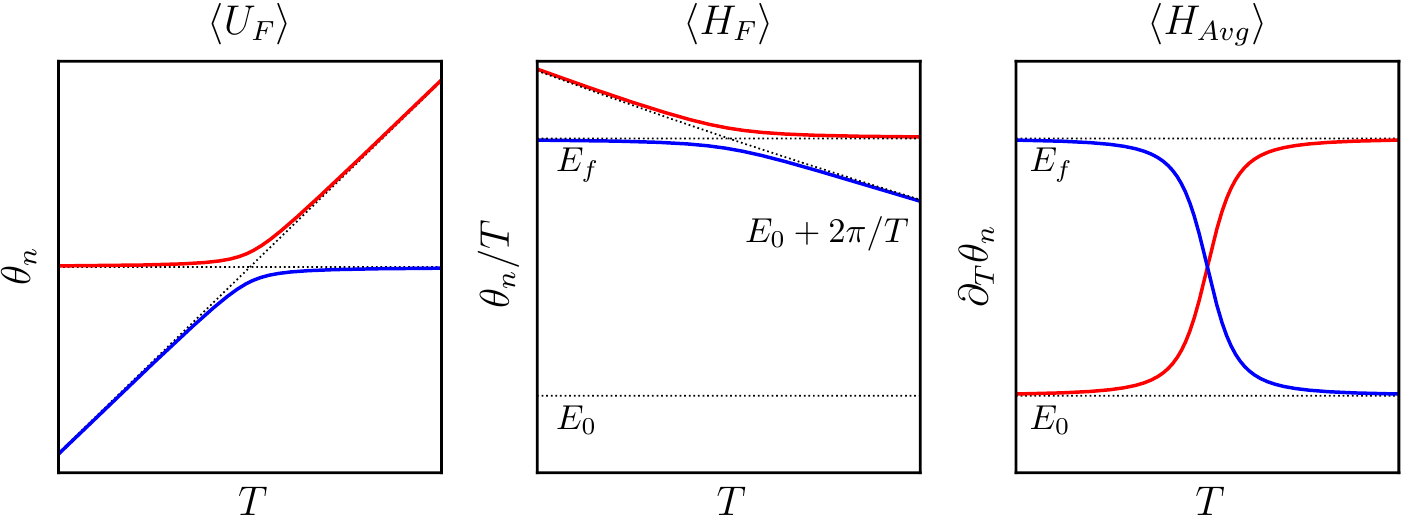}
\caption{Illustration of many-body resonances starting from eigenstates of the time-averaged Hamiltonian with energies $E_0$ and $E_f$. \label{fig:floq:resonances}\vspace{-\baselineskip}}
\end{center}
\end{figure}

These can be easily understood by applying degenerate perturbation theory to the Floquet operator. In the same way that a small perturbation on a stationary Hamiltonian can strongly couple degenerate eigenstates of the unperturbed Hamiltonian, small deviations of $U_F = e^{-i H_F T}$ from $e^{-i H_{Avg}T}$ can strongly couple quasi-degenerate eigenstates of $H_{Avg}$. Consider two eigenstates of the time-averaged Hamiltonian $\ket{\phi_0}$ and $\ket{\phi_f}$ with eigenvalues $E_0$ and $E_f$, and take $\Delta$ to be some small parameter controlling the deviation of $U_F$ from $e^{-i H_{Avg}T}$. In this (many-body) basis the Floquet operator can be expressed as
\begin{equation}
U_F \approx \begin{bmatrix}
e^{-i E_0 T} & 0 \\
0 & e^{-i E_f T}
\end{bmatrix}
 +  e^{-i(E_0+E_f)T/2} 
\begin{bmatrix}
0 & i\Delta^* \\
i\Delta & 0
\end{bmatrix},
\end{equation}
where the additional phase in the perturbation is necessary to guarantee unitarity. The resulting eigenvalues and quasi-energies are given in Figure \ref{fig:floq:resonances} at different values of the driving period.

Away from resonance, the eigenphases are approximately $e^{-iE_0 T}$ and $e^{-i E_f T}$ and the eigenstates approximately $\ket{\phi_0}$ and $\ket{\phi_f}$. However, at resonance the unperturbed eigenstates are quasi-degenerate since $E_f-E_0 =  2\pi/T$. This then leads to a degeneracy in the unperturbed evolution operator since $e^{-iE_0T} = e^{-iE_fT}$, and any non-zero $\Delta$ results in an avoided crossing in the spectrum of both $U_F$ and $H_F$. Note that one of the quasi-energies has been shifted over $2\pi/T$ in order to make the avoided crossing clear. At resonance the eigenstates follow as $\frac{1}{\sqrt{2}} \left(\ket{\phi_0} + e^{i \theta} \ket{\phi_f} \right)$ for $\Delta= |\Delta|e^{i \theta}$, and it is precisely these states that are known as many-body resonances. While these immediately follow from a small perturbation on the evolution operator, this does not translate to a small perturbation on the Floquet Hamiltonian. The interactions necessary to obtain such many-body resonances as eigenstates of a Hamiltonian are localized in energy space and generally do not follow from any truncated Magnus expansion \cite{weinberg_adiabatic_nodate}. However, the Magnus expansion still provides a Hamiltonian which can accurately model the off-resonant eigenstates, whereas the many-body resonances at quasi-degeneracies need to be included at a later step.

\subsection{Different regimes}

Despite the involved technical difficulties, the dependence of the Floquet Hamiltonian on the driving frequency has become well-understood in recent years \cite{dalessio_long-time_2014,bukov_heating_2016,weinberg_adiabatic_nodate}. At high enough driving frequencies no quasi-degeneracies are present and the Floquet Hamiltonian can be accurately approximated by an effective Hamiltonian, leading to strongly suppressed heating \cite{mori_rigorous_2016,kuwahara_floquetmagnus_2016,abanin_effective_2017,abanin_rigorous_2017}. This effective Hamiltonian can be obtained from the Magnus expansion, where the time-averaged Hamiltonian $H_{Avg} = \eta H_1 + (1- \eta) H_2$  presents a first-order approximation \cite{eckardt_high-frequency_2015,mikami_brillouin-wigner_2016,klarsfeld_baker-campbell-hausdorff_1989,blanes_magnus_2009}. Lowering the driving frequency $2 \pi/T$, many-body resonances are introduced where quasi-degenerate eigenstates of this effective Hamiltonian interact strongly and hybridize \cite{dalessio_long-time_2014,bukov_heating_2016,weinberg_adiabatic_nodate}.  When many-body resonances remain rare, they where shown to dominate the time evolution and lead to slow nonthermalizing time evolution \cite{bukov_heating_2016}. Further lowering the driving frequency, these many-body resonances multiply and lead to so-called `infinite-temperature states'. At low enough driving frequencies the majority of states are approximately quasi-degenerate, and the  Floquet eigenstates reduce to superpositions of a macroscopically large amount of quasi-degenerate eigenstates, behaving essentially as a random (`infinite temperature') state. This then immediately leads to thermalization to an infinite temperature state when considering time evolution.

However, at all points it should be kept in mind that this only holds in finite systems. The unbounded spectrum in infinitely large systems immediately leads to a proliferation of many-body resonances and infinite-temperature Floquet eigenstates at all possible driving frequencies. However, in large but finite driven systems, it remains possible to isolate and target specific resonances.

\section{Driving the central spin model}

This will now be applied to the central spin Hamiltonian as given by (see Section \ref{RG:subsec:cs}) 
\begin{equation}
{H} = B_z S_0^z + \sum_{j=1}^{L} A_j \vec{S}_0 \cdot \vec{S}_j,
\end{equation}
where $S_0^{\alpha}$ and $S_j^{\alpha}$ are the spin operators of the central spin and the bath spins respectively\footnote{The total system size now corresponds to $L+1$.}. In the following, these spins are taken to be spin-$1/2$ particles, and the coupling constants are taken to be $A_j = \exp\left[-(j-1)/L\right]$, corresponding to a quantum dot in a 2D Gaussian envelope \cite{coish_hyperfine_2004}. However, the integrability of the central spin model is versatile enough that our proposed method holds for arbitrary spins and parametrizations. Defining $\epsilon_j = -A_j^{-1}$ and $\epsilon_0=0$, the exact eigenstates are given by Bethe ansatz states (where the dependence on the magnetic field $B_z$ is made explicit)
\begin{equation}
\ket{B_z;v_1 \dots v_N} = \prod_{a=1}^N \left(\sum_{j=0}^{L}\frac{S_j^+}{\epsilon_j-v_{a}}\right)\ket{\downarrow \dots \downarrow},
\end{equation}
with rapidities $\{v_1 \dots v_N\}$ satisfying the Bethe equations (\ref{RG:cs:BAE}).

This has two major advantages when considering periodic driving. First, these equations can be efficiently solved in a computational time scaling polynomially with system size. This should be contrasted with the conventional construction and diagonalization of the Hamiltonian matrix in an exponentially large Hilbert space and allows for exact results for large system sizes. Second, this also allows for the targeting of eigenstates in a systematic way, since eigenstates can be obtained by targeting specific solutions to the Bethe equations without immediately calculating all possible eigenstates. Furthermore, the key to our proposed approach is that overlaps between eigenstates of central spin Hamiltonians with different magnetic fields $\braket{B_{z,1};v_1 \dots v_N|B_{z,2};w_1 \dots w_N}$ can be efficiently calculated numerically\footnote{The crucial realization is that \emph{on-shell} states at a given value of $B_z$ can be interpreted as \emph{off-shell} states at a different value of $B_z$, since the dependence on the magnetic field is only implicit in the Bethe states due to the dependence of the rapidities on the Bethe equations.} (see Chapter \ref{chap:innerproducts}).

\begin{figure}
\begin{center}
\includegraphics{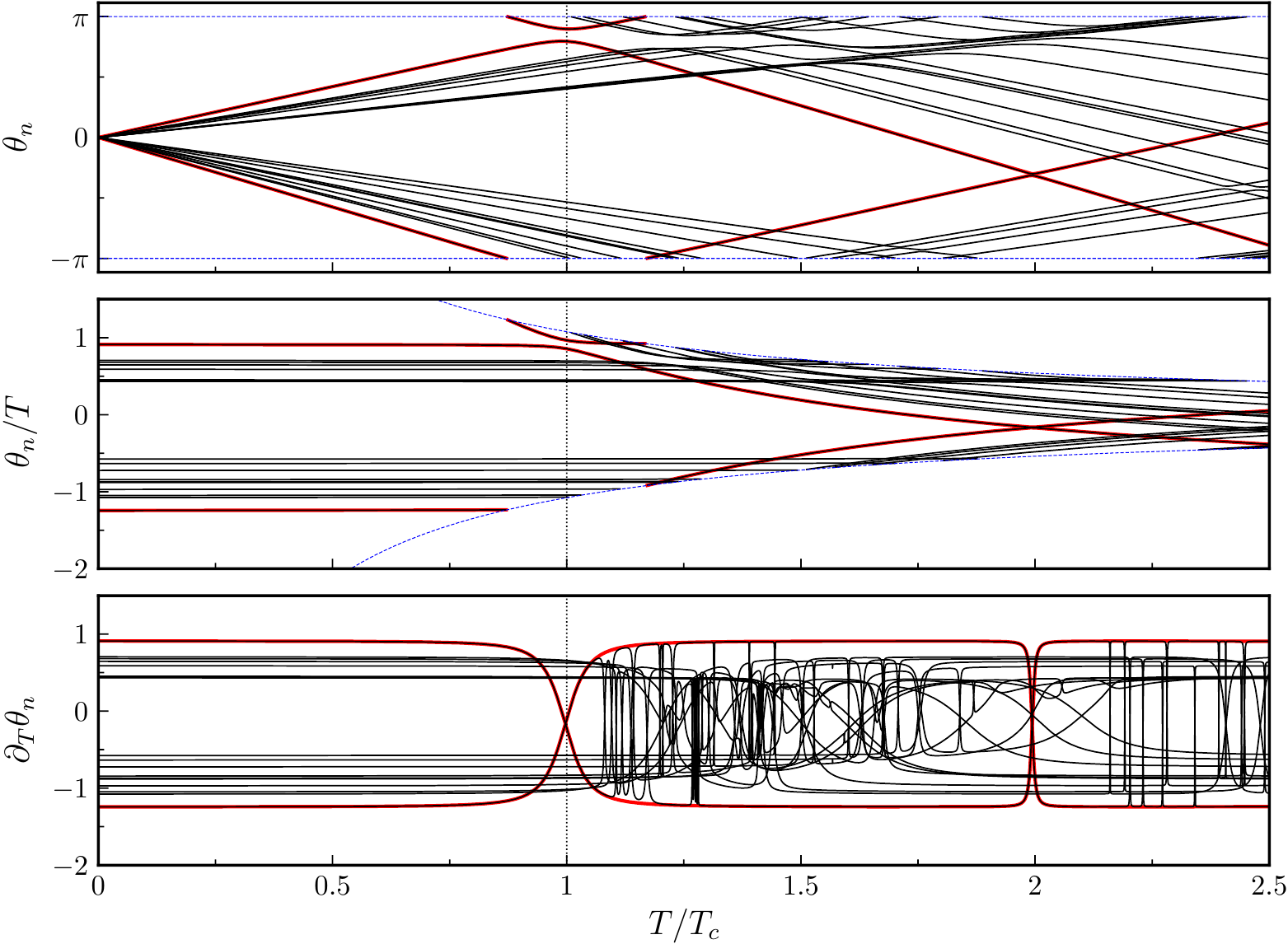}
\caption{Phase spectrum of the Floquet operator, quasi-energies and dynamical energies for a periodically driven central spin Hamiltonian at different driving periods $T$. The dotted (blue) lines mark the edges of the Brillouin zone $\pm \pi$ $(\theta_n)$ and $\pm \pi/T$ $(\theta_n/T)$ while the vertical dotted line denotes $T_c=2\pi/W$. The ground state and highest excited state are highlighted red using the approximative results from integrability (see Section \ref{sec:floq:model}). \vspace{-\baselineskip}\label{spec_centralspin}}
\end{center}
\end{figure}

Returning to Floquet dynamics, a protocol is considered where $B_z$ is periodically switched between two values $B_{z,1}$ and $B_{z,2}$. To fix ideas, the eigenphases of the Floquet operator have been given in Figure \ref{spec_centralspin} for different driving periods $T$, with total spin projection 0, $\eta=0.5$ and $B_z$  switched between $1.2$ and $0.8$. These calculations have been performed using exact diagonalization on a small system with $L=5$ in order to provide a clear graphical representation, but are representative for larger system sizes. Note that no techniques from integrability were used so far, since the dimension of the Hilbert space is ${6 \choose 3} = 20$ and the Floquet operator can be straightforwardly constructed and diagonalized. Next to the spectrum of the Floquet operator, the two different energy measures of a Floquet state $\ket{\phi_n}$ have been given as $\theta_n/T$ and $\partial_T \theta_n$. This second quantity is convenient for the visualization of avoided crossings in the spectrum of the Floquet Hamiltonian, highlighting the absorption/emission of energy.

At small driving periods (high frequencies), the spectrum of $H_F$ reduces to that of ${H}_{Avg}$ and both energies coincide. The onset of many-body resonances can be observed at $T_c = 2\pi/W$, with $W=E^{Avg}_{max}-E^{Avg}_{min}$ the bandwidth of ${H}_{Avg}$. At this critical frequency, the energy difference between the ground state and the highest excited state exactly matches the driving frequency. These states are then quasi-degenerate and will interact resonantly, which can be clearly observed in the avoided crossing between their respective quasi-energies in $\braket{H_{F}}=\theta_n/T$ and the crossing between their respective energies in $\braket{H_{Avg}}=\partial_T \theta_n$. Note that this quasi-degeneracy again necessitates the shift of one of the two states into the first Brillouin zone. Further increasing the driving period, more and more resonances are introduced. Remarkably, in the crossover regime the off-resonant parts of the spectrum can still be accurately approximated using the time-averaged Hamiltonian, a feature which was already noted in Refs. \cite{bukov_heating_2016,russomanno_floquet_2017}.

\subsection{Resonant transitions}
Following the discussion on many-body resonances, these resonances have a major influence on the concept of adiabaticity, with distinct effects on the eigenstates of the Floquet Hamiltonian and the time-averaged Hamiltonian \cite{breuer_adiabatic_1989,young_adiabatic_1970,eckardt_superfluid-insulator_2005,russomanno_kibble-zurek_2016,russomanno_floquet_2017,russomanno_spin_2017}. Starting from an eigenstate of the Floquet Hamiltonian and adiabatically changing the driving frequency\footnote{Provided the micromotion operator does not change, see Ref. \cite{weinberg_adiabatic_nodate}.}, the initial state will adiabatically follow the eigenstate of the Floquet Hamiltonian at stroboscopic times. Following Section \ref{sec:floq:mbr}, this would lead to a superposition of quasi-degenerate states at the point of resonance, while a transition across resonance would lead to a transition from e.g. the ground state to a highly excited state of the time-averaged Hamiltonian, since these are adiabatically connected through eigenstates of the Floquet Hamiltonian.

\begin{figure}
\includegraphics{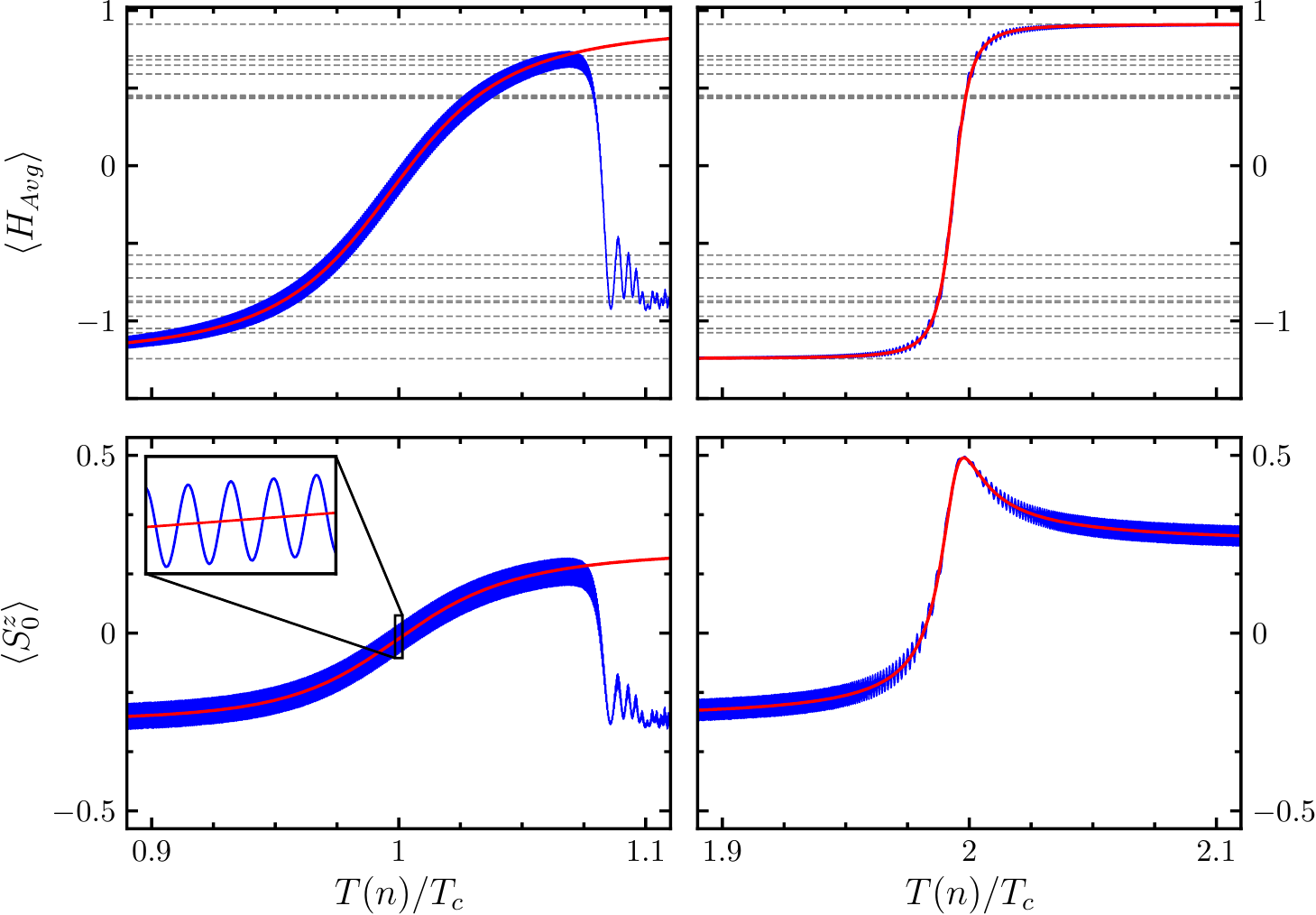}
\caption{Expectation values of ${H}_{Avg}$ and $S^z_0$ w.r.t. the adiabatic ground state of the Floquet Hamiltonian when the driving period is slowly increased from $0.8 \ T_c$ to $1.2 \ T_c$ (first column) or from $1.8 \ T_c$ to $2.2 \ T_c$ (second column) with $T_{i+1}-T_i = 10^{-4}$. Blue lines are exact results, while the red line is the approximation from integrability (see further). The dashed lines indicate the spectrum of $H_{Avg}$. \label{transition_L=6}}
\end{figure}

Focusing on the ground and highest excited state and adiabatically increasing the driving period across resonance, starting from the ground state $\ket{\phi_0(\overline{B}_z)}$ of $H_{Avg}$ leads to 
\begin{equation}
U(T_n) \dots U(T_2) U(T_1) \ket{\phi_0(\overline{B}_z)},
\end{equation}
with $T$ slowly increased from $T_1$ to $T_n$. We will refer to this state as the `adiabatic ground state', which is expected to adiabatically follow the corresponding eigenstate of the Floquet Hamiltonian, leading to a transition between the resonant states. For the small system with $L=5$, such transitions are shown in Figure \ref{transition_L=6} for the first $(T \approx T_c)$ and second-order ($T \approx 2 T_c$) resonance. 
When slowly increasing the driving period, the system ends up in the highest excited state of ${H}_{Avg}$ in the second resonance, while it undergoes another resonance in the first-order transition before the highest excited state can be reached. Still, it is clear that the ground state adiabatically follows the eigenstates of the Floquet Hamiltonian if the driving period is varied adiabatically. As evidenced by the oscillations in the expectation values (see inset in Figure \ref{transition_L=6}), the ground state of $H_{Avg}$ only provides a (good) approximation and is not an exact eigenstate of $H_F$. The polarization generally increases from a negative to a positive value, as could be expected. However, this need not happen monotonously. Note how $\langle S_0^z \rangle$, as shown in the lower panel of Figure \ref{transition_L=6}, nears its maximal value of $1/2$ in the second resonance, which will be extensively investigated in the following.


\subsection{Modeling the resonant transition}
\label{sec:floq:model}

All adiabatic transitions and many-body resonances are encoded in the spectrum of the Floquet Hamiltonian, which fully governs the periodic dynamics. However, due to the exponential scaling of the Hilbert space and the inherently non-diagonal nature of time evolution operators it is generally impossible to obtain this Hamiltonian in interacting systems of a realistic size. Modeling this transition would require constructing the evolution operators for both driving Hamiltonians at each value of the driving period, and constructing and subsequently diagonalizing the resulting Floquet operator. Each of these steps involves the full Hilbert space, making such calculations unfeasible for realistic system sizes. However, knowledge acquired from quantum quenches can be transferred to the present situation under one key assumption, allowing for the study of resonances in the relevant part of the Floquet spectrum. 

Namely, it is assumed that each many-body resonance can be accurately modeled by treating it as a two-level system including only the corresponding quasi-degenerate eigenstates of $H_{Avg}$. This assumes that quasi-degenerate states do not interact strongly with off-resonant states or other quasi-degenerate states with a different quasi-energy, similar in spirit to degenerate perturbation theory. This approximation can be validated through the Floquet-Magnus expansion and is expected to hold if the deviations of the driving Hamiltonians are small w.r.t. the time-averaged Hamiltonian \cite{mikami_brillouin-wigner_2016,klarsfeld_baker-campbell-hausdorff_1989,blanes_magnus_2009,kuwahara_floquetmagnus_2016}. The Floquet operator can then be constructed in the $2$-dimensional basis $\{\ket{\phi_0(\overline{B}_z)},\ket{\phi_f(\overline{B}_z)}\}$ spanned by the relevant quasi-degenerate eigenstates of the time-averaged Hamiltonian 
\begin{equation}
{U}_F = 
\begin{bmatrix}
\braket{\phi_0(\overline{B}_z)|{U}_F|\phi_0(\overline{B}_z)} & \braket{\phi_0(\overline{B}_z)|{U}_F|\phi_f(\overline{B}_z)} \\
\braket{\phi_f(\overline{B}_z)|{U}_F|\phi_0(\overline{B}_z)} & \braket{\phi_f(\overline{B}_z)|{U}_F|\phi_f(\overline{B}_z)} 
\end{bmatrix}.
\end{equation}
Explicitly writing out the Floquet operator and expanding in the eigenstates of the two driving Hamiltonians, each of the $4$ matrix elements of the Floquet operator can be written as
\begin{align}\label{UF_mat}
&\braket{\phi_i(\overline{B}_z)| {U}_F|\phi_j(\overline{B}_z)} = \sum_{m,n} e^{-i(1-\eta) E_m(B_{z,2}) T }e^{-i \eta E_n(B_{z,1}) T } \nonumber \\
&\qquad \qquad \qquad \qquad \times \braket{\phi_i(\overline{B}_z)|\phi_m(B_{z,2})}\braket{\phi_m(B_{z,2})|\phi_n(B_{z,1})}\braket{\phi_n(B_{z,1})|\phi_j(\overline{B}_z)}.
\end{align}
The calculation of each matrix element involves a double summation over the Hilbert space of overlaps and energy matrix elements, which in turn involve sums over the Hilbert space. This is impossible to calculate in general, but integrability already provides us with numerically efficient expressions for the energies and the overlaps. However, Eq. (\ref{UF_mat}) still involves a double summation over the full Hilbert space. As noticed in the context of quantum quenches, an important feature of Bethe states is that they offer an optimized basis in which only a very small minority of eigenstates carry substantial correlation weight, allowing such summations to be drastically truncated \cite{faribault_bethe_2009,faribault_quantum_2009,caux_correlation_2009}.

\begin{figure}
\begin{center}
\includegraphics[width=0.75\textwidth]{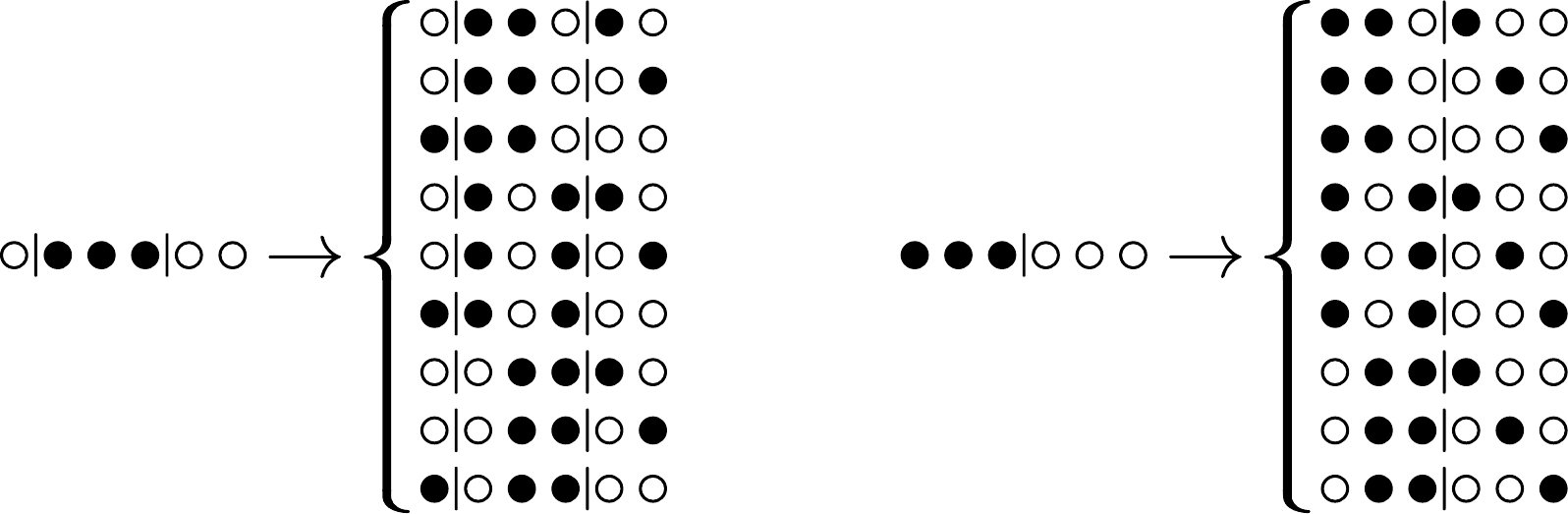}
\end{center}
\caption{Graphical representation of relevant states for $L=5$ and $N=3$ starting from the ground state and the highest excited state. \label{excitations}}
\end{figure}

The targeting of relevant eigenstates for the summation can be represented in a graphical way by making the connection with the $B_z \to \infty$ limit. In this limit, the rapidities behave as $v_a = \epsilon_{i(a)} + \mathcal{O}(B_z^{-1})$, where $i(a)$ associates a spin index with each rapidity such that the eigenstates reduce to
\begin{equation}
\ket{B_z \to \infty;v_1 \dots v_N} \propto \prod_{a=1}^N S^+_{i(a)}\ket{\downarrow \dots \downarrow}.
\end{equation}
In this limit, all states reduce to simple product states defined in terms of a set occupied spin levels $\{i(1) \dots i(N)\}$. For the given parametrization, the ground state reduces to the state where the central spin is unoccupied and the states interacting most strongly with the central spin are occupied. The highest-excited state will correspond to the state where the central spin is occupied and the states interacting most strongly with the central spin are similarly occupied. This can be represented graphically (for e.g. $L=5$ and $N=3$) as
\begin{align}
\ket{\phi_0(B_z \to \infty)} &\propto S_1^+ S_2^+ S_3^+  \ket{\downarrow \dots \downarrow}  = \ket{\circ | \bullet\bullet\bullet | \circ\circ},  \\
\ket{\phi_f(B_z \to \infty)} &\propto S_0^+ S_1^+ S_2^+  \ket{\downarrow \dots \downarrow}\nonumber = \ket{\bullet\bullet\bullet | \circ\circ\circ}.
\end{align}
When expanding an eigenstate at fixed $B_z$ into eigenstates of a model at different $B_z$, it is often sufficient to restrict the expansion to states which can be related to the initial state by simple spin-flip excitations in the limit $B_z \to \infty$. This results in a set of $N(L+1-N)$ states for which the overlap needs to be calculated, which are again represented graphically in Figure \ref{excitations}. When expanding an initial state $\ket{\phi_i(\bar{B}_z)}$ in such a restricted basis, the error induced by this truncation can easily be checked from sum rules. If this initial truncation would prove to be insufficient and the error exceeds a certain threshold (when e.g. there are large differences between the time-averaged Hamiltonian and the driving Hamiltonians), this summation can be extended in a systematic way by including higher-order spin-flip excitations. In practice, this truncation scheme allows for a numerically exact construction of the matrix elements of the Floquet operator. 
\begin{figure}[t]
\begin{center}
\includegraphics{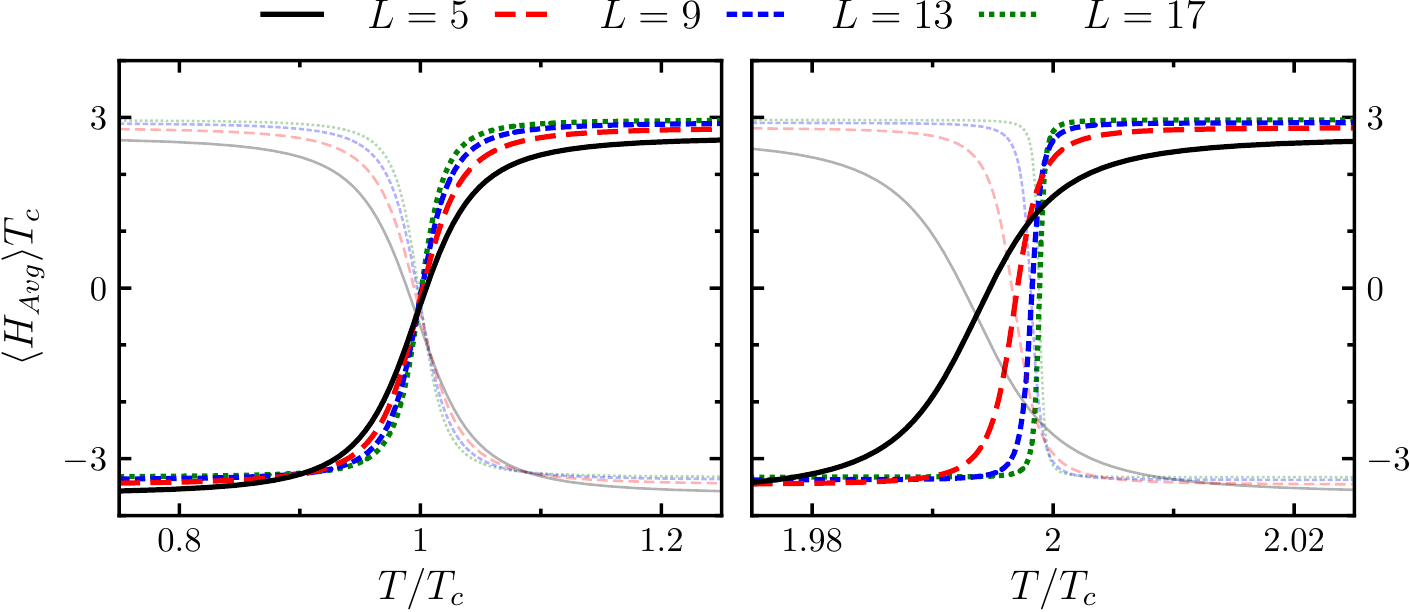}
\caption{Expectation value of the time-averaged Hamiltonian in the adiabatic ground (and highest excited) state of the Floquet Hamiltonian with driving  $B_z=1 \pm 0.2$ and $\eta=1/2$ for different system sizes $L$. \label{compareL}}
\end{center}
\end{figure}

The resulting $2 \times 2$ operator can then be easily diagonalized, and integrability allows for an efficient calculation of expectation values from the resulting wave function. The main approximation introduced in this scheme is the restriction of the Hilbert space to a $2$-dimensional space, but within this space the Floquet operator is numerically exact. The error introduced in this way can be quantified by calculating $\lVert {U}_F^{\dagger}{U}_F - \mathbbm{1} \rVert $, which should reduce to zero if the approximation is exact. In this case, the time evolution within a single period only couples the states within the restricted basis. 

\begin{figure}[t]
\begin{center}
\includegraphics{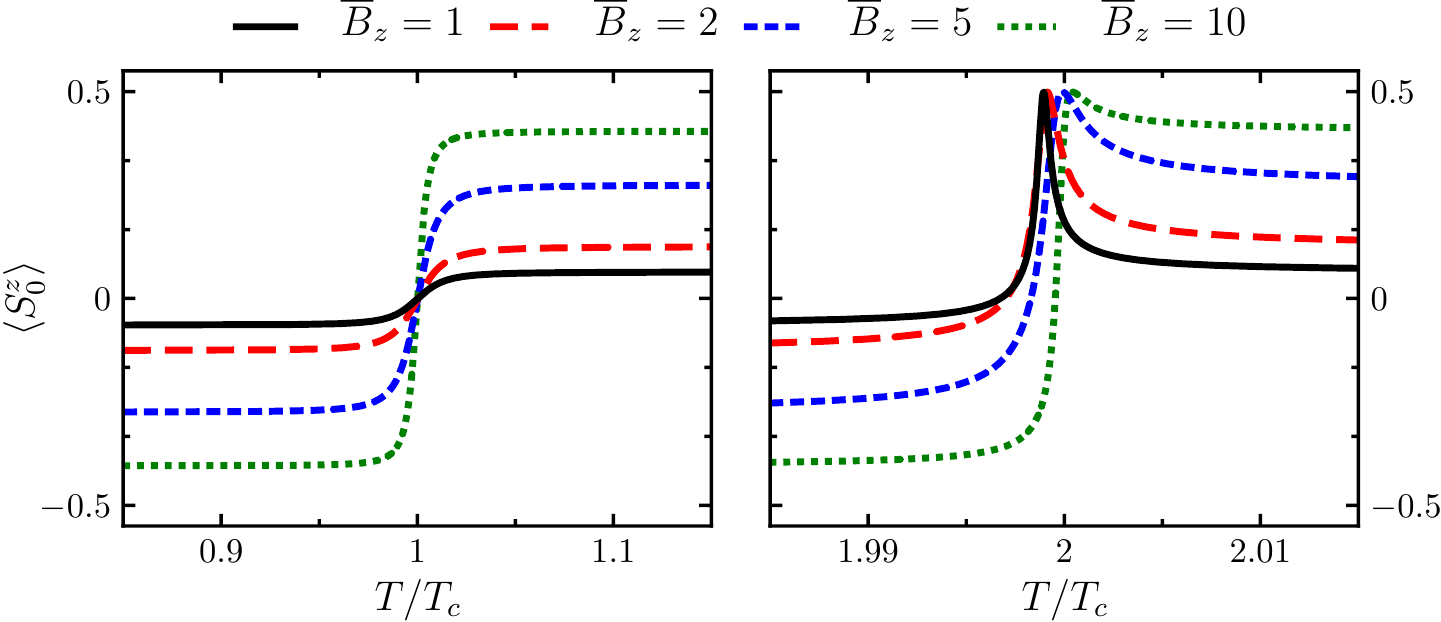}
\caption{Magnetization of the central spin in the adiabatic ground state of the Floquet Hamiltonian at different driving periods with $L=25$ and driving $B_z=\overline{B}_z \pm 0.2$. \label{compareB}}
\end{center}
\end{figure}

It is possible to include an increasing number of states in this basis in order to systematically reconstruct part of the Floquet spectrum, but we choose to focus on the interaction between the ground state and the highest excited state only. The accuracy of this approximation can already be appreciated in Figures \ref{spec_centralspin} and \ref{transition_L=6}, where the avoided crossings near the resonances are well approximated but fail to take into account the interactions with other states. The results are extended in Figures \ref{compareL} to different system sizes and in Figure \ref{compareB} to different values of the average magnetic field, where the same qualitative behaviour can be observed. 

\subsection{Perturbation expansion}

While the expectation value of $H_{Avg}$ varies smoothly from the initial to the final value, the behaviour of the central spin is highly dependent on the order of the resonance. The magnetization $\braket{S_0^z}$ vanishes at the first resonance, while it nears the maximal value $1/2$ at the second resonance. Clearly, such a protocol could then be used to realize a state with magnetization exceeding that of both states, uncompatible with any stationary central spin Hamiltonian, since a maximal value of $1/2$ implies a pure state decoupled from its environment.

A first step toward understanding this behavior can be found in the structure of the relevant eigenstates. In both the ground state and the highest excited state the bath spins tend to align, either anti-parallel or parallel to the central spin, and can be approximated by treating the bath spins as a single collective spin. In this space the Hamiltonian can be simplified to
\begin{equation}
{H} \approx \overline{B}_z S_0^z + A_b\vec{S}_b \cdot \vec{S}_0.
\end{equation}
Although it is not a necessary approximation (see Appendix \ref{app:PT}), some intuition can be gained by assuming $|\overline{B}_z| \ll |A_b S_b|$, when the relevant eigenstates can be approximated as angular momentum-recoupled states
\begin{equation}
\ket{\phi_{\pm}}\approx\frac{1}{\sqrt{2}} \left(\ket{\tfrac{1}{2},\tfrac{1}{2}}_0\ket{S_b,-\tfrac{1}{2}}_b \pm \ket{\tfrac{1}{2},-\tfrac{1}{2}}_0\ket{S_b,\tfrac{1}{2}}_b\right).
\end{equation}
At resonance, the Floquet states are approximately given by
\begin{equation}
\ket{\phi} = \frac{1}{\sqrt{2}} \left(\ket{\phi_{+}} \pm e^{i \theta} \ket{\phi_-}\right),
\end{equation}
where the relative phase $\theta$ is a priori unknown. However, the magnetization of the central spin depends on this relative phase as 
\begin{equation}
\braket{\phi | S_0^z | \phi} = \frac{1}{2}\cos(\theta).
\end{equation}
The different magnetizations hence correspond to different relative phases acquired by these states. This relative phase can be deduced from second-order perturbation theory, expanding the matrix elements of the Floquet operator (\ref{UF_mat}) at resonance for small deviations from the average magnetic field $(B_{z}-\overline{B}_z)$ (see Appendix \ref{app:PT}). Evolving either state over a full driving cycle will lead to a global phase and introduce off-diagonal corrections on the initial state, which are shown to either interfere constructively or destructively depending on the order of the resonance. This is reflected in the dependence of the off-diagonal elements on the order of the resonance $k$ through terms $e^{\pm i \eta k 2 \pi}$, and perturbation theory leads to relative phases $\pi/2$ and $3 \pi/2$ in the first resonance, while it leads to relative phases $0$ and $\pi$ in the second resonance. These explain the observed magnetization $\braket{S_0^z}=0$ or $\pm 1/2$, leading to the decoupling and polarization of the central spin in the second-order resonance. This behaviour extends towards higher-order resonances, where the polarization occurs at even-order resonances but vanishes at odd-order resonances. However, there is no guarantee that such resonances will be isolated and hence observable.

\section{Conclusions}

In this chapter, we investigated how many-body resonances and adiabatic transitions in the Floquet Hamiltonian when varying the driving frequency can lead to a transition between the ground state and the highest excited state in the time-averaged Hamiltonian. Integrability-based techniques were shown to be able to model this transition, which allows for an investigation of larger system sizes and presents a first step in applying techniques from integrability to interacting integrable systems subjected to periodic driving. This scheme was then applied to a central spin model on which a periodically varying magnetic field is applied. Physically, transitions across resonance result in a transfer of polarization from the environment to the central spin. At resonance this can be used to construct coherent superpositions of eigenstates of the time-averaged Hamiltonian, where the relative phase was shown to depend highly on the order of the resonance, leading either to a vanishing magnetization or to a spin state exactly aligned with the magnetic field. Remarkably, the latter effectively corresponds to a decoupling of the central spin from its environment. A natural extension would then be to combine this approach with the integrability-based method in Ref. \cite{van_den_berg_competing_2014}, approximating time evolution using a Suzuki-Trotter decomposition evaluated in a basis of Bethe states, allowing for the description of more general driving protocols.

\begin{subappendices}
\makeatletter
\@openrightfalse
\makeatother
\chapter*{Appendices}

\section{Energies in a Floquet system}
\label{app:floquetenergies}
In this Appendix, the different definitions of energy are highlighted in a Floquet system for a two-step driving protocol 
\begin{equation}\label{app:timedepHam}
{H}(t) = 
  \begin{cases}
 {H}_1 &\text{for}\qquad 0 < t < \eta T, \\
  {H}_2    & \text{for}\qquad \eta T < t < T,
  \end{cases}
\end{equation}
with ${H}(t+T) = {H}(t)$. The Floquet operator and Floquet Hamiltonian are subsequently defined as
\begin{equation}\label{app:FloquetHam}
{U}_{F} \equiv  e^{-i {H}_F T} \equiv e^{-i(1-\eta){H}_2 T} e^{-i \eta {H}_1 T},
\end{equation}
and can be simultaneously diagonalized as
\begin{align}
{H}_F = \sum_n \epsilon_n \ket{\phi_n}\bra{\phi_n}, \ \
{U}_F = \sum_n e^{-i \theta_n} \ket{\phi_n}\bra{\phi_n},
\end{align}
where $\epsilon_n = \theta_n/T$. In Ref. \cite{grifoni_driven_1998}, it was shown how
\begin{equation}\label{app:intHam}
\frac{\partial \theta_n}{\partial T} = \frac{1}{T} \int_{0}^T \braket{\phi_n(t) | {H}(t) | \phi_n(t)} \mathrm{d}t ,
\end{equation}
which is the average energy of a Floquet state during one driving cycle and was shown there to act as a dynamical contribution to the Floquet phase $\theta_n$. For the Floquet operator (\ref{app:FloquetHam}), this can be further simplified, combining
\begin{equation}\label{app:idtUF}
i \partial_T {U}_F = (1-\eta){H}_2 {U}_F + \eta {U}_F  {H}_1,
\end{equation}
with the Hellmann-Feynman theorem,
\begin{equation}
 \frac{\partial \theta_n}{\partial T} = i \frac{\partial_T \braket{\phi_n | {U}_F |\phi_n}}{\braket{\phi_n | {U}_F |\phi_n}} = \frac{ \bra{\phi_n} i \partial_T {U}_F\ket{\phi_n}}{\bra{\phi_n}{U}_F\ket{\phi_n}}.
\end{equation}
Making use of Eq. (\ref{app:idtUF}) and ${U}_F \ket{\phi_n} = e^{-i \theta_n}\ket{\phi_n}$, this simplifies to
\begin{equation}
\frac{\partial \theta_n}{\partial T} = \bra{\phi_n}(1-\eta){H}_2+ \eta {H}_1 \ket{\phi_n} = \bra{\phi_n} {H}_{Avg} \ket{\phi_n}.
\end{equation}
Alternatively, this also follows from Eq. (\ref{app:intHam}) by considering the explicit time evolution of the Floquet states $\ket{\phi_n} \equiv \ket{\phi_n(t=0)}$, as governed by 
\begin{equation}
\ket{\phi_n(t)} = {P}(t) \ket{\phi_n} = {U}(t) e^{i {H}_F t} \ket{\phi_n},
\end{equation}
where the evolution operator follows from Eq. (\ref{app:timedepHam}) as
\begin{equation}
{U}(t) = 
  \begin{cases}
 e^{-i {H}_1 t} &\text{for}\ \ 0 < t < \eta T, \\
e^{-i {H}_2 (t-\eta T) } e^{-i \eta {H}_1 T}     & \text{for}\ \  \eta T < t < T.
  \end{cases}
\end{equation}
The two-step driving again allows for a simplification as
\begin{equation}
{U}(t) = 
  \begin{cases}
 e^{-i {H}_1 t} &\text{for}\ \ 0 < t < \eta T, \\
e^{i {H}_2 (T-t) } e^{-i {H}_F T}     & \text{for}\ \  \eta T < t < T.
  \end{cases}
\end{equation}
The kick operator is subsequently given by
\begin{equation}
{P}(t) = 
  \begin{cases}
 e^{-i {H}_1 t} e^{i {H}_F t}&\text{for}\ \ 0 < t < \eta T, \\
e^{i {H}_2 (T-t) } e^{-i {H}_F (T-t)}    & \text{for}\ \  \eta T < t < T,
  \end{cases}
\end{equation}
and the time-evolved eigenstates of the Floquet operator by
\begin{equation}
\ket{\phi_n(t)} = 
  \begin{cases}
 e^{i \epsilon_n t } e^{-i{H}_1t} \ket{\phi_n} &\text{for}\ \ 0 < t < \eta T, \\
e^{-i \epsilon_n (T-t)} e^{i{H}_2(T-t)} \ket{\phi_n}     & \text{for}\ \  \eta T < t < T.
  \end{cases}
\end{equation}
This has a clear interpretation because of the simplicity of the driving protocol. In order to obtain the state in the first part of the period ($0 < t < \eta T$), it is possible to evolve the state forward in time from $t=0$ using only ${H}_1$. For the second half of the period ($\eta T < t < T$), it is possible to evolve the state back in time starting from $t=T$ using only ${H}_2$. This then results in
\begin{equation}
\braket{\phi_n(t)|{H}(t)|\phi_n(t)} = \braket{\phi_n|{H}(t)|\phi_n},
\end{equation}
where inserting this equality in Eq. (\ref{app:intHam}) again returns the time-averaged Hamiltonian.

Given the Floquet phases $\theta_n$, these thus allow for two different measures of the energy of a Floquet state,
\begin{equation}
\frac{\theta_n}{T}=\bra{\phi_n} {H}_F \ket{\phi_n}, \qquad \frac{\partial \theta_n}{\partial T} = \bra{\phi_n}{H}_{Avg} \ket{\phi_n}.
\end{equation}
The derivatives of the phases w.r.t. the period have an interpretation as the average energy of a Floquet state during one driving cycle. These follow from the expectation values of the time-averaged Hamiltonian, and are as such uniquely defined and bounded by the extremal eigenvalues of ${H}_{Avg}$. These can be contrasted to the quasienergies, which are the ratio of the phases and the period, only defined modulo $\frac{2 \pi}{T}$, and commonly taken to be restricted to a single Brillouin (Floquet) zone $\left[-\frac{\pi}{T}, \frac{\pi}{T}\right]$.

\section{Perturbation expansion of the Floquet operator}
\label{app:PT}

In order to better understand the behaviour of the Floquet eigenstates near resonance, we perform a perturbative expansion of the Floquet operator when the model is being driven in such a way that there are only small deviations from the average magnetic field. Then within each matrix element all non-diagonal overlaps in the summation will be of order $B_{z,i} - \overline{B}_z \equiv \mathcal{O}(\Delta)$, allowing the summation to be severely restricted. For corrections up to $\mathcal{O}(\Delta^2)$, only the initial and final state are relevant as intermediate states in the summation (\ref{UF_mat}). The diagonal elements can easily be found as
\begin{align}
\braket{\phi_0(\overline{B}_z)|{U}_F|\phi_0(\overline{B}_z)} &= e^{-i k E_0(\overline{B}_z)T_c} + \mathcal{O}(\Delta^2), \\
\braket{\phi_f(\overline{B}_z)|{U}_F|\phi_f(\overline{B}_z)} &= e^{-i k E_f(\overline{B}_z)T_c} + \mathcal{O}(\Delta^2),
\end{align}
which holds for arbitrary values of the driving period. In the first element, the summation has been restricted to $(m,n)=(0,0)$, while it has been restricted to $(m,n)=(f,f)$ in the second element. Here $0$ and $f$ label the resonant states, which we will take to be the ground (initial) state and highest excited (final) state. All other terms involve at least two off-diagonal overlaps and are as such of $\mathcal{O}(\Delta^2)$. Due to the expansion around $\overline{B}_z$, the first-order corrections on the phases can also be shown to vanish. Note that we discard a summation of $\mathcal{O}(\Delta^2)$ over an exponentially large Hilbert space, but we can assume that the phases do not add coherently and as such this summation can still be neglected. If this would not be the case then the resonant states have a strong interaction with off-resonant states and our 2-level approximation would also prove to be insufficient. 

The off-diagonal elements contain at least one off-diagonal overlap, restricting the summation to $(m,n) = (0,0)$, $(f,f)$ and $(0,f)$ or $(f,0)$, leading to
\begin{align}
\braket{\phi_0(\overline{B}_z)|{U}_F|\phi_f(\overline{B}_z)} &=  e^{-i(1-\eta) E_0(\overline{B}_z) T }e^{-i \eta E_0(\overline{B}_z) T } (B_{z,1}-\overline{B}_z)\braket{\partial_{B_z}\phi_0|\phi_f} \nonumber \\
& + e^{-i(1-\eta) E_f(\overline{B}_z) T }e^{-i \eta E_f(\overline{B}_z) T } (B_{z,2}-\overline{B}_z) \braket{\phi_0|\partial_{B_z}\phi_f} \nonumber \\
& + e^{-i(1-\eta) E_0(\overline{B}_z) T }e^{-i \eta E_f(\overline{B}_z) T } \nonumber \\
& \ \ \times \left[(B_{z,2}-\overline{B}_z)\braket{\partial_{B_z}\phi_0|\phi_f} + (B_{z,1}-\overline{B}_z)\braket{\phi_0|\partial_{B_z}\phi_f}\right] \nonumber \\
& +\mathcal{O}(\Delta^2),
\end{align}
and similar for $\braket{\phi_0(\overline{B}_z)|{U}_F|\phi_f(\overline{B}_z)}$. Here the inner products have been expanded as e.g. $\ket{\phi_{0}({B}_{z,i})} =\ket{\phi_{0}(\overline{B}_{z})} + (B_{z,i}-\overline{B}_z)\ket{\partial_{B_z}\phi_{0}(\overline{B}_z)} + \mathcal{O}(\Delta^2)$, where the dependence on $\overline{B}_z$ has been made implicit in the final expressions. Evaluating these at $T= k \cdot T_c, k \in \mathbb{N}$ and explicitly setting $E_f = E_0 + 2\pi/T_c$, the matrix elements can (up to $\mathcal{O}(\Delta^2)$) be rewritten as
\begin{align*}
\braket{\phi_0(\overline{B}_z)|{U}_F|\phi_0(\overline{B}_z)} &= e^{-i k E_0(\overline{B}_z)T_c}, \\
\braket{\phi_f(\overline{B}_z)|{U}_F|\phi_f(\overline{B}_z)} &= e^{-i k E_F(\overline{B}_z)T_c}= e^{-i k E_0(\overline{B}_z)T_c}, \\
\braket{\phi_0(\overline{B}_z)|{U}_F|\phi_f(\overline{B}_z)} &= e^{-i k E_0(\overline{B}_z)T_c} (B_{z,1}-B_{z,2})  \braket{\partial_{B_z} \phi_0|\phi_f} (1-e^{-i \eta k 2 \pi})   , \\
\braket{\phi_f(\overline{B}_z)|{U}_F|\phi_0(\overline{B}_z)} &= e^{-i k E_0(\overline{B}_z)T_c} (B_{z,1}-B_{z,2}) \braket{\partial_{B_z} \phi_f|\phi_0} (1-e^{+ i \eta k 2 \pi}).
\end{align*}
For the central spin model, it is known that all eigenstates are purely real and hence $\braket{\partial_{B_z} \phi_0|\phi_f}=-\braket{\partial_{B_z} \phi_f|\phi_0}$. Taking $k=1$ and $\eta=1/2$ then results in 
\begin{align}
&U_F = \ e^{-i k E_0(\overline{B}_z)T_c} \left(\mathbbm{1} + 2 (B_{z,1}-B_{z,2}) \braket{\partial_{B_z} \phi_0|\phi_f}
\begin{bmatrix}
0 & 1\\
-1 & 0
\end{bmatrix}
\right) 
+ \mathcal{O}(\Delta^2),
\end{align}
returning the first-order correction proposed in the main text. It can similarly be shown that the second-order correction is purely imaginary and has to be anti-hermitian because of the unitarity of $U_F$. For $\eta=1/2$ this is a direct consequence of the fact that exchanging $B_{z,1}$ and $B_{z,2}$ and taking the transpose leaves $U_F$ invariant, while it changes $\Delta \rightarrow - \Delta$. The argument presented in the main text can now be extended to systems where the condition $|B_z| \ll |A_b S_b|$ does not hold, since the second-order resonance will interpolate between two real states with relative phase approximately $0$ or $\pi$. Because of the orthogonality of the initial and final states, this necessitates an intermediate state where only a pure state remains, resulting in a decoupled central spin. While this might not always be exactly at the point of resonance where $T=2T_c$, this will generally occur close to resonance because of the second-order nature of the perturbation. 

As supported by numerical results, perturbation theory then generally predicts odd-order resonances where the magnetization is approximately zero and even-order resonances near which the magnetization is either maximal or minimal.

\end{subappendices}
\makeatletter
\@openrighttrue
\makeatother

\chapter{Conclusions} 
\label{chap:concl}

\setlength\epigraphwidth{.4\textwidth}
\epigraph{\emph{...the end is simply the beginning of an even longer story.}}{{Zadie Smith}}

Despite being seemingly counterintuitive, the rules of quantum mechanics are by now quite well understood. Given a physical system, all interactions can be captured in a Hamiltonian, and the wave function provides a complete description of this system. Following the Schr\"odinger equation, this wave function can in principle be obtained through a diagonalization of the Hamiltonian. While this might be feasible in few-body systems, things get complicated in many-body systems. Although the equations governing the fundamental laws of physics remain the same, the exponentially large dimension of the Hilbert space prevents a straightforward solution to the Schr\"odinger equation. One way of circumventing this exponential scaling is through the notion of \emph{integrability}.

In this work, the structure and applications of Richardson-Gaudin integrable models are investigated. Integrable systems are characterized by two properties going hand in hand -- they support a large amount of conserved quantities, and their eigenstates can be exactly obtained using Bethe ansatz techniques. The Bethe ansatz wave function circumvents the exponential scaling of the Hilbert space, allowing exact eigenstates to be obtained by solving a set of non-linear equations, which remains feasible even for large system sizes. As introduced in Chapter \ref{Chap_RGmodels}, Richardson-Gaudin models present a specific class of integrable models, closely connected to the class of (trivially integrable) free models. This thesis then investigates ways of applying the properties of Richardson-Gaudin models in various contexts both in and out of integrability.

First, a framework was presented for the numerical and theoretical treatment of the Bethe ansatz in Richardson-Gaudin models. Starting from Chapter \ref{chap:EVB}, it was shown how any eigenstate of a Richardson-Gaudin integrable model can be characterized in two distinct ways, either through a set of rapidities parametrizing the Bethe wave function, or through a set of variables parametrizing the eigenvalues of the conserved charges. While the rapidities allow for a straightforward construction of the Bethe states, the Bethe equations that need to be solved are highly non-linear and exhibit singular behaviour, hampering (numerical) solutions. Alternatively, the eigenvalue-based variables can be easily obtained numerically, but it is not immediately clear in what way these determine the Bethe state. 

However, following Chapter  \ref{chap:innerproducts}, determinant expressions for inner products and correlation coefficients can be obtained without explicit knowledge of the rapidities, showing how the eigenvalue-based variables fully characterize the Bethe state. Again, such determinant expressions circumvent the usual exponential scaling of the Hilbert space, allowing for an efficient application of the Bethe ansatz. These two different approaches can be related to two dual ways of calculating these determinants, as made explicit by connecting both to the structure of Cauchy matrices. This holds for integrable models describing spin models as well as integrable models containing a bosonic degree of freedom, where the connection between both can be made through a pseudo-deformation connecting the relevant algebras, as done in Chapter \ref{chap:contraction}.

Second, it was investigated through this framework how the Bethe ansatz can be applied in different settings. Building on an integrable model, it was shown in Chapter \ref{chap:readgreen} how topological superconductivity and the exchange of particles with an environment can be modelled, where particle-exchange destroys the topological phase transition and instead leads to Read-Green resonances in low-energy and -momentum levels. The structure of the Bethe ansatz could then be used to shed light on the underlying mechanism. Moving towards non-integrable models, Chapter \ref{chap:VarRG} applied the toolbox of integrability for the use of the Bethe ansatz as an approximate wave function through a variational approach. Once a first approximation to the ground state of a non-integrable system is obtained, excitations can be added, and the resulting method was termed the Richardson-Gaudin Configuration-Interaction method. This was applied to nuclear pairing models, where it was found that Bethe states are able to accurately model the collective ground state and low-lying excited states. This concerned stationary Hamiltonians on which perturbations have been added, explicitly breaking integrability. However, another way of breaking integrability is through the introduction of dynamics and periodic driving. Such driven systems are characterized by a non-integrable Floquet Hamiltonian, which is generally impossible to construct due to the exponential scaling of the Hilbert space. In Chapter \ref{chap:floquet}, it was then shown how techniques from integrability can still be used in order to model this Hamiltonian, with special attention paid to many-body resonances of Bethe states. The presented techniques in these chapters can then be applied to more general Richardson-Gaudin models.

As hopefully made apparent throughout this work, the clear-cut structure and relatively large freedom in Richardson-Gaudin models makes them ideal for an investigation of the general principles of integrability, where many of the results in this work are expected to have analogues in more general integrable models, as well as being a perfect testing ground for the development of new quantum many-body techniques beyond integrability.


\bibliography{MyLibrary}


\chapter*{Nederlandstalige samenvatting}
\addcontentsline{toc}{chapter}{Nederlandstalige samenvatting}
\markboth{{Nederlandstalige samenvatting}}{{Nederlandstalige samenvatting}}

\setlength\epigraphwidth{.4\textwidth}
\epigraph{\emph{We call upon the author to explain.}}{{Nick Cave}}

Kwantumfysica is de tak van de wetenschap die het gedrag van deeltjes op de kleine schaal beschrijft. Deze theorie heeft een ongelooflijk succes gekend in het beschrijven en voorspellen van experimenten, ondanks het feit dat de basisregels van de kwantumfysica vaak lijken in te druisen tegen onze intu\"itie. Kwantumfysisch is elk system volledig bepaald door zijn zogenaamde \emph{golffunctie}. E\'en van de bouwblokken van de kwantumfysica is dan de Schr\"odinger vergelijking, die toelaat om de precieze wiskundige vergelijkingen op te stellen die deze golffunctie bepalen. Een belangrijke taak van de fysica is dan om het gedrag van deeltjes te voorspellen vertrekkende van de Schr\"odinger vergelijking. Hier worden we echter geconfronteerd met het \emph{kwantum veeldeeltjesprobleem}. De resulterende wiskundige vergelijkingen kunnen misschien wel opgelost worden voor \'e\'en of twee deeltjes, maar zodra een systeem een groot aantal deeltjes bevat neemt de complexiteit van de relevante vergelijkingen exponentieel toe. Exacte oplossingen van de Schr\"odinger vergelijking worden onmogelijk.

E\'en van de manieren om deze exponenti\"ele complexiteit te omzeilen is het gebruik van \emph{integreerbare modellen}. Deze modellen worden gekarakteriseerd door twee fundamentele eigenschappen -- enerzijds leiden integreerbare modellen tot een groot aantal behoudswetten en behouden grootheden, anderzijds kunnen de golffuncties in deze modellen exact bekomen worden via de \emph{Bethe ansatz} golffunctie. Exacte golffuncties kunnen hier bekomen worden door een aantal niet-lineaire vergelijkingen op te lossen, wat mogelijk blijft voor relatief grote systemen in verschillende fysische contexts (zoals supergeleiding, kernfysica, de interactie tussen licht en atomen,...). Deze thesis behandelt dan de specifieke klasse van \emph{Richardson-Gaudin} integreerbare modellen, en de toepassingen van de Bethe ansatz in zowel integreerbare als niet-integreerbare modellen. In Hoofdstuk \ref{Chap_Intro} wordt een inleiding gegeven tot kwantummechanica en het veeldeeltjesprobleem, waarna integreerbaarheid en de Bethe ansatz ge\"introduceerd worden in Hoofdstuk \ref{Chap_RGmodels}.

In de eerste helft van deze thesis wordt dan een theoretisch kader opgebouwd voor de numerieke en theoretische behandeling van deze modellen. In Hoofdstuk \ref{chap:EVB} wordt aangetoond hoe elke Bethe golffunctie op twee verschillende manieren beschreven kan worden, oftewel via een set rapiditeiten die de golffunctie parametriseren, oftewel via een set variabelen die de behouden grootheden karakteriseren. De Bethe vergelijkingen die de rapiditeiten bepalen vertonen singulier gedrag en kunnen niet eenvoudig opgelost worden, terwijl de behouden grootheden relatief eenvoudig bepaald kunnen worden. Hier gaat de directe link met de Bethe golffunctie wel verloren, die duidelijk blijft voor de rapiditeiten.

In Hoofdstuk \ref{chap:innerproducts} wordt echter aangetoond hoe determinant-uitdrukkingen voor de berekening van inproducten bekomen kunnen worden vertrekkende van de behouden grootheden, en de twee verschillende beschrijvingen worden gelinkt aan twee verschillende manieren om dergelijke inproducten te berekenen. Dit toont aan dat beide beschrijvingen equivalent zijn, en zo'n determinant-uitdrukkingen omzeilen opnieuw de exponenti\"ele complexiteit van het veeldeeltjesprobleem. Dit geldt zowel voor integreerbare modellen die spin-systemen beschrijven als integreerbare modellen die een bosonische vrijheidsgraad bezitten, zoals aangetoond in Hoofdstuk \ref{chap:contraction} via een deformatie van de onderliggende algebra.

De tweede helft van deze thesis past dan dit kader toe op verschillende fysische systemen. Allereerst wordt in Hoofdstuk \ref{chap:readgreen} de Bethe ansatz golffunctie uitgebreid en toegepast op een (topologische) supergeleider. Het resulterende model is integreerbaar en het fasediagram van dit model wordt besproken en gelinkt aan de symmetrie in de Bethe ansatz golffunctie. De vele voordelen van de exacte oplossing van integreerbare modellen worden enigszins gecompenseerd door het feit dat de interacties in een model precies afgesteld moeten worden vooraleer een model integreerbaar is. Zelfs bij kleine storingen op het fysisch model wordt integreerbaarheid gebroken en valt het volledig theoretisch kader van de Bethe ansatz weg. Hoewel een exacte oplossing niet langer mogelijk is in niet-integreerbare modellen kan de Bethe ansatz nog gebruikt worden als een benaderende golffunctie, en dit wordt besproken in Hoofdstukken \ref{chap:VarRG} en \ref{chap:floquet}. In Hoofdstuk \ref{chap:VarRG} wordt aangetoond hoe een beschrijving van de grondtoestand en het lage-energie spectrum van een expliciet niet-integreerbaar model bekomen kan worden aan de hand van benaderende Bethe ansatz golffuncties. Integreerbaarheid kan ook impliciet gebroken worden door externe krachten te laten inwerken op een integreerbaar model, zoals in periodiek aangedreven systemen. In Hoofdstuk \ref{chap:floquet} wordt dan aangetoond hoe de Bethe ansatz en integreerbaarheid gebruikt kunnen worden om aangedreven systemen te beschrijven, waar veeldeeltjesresonanties gemodelleerd worden met Bethe golffuncties. De technieken ge\"introduceerd in deze hoofdstukken zijn algemeen van toepassing voor verschillende Richardson-Gaudin modellen.

Het hoofddoel van deze thesis is dan om duidelijk te maken dat Richardson-Gaudin modellen een duidelijke structuur en relatief veel vrijheid bezitten, waardoor deze ideale kandidaten zijn om zowel het theoretisch kader van integreerbaarheid te onderzoeken, waar verwacht wordt dat veel van de resultaten in deze thesis uitgebreid kunnen worden naar algemenere integreerbare systemen, als om nieuwe veeldeeltjestechnieken te ontwikkelen wanneer integreerbaarheid gebroken wordt.

\newpage
\thispagestyle{empty}
\hbox{}
\newpage
\thispagestyle{empty}
\hbox{}

\end{document}